\documentclass[aar]{svjour}
\usepackage{latexsym}
\usepackage{graphics}
\usepackage[]{natbib}
\usepackage{times}
\usepackage{mathptmx}
\bibpunct[]{(}{)}{;}{a}{}{,}  
\def\ga{\;\rlap{\lower 2.5pt\hbox{$\sim$}}\raise 1.5pt\hbox{$>$}\;}
\def\la{\;\rlap{\lower 2.5pt\hbox{$\sim$}}\raise 1.5pt\hbox{$<$}\;}
\journalname{Astronomy and Astrophysics Review}
\begin{document}
\title{X-ray astronomy of stellar coronae
\thanks{This article is dedicated to the late Rolf Mewe, a prominent 
       astrophysicist who contributed major work to the field of stellar 
       X-ray astronomy and spectroscopy. He died on May 4, 2004.}
}
\author{Manuel G\"udel
}                     
%
%
\institute{Paul Scherrer Institut, W\"urenlingen and Villigen, CH-5232 
           Villigen PSI, Switzerland}
\date{Received: 27 February 2004 / accepted 27 May 2004}
%
\maketitle
\begin{abstract}
X-ray emission from stars in the cool half of the Hertzsprung-Russell diagram is generally attributed
to the presence of a magnetic corona that contains plasma at temperatures exceeding 1 million K.
Coronae are ubiquitous among these stars, yet many fundamental mechanisms operating
in their magnetic fields still elude an interpretation through a detailed physical description. 
Stellar X-ray astronomy is therefore contributing
toward a deeper understanding of the generation of  magnetic fields in magnetohydrodynamic dynamos,
the release of  energy in tenuous astrophysical plasmas through various plasma-physical processes, 
and the interactions of high-energy radiation with the stellar environment. Stellar X-ray emission also provides 
important diagnostics to study the structure and evolution of stellar magnetic fields from the first days 
of a protostellar life to the latest stages of stellar evolution among giants and supergiants. 
The discipline of stellar coronal X-ray astronomy has now reached a level of sophistication that makes 
tests of advanced theories in stellar physics possible. This
development is based on the  rapidly advancing instrumental possibilities that today 
allow us to obtain images with sub-arcsecond resolution
and spectra with resolving powers exceeding 1000. High-resolution X-ray spectroscopy has, in fact, opened 
new windows into astrophysical sources, and has played a fundamental role in coronal
research.

The present article reviews the development and current status of various 
topics in the X-ray astronomy of stellar coronae, focusing on observational results and on theoretical 
aspects relevant to our understanding of coronal magnetic structure and evolution.

\end{abstract}

\vskip 0.2truecm
\hskip 4.6cm\begin{minipage}{7cm}
{\it For my part I know nothing with any certainty, but the sight of the stars makes me dream.

\hfill Vincent van Gogh (1853-1890)}
\end{minipage} 
\vfill


\section{Introduction}\label{intro}

More than half a century ago, the presence of a very hot, tenuous gas surrounding the Sun,
the {\it X-ray corona}, was inferred indirectly from optical coronal lines of highly ionized 
species \citep{grotrian39, edlen42} and more directly by detecting X-ray photons in the course of  a rocket 
flight \citep{burnight49}\footnote{T. Burnight wrote, 
{\it ``The sun is assumed to be the source of this radiation although radiation of wave-length shorter 
than 4 angstroms would not be expected from theoretical estimates of black body radiation from the solar 
corona.''} The unexpected has prevailed all through the history of coronal physics indeed!}. 
Around the same time, radio observations revealed  a {\it radio corona} as well
\citep{hey46}. Of course,
the sheer beauty of the solar corona has been admired in scattered visible light ever since
humans first wondered about solar eclipses, but only during the last very few decades
have we started to seriously grasp solutions to some of the fundamental
astrophysical problems within the field of solar and stellar astronomy. 

Magnetic fields
have come to the center of our attention in this endeavor. They seem to be ubiquitous among
stars, but neither do we understand precisely why, nor have we fully understood the
bewildering variety of plasma physical mechanisms that act in stellar environments. 
We have found magnetic fields on stars that ought to have none so long as we appeal to our
limited understanding of magnetic field production and amplification. 
We witness various processes of energy transport and energy release intimately related
to those very magnetic fields;  the fields not only guide mass and energy flows, 
they are the sources of energy themselves. But our understanding of energy dissipation has
remained patchy, in particular in magnetically active stars. 

The coronal magnetic fields
reach into the stellar environment, structuring it and governing heating and
particle acceleration. Nevertheless, except in the case of the Sun, we have  very little, and
usually only indirect evidence of the topology of magnetic fields. In very young stars,
magnetic fields may reach out to the circumstellar accretion disks. Again, their role
is manifold: they transport angular momentum and thus control the spin rate of the star.
They guide mass flows, thus take a leading role in the mass accretion process.
They release energy and thus ionize the stellar molecular environment, possibly
altering the physics and chemistry of accretion disks and thereby influencing the
formation of planets. At later stages, they control stellar rotation through 
angular momentum transport via a stellar wind and thus engage in a feedback loop
because the magnetic field production is rooted in precisely this
rotation. 

The magnetic field thus plays a fundamental role in the evolution of the
radiative environment of a star, with far-reaching consequences for the chemical development
of planetary atmospheres and, eventually, the formation of life. These and many 
further challenges have stimulated our field of research both in theory and 
observation, producing a rich treasure of ideas and models from stellar evolution to elementary 
plasma physical processes that reach way beyond specific coronal physics problems.

Yet, it has proven surprisingly difficult to study these magnetic fields in the 
outer stellar atmospheres and the stellar environments. Thus, the essence of a stellar corona is not yet accessible.
It is the mass loading of magnetic fields that has given us specific diagnostics  
for magnetic activity to an extent that we often consider the hot plasma and 
the accelerated high-energy particles themselves to be our primary subjects.  

Although often narrowed down to some specific energy ranges, coronal emission is 
intrinsically a multi-wavelength phenomenon revealing itself from the meter-wave radio range
to gamma rays. The most important wavelength regions from which we have learned {\it diagnostically} on 
{\it stellar} coronae include the radio (decimetric to centimetric) range and the X-ray domain.
The former is sensitive to accelerated 
electrons in magnetic fields, and that has provided the only direct means of imaging 
stellar coronal structure, through very long baseline interferometry. 

The most productive
spectral range for stellar coronal physics has, however, been the soft X-ray domain where the mysteriously 
heated bulk plasma trapped in the coronal fields radiates. The X-ray diagnostic power 
has been instrumental for our understanding of physical processes in coronae, and the recent
advent of high-resolution X-ray spectroscopy with the {\it Chandra} and {\it XMM-Newton} X-ray
observatories is now accessing physical parameters of coronal plasma  directly.

This review is predominantly concerned with the soft X-ray domain of stellar coronal
physics, and partly with the closely related extreme ultraviolet range. Radio aspects of stellar 
coronae have been  reviewed elsewhere \citep{guedel02d} and
will be occasionally addressed when they provide complementary information to our present subject.

Notwithstanding the importance of these two wavelength regimes, I emphasize that further
diagnostics are available at other photon energies. Although optical and ultraviolet spectroscopy
refers predominantly to cooler layers of stellar atmospheres, a few {\it coronal} emission lines
detected in this wavelength range  promise some complementary diagnostics
in particular through the very high spectral resolving power available. And second, the hard X-ray 
and $\gamma$-ray range, recognized as the fundamental source of information for energy release
physics in the solar corona, will be of similar importance for stars although, at the time of
writing, it remains somewhat of a stellar {\it terra incognita}, waiting for more sensitive instruments
to detect these few elusive photons.

A review of a field that has accumulated massive primary literature from three decades of continuous 
research based on numerous satellite observatories, necessarily needs to focus on selected aspects. 
The review in hand is no exception. While trying to address issues and problems across the field of
stellar X-ray astronomy, I have chosen to put emphasis on physical processes and diagnostics
that will help us understand mechanisms not only in stars but in other astrophysical
environments as well. 
Understanding energy-release physics, magnetic-field generation mechanisms, and evolutionary
processes of magnetic structures from protostars to giants will eventually contribute to our understanding
of the physics in other astrophysical objects. Examples are accretion-driven mechanisms in disks 
around active galactic nuclei, the physics of large-scale galactic magnetic fields, or the heating 
and cooling of galaxy cluster gas. 

In the
course of this review, a number of controversial issues and debates will deliberately be exposed - this
is where more investment is needed in the future.
Also, while I will address some topics relatively extensively, I will touch upon others in
a somewhat more cursory way. My hope is that various previous reviews 
help close the gaps. They themselves are far too numerous to name individually. I would recommend, 
among others, the following reviews as entry points to this field:  
The extreme-ultraviolet domain has recently been summarized extensively by \citet{bowyer00}. 
\citet{haisch91a} summarized the multi-wavelength view of stellar and solar flares.
Two early comprehensive reviews of X-ray astrophysics of stellar coronae were written by
\citet{rosner85} and \citet{pallavicini89}, and very recently \citet{favata03b} have presented a comprehensive
observational overview of stellar X-ray coronae. 

An interesting early summary of {\it nonradiative}
processes in outer stellar atmospheres that comprises and defines many of our questions was given
by \citet{linsky85}. \citet{mewe91} reviewed X-ray spectroscopic methods for  stellar coronae.
High-energy aspects in the pivotal domain of star formation and early
stellar evolution were comprehensively summarized by \citet{feigelson99}.

Topics that are - 
despite their importance for stellar astronomy - predominantly the subject
of solar physics are not discussed here. In particular, plasma-physical mechanisms of coronal heating, the acceleration
of (solar and stellar) winds, and the operation of internal dynamos will only be touched upon
in so far  as stellar observations are contributing specifically to our knowledge. I also
note that insights relating to our subject provide diagnostics that reach out to entirely
different fields, such as studies of the rotational history of stars, the ionization structure and
large-scale evolution of star-forming molecular clouds, the structure and  the
composition of galactic stellar populations, and dynamo theory for various types of stellar systems. Beyond
what I can address in this summary, I need to refer the reader to the more specialized literature.  

\section{The study of stellar coronae} 

The study of stellar coronae, of course, starts with the Sun. This provides a rather important advantage
for stellar astronomers: that of having a nearby, bright standard example 
available at high spatial and spectral resolution. What, then, should we expect from the study of stellar 
coronae?

\begin{figure} 
\centerline{\resizebox{1\textwidth}{!}{\includegraphics{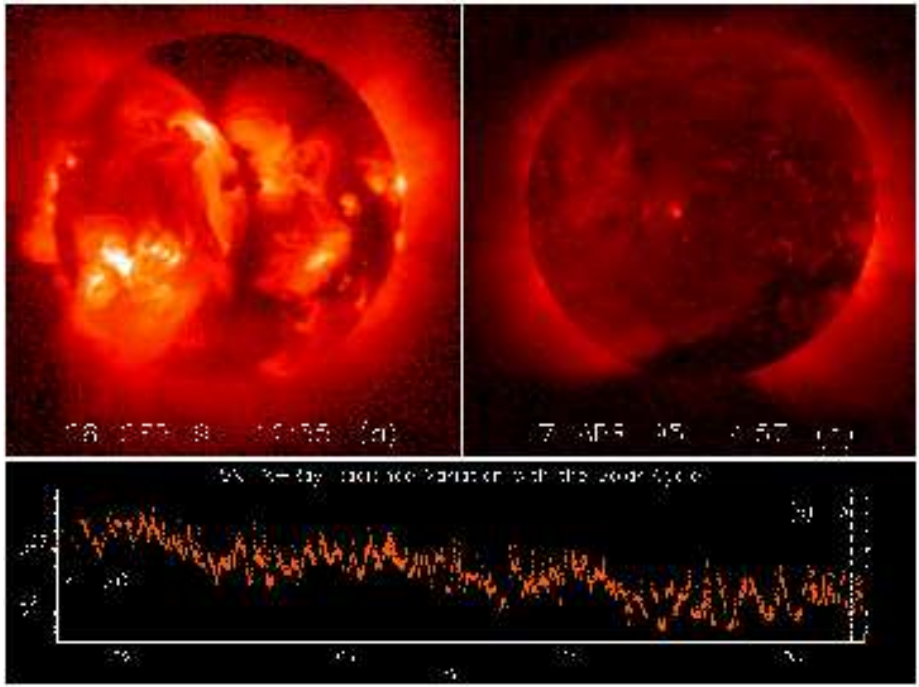}}}
\caption{{\it Yohkoh} images of the Sun during activity maximum (left, in 1991) and minimum (right, in 1995). The light curve
          in the lower panel illustrates the long-term ``cyclic'' variability of the solar
	    soft X-ray output. }\label{solarminmax}
\end{figure}

In the context of the solar-stellar connection, stellar X-ray astronomy has introduced a range of
stellar rotation periods, gravities, masses, and ages into the debate on the magnetic dynamo. Coronal magnetic structures
and heating  mechanisms may vary together with variations of these parameters. Parameter studies
could provide valuable insight for constraining relevant theory. Different topologies and sizes
of magnetic field structures lead to different wind mass-loss rates, and this will regulate the stellar spin-down
rates differently. As is now clear, on the other hand, rotation
is one of the primary determinants of the magnetic dynamo. This point could not be demonstrated by observing the Sun: 
The Sun's magnetic activity is in fact strongly modulated  (due to the 11-year magnetic spot cycle, Fig.~\ref{solarminmax}),
but this effect is not directly dependent on the rotation period. Conversely, the well-studied solar activity cycle 
motivates us to investigate similar magnetic modulations in stars in order to confine the underlying  dynamo mechanism.

Stars allow us to study long-term evolutionary effects by observing selected samples with known and largely differing
ages. While models of the solar interior and its evolution  predict that the young Zero-Age Main-Sequence (ZAMS) 
Sun was fainter by $\approx 25$\% than at present, stellar X-ray astronomy has revealed that the solar high-energy
emission was likely to be hundreds of times more intense at such young ages. The increased level of ionizing
and UV radiation must then have had an important impact on the formation and chemistry of planetary atmospheres.

At younger stages still, the radiation at short wavelengths may have been pivotal for ionizing the circumstellar accretion 
disk. This must have resulted in at least five significant effects that determine the further process of planet formation and 
stellar evolution.  First, stellar magnetic fields
couple  to the inner disk and thus guide mass accretion onto the star; second, and at the same time, torques mediated by
the magnetic fields regulate the rotation period of the star; third, winding-up magnetic fields between star and disk
may release further magnetic energy that irradiates the disk and may lead to jet (and indirectly, to molecular outflow) activity;
fourth, permeation of the weakly ionized disk by magnetic fields induces instabilities (e.g., the Balbus-Hawley instability)
that are relevant for the accretion process and planet formation; and fifth, the X-ray irradiation may directly
influence the disk chemistry and thus the overall evolution of the dust disk (e.g., \citealt{glassgold00}).

Including stars into the big picture of coronal research has also widened our view of coronal plasma physics.
While solar coronal plasma resides typically at $(1-5)\times 10^6$~K with temporary excursions to $\approx 20$~MK
during large flares, much higher temperatures were found  on some active stars, with steady plasma temperatures
of several tens of MK and flare peaks beyond 100~MK. Energy release in stellar flares involves up to $10^5$ times
more thermal energy than in solar flares, and pressures that are not encountered in the solar corona.

\section{The early days of stellar coronal X-ray astronomy} 

While this review is entirely devoted to (non-solar) stellar studies, an important anchor point would 
be missing if the success of the {\it Skylab} mission in the early seventies were not mentioned. The 
high-quality images of the full-disk Sun in X-rays formed, together with data from previous rocket flights,
our modern picture of the solar and therefore stellar coronae. The solar X-ray corona is now understood as a dynamic 
ensemble of magnetic loops that contain hot plasma in ever-changing constellations, yet  always related to
the underlying photospheric magnetic field. Interacting loops may episodically evolve into flares that release much
of their energy as soft X-rays. Coronal holes, characterized by low X-ray emission and predominantly open magnetic field lines
along which the solar wind escapes at high speed, fill volumes between bright coronal active regions. From a stellar astronomer's
point of view, the early solar coronal studies culminated in the formulation of scaling laws for coronal
structures. Best known is the seminal paper by \citet{rosner78} in which several of the basic ideas of  coronal 
structure and static loops were developed or extended. These concepts paved the way to interpreting stellar coronae 
without requiring the imaging capabilities that have been so central to solar studies.

The field of stellar coronal X-ray astronomy was in fact opened around the same time, in 1975 when
\citet{catura75} reported the detection of Capella as the first stellar coronal X-ray source on the occasion of 
a rocket flight. They estimated the X-ray luminosity at $10^{31}$~erg~s$^{-1}$ -- four orders of magnitude above 
the Sun's -- and the plasma electron temperature at about $8\times 10^6$~K, again several times higher than the Sun's.  
I note in  passing that the latter measurement is in quite close agreement with modern values. This
result was confirmed by \citet{mewe75} from observations with the {\it ANS} satellite; they 
were the first to interpret the soft X-rays as solar-like coronal emission at an enhanced level. Around the same time, 
\citet{heise75} monitored the first stellar coronal X-ray flares (on YZ CMi and UV Cet) with 
{\it ANS}; one of the flares was recorded simultaneously with an optical burst. The possible 
contributions of stellar X-ray flares to the diffuse galactic soft X-ray background and of 
associated  particles to the cosmic-ray particle population were immediately recognized and discussed. Numerous 
flare observations followed, opening up new avenues of research on energy release physics familiar
from the Sun. For example, \citet{white78} related a soft X-ray flare on HR~1099 with a simultaneous radio 
burst.

Further detections followed suit. Algol was next in line, defining another new class of stellar
X-ray sources \citep{schnopper76, harnden77}. The initial discussion related the X-rays 
to an accretion mass stream, however, while modern interpretation ascribes them  
to coronal structures on the K-type secondary. A series of  detections with {\it HEAO 1}, 
namely of RS CVn and HR~1099  \citep{walter78a}, Capella \citep{cash78}, and UX Ari \citep{walter78b} 
established the RS CVn binaries as a class of coronal X-ray sources that may also  be significant 
contributors to the galactic soft X-ray background. The unusually high X-ray production of 
this class was confirmed in the survey by \citet{walter80a}, also based on {\it HEAO 1}. 

Main-sequence (MS) stars were brought on stage with the first discovery of Proxima Centauri in
the extreme ultraviolet range by \citet{haisch77}, while \citet{nugent78} identified the ``solar twin'' 
$\alpha$ Cen as an inactive coronal source at a luminosity level similar to the Sun. Several further single 
MS  stars at the high end of the activity scale were found, putting our own Sun into a new
perspective as a rather modest X-ray star \citep{walter78c, cash79a, walter80b}. Discoveries soon
extended the coronal range into the A spectral class \citep{mewe75, topka79} although these early detections have not been
confirmed as coronal sources, from the present-day point of view. (\citealt{mewe75} suggested
the white dwarf in the Sirius system to be the X-ray source.) 

An important new chapter was opened with 
the introduction of medium-resolution X-ray spectroscopy. \citet{cash78} used {\it HEAO 1} to obtain the 
first coronal X-ray spectrum of Capella. They correctly interpreted 
excess emission between 0.65 and 1~keV as being due to the Fe\,{\sc xvii/xviii} complex. 
The earliest version of a (non-solar) stellar coronal emission measure distribution can be 
found in their paper already! They noticed that the high temperature of Capella's 
corona requires magnetic confinement unless we are seeing a free-flowing coronal 
wind. Another element of modern coronal X-ray interpretation was introduced by \citet{walter78b} 
when they realized that subsolar abundances were required to fit their medium-resolution spectrum of UX Ari. 
The first explicit X-ray  spectra of (non-solar) stellar flares were obtained by \citet{kahn79}. The thermal 
nature of the emission was confirmed from the detection of the 6.7~keV Fe K$\alpha$ line.

The {\it Einstein} satellite revolutionized the entire 
field of coronal X-ray astronomy, transforming it from a domain of mostly exotic and extreme 
stars to a research area that eventually addressed X-ray emission from all stars across 
the Hertzsprung-Russell diagram, with no lack of success in detecting them either
as steady sources \citep{vaiana81} or during large outbursts \citep{charles79}. The field of star formation entered
the scene when not only star formation regions as a whole, but individual young stars such as T Tau stars
were detected as strong and unexpectedly variable X-ray sources \citep{ku79, ku82, gahm80, walter81c}, including 
the presence of strong flares \citep{feigelson81a}.

The solid-state spectrometer on board {\it Einstein} also provided spectroscopic access to 
many coronal X-ray sources and identified individual emission-line blends of various elements 
\citep{holt79}. These spectra permitted for the first time multi-temperature, variable-abundance 
spectral fits that suggested the co-existence of cool and very hot ($\ga 2\times 10^7$~K) plasma 
in RS  CVn binaries \citep{swank81, agrawal81}.

Lastly, grating spectroscopy started to  resolve individual 
spectral lines or blends in coronal X-ray spectra. \citet{mewe82} described spectroscopic 
observations of Capella that separated Fe and O lines in the 5--30~\AA\ region using the {\it Einstein} 
objective grating spectrometer with a resolving power up to at least 30.  
Only few high-resolution stellar spectra were obtained with the Focal Plane
Crystal Spectrometer (FPCS) on {\it Einstein} and the Transmission Grating Spectrometer (TGS)
on {\it EXOSAT}. The instruments offered spectral resolving powers of 50--500 in the
X-ray band (FPCS) and of $\la 60$  in the EUV band (TGS), respectively. The former instrument was used in narrow
wavelength bands only; examples include observations of $\sigma^2$ CrB \citep{agrawal85} and
Capella \citep{vedder83}. While these instruments marked a breakthrough in X-ray spectroscopy at the time, 
the S/N achieved was not sufficient to derive detailed information on emission measure (EM) distributions.
Nevertheless, rough models were derived that  grossly resemble the results from modern high-sensitivity
spectroscopy \citep{vedder83}. Later, the TGS was used to derive information on the EM
distribution of bright X-ray sources \citep{lemen89} and to study loop models \citep{schrijver89b}.

The ultimate breakthrough in stellar coronal physics came
with the initial {\it Einstein} survey that led to three significant insights. First,
X-ray sources abound among all types of stars, across the Hertzsprung-Russell diagram and across
most stages of evolution \citep{vaiana81}. {\it Stars became one of the most prominent classes of cosmic X-ray
sources.}  Second, the X-ray luminosities and their distribution now uncovered along the
main sequence could not be in agreement with the long-favored acoustic heating theories;
the X-ray emission was now interpreted as the effect of {\it magnetic coronal heating.}  
And third,  stars that are otherwise similar reveal large differences in their X-ray output
if their rotation period is different \citep{pallavicini81, walter81a}. These systematics have been in the 
center of {\it dynamo theory} up to the present day, and it is fair to say that no dynamo theory will be 
deemed fully successful without addressing the latter two points in some detail.

The initial findings were rapidly consolidated (e.g., \citealt{johnson81, ayres81a}) but  cool-star X-ray 
astronomy has remained an active research area to the present day, with no lack of debate. The following
chapters are devoted to our still exciting era of stellar X-ray astronomy.

\section{A walk through the X-ray Hertzsprung-Russell diagram}

\begin{figure} 
\center{\resizebox{0.86\hsize}{!}{\includegraphics{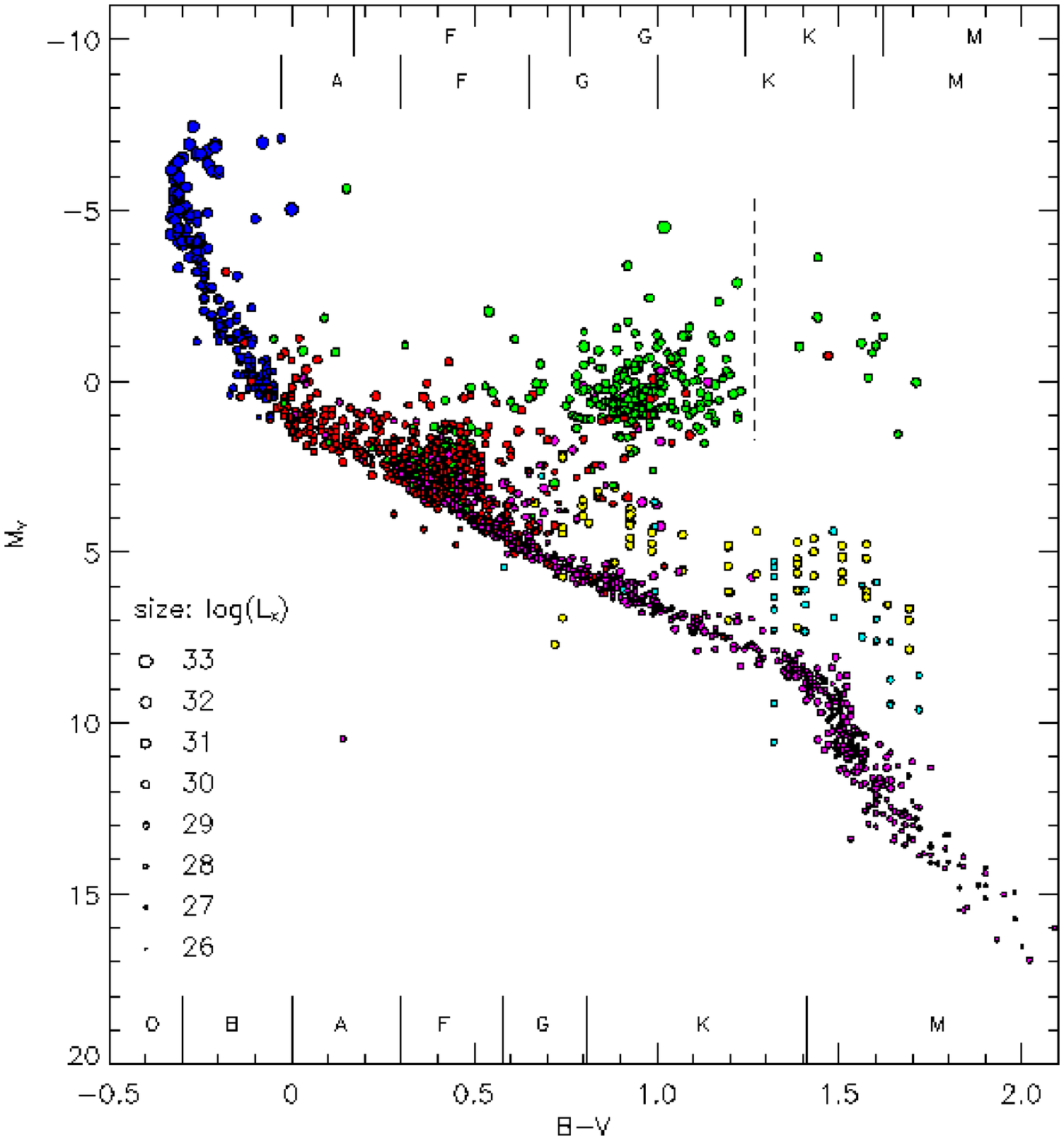}}}    
\caption{Hertzsprung-Russell diagram based on about 2000 X-ray detected stars extracted from the
catalogs by \citet{berghoefer96} (blue), \citet{huensch98a, huensch98b} (green and red, respectively),  
and \citet{huensch99} (pink). 
Where missing, distances from the Hipparcos catalog \citep{perryman97} were used to calculate the relevant 
parameters. The low-mass pre-main sequence stars
are taken from studies of the Chamaeleon I dark cloud (\citealt{alcala97, lawson96}, yellow and cyan, respectively)
and are representative of other star formation regions. 
The size of the circles
characterizes log\,$L_\mathrm{X}$ as indicated in the panel at lower left. The ranges for the spectral classes
are given at the top (upper row for supergiants, lower row for giants), and at the bottom of the figure (for 
main-sequence stars). }\label{hrd}
\end{figure}

A look at the Hertzsprung-Russell diagram (HRD) of detected X-ray 
stars in Fig.~\ref{hrd}, compiled from selected catalogs of
survey programs \citep{alcala97, berghoefer96, huensch98a, huensch98b, huensch99, lawson96}, 
shows all basic features 
that we know from an optical HRD (we plot each star at the locus of the optically determined
absolute magnitude $M_\mathrm{V}$ and the color index \textit{B}$-$\textit{V} regardless of possible 
unresolved binarity).  Although the samples used for the figure are in no 
way ``complete'' (in volume or brightness), the main sequence is clearly evident, and so 
is the giant branch.  The top right part of the diagram, comprising cool giants, is almost devoid of
detections, however. The so-called corona vs. wind dividing line  (dashed in Fig.~\ref{hrd}; after Linsky \& Haisch 
1979) separates coronal giants and supergiants to its left from stars with massive winds
to its right. It is unknown whether the wind giants possess magnetically
structured coronae at the base of their winds -- the X-rays may simply be absorbed 
by the overlying wind material (Sect.~\ref{dividing}). The few residual detections may at least partly be attributed to 
low-mass companions. The large remaining area from spectral class~M up to at
least mid-F comprises stars that are -- in the widest sense -- solar-like and that define the 
subject of this review. I now turn to a few selected domains within the HRD that have attracted special 
attention. The domain of star formation and pre-main sequence evolution will be discussed in a wider
context toward the end of this review (Sect.\ref{starformation}).

\begin{table}[t!] 
\caption{Symbols and units used throughout the text} 
\label{symbols}        
\begin{tabular}{ll} 
\hline\noalign{\smallskip}
Symbol, acronym	          	    & Explanation	       \\
\noalign{\smallskip}\hline\noalign{\smallskip}
$R_*$             	                    & Stellar radius [cm]		      \\    
$R_{\odot}$             	            & Solar radius [$7\times 10^{10}$~cm]		      \\    
$M_{\odot}$             	            & Solar mass [$2\times 10^{33}$~g]		      \\    
$P_{\rm rot}$ or $P$                        & Rotation period [d]              \\
$p$             	                    & Electron pressure [dyne~cm$^{-2}$]		      \\    
$L$             	                    & Coronal loop semi-length [cm]		      \\    
$T$             	        	    & Coronal electron temperature [K]	  \\	
$n_e$             	        	    & Electron density [cm$^{-3}$]	  \\	
$n_{\rm H}$             	            & Hydrogen density [cm$^{-3}$]	  \\	
$B$             	        	    & Magnetic field strength [G]	  \\	
$f$             	        	    & Surface filling factor [\%]	  \\	
$\Gamma$             	        	    & Loop area expansion factor (apex to base)	  \\	
$F_X$             	                    & X-ray surface flux  [erg~s$^{-1}$~cm$^{-2}$]		      \\    
$L_X$             	                    & X-ray luminosity 	[erg~s$^{-1}$]		      \\    
$L_{\rm bol}$             	            & Stellar bolometric luminosity 	[erg~s$^{-1}$]		      \\    
$\Lambda = \Lambda_0T^{\gamma}$             & Cooling function  [erg~s$^{-1}$~cm$^{3}$]   \\    
$Ro$             	        	    & Rossby number	  \\	
RTV loop          	                    & Constant cross-section loop after \citet{rosner78}\\    
VAU loop          	                    & Expanding cross-section loop after \citet{vesecky79}\\    
2-R flares          	                    & Two-Ribbon flares\\    
HRD          	                	    & Hertzsprung-Russell Diagram			 \\   
EM             	                	    & Emission Measure  		     \\    
$Q$, DEM             	        	    & Differential Emission Measure  Distribution		      \\    
EMD             	        	    & (discretized, binned) Emission Measure Distribution      \\    
(ZA)MS             	        	    & (Zero-Age) Main Sequence		      \\    
PMS                                         & Pre-Main Sequence     \\
(W, C)TTS                                   & (Weak-lined, classical) T Tauri Star \\
BD                                          & Brown Dwarf   \\
\noalign{\smallskip}\hline
\end{tabular}
\end{table}

\subsection{Main-sequence stars}

The main sequence (MS henceforth) has arguably played the most fundamental role in the interpretation
of stellar magnetic activity. It is here that we find a relatively clear correspondence between mass, radius, and color. On the other
hand, evolutionary processes map poorly on the MS, providing us with a separate free parameter,
namely age or, often equivalently, rotation rate. Current wisdom has it that the most massive coronal MS stars  
are  late-A or early F stars, a conjecture that is supported both by observation and
by theory. Theory predicts the absence of a magnetic dynamo, given the lack of a significant outer
convection zone in A-type stars. (In earlier-type stars of spectral type O and B, shocks developing
in unstable winds are the likely  source of X-rays.)

MS stars define by far the largest stellar population for systematic survey studies \citep{maggio87, fleming88, 
schmitt90a, barbera93}. The {\it ROSAT} All-Sky Survey (RASS) from which the samples shown in Fig.~\ref{hrd} 
were drawn has contributed invaluably to our science by providing {\it volume-limited} samples including 
stars down to the end of the MS and to the minimum levels of X-ray activity. Comprehensive surveys of
cool MS stars, some of them complete out to more than 10~pc, were presented by \citet{schmitt95}, \citet{fleming95}, 
\citet{schmitt97} and \citet{huensch99}, with detection rates as high as 95\% per spectral class, except for the 
intrinsically faint A stars. The survey sensitivities were sufficient to suggest a {\it lower limit} to the MS X-ray 
luminosity probably around a few times $10^{25}$~erg~s$^{-1}$ \citep{schmitt95} which translates to 
a lower limit to the surface X-ray flux that is similar to that of solar coronal holes \citep{schmitt97}.

Such studies have  contributed  much to our current understanding of coronal physics, in particular with regard to
the dependence of magnetic activity on rotation, the ingredients controlling the coronal
heating efficiency, and the feedback loop between activity and evolution, subjects broadly
discussed across this review. Before moving on to giants and binaries, I now specifically address three fundamental 
issues within the main-sequence domain:
that of very low-mass stars, brown dwarfs, and A stars with very shallow convective zones.

\subsection{The coolest M dwarfs}\label{coolm}

Beyond spectral type M5, the internal structure of dwarf stars changes significantly as they
become fully convective. The classical $\alpha\omega$ dynamo can thus no longer operate.
On the other hand, a distributed (or $\alpha^2$) dynamo may become relevant (e.g., \citealt{giampapa96} and 
references therein). One would then naturally expect that both the magnetic flux on the surface and the topology of the magnetic 
fields in the corona systematically change across this transition, perhaps resulting in
some discontinuities in the X-ray characteristics around spectral class dM5.

\begin{figure} 
\centerline{\resizebox{0.76\textwidth}{!}{\rotatebox{270}{\includegraphics{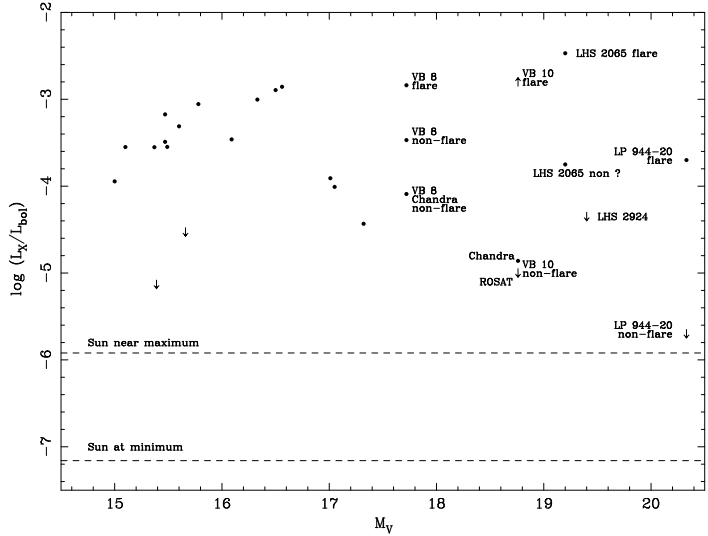}}}}
\caption{Diagram showing $L_X/L_{\rm bol}$ for lowest-mass stars later than spectral type
         M5. Flare and non-flaring values for several detections are marked (figure
	 courtesy of T. Fleming and M. Giampapa, after \citealt{fleming03}).}\label{lowmass}
\end{figure}

Observations do not seem to support this picture, however. The long-time lowest-mass X-ray detection, 
VB~8 (M7e~V) has shown steady emission at levels of $L_X \approx 10^{26}$~erg~s$^{-1}$ 
\citep{johnson81, fleming93, drake96} and flares up to an order of magnitude higher 
\citep{johnson87, tagliaferri90, drake96}. If its X-ray luminosity $L_X$ or the ratio of $L_X/L_{\rm bol}$ are
compared with other late M dwarfs, a rather continuous trend becomes visible (\citealt{fleming95}, Fig.~\ref{lowmass},
although there have been scattered claims to the contrary, see, e.g., \citealt{barbera93}). The maximum levels
attained by these stars  ($L_X \approx 10^{-3}L_{\rm bol}$) remains the same across spectral class M. If a
change from an $\alpha\omega$  to a distributive dynamo indeed does take place, then the efficiencies of
both types of dynamos must be very similar, or the transition must be very smooth and gradual, with the
two regimes possibly overlapping \citep{fleming93, weiss93, drake96}.

The same question also arose in the context of the coolest M dwarfs, namely relating to the boundary toward 
the substellar regime (around masses of 0.07$M_{\odot}$).  A change in the magnetic behavior is suggested there, 
for the following reason. The photospheres
of such stars are dominated by molecular hydrogen, with a very low ionization degree of approximately $10^{-7}$.
Electric currents flow parallel to the coronal magnetic field lines in the predominant non-flaring force-free
configuration, but since currents cannot flow into the almost neutral photosphere, 
any equilibrium coronal configuration will be potential, that is,  not capable of liberating energy for heating
(\citealt{fleming00} and references therein). A precipitous drop of $L_X$ would thus be expected.  
Heating could be due to episodic instabilities in more complex magnetic configurations that then produce prominent 
flares.  Indeed, stars at the bottom of the main-sequence \citep{fleming00, schmitt02a} and evolved brown dwarfs (see below) have
been detected in X-rays during flares but not generally at steady levels, similar to what is seen in H$\alpha$
observations (see references in \citealt{fleming03}). This picture has become somewhat questionable with 
the X-ray detection VB~10 (M8e) during  3.5~hrs at a level of $L_X \approx  
2.4\times 10^{25}$~erg~s$^{-1}$and log\,$L_X/L_{\rm bol} \approx -4.9$ by \citet{fleming03} who claim
this emission to be non-flaring. It is, however, inherently difficult to identify a steady process in data with
very low signal-to-noise ratios  (Sect.~\ref{stochvariability}), so that the last word on the emission type 
in those stars may not have been spoken. A most productive strategy is to
push the limit further toward lower-mass objects, as discussed below.

\subsection{Brown dwarfs (and planets?)}\label{BD}

Below the stellar mass limit at 0.07$M_{\odot}$, the realm of brown dwarfs (BD) has attracted immense
attention in recent years. The field is, at the time of  writing, still quite poorly explored in X-rays. 
Whereas young X-ray emitting BDs have now amply been detected in star forming regions, these objects
behave like contracting T Tauri stars rather than evolved, low-mass MS stars. 
Recent findings on young BDs are therefore summarized in Sect.~\ref{youngBD} in the context of star formation.
At some quite early point in their evolution (before the age of 100~Myr), they must drop to quite
low activity levels. \citet{krishnamurthi01} could not detect any BD in their Pleiades field down to 
a sensitivity limit of $L_X \approx 3\times 10^{27}$~erg~s$^{-1}$.

First X-ray detections among older, contracted BDs now exist, but a number of anomalies come to light. 
The first X-ray detected evolved brown dwarf, LP 944-20, was recorded exclusively during a flare, with $L_X \approx 1.2\times 
10^{26}$~erg~s$^{-1}$ and with a decay time of $\approx 5400$~s \citep{rutledge00}. Although this star is old 
($\approx 500$~Myr), it has not spun down ($P_{\rm rot} \le 4.4$~hrs). While it should thus be in a 
``supersaturated'' regime (Sect.~\ref{rotation}), it shows no detectable steady X-ray emission, quite 
in contrast to its strong and also flaring radio radiation  \citep{berger02}. \citet{tsuboi03}
found the BD companion of TWA~5 (with an age of 12~Myr) at levels of $L_X \approx 4\times 10^{27}$~erg~s$^{-1}$
and, again, below the empirical saturation limit, with a rather soft spectrum ($kT = 0.3$~keV). Finally,
\citet{briggs04} reported a weak detection of Roque 14 in the Pleiades (age 100~Myr), with $L_X \approx 3\times 
10^{27}$~erg~s$^{-1}$. This time, the emission is not compatible with a flare if the decay time is
shorter than 4~ks.

Ultimately, X-ray observations offer the possibility to detect (X-ray dark) planets as they eclipse part of the 
corona of their parent star while in transit. Such methods are particularly promising for low-mass stars 
as a Jupiter-like planet could eclipse a rather significant coronal area. \citet{briggs03} 
presented an example of a Pleiades member star that showed a significant dip in a flaring light curve (Sect.~\ref{eclipseflare}). The 
requirements for such a flare eclipse to occur are somewhat implausible, however, unless enhanced magnetic activity
is induced by star-planet interactions  (see \citealt{briggs03}).

\subsection{A-type stars}\label{astars}

The outer convection zones of stars become very shallow toward early 
F stars and disappear in A-type dwarfs. These stars are therefore not capable of operating a
classical $\alpha\omega$-type dynamo. Even if magnetic fields existed in early A-type stars, efficient coronal energy 
release is not expected because no  strong surface convective motions are present to transport
energy into non-potential coronal fields. Coincidentally, the acoustic flux from the interior reaches a maximum 
for late A and early F stars (\citealt{vaiana81}, \citealt{schrijver93} and references therein), a fact that has provoked
several early survey programs to look for strong X-ray emission in these stars \citep{vaiana81, pallavicini81,
topka82, walter83, schmitt85a}. Acoustic heating has been widely ruled out as the principal 
coronal heating mechanism along the main sequence (\citealt{vaiana81, stern81}, although the issue re-surfaces from time
to time, e.g., \citealt{mullan94a} and \citealt{mullan96}, for M dwarfs). However,  a significant ``base'' contribution of
acoustic waves in particular to the heating of the lower atmospheres of  A-F stars remains a viable possibility 
\citep{schrijver93, mullan94b}.

Investigations of magnetic activity in A-type stars have proceeded along three principal
lines: i) Search for genuine magnetic activity in single, normal A-type and early F stars; ii) 
study of magnetic activity in chemically peculiar Ap/Bp stars; iii) search for signatures of magnetic 
fields in very young, forming A-type stars (e.g., Herbig Ae/Be stars). Some selected results are briefly summarized below.

\subsubsection{X-ray emission from normal A stars}

Volume-limited stellar samples reveal a rather abrupt onset of X-ray emission around spectral
type A7-F0, with a large range of luminosities developing across spectral class F \citep{schmitt97}. 
The drop in X-rays toward earlier stars is appreciable: The definitive X-ray detection of 
Altair (A7~V) shows a very soft spectrum (at low luminosity, log\,$L_X \approx 27.1$, log\,$L_X/L_{\rm bol} \approx -7.5$, 
\citealt{schmitt85a}) with a temperature of only $\approx 1$~MK \citep{golub83, schmitt90a}. 
Quite in general,  ostensibly  single A-to-early F stars show distinctly soft spectra if detected
in X-rays  \citep{panzera99}.

Optically selected samples have produced  a number of additional detections up to early A stars, 
among them quite luminous examples. However, there are several reasons to believe that unidentified 
cooler companions are responsible for the X-rays. Given the rapid evolution of A-type stars, a companion of,
say, spectral type K or M would still be quite active. The companion hypothesis is thus particularly
likely for X-ray luminous examples with a hard spectrum \citep{golub83}. Late-B and A-type stars in the Pleiades,
for example, have X-ray properties that are indeed consistent with the presence of a cooler coronal
companion \citep{daniel02, briggs03}. Another indication
is the break-down of any correlation between rotation and $L_X$ or rotation and $B-V$ among these stars, once
again indicating that the X-rays may not actually be related to the A star \citep{simon95, panzera99}, although
F stars are considered to be coronal sources in any case (see below).

Illustrative examples of the complications involved in A-star X-ray astronomy are nearby A star binaries. 
The X-ray detections of the A-type binary Castor, even as a strong flaring source 
\citep{pallavicini90b, schmitt94b, gotthelf94}, opened up new speculations on A-star coronae.
Spatially resolved observations \citep{guedel01a, stelzer03} with {\it XMM-Newton} and
{\it Chandra} showed that both components are X-ray active. However, Castor is in fact a  
hierarchical quadruple system consisting of two A stars that are each surrounded by a low-mass 
(K-M type, \citealt{guedel01a} and references therein) companion. They are additionally accompanied by
the well-known X-ray strong M dwarf binary YY Gem. Both Castor components are frequently flaring \citep{guedel01a, 
stelzer03} and reveal X-ray spectra and fluxes that are quite similar to M dwarfs.

Restricting our attention now to the few genuine late A- or early F-type coronal emitters, we find that their weak dynamo operation 
is generally not able to brake the rapidly spinning star considerably during their short lifetime \citep{schmitt85a}.
One step further, several authors \citep{pallavicini81, walter83, simon91} have questioned the presence of any 
activity-rotation relation from spectral class A to F5 (beyond which it holds) altogether.  These coronae are also
conspicuous by their severe deficit of X-ray emission compared to chromospheric and transition region
fluxes; the latter can be followed up to mid-A type stars at quite high levels \citep{simon89, simon93, simon91, simon97}. 
Whether or not these atmospheres are indeed heated acoustically and drive an ``expanding'', weak and cool corona \citep{simon89}
or whether they are heated magnetically, the X-ray deficit and the low coronal temperatures clearly
attest to the inability of these stars to maintain substantial, hot coronae in any way comparable to
cooler active stars, their appreciable chromospheres notwithstanding.
 
\subsubsection{Chemically peculiar A stars}

Magnetic chemically peculiar stars of spectral type Bp or Ap  are appreciable
magnetic radio sources \citep{drakes87}, but they  have produced quite mixed results in X-rays. While a number
of detections were reported early on \citep{cash79b, cash82, golub83}, most Bp/Ap stars remained undetected,
and only few of them can be identified as probably single stars \citep{drakes94b}. When detected, their X-ray
luminosities  are quite high (log\,$L_X \approx 29.5-30$, log\,$L_X/L_{\rm bol} \approx -6$) and do not follow
the systematics of earlier-type stars. Given the strong surface magnetic fields in Bp/Ap stars, 
the currently favored models involve dipolar magnetospheres either featuring equatorial reconnection
zones that heat plasma \citep{drakes94b} or winds that are magnetically guided to the equatorial plane where 
they collide and heat up \citep{babel97}. 

A suspiciously high detection rate of CP stars was reported for the open cluster NGC 2516 by 
\citet{dachs96}, \citet{jeffries97}, and \citet{damiani03} (6 or 7 detected out of 8 observed CP 
stars in the latter study, amounting to one half of all detected stars optically identified as A-type). 
Such statistical samples may argue in favor of some of the Ap stars in fact being the sites of 
the X-ray emission. As for nonmagnetic Am stars, only scattered evidence is present that some may be X-ray sources 
\citep{randich96b, panzera99}, but again the caveats with undetected cooler companions apply.

\subsubsection{Herbig Ae/Be stars}

The nature of strong X-ray emission from pre-main sequence Herbig Ae/Be stars has remained
rather controversial. Models include unstable stellar winds, colliding winds, magnetic coronae, 
disk coronae, wind-fed magnetospheres, accretion  shocks, the operation of a shear dynamo, and the presence of
unknown late-type companions. Some X-ray properties are reminiscent of hot stars \citep{zinnecker94}, 
others point to coronal activity as in cool stars, in particular the presence of flares \citep{hamaguchi00, giardino04} 
and very high temperatures ($T \ga 35$~MK, \citealt{skinner04}). For reviews -
with differing conclusions - I refer the reader to the extensive critical discussions in  
\citet{zinnecker94}, \citet{skinner96}, and \citet{skinner04}.

\subsection{Giants and supergiants}

The evolution of X-ray emission changes appreciably  in the domain
of giants and supergiants. The area of 
red giants has attracted particular attention  because hardly any X-rays are found there. The cause of
the X-ray deficiency is unclear. It
may involve a turn-off of the dynamo, a suppression by competing wind production, or
simply strong attenuation by an overlying thick chromosphere (Sect.~\ref{dividing}).
This region of the HRD  was  comprehensively surveyed by \citet{maggio90} using
{\it Einstein}, and by \citet{ayres95}, \citet{huensch98a},  and \citet{huensch98b}  using the {\it ROSAT} 
All-Sky Survey.  A volume-limited sample was discussed by \citet{huensch96b}. 
Several competing effects influence the dynamo during this evolutionary phase especially
for stars with masses $\ga 2M_{\odot}$: While the growing 
convection zone enhances the dynamo efficiency, angular momentum loss via a magnetized wind tends to
dampen the dynamo evolution. These processes occur during the rather rapid crossing of 
the Hertzsprung gap toward M type giants and supergiants. The systematics of the X-ray 
emission are still not fully understood.

A rather exquisite but very small family of stars is defined by the so-called FK Com stars,
giants of spectral type K with an unusually rapid rotation and signs of extreme activity.
Their X-ray coronae are among the most luminous ($L_X \ga 10^{32}$~erg~s$^{-1}$) and hottest
known (with dominant temperatures up to 40~MK; e.g., \citealt{welty94, gondoin02, gondoin04b, audard04}).
These stars are probably descendants of rapidly rotating B-A MS stars
that, during the fast evolution across the Hertzsprung gap, have been able to
maintain their rapid rotation as  the convection zone deepened, while now being in a stage 
of strong magnetic braking due to increased magnetic activity \citep{gondoin02, gondoin03a}. 
The leading hypothesis, however, involves a merger of a close binary system, in which the orbital angular 
momentum of the companion is transferred to the primary \citep{bopp81}.

\subsection{Close binaries}

Close, tidally interacting binaries keep their fast rotation
rates often throughout their MS life and possibly into the subgiant
and giant evolution. Their rotation-induced dynamos maintain high magnetic activity
levels throughout their lifetimes, making them ideal laboratories for the  study
of magnetic dynamo theory. 

The most common binary systems available for study are RS CVn-type systems that
typically contain a G- or K-type giant or subgiant with a late-type subgiant or
MS companion. The similar class of BY Dra-type
binaries contain two late-type MS stars instead. If their separation is sufficiently small,
the two components may come into physical contact, defining the class
of W UMa-type contact systems (see Sect.~\ref{contact} below). And finally, Algol-type binaries are
similar to RS CVn systems, but the MS component is of early type (typically a B star).
The cool subgiant fills its Roche lobe, and mass transfer may be possible.

Extensive X-ray surveys of RS CVn-type binaries were presented by \citet{walter81a}, 
\citet{drakes89}, \citet{drakes92}, \citet{dempsey93a}, \citet{dempsey93b}, and \citet{fox94}.
Comparative studies suggested that the secondary star plays no role in determining the activity 
level of the system other than providing the mechanism to maintain rapid rotation \citep{dempsey93a}. 
However, the surface X-ray activity does seem to be enhanced  compared to single stars
with the same rotation period \citep{dempsey93a}.
The X-ray characteristics of BY Dra binaries are essentially
indistinguishable from RS CVn binaries so that they form a single
population for statistical studies \citep{dempsey97}.

The X-ray emission of Algols was surveyed by \citet{white83}, \citet{mccluskey84} and
\citet{singh95}. \citet{white80} were the first to indicate that X-rays from Algol-type binaries 
are also coronal (with X-ray sources located on the late-type secondary),
and that they resemble RS CVn-type binaries in that respect.
However, Algols are underluminous by a factor of 3-4 compared to similar RS CVn binaries \citep{singh96b} . 
It is therefore rather unlikely that possible accretion streams  contribute significantly to
the X-ray emission in Algol.

\citet{ottmann97} presented the first survey of Population 
II binaries. They concluded that their overall X-ray emission is weaker  than what is typical
for similar Pop I RS CVn binaries. Here, part of the trend may, however, be explained 
by the Pop II sample containing fewer evolved stars. On the other hand, the reduced 
metallicity may also inhibit efficient coronal radiation. 

\subsection{Contact binary systems}\label{contact}

W UMa systems are contact binaries of spectral type F-K with rotation periods from 0.1--1.5~d. 
They  were first detected in X-rays by \citet{carroll80} and surveyed by \citet{cruddace84}.
Their rapid rotation periods suggest enhanced activity, and this is indeed confirmed by more recent comprehensive 
surveys \citep{stepien01}. However, early work already found an order-of-magnitude deficiency in $L_X/L_{\rm bol}$ 
when compared to similar detached systems \citep{cruddace84, vilhu83, vilhu84}. This 
phenomenon is  also known as ``supersaturation''
(see Sect.~\ref{rotation}). A survey by \citet{mcgale96} with {\it ROSAT} essentially confirmed the 
luminosity deficit in all targets and additionally reported somewhat lower maximum temperatures and 
a smaller amount of very hot plasma when compared with detached RS CVn binaries.

The related  {\it near-}contact binaries do not share a 
common envelope but may in fact be evolutionary precursors of contact
systems. They were studied by \citet{shaw96} who found luminosities similar to those
of contact systems but again significantly lower than those of RS CVn binaries.

\section{X-ray activity and rotation}\label{rotation}

\subsection{Rotation-activity laws} 

Stellar rotation and magnetic activity  operate
in a feedback loop; as a single low-mass MS star ages, it sheds a magnetized wind, thus spinning 
down  due to angular momentum transport away from the star. This, in turn, weakens the internal 
dynamo and thus reduces magnetic activity (e.g., \citealt{skumanich72}). 
This negative feedback loop tends to converge toward a definitive rotation period $P$  
that depends only on mass and age once the star has evolved for a few 100~Myr (e.g., \citealt{soderblom93}). 
It is thus most likely rotation, and only indirectly age, that determines the level of magnetic activity,
a contention confirmed in recent studies by \citet{hempelmann95}. X-rays offer an ideal and sensitive
tool to test these dependencies, and corresponding results were found from the initial 
X-ray survey with {\it Einstein}.  \citet{pallavicini81}  suggested a relation
between X-ray luminosity and projected rotational velocities  $v{\rm sin}i$ (where $v$ is measured 
in km~s$^{-1}$)
\begin{equation}\label{lxprot}
L_X \approx 10^{27}(v{\rm sin}i)^2 \quad{\rm [erg~s^{-1}]} 
\end{equation}
(although the stellar sample included saturated stars, which were recognized
only later). A similar trend was visible in a sample shown by \citet{ayres80}. 
The overall relation was subsequently widely confirmed, e.g., by \citet{maggio87} for F-G MS and subgiant stars,
or  by   \citet{wood94} for a large sample of nearby stars based on EUV measurements
(with somewhat smaller indices of 1.4--1.6). For the surface flux $F_X$ or the ratio $L_X/L_{\rm bol}$, 
a relation like~(\ref{lxprot}) implies
\begin{equation}
F_X, {L_X \over L_{\rm bol}} \propto  \Omega^2 \propto P^{-2}
\end{equation}
where $\Omega$ is the angular rotation velocity and we have, for the moment, ignored the temperature
term distinguishing the two measures on the left-hand side. \citet{walter81b} reported 
$L_X/L_{\rm bol} \propto \Omega$ but his sample
included saturated stars (not recognized as such at that time). He later introduced broken
power-laws and thus in fact corrected for a saturation effect in rapid rotators (\citealt{walter82}, see below).
\citet{schrijver84}  included other determining factors,
concluding, from a common-factor analysis, that the specific emission measure $\zeta$ (total 
EM divided by the stellar surface area) is related to $P$ and the (dominant) coronal temperature as
\begin{equation}\label{schrijver_em}
\zeta = 10^{28.6\pm 0.2}T^{1.51\pm 0.16}P^{-0.88\pm 0.14}
\end{equation}
where here $T$ is given in MK, $P$ in d, and $\zeta$ in cm$^{-5}$. 
Since the dynamo efficiency also depends
on the convection zone depth, \citet{noyes84} and \citet{mangeney84} 
introduced the Rossby number $Ro$ as the ratio of the two relevant time scales of rotation and
convection ($Ro = P/\tau_c$,
where $\tau_c$ is the convective turnover time). The most general rotation-activity
diagrams that may include stars of various spectral classes and radii are now conventionally
drawn for the variables  $Ro$ and $L_X/L_{\rm bol}$ \citep{dobson89} although there has been considerable 
discussion as to which parameters are to be preferred \citep{rutten87, basri87}. 
A critical appraisal of the use of $Ro$ for activity-rotation 
relations was given, for example, by \citet{stepien94} who described some limitations also with 
regard to the underlying theoretical concepts.

\begin{figure} 
\hbox{\hskip -0.3cm\rotatebox{270}{\resizebox{0.7\hsize}{!}{\includegraphics{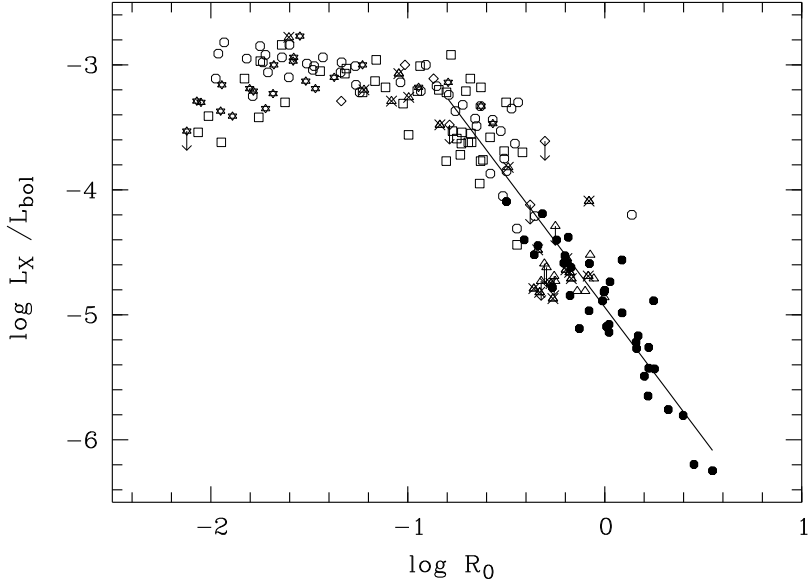}}}}    
\caption{Activity-rotation relationship compiled from several samples of open cluster stars. 
         Key to the symbols: circles: Pleiades; squares: IC 2602 and IC 2391: stars: $\alpha$ Per;
	 triangles: single Hyades stars; crossed triangles: Hyades binaries; diamonds: IC 4665;
	 filled symbols: field stars  (figure courtesy of S. Randich, after \citealt{randich00b},
	 by the kind permission of the Astronomical Society of the Pacific).}\label{activityrotation}
\end{figure}

The overall rotation-activity relation was perhaps best clarified by using large samples of
stars from stellar clusters.  
The comprehensive diagram in Fig.~\ref{activityrotation} clearly shows a regime where $L_X/L_{\rm bol} \propto
Ro^{-2}$ for intermediate and slow rotators (from \citealt{randich00b}).
However, in fast rotators $L_X$ appears to become a unique function of $L_{\rm bol}$, 
$L_X/L_{\rm bol} \approx 10^{-3}$ regardless of the rotation period \citep{agrawal86a, fleming88, pallavicini90a}. 
The tendency for a corona to  ``saturate'' at this level  once the rotation period
(or the Rossby number) is sufficiently small, or $v$ sufficiently large, 
was identified and described in detail by \citet{vilhu83}, \citet{vilhu84}, \citet{vilhu87}, and
\citet{fleming89}. It is valid for all 
classes of stars but the onset of saturation varies somewhat depending on the spectral type. 
Once MS coronae are saturated, $L_X$ {\it also becomes a function of mass, color, or radius simply owing to
the fundamental properties of MS stars.} 

\subsection{Activity and rotation in stars with shallow convection zones} 

The shallow convection zones toward early F stars express themselves through two effects.
First,  the maximum $L_X/L_{\rm bol}$ ratio or the maximum surface flux $F_X$ 
decrease when compared to the nearly constant values for cooler stars (\citealt{walter83, vilhu87, gagne95b}, also 
\citealt{wood94} for EUV emission). In other words, {\it the dynamo becomes less efficient}. 
And second, little or no dependence of $L_X$ on rotational parameters is found in early F stars 
\citep{pallavicini81, walter83, simon89, simon93, stauffer94}.
These results were initially taken to suggest a change in the dynamo mode (possibly from non-solar-like 
to solar-like  at a spectral type of $\approx$F5~V), or a change in the surface magnetic field configuration as the 
convection zone deepens \citep{schrijver96}, or a change from acoustic to magnetic coronal
heating \citep{simon89}.  

Comprehensive surveys by \citet{mangeney84}, \citet{schmitt85a}, and \citet{dobson89}, however, put
the absence of an activity-rotation relation into question. If activity is correlated with the Rossby number,
then in fact the same dependence is recovered for early F-type stars as for later stars. Consideration of 
the convective turnover time may indeed be important for these stars because, in contrast to the range of 
cooler stars, $\tau_c$ varies largely across the F spectral class (e.g., \citealt{stauffer94, randich96a}).
Once a ``basal'' flux is subtracted from the observed chromospheric flux, one also finds  that all
convective stars, including late-A and F stars, follow the same coronal-chromospheric flux-flux relations
(Sect.~\ref{fluxfluxrel}; \citealt{schrijver93}). It is essentially the  
{\it dynamo efficiency} (i.e., the surface magnetic flux density for a given 
rotation rate)  that decreases toward earlier F stars while the basic dynamo physics may be identical in
all convective main-sequence stars (except, possibly, for late-M dwarfs; Sect.~\ref{coolm}). 
At  the same time, the magnetic braking efficiency is reduced in early-F stars. The dynamo and, as a consequence, the 
magnetic flux production are thus never sufficient in these stars to ``saturate'' the way cooler stars do.

\subsection{Rotation and saturation; supersaturation} 

The issue of which parameters most favorably represent the activity-rotation relation 
was studied in great detail by \citet{pizzolato03} for MS stars in the context of saturation. They reported the following results
(see also \citealt{micela99a} for a qualitative description): i) The rotation period
{\it is} a good activity indicator for {\it non-saturated} stars for which it correlates with the unnormalized 
$L_X$ regardless of mass. The slope of the power law is --2. ii) The period at which saturation is reached 
increases with decreasing mass ($\approx 1.5$~d for an $1.05M_{\odot}$ star, $\approx 3.5$~d for a $0.7M_{\odot}$ 
star, and rapidly increasing further for lower masses), therefore reaching to progressively higher maximum $L_X$. 
iii) In the saturation  regime, $L_X/L_{\rm bol}$ becomes strictly 
independent of rotation and mass with the possible exception of stars with masses $> 1.1M_{\odot}$.
iv) A modified, empirical convective turnover time (hence a modified Rossby number $Ro^{\prime}$) can be derived as a function of stellar mass with
the goal of defining a universal function $L_X/L_{\rm bol} = f(Ro^{\prime})$  that is valid for all cool stars irrespective
of mass. The empirical turnover time is found to be similar to the calculated $\tau_c$, and it scales with $L_{\rm bol}^{-1/2}$. 
As a consequence, the two descriptions, $L_X$ vs. $P$ and $L_X/L_{\rm bol}$ vs. $Ro^{\prime}$ become fully equivalent for non-saturated stars.
v) The saturation is triggered at a fixed  $Ro^{\prime} \approx 0.1$, and the critical period where
saturation starts is $P_{\rm sat}$. The full description of the rotation-activity relation in this picture is, then,
\begin{equation} 
\left. 
   \begin{array}{ll}
     {\displaystyle{L_X \over L_{\rm bol}}} \propto Ro^{\prime-2}  \quad {\rm and}\quad  L_X \propto P^{-2}  &\quad\quad {\rm for}\quad\  P \ga  \\
     {\ } \\
     {\displaystyle{L_X \over L_{\rm bol}}} \approx 10^{-3}                                                 &\quad\quad {\rm for}\quad\ P \la        
   \end{array} 
   \right\} 
   P_{\rm sat} \approx 1.2\left({L_{\rm bol}\over L_{\odot}}\right)^{-1/2}
\end{equation}
As rotational equatorial velocities exceed $\approx 100$~km~s$^{-1}$, the $L_X/L_{\rm bol}$ values
begin to slightly decrease again. This ``supersaturation'' phenomenon (\citealt{prosser96, randich96a,
james00}, Fig.~\ref{activityrotation})  may be ascribed to a fundamental change of the dynamo action or to a decrease of the surface 
coverage with active regions (see Sect.~\ref{saturationcause}).

\subsection{Physical causes for saturation and supersaturation} \label{saturationcause}

Considering all aspects described above, it seems fair to say that all spectral classes between F and M 
are capable of maintaining coronae up to a limit of log\,$(L_X/L_{\rm bol}) \approx -3.0$. 
Some decrease of the maximum (``saturation'') level
toward earlier F and late-A stars may be real because of the shallowness of the convection zone
in these stars. The physical causes of saturation and supersaturation are not well understood. Ideas include the
following:
\begin{description}

\item [1.] The internal dynamo saturates, i.e., it produces no more magnetic flux if
      the rotation period increases \citep{gilman83, vilhu87}.

\item [2.] The surface filling factor of magnetic flux approaches unity at saturation \citep{vilhu84}.
      This is also  motivated by a strong correlation between
      saturated $L_X$ and radius rather than $L_X$ and surface temperature \citep{fleming89}.
      However, if the entire solar surface were filled with normal active regions,
      its X-ray luminosity would amount to only $\approx (2-3)\times 10^{29}$~erg~s$^{-1}$ \citep{vaiana78, wood94}, 
      with $L_X/L_{\rm bol} \approx 10^{-4}$ \citep{vilhu84}, 
      short of the empirical saturation value by one order of magnitude. To make up for this deficiency,
      one requires enhanced densities, larger coronal heights, or different mechanisms such as continuous flaring 
      (see Sect.~\ref{stochasticflares}).
      The detection of rotational modulation in some saturated or
      nearly saturated stars (\citealt{guedel95, kuerster97, audard01a}, see Sect.\ref{rotmod}) casts some doubt on a stellar 
      surface that is completely covered with an ensemble of similar active regions.      
       
\item [3.] \citet{jardine99} argued that the radius where centrifugal forces balance gravity (the co-rotation radius) 
      approaches the outer X-ray coronal radius in rapid rotators. As the rotation rate increases, centrifugal 
      forces lead to a rise in pressure in the outer parts of the largest loops. Once the corotation radius
      drops inside the corona, the local gas pressure may increase sufficiently to  blow open the magnetic structures, thus
      leading to open, X-ray dark volumes. This mechanism confines the coronal height. This coronal 
      ``stripping'' overcomes effects due to increased pressure, leading to approximately constant emission in 
      the saturation regime. As the rotation rate increases further and the corona shrinks, a more structured 
      low corona is left behind that is less luminous (``supersaturated'') and
      is more prone to rotational modulation \citep{jardine04}. Deep rotational modulation has indeed been 
      found in the supersaturated young G star VXR45 (\citealt{marino03a}, see Sect.~\ref{rotmod}).

\item [4.] An alternative explanation
      was  given by \citet{stepien01} who conjectured that rapid rotation produces, through a strong
      centrifugally induced gradient of the effective gravity from the equator to the pole, a heat 
      flux excess toward the poles in the stellar interior. Consequently, an excess convective updraft develops
      at the poles, accompanied by poleward circulation flows in the lower part of the convection zone,
      and equatorward surface return flows. This circulation system sweeps magnetic fields from the generation 
      region in the lower convection zone toward the poles. This effect strongly
      amplifies with rotation rate, thus leaving progressively more of the equatorial region free of
      strong magnetic fields. Therefore,  the filling factor decreases. The effect is particularly strong
      in W UMa-type contact binaries although an additional suppression, presumably due to equatorial
      flows between the components, is found. The suppression of equatorial activity has an interesting
      side effect in that loss of angular momentum through a wind is strongly suppressed \citep{stepien01}.
      
\end{description}

\subsection{Rotation and activity in pre-main sequence stars, giants and binaries} 

Among giants and supergiants, the dependence between rotation and activity becomes  
much less evident \citep{maggio90}.  Whereas cooler giants follow the same dependence as 
MS stars, this does not hold for warmer giants \citep{ayres98}. The evolution across the Hertzsprung gap
features two competing effects, namely a deepening convection zone that strengthens the dynamo, and
rapid spin-down that weakens it. It is likely that the rapid evolution through this regime
does not leave sufficient time for the stars to converge to a unified rotation-activity relation.

There are also mixed results from pre-main sequence stars. Whereas
the standard behavior including saturation applies to some star-forming regions
such as Taurus, other regions show all stars  in a saturation regime, up to rotation periods
of 30~d. This effect could be related to the long convective turnover time in
these stars, as discussed by \citet{flaccomio03c} and \citet{feigelson03} (see Sect.~\ref{ttau} 
for further details).

Close binary systems are interesting objects to study the effect of 
rapid rotation that is maintained due to tidal interactions with the orbiting
companion.  \citet{walter81a} found
\begin{equation}
 {L_X \over L_{\rm bol}} \propto \Omega
\end{equation}
(and no dependence on $v$) although this relation contains much scatter and may
in fact be a consequence of a relation between stellar radius and orbital
period in close binaries, larger stars typically being components of longer-period systems and
being bolometrically brighter \citep{walter81a, rengarajan83, majer86, dempsey93a};
this explanation was however questioned again by \citet{dempsey97}. In general, caution is
in order with regard to activity-rotation relationships in these binaries because many of them 
are at or close to the saturation limit.

\section{Flux-flux relations}\label{fluxfluxrel}

\subsection{Chromosphere-transition region-corona}

\begin{figure} 
\centerline{
\resizebox{0.7\textwidth}{!}{\includegraphics{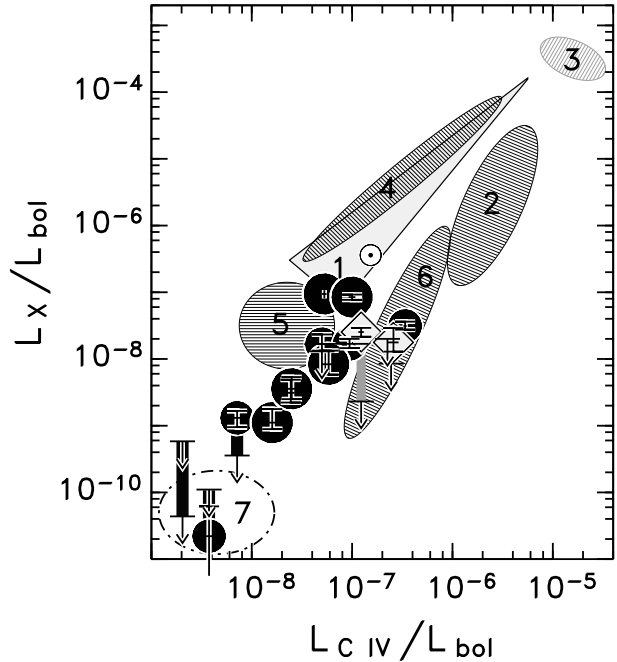}}
}
\caption{``Flux-flux'' diagram for $L_X$ and the C\,{\sc iv} luminosity.  Different groups of stars 
are schematically represented by different shading, labeled as follows: 1 standard relation for
  main-sequence stars; 2 X-ray deficient G supergiants; 3, 4 G-K0~III giants; 5 low-activity
  K0-1 III giants, 6 X-ray deficient F-G0 Hertzsprung gap stars; 7 probable region of red giants (Aldebaran, Arcturus);
   the black circles mark ``hybrid'' stars   (figure courtesy of T. Ayres, after \citealt{ayres04}).}\label{fluxflux}
\end{figure}

Flux-flux (or luminosity-luminosity) relations from the chromosphere to the corona contain telltale signatures
of the overall heating process and of systematic deficiencies at any of the temperature layers.
The standard relation between normalized luminosities from transition-region emission lines such as C\,{\sc iv} and 
coronal X-ray luminosities is non-linear, with a power-law slope of about 1.4--1.5 \citep{ayres81b, vilhu84, agrawal86a, 
haisch90c, ayres95};  the power-law becomes steeper if chromospheric lines are used, e.g., Mg\,{\sc ii}; thus
\begin{equation}
{L_X\over L_{\rm bol}} \approx \left({L_{\rm C~IV}\over L_{\rm bol}}\right)^{1.5}; \quad\quad
{L_X\over L_{\rm bol}} \approx \left({L_{\rm Mg~II}\over L_{\rm bol}}\right)^{3}
\end{equation}
(see Fig.~\ref{fluxflux}). These relations hold for RS CVn-type binaries as well \citep{dempsey93a}, although
\citet{mathioudakis89} reported a near-linear correlation between the Mg~{\sc ii} and and X-ray surface fluxes in
dMe and dKe dwarfs. Schrijver and co-workers suggested that a color-dependent {\it basal} component be subtracted from the 
chromospheric (and partly transition region) stellar line fluxes to obtain the magnetically induced
excess flux $\Delta_{HK}$ in the Ca\,{\sc ii} H\&K  lines (\citealt{schrijver83, schrijver87, schrijver92, rutten91} and further 
references cited therein). The justification for this procedure is that the flux-flux 
power-law relations tighten and become color-independent. They  flatten somewhat but  remain
non-linear between X-rays and chromospheric fluxes: $F_X \propto \Delta_{HK}^{1.5-1.7}$. 
On the other hand, this procedure results in
little change for the X-ray vs. transition region flux correlation \citep{rutten91, ayres95}. The basal chromospheric
flux is then independent of activity and was suggested to be the result of steady (non-magnetic) acoustic heating 
\citep{schrijver87}. {\it No} such basal flux is found for X-rays \citep{rutten91} which may imply that
any lower limit to the average X-ray surface flux may be ascribed to a genuine minimum magnetic heating.
If done so, the lower limit to the X-ray surface flux empirically found by \citet{schmitt97}  implies,
however, a minimum {\it magnetic} chromospheric flux that is still much in excess of the 
basal fluxes for G-M stars, thus putting into question whether truly basal stars are realized, except
possibly for F-type stars (see also \citealt{dempsey97}). 

The flux-flux relations must be rooted in the magnetic flux on the stellar surface. Whereas
{\it chromospheric} fluxes appear to depend non-linearly on the average photospheric magnetic 
flux density $fB$  ($f$ is the surface filling factor of the magnetic fields) 
both for solar features and for entire stars \citep{schrijver89a, schrijver89c}, the 
stellar {\it coronal} correlation between X-ray surface flux and $fB$  becomes nearly linear 
\citep{schrijver89a}. This suggests that an important cause of the non-linearities
in the chromospheric-coronal flux-flux correlations is actually rooted in the behavior
of the chromospheric radiative losses. On the other hand, \citet{ayres96} explained the non-linearity in
the corona-transition region flux-flux relation by the increasing coronal temperatures with increasing
activity (Sect.~\ref{templx}), bringing a progressively larger fraction of the emission into the X-ray band. 

For various solar features (active regions, bright points, etc),
a strong {\it linear} correlation was reported between $L_X$ and total unsigned magnetic flux \citep{fisher98};
the principal determining factor is the surface  area of the feature. 
Interestingly, this correlation extends linearly over orders of magnitude to magnetically 
active stars. Because  the entire stellar coronae may be made up of various solar-like 
features, the overall correlation  suggests that a common heating mechanism is present 
for all solar and stellar coronal structures \citep{pevtsov03}.  Empirically, a somewhat different 
relation was suggested by \citet{stepien94} if the stellar color (or the photospheric effective 
temperature) was also taken into account, namely $F_X \propto T^{8.3}B^{1.9}$ where $B$ is the surface 
magnetic field strength.

There are interesting deviations from the flux-flux correlations, in particular in
the giant and supergiant domain. These are briefly addressed in the context of evolution and 
coronal structure in Sect.~\ref{giants} and \ref{hybrid}. Also, several flux-flux relations
have been reported between flaring and low-level emission. I describe those
in Sect.~\ref{flarequicorr}.

\subsection{Radio--X-ray correlations}

There is considerable interest in correlating emissions that  connect
different parts of causal chains. As I will discuss further
in Sect.\ref{flaremodel}, standard flare models involve electron beams (visible
at radio wavelengths) that heat the chromospheric plasma to X-ray emitting 
temperatures. The heating processes of the quiescent coronal plasma may be entirely
different, however. Nevertheless, average radio and X-ray luminosities
are nearly linearly correlated in magnetically active stars \citep{guedel93},
$L_X/L_R \approx 10^{15.5}$~[Hz]\footnote{Note that the radio luminosity $L_R$ is conventionally
derived from a {\it flux density} measured at a fixed frequency such as 4.9~GHz or 8.4~GHz, per unit 
frequency interval.} with some scatter, and this correlation appears to mirror
the behavior of individual  solar flares \citep{benz94}. This flux-flux relation
applies to quite different classes of active stars such as RS CVn binaries
\citep{drakes89, drakes92, dempsey93a, dempsey97, fox94} and active M dwarfs
\citep{guedel93b}. It is not entirely clear what the underlying cause is, but
the most straightforward interpretation is in terms of chromospheric
evaporation of frequent, unresolved flares that produce the
observed radio emission and at the same time heat the plasma to
coronal temperatures.

\section{Thermal structure of stellar coronae}

\subsection{Thermal coronal components}\label{thermalcomp}

The large range of temperatures measured in stellar coronae has been a challenge for 
theoretical interpretation from the early  days of coronal research. Whereas much 
of the solar coronal plasma can be well described by a  component of a few
million degrees, early investigations of RS CVn binaries with low-resolution  detectors already recognized
that active stellar coronae cannot be (near-)isothermal but require a parameterization
in terms of at least two largely different temperature components, one around $4-8$~MK and the other around 
$20-100$~MK (\citealt{swank81}, also \citealt{holt79}, \citealt{white80}, and \citealt{agrawal81}
for individual cases). First estimates based on static loop models showed that a simple corona
requires either very high pressures (of order 100 dynes~cm$^{-2}$), implying very compact  
sources with small filling factors, or extremely extended magnetic loops, with a possibility
to connect to the binary companion \citep{swank81}. The solar analogy had thus immediately reached its 
limitation for a proper interpretation of stellar data. This theme and its variations
have remained of fundamental interest to the stellar X-ray community ever since.

Although even low-resolution devices provide meaningful temperature measurements, there has been 
a long-standing debate on the interpretation of ``1-$T$'' or ``2-$T$'' models. 
Historically, the opinions were split; some argued that the individual temperature components represent
separate plasma features \citep{schrijver84, mewe86, singh87, pallavicini88, 
lemen89, pasquini89, schrijver89b, dempsey93b, singh95, singh96a, singh96c, rodono99};
others suggested that they parameterize a continuous distribution of EM in temperature  and thus represent
a continuum of source types 
\citep{majer86, schmitt87, schmitt90a, schmitt97, drake95b, drake01}. 
As we have been learning from high-resolution spectroscopy, the correct solution may be a diplomatic one. There is little
doubt (also from the solar analogy and simple physical models) that coronae display truly continuous EM distributions, but there
are a number of superimposed features that may trace back to individual physical coronal structures.
The differential emission measure distribution (DEM) thus became an interesting diagnostic tool for coronal 
structure and heating (see, for  example, \citealt{dupree93, brickhouse95, kaastra96, guedel97a, favata97c, griffiths98}, to 
mention a few).

The temperatures determined from low-resolution spectral devices may in fact also be driven by detector
characteristics, in particular the accessible energy range as well as the spectral behavior of the detector's 
effective area \citep{majer86, schmitt87, pasquini89, schmitt90a, favata97c}. 
Overall, there is little doubt that the gross temperature determinations of low-resolution devices are correct, but
a comprehensive description of the  emission measure distribution requires high-resolution
spectra that allow for more  degrees of freedom and thus independent parameters, although the accuracy of 
the spectral inversion remains  limited on principal mathematical grounds (see Sect.~\ref{demmeth}).
I will in the following  focus on more recent results that are based on reconstructions of full DEMs 
mainly from high-resolution devices, first reviewing some general data and basic definitions.

\section{High-resolution X-ray spectroscopy}

With the advent of  {\it Chandra} and {\it XMM-Newton}, high-resolution X-ray spectroscopy  has opened
a new window to stellar coronal research. The {\it Chandra} High-Energy Transmission Grating Spectrometer (HETGS), the 
Low-Energy Transmission Grating Spectrometer (LETGS) as well as the two {\it XMM-Newton} Reflection Grating Spectrometers 
(RGS) cover a large range of spectral lines that can be separated and  analyzed in detail. The spectra contain 
the features required for deriving emission measure distributions, abundances, densities, and opacities as discussed 
throughout this paper. Here, I give only a brief description of sample spectra and some distinguishing properties 
that are directly related to the thermal structure.
 
\begin{figure} 
\centerline{\resizebox{1.0\textwidth}{!}{\includegraphics{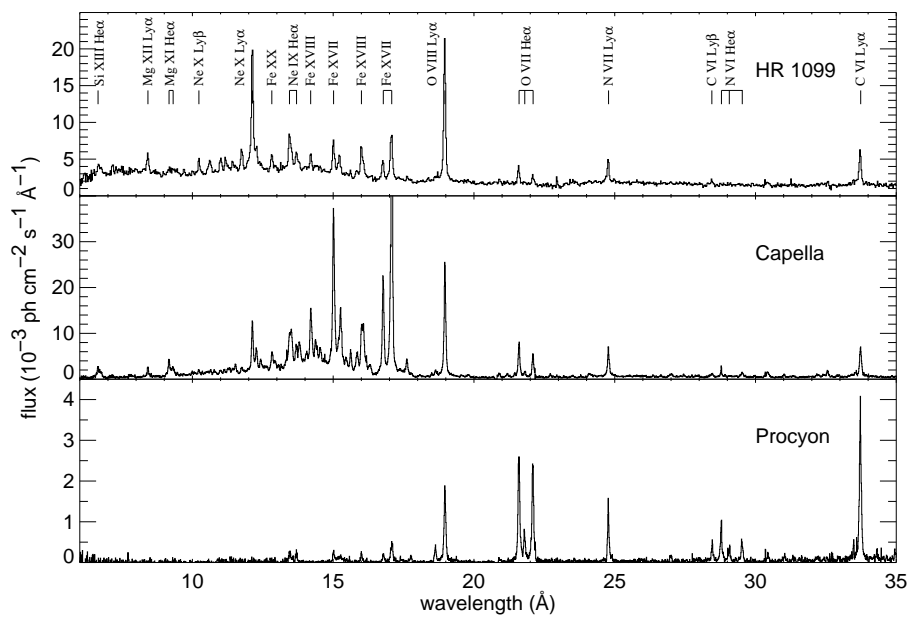}}}
\caption{Three high-resolution X-ray spectra of stars with largely differing activity levels: HR~1099,
Capella, and Procyon. Data from {\it XMM-Newton} RGS.}\label{spectra1}
\end{figure}

\begin{figure} 
\centerline{\resizebox{1.0\textwidth}{!}{\includegraphics{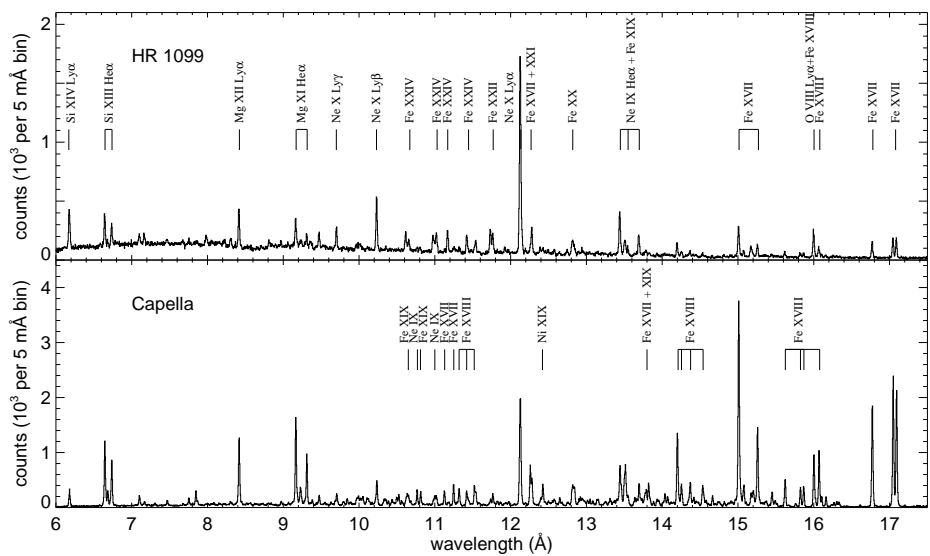}}}
\caption{Extracts of two high-resolution X-ray spectra from  HR~1099 
and Capella, showing the region of the Fe L-shell transitions. Strong  lines in the Capella spectrum without
identification labels correspond to those labeled in the HR~1099 spectrum  (data from {\it Chandra} HETGS, 
courtesy of N. Brickhouse).}\label{spectra2}
\end{figure}

Figures \ref{spectra1} and \ref{spectra2} show examples of spectra obtained by {\it XMM-Newton} and {\it Chandra}, respectively.
The stars cover the entire range of stellar activity: HR~1099 representing a very active RS CVn system, Capella 
an intermediately active binary, and Procyon an inactive F dwarf. The spectrum of HR~1099 reveals a considerable
amount of continuum and comparatively weak lines, which is a consequence of the very hot plasma in this corona 
($T \approx 5-30$~MK). Note also the unusually strong Ne\,{\sc ix}/Fe\,{\sc xvii} and Ne\,{\sc x}/Fe\,{\sc xvii} 
flux ratios if compared to the other stellar spectra. These anomalous ratios are in fact due to 
an abundance anomaly discussed in Sect.~\ref{abunnew}. The spectrum of Capella is dominated by Fe\,{\sc xvii} and Fe\,{\sc xviii} 
lines which are preferentially formed in this corona's plasma at $T \approx 6$~MK. Procyon, in contrast, shows
essentially no continuum and only very weak lines of Fe. Its spectrum is dominated by the H- and He-like
transitions of C, N, and O formed  around 1--4~MK.  The flux ratios between H- to He-like transitions are also convenient 
temperature indicators: The O\,{\sc viii}~$\lambda 18.97$/O\,{\sc vii}~$\lambda 21.6$ flux ratio, for example, is very large for HR~1099 but drops 
below unity for Procyon.

\section{The differential emission measure distribution}

\subsection{Theory}

The flux $\phi_j$ observed in a line from a given atomic transition can be written as
\begin{equation}\label{fluxdem}
\phi_j = {1\over 4\pi d^2}   \int AG_j(T) {n_en_H dV\over d{\rm ln}T}d{\rm ln}T
\end{equation}
where $d$ denotes the distance, and $G_j(T)$ is the ``line cooling function'' (luminosity 
per unit EM) that contains 
the atomic physics of the transition as well as the ionization  fraction for  the
ionization stage in question, and $A$ is the  abundance of the element with respect
to some basic tabulation used for $G_j$. For a fully ionized plasma
with cosmic abundances, the hydrogen density  $n_H \approx 0.85n_e$. The expression
\begin{equation}\label{demdef}
Q(T)  = {n_en_HdV\over d{\rm ln}T}
\end{equation}
defines the {\it differential emission measure distribution} (DEM). I will use this definition
throughout but note that some authors define $Q^{\prime}(T) = n_en_HdV/dT$  which is 
smaller by one power of $T$. For a plane-parallel  atmosphere with surface area $S$,
Eq.~(\ref{demdef}) implies 
\begin{equation}
Q(T) = n_en_H SH(T), \quad\quad H(T) = \left|{1\over T}{dT\over ds}\right|^{-1}
\end{equation} 
where  $H$ is the {\it temperature scale height.}

\subsection{Interpretation}\label{deminterpret}

Equations~(\ref{fluxdem}) and (\ref{demdef}) introduce the  DEM as the basic interface between the stellar X-ray observation and the model 
interpretation of the thermal source. It contains information on the plasma temperature and 
the density-weighted plasma mass that emits X-rays at any given temperature. Although a DEM is often a  
highly degenerate description of a complex real corona, it provides important constraints
on heating theories and on the range of coronal structures that it may describe.
{\it Solar} DEMs can, similarly to the stellar cases, often be approximated
by two power laws $Q(T)  \propto T^s$, one on each side of its peak. \citet{raymond81} reported
low-temperature power-law slopes of $s = 0.9$ for the coronal hole network, $s = 2.1$ for
the quiet Sun, and $s = 3.1$ for flares. The Sun has given considerable guidance in physically
interpreting the observed stellar DEMs, as the following subsections summarize.

\subsubsection{The DEM of a static loop}\label{demstatic}

\begin{figure} 
\hbox{
\resizebox{0.526\textwidth}{!}{\includegraphics{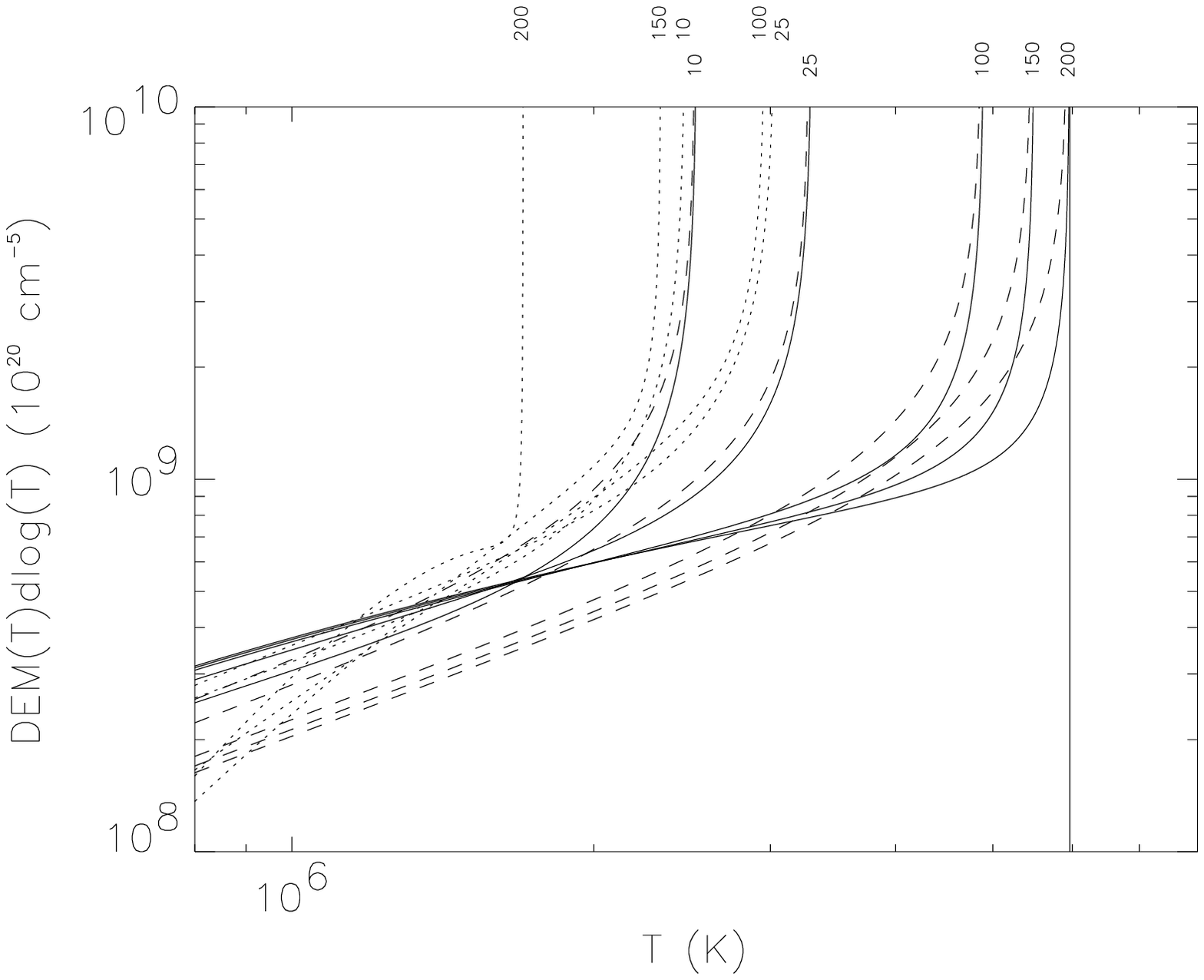}}
\resizebox{0.474\textwidth}{!}{\includegraphics{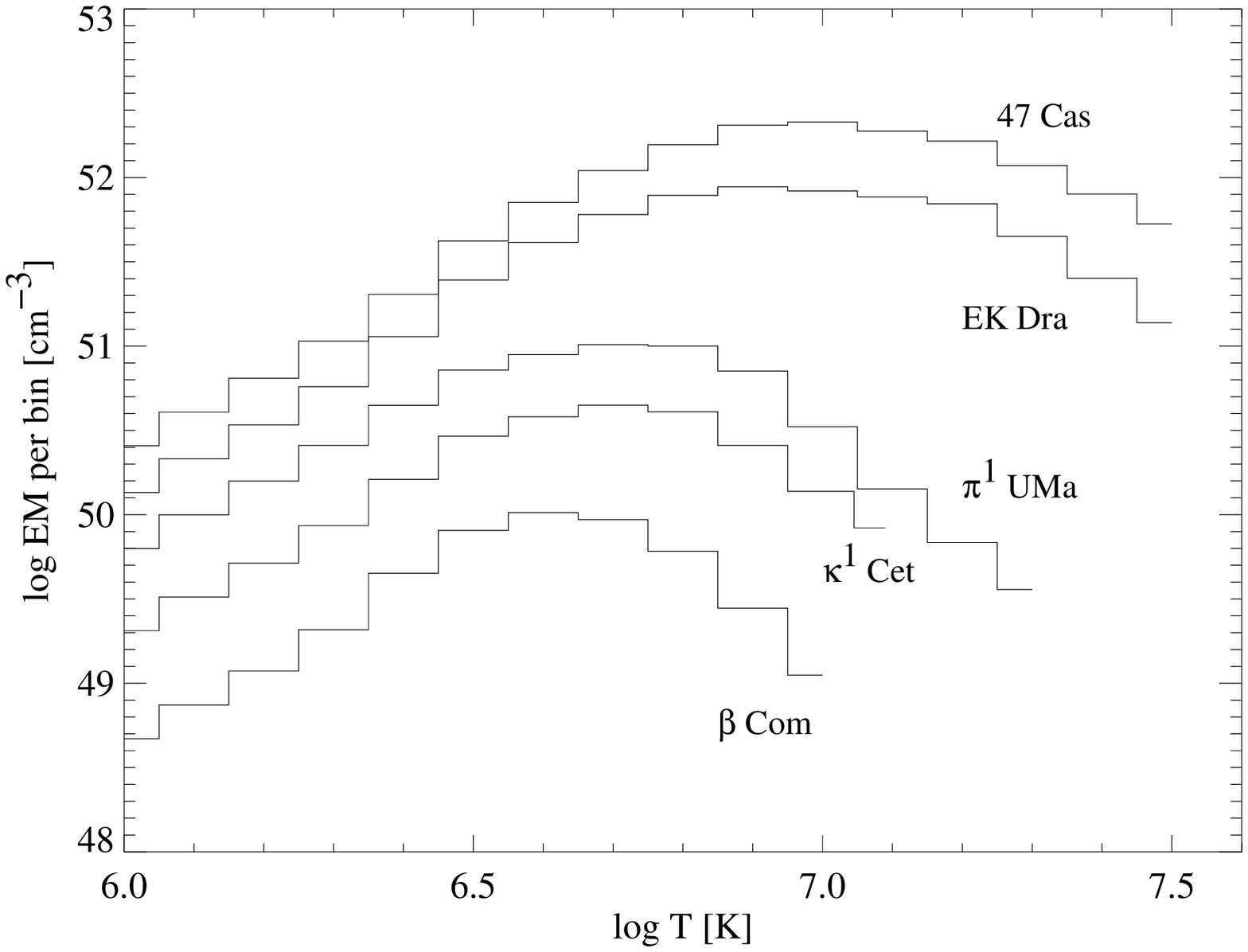}}
}
\caption{{\it Left:} Calculated differential emission measures of individual
static loops. The solid curves refer to uniform heating along the loop and some
fixed footpoint heating flux, for different loop half-lengths labeled above the 
figure panel in megameters. The dashed curves illustrate the analytical
solutions presented by \citet{rosner78} for uniform heating. The dotted
lines show solutions assuming a heating scale height of $2\times 10^9$~cm (figure
courtesy of K. Schrijver, after \citealt{schrijver02}).
{\it Right:} Examples of discrete stellar emission measure distributions derived
from spectra of solar analogs. The slopes below log\,$T \approx 6.5$ are only approximate 
(figure courtesy of A. Telleschi).
}\label{calcdem}
\end{figure}

The DEM  of a hydrostatic, constant-pressure loop was discussed by 
\citet{rosner78} (= RTV), \citet{vesecky79} (= VAU),
and \citet{antiochos86b}. Under the conditions of negligible gravity, i.e., constant 
pressure in the entire loop, and negligible thermal conduction at the footpoints,
\begin{equation}\label{staticem}
Q(T) \propto pT^{3/4-\gamma/2+\alpha}  {1 \over \left( 1 - \left[T/T_{\rm a}\right]^{2-\gamma+\beta}\right)^{1/2}}
\end{equation}
\citep{bray91} where $T_{\rm a}$ is the loop apex temperature,  and $\alpha$ and $\beta$ are power-law 
indices of, respectively,  the loop cross section area $S$ and the heating power $q$ as a function of $T$:
$S(T) = S_0T^{\alpha}$,  $q(T) = q_0T^{\beta}$, and 
$\gamma$ is the exponent in the cooling function over the relevant temperature range: 
$\Lambda(T) \propto T^{\gamma}$. If $T$ is not close to $T_{\rm a}$ and the loops have constant cross
section ($\alpha = 0$), we have $Q(T) \propto T^{3/4 -\gamma/2}$,
i.e., under typical coronal conditions for non-flaring loops ($T < 10$~MK, $\gamma \approx -0.5$), the DEM 
slope is near unity \citep{antiochos86b}. If strong thermal conduction is included at the footpoints, then
the slope changes to +3/2 if not too close to $T_{\rm a}$ \citep{vdoord97}, but note that the exact slope
again depends on $\gamma$, i.e., the run of the cooling function over the temperature range
of interest. The single-loop DEM sharply increases at $T \approx T_{\rm a}$ (Fig.~\ref{calcdem}). 

Such models may already resemble some stellar DEMs \citep{ayres98}, and they are close to
observed solar full-disk DEMs that indicate $Q \propto T^{3/2}$ \citep{jordan80, laming95, peres01}.  
However, the DEMs of many active stars are much steeper (see below; Fig.~\ref{calcdem}). Loop expansion
($\alpha > 0$) obviously steepens the DEM. Increased heating at the loop footpoints (instead of 
uniform heating) makes the $T$ range narrower and will also increase the slope of the DEM 
\citep{bray91, argiroffi03}. Further, if the heating is non-uniform, as for example in loops
that are predominantly heated near the footpoints, the DEM becomes steeper as well (see 
numerical calculations of various loop examples by \citealt{schrijver02} and \citealt{aschwanden02}). 
Examples are illustrated in Fig.~\ref{calcdem} together with discrete emission measure distributions derived from
stellar spectra. Comprehensive numerical hydrostatic energy-balance loop models undergoing steady 
apex heating have been computed by \citet{griffiths99}, with applications to observed DEMs, and by the
Palermo group (see Sect.~\ref{loopmodels}).

If the loops are uniformly distributed in $T_{\rm a}$, and one assumes a heating rate proportional
to the square of the magnetic field strength, $B^2$, then  $Q$ {\it is dominated by the hottest
loops} because these produce more EM at any given $T$ than cooler loops \citep{antiochos86b}. In the more
general case, the descending, high-$T$ slope is related to the statistical distribution of the loops in $T_{\rm a}$; 
a sharp decrease of the DEM indicates that only few loops are present with a temperature exceeding the 
temperature of the DEM peak \citep{peres01}. 

\citet{lemen89} found that EM is concentrated at temperatures  where the cooling
function $\Lambda(T)$ has a positive slope or is flat; these are regions insensitive to heating fluctuations. 
This idea was further discussed by \citet{gehrels93} who found that most 2-$T$ fits of low-resolution RS CVn spectra
show plasma in two regions of relative stability, namely at $5-8$~MK and above 25~MK.

\subsubsection{The DEM of flaring structures}\label{demflaring}

\citet{antiochos80} (see also references therein) discussed DEMs of flaring loops 
that cool by i) static conduction (without flows), ii) evaporative conduction (including flows), and
iii) radiation. The inferred DEMs scale, in the above order, like
\begin{equation}
Q_{\rm cond} \propto  T^{1.5}, \quad\quad  Q_{\rm evap} \propto T^{0.5}, 
     \quad\quad Q_{\rm rad} \propto T^{-\gamma + 1}. 
\end{equation}
Since $\gamma \approx 0\pm 0.5$ in the range typically of interest for stellar flares
($5-50$~MK), all above DEMs are relatively flat (slope $1\pm 0.5$). If multiple loops with equal slope 
but different peak $T$ contribute, then the slope up to the first DEM peak can only become smaller.
Non-constant loop cross sections have a very limited influence
on the DEM slopes. 

Stellar flare observations are often not of sufficient quality to derive temperature and EM characteristics for 
many different time bins. An interesting diagnostic was presented by \citet{mewe97} who calculated
the time-integrated (average) DEM of a flare that decays quasi-statically. They find
\begin{equation}\label{demflareint}
Q \propto T^{19/8}
\end{equation}
up to a maximum $T$ that is equal to the temperature at the start of the decay phase.

\citet{sturrock90} considered episodic flare heating. In essence, they showed how loop cooling together with 
the rate of energy injection as a function of $T$ may form the observed solar ``quiescent'' DEM, i.e., 
the latter would be related to the shape of the cooling function $\Lambda(T)$: the negative slope of $\Lambda$
between $10^5$~K and a few times $10^6$~K results in an increasing $Q(T)$. Systems of this kind were 
computed semi-analytically by \cite{cargill94}, using analytic approximations for conductive and radiative 
decay phases of the flares. Here, the DEM is defined not by the internal loop structure but
by the time evolution of a flaring plasma (assumed to be isothermal). Cargill argued that for radiative 
cooling, the (statistical) contribution of a flaring loop to the DEM is, to zeroth order,
inversely proportional to the radiative decay time, which implies
\begin{equation}
Q(T) \propto T^{-\gamma+1}
\end{equation}
up to a maximum $T_m$, and a factor of $T^{1/2}$ less if subsonic draining of the cooling loop is 
allowed. Simulations with a uniform distribution of small flares within a limited energy range agree with
these rough predictions, indicating a time-averaged DEM that is relatively flat below $10^6$~K but 
steep ($Q[T] \propto T^4$) up to a few MK, a range in which the cooling function drops rapidly. 
\citet{guedel97c}  used a semi-analytical hydrodynamic approach formulated by \citet{kopp93} to compute the 
time-averaged DEM for typical stellar conditions up to several $10^7$~K for a power-law distribution of the flare energies
(see also Sect.~\ref{obsstochflare}), finding two DEM peaks that are dominated by the large number of small, 
cool flares (``microflares'' producing the cooler plasma) and by the much less frequent energetic, hot flares, respectively. 

Let us next assume - in analogy to solar flares - that the occurrence rate of flares is distributed in energy as a 
power law with an index $\alpha$  ($dN/dE \propto E^{-\alpha}$) and that the peak emission measure EM$_p$ of a flare is a 
power-law function of its peak temperature $T_p$ at least over a limited range of temperatures: 
EM$_p \propto T_p^b$ as found by \citet{feldman95}  
(see also Sect.~\ref{flaretemp} for a larger flare sample).  Then, an analytic expression can be derived for the 
time-averaged DEM of such a flare ensemble, i.e., a ``flare-heated corona'', revealing a power law on each side of 
the DEM peak \citep{guedel03a}:
\begin{equation} 
Q(T) \propto \left\{ 
   \begin{array}{ll}\label{demflare}
        T^{2/\zeta}                                & \mbox{\quad for\quad  $T < T_m$ } \\
        T^{-(\alpha-2)(b+\gamma)-\gamma}           & \mbox{\quad for\quad  $T > T_m$ } 
   \end{array} 
   \right. 
\end{equation}
where we have assumed the same luminosity decay time scale for all flares.
Here, $T_m$ (a free parameter) is the temperature of the DEM peak, and $b \approx 4.3\pm 0.35$ (Sect.~\ref{flaretemp}) 
in the temperature range of interest for active-stellar conditions. The parameter $\zeta \equiv \tau_n/\tau_T$,
$\tau_n$ and $\tau_T$ being the e-folding decay times of density and temperature, respectively, is found to vary
between $\zeta = 0.5$ (strong heating during the decay) and $\zeta = 2$ (no heating, see \citealt{reale97}, 
Sect.~\ref{contheat}). This model produces DEM slopes below $T_m$ that are steeper than unity and range
up to a maximum of four.

\subsection{Reconstruction methods and limitations}\label{demmeth}

Deriving DEMs or their discretized, binned equivalents, the ``emission measure 
distributions'' [EMD] in log\,$T$ from X-ray spectra has been one of the central issues in observational
stellar X-ray astronomy. As implied by Eq.~(\ref{fluxdem}), it is also of considerable 
importance in the context of determining the coronal composition (see Sect.~\ref{composition}).
Although a full description of the methodology of spectral inversion is beyond
the scope of this review, I will briefly outline the available strategies as well as
the current debate on optimizing results. This may serve as an introduction and
guide to the more technical literature.

If a spectrum of an isothermal plasma component with unit EM
is written in vector form as ${\bf f}(\lambda, T)$, then 
the observed spectrum is the weighted sum 
\begin{equation}\label{DEM}
{\bf g}(\lambda) = \int {\bf f}(\lambda, T)Q(T) d{\rm ln}T \equiv {\bf F\cdot Q}.
\end{equation}
In discretized form for bins $\Delta {\rm log}\,T$, $\mathbf{F}$ is a rectangular matrix 
(in $\lambda$ and $T$). Eq.~(\ref{DEM}) constitutes a Fredholm equation of the first 
kind for {\bf Q}. Its inversion aiming at solving for Q is an
ill-conditioned problem with no unique solution unless one imposes
additional constraints such as positivity, smoothness, or functional form, most of which may not be physically founded.
A formal treatment is given in \citet{craig76}. The problem is particularly
serious due to several sources of unknown and systematic uncertainties, such as inaccurate atomic
physics parameters in the spectral models, uncertainties in the instrument calibration and imprecise flux determinations, 
line blends (see detailed discussion in \citealt{vdoord97} and \citealt{kashyap98} and references therein) and, in 
particular, unknown element abundances. The latter need to be determined from the same spectra. They  are usually assumed 
to be constant across the complete DEM although this hypothesis is not supported by solar investigations \citep{laming95, jordan98}. 
The following {\it constrained inversion} techniques have turned out to be convenient:\footnote{I  henceforth avoid  expressions 
such as ``global'' or ``line-based  methods'' that have often been used in various, ill-defined contexts. Spectral inversion methods
should be distinguished by i) the range and type of the data to be fitted, ii) the parameters to be determined (model assumptions), 
iii) the iteration scheme for the fit (if an iterative technique is applied),
iv) the convergence criteria,  v) the constraints imposed on the solution (e.g., functional form of DEM, smoothness, positivity, etc),
and vi) the atomic database used for the interpretation.
Several methods described in the literature vary in some or all of the above characteristics. 
Most of the methods described here are not inherently tailored to a specific spectral resolving power.
What does require attention are the possible biases that the selected iteration scheme and the constraints imposed on the 
solution may introduce, in particular because the underlying atomic physics tabulations are often inaccurate or
incomplete (``missing lines'' in the codes).}
\begin{description}

\item [1.] {\it Integral inversion with regularization}. A matrix inversion of Eq.~(\ref{DEM})
       is used with the additional constraint that the 
       second derivative of the solution $Q$(log\,$T$) is as smooth as statistically allowed by the data.
       Oscillations in the data that are due to data noise are thus damped out. This method is
       appropriate for smooth DEMs, but tends to produce artificial wings in sharply peaked DEMs
       \citep{mewe95, schrijver95, cully97}. A variant using singular value decomposition for
       a series of measured line fluxes was discussed by  \citet{schmitt96b} .

\item [2.] {\it Multi-temperature component fits}. This approach uses a set of elementary DEM building blocks such as 
       Gaussian DEMs centered at various $T$ but is otherwise similar to the traditional multi-component fits applied
       to low-resolution data (examples were given by \citealt{kaastra96} and \citealt{guedel97b}). 

\item [3.] {\it Clean algorithm}. This is both a specific iteration scheme and
      a special case of (2) that uses delta functions as building blocks.  The observed spectrum 
      (or part of it)  is correlated with  predictions from
      isothermal models. The model spectrum with the highest
      correlation coefficient indicates the likely dominant $T$
      component. A fraction of this spectral component is subtracted from the
      observation, and the corresponding model EM is saved. This process
      is iterated until the residual spectrum contains only noise. The summed
      model EM tends to produce sharp features while positivity is ensured \citep{kaastra96}. 
       
\item [4.] {\it Polynomial DEMs.} The DEM is approximated by
       the sum of Chebychev polynomials $P_k$. For better convergence, the logarithms of the 
       EM and of $T$ are used:  
       log${[Q(T)d{\rm log}\,(T)]} = \sum_{k=0}^{N-1}a_kP_k({\rm log}\,T)$ which ensures 
       positivity. The degree $N$ of the polynomial fit can be adjusted to
       account for broad and narrow features  \citep{lemen89, kaastra96, schmitt04, audard04}.

\item [5.] {\it Power-law shaped DEMs} of the form $Q(T) \propto (T/T_{\rm max})^{\alpha}$ up to
      a cutoff temperature $T_{\rm max}$ are motivated by the approximate DEM shape of a single
      magnetic loop \citep{pasquini89, schmitt90a}. 
            
\end{description}
The {\it ranges and types} of data may vary depending on the data in use. Low-resolution spectra are commonly
inverted as a whole because individual features cannot be isolated. If a high-resolution spectrum is 
available,  then inversion methods have been applied either to the entire spectrum, to selected features 
(i.e., mostly bright lines), or to a sample of extracted line fluxes.

As for {\it iteration schemes}, standard optimization/minimization techniques are available.
Various methods have been developed for fits to samples of line fluxes (e.g., \citealt{lemen89, huenemoerder01, 
huenemoerder03, osten03, sanz03, telleschi04}), with similar principles: 
\begin{description}
\item [1.]  The DEM {\it shape} is iteratively 
derived from line fluxes of one element only, typically Fe ({\sc xvii-xxvi} in X-rays, covering $T$ up 
to $\approx$100~MK), e.g., by making use of one of the above inversion schemes tailored to a sample of line fluxes.
Alternatively, one can use $T$-sensitive but abundance-independent flux ratios between He-like 
and H-like transitions of various elements to construct
the DEM piece-wise across a temperature range of $\approx 1-15$~MK \citep{schmitt04}.

\item [2.] The Fe abundance (and thus the DEM normalization) is found by requiring 
that the continuum (formed mainly by H and He) agrees with the observations. 

\item [3.] The abundances
of other elements are found by comparing their DEM-predicted line fluxes (e.g., assuming solar
abundances), with the observations.  

\end{description}
The advantage of such schemes is that they treat the DEM inversion
and the abundance determination sequentially and independently. \citet{huenemoerder01} and \citet{huenemoerder03} used an 
iteration scheme that fits DEM and abundances simultaneously based on a list of line fluxes
plus a continuum. \citet{kashyap98} further introduced an iteration scheme based on Markov-chain 
Monte Carlo methods for a list of line fluxes. This approach was applied to stellar data by \citet{drake01}.
Genetic algorithms have also, albeit rarely, been used as iteration schemes (\citealt{kaastra96} for low-resolution
spectra). 

There has been a lively debate in the stellar community on the ``preferred'' spectral inversion approach.  
Some of the pros and cons for various strategies are: 
Methods based on full, tabulated spectral models or on a large number of individual line fluxes
may be compromised by inclusion of transitions with poor atomic data such as emissivities or wavelengths. 
On the other hand, a large line sample may smooth out the effect of such uncertainties. Consideration of all tabulated 
lines further leads to a treatment of line blends that is self-consistent within the limits given by the
atomic physics uncertainties. A most serious problem 
arises from weak lines that are not tabulated in the spectral codes while they contribute to the spectrum in
two ways: either in the form of excess flux that may be misinterpreted as a continuum, thus modifying the
DEM; or in the form of unrecognized line blends, thus modifying individual line fluxes and the pedestal
flux on which individual lines are superimposed. A careful selection of spectral regions and lines for 
the inversion is thus required (see discussion and examples in \citealt{lepson02} for the EUV range).

If DEMs and abundances are iterated simultaneously,  numerical cross-talk between abundance and 
DEM calculation may be problematic, in particular if multiple solutions exist.
Nevertheless, each ensemble of line flux ratios of one element determines the same 
DEM and thus simultaneously enforces agreement.  If a list of selected line fluxes is used, e.g., for one element 
at a time, DEM-abundance cross talk can be avoided, and the influence of the atomic physics uncertainties can
be traced throughout the reconstruction process. But there may be a strong dependence of the 
reconstruction on the atomic physics uncertainties and the flux measurements of a few lines. The lack of a priori 
knowledge on line blends affecting the extracted line fluxes will introduce systematic uncertainties as well. 
This can, however, be improved if tabulated potential line blends are iteratively included.

The presence of systematic uncertainties also requires a careful and conservative choice
of {\it convergence criteria} or smoothness parameters to avoid introduction of spurious
features in the DEM. The result is {\it a range of solutions that acceptably describe the data} based
on a goodness-of-fit criterion, in so far as the data can be considered to be represented by the 
spectral database in use.
Within this allowed range, ``correctness'' cannot be judged on by purely statistical arguments. The spectral 
inversion is non-unique because the mathematical problem is ill-posed --
{\it the atomic data  deficiencies cannot be overcome by statistical methodology but require external 
information.}

Direct comparisons of various methods, applied to the {\it same} data,  are needed.
\citet{mewe01} presented an EMD  for Capella based on selected Fe lines that compares
very favorably with their EMD derived from a multi-$T$ approach for the complete spectra, and these results also 
seem to agree satisfactorily with previously published EMDs from various methods and various  data sets. 
EMDs of the active HR~1099 found from spectra of {\it XMM-Newton} RGS \citep{audard01a} and
from {\it Chandra} HETGS \citep{drake01} agree in their principal features, notwithstanding
the very different reconstruction methods applied and some discrepancies in the abundance determinations. 
\citet{telleschi04} determined
EMDs and abundances of a series of solar analogs at different activity levels from
polynomial-DEM fits to selected spectral regions and from
an iterative reconstruction by use of  extracted line-flux lists. The resulting EMDs
and the derived abundances of various elements are in good agreement. 
\citet{schmitt04}
compared two approaches within their polynomial DEM reconstruction method, again concluding that
the major discrepancies result from the uncertainties in the atomic physics, in particular when
results from EUV lines are compared with those from X-ray lines,
rather than from the reconstruction approach. A somewhat different conclusion was reported by
\citet{sanz03} from an iterative analysis of Fe-line fluxes of AB Dor; nevertheless, their abundance distribution 
is in fact quite similar to results reported by \citet{guedel01b} who fitted a complete spectrum. 

The evidence hitherto 
reported clearly locates the major obstacle not in the inversion method but in the incompleteness of, 
and the inaccuracies in, the atomic physics  tabulations. \citet {brickhouse95} gave a critical 
assessment of the current status of Fe line emissivities and their discrepancies in the EUV range, 
together with an analysis of solar and stellar spectra.
The effects of missing atomic transitions in the spectral codes were
demonstrated by \citet{brickhouse00} who particularly discussed the
case of Fe transitions from high $n$ quantum numbers, i.e., of transitions that have only
recently been considered.

\subsection{Observational results}\label{demresults}

\begin{figure} 
\vskip 0.4truecm
\centerline{\resizebox{0.75\textwidth}{!}{\includegraphics{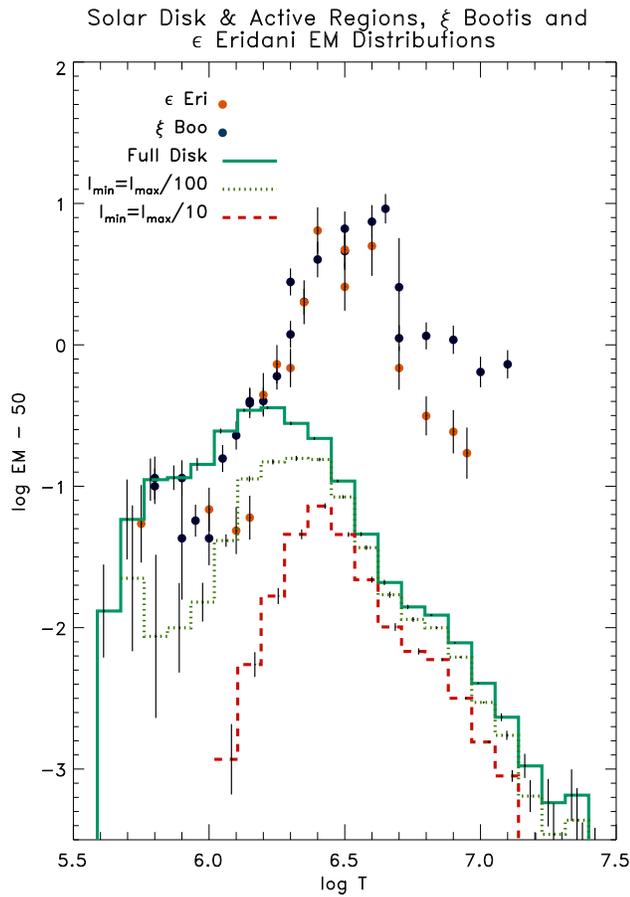}}}
\caption{Emission measure distributions  of two intermediately active stars and the Sun. The axes are logarithmic, 
with  the EM given in units of $10^{50}$~cm$^{-3}$, and the temperature in K. The blue and
red circles refer to  EMDs for $\xi$ Boo and $\epsilon$ Eri, 
respectively. The histograms refer to full-disk solar EMDs derived from {\it Yohkoh} images at solar maximum, 
including also two versions for different lower cutoffs for the intensities in {\it Yohkoh} images. Note the similar low-$T$  shapes
but the additional high-$T$ contributions in the stars that reveal EMDs similar to bright
active regions (figure courtesy of J. Drake, after \citealt{drake00}).}\label{demsolarstellar}
\end{figure}

If the caveats and the principal mathematical limitations of the present state of the art in 
the derivation of EMDs discussed above are properly taken into account, then the physical implications 
of some of the more secure results offer access to the underlying coronal physics.
Most EMDs have generally been found to be singly or doubly peaked 
\citep{mewe95, mewe96, mewe97, drake95b, drake97, rucinski95, schrijver95, kaastra96,
schmitt96b, guedel97a, guedel97b} and confined on either side approximately by power laws (e.g., 
\citealt{mathioudakis99, guedel03a}). Examples are shown in Figs.~\ref{calcdem} and \ref{demsolarstellar}, where
stellar and solar EMDs are compared. These power laws open up interesting ways of interpretation 
as discussed in Sect.~\ref{demstatic} and \ref{demflaring}.

It is notable that the complete EMD shifts 
to higher temperatures with increasing stellar activity (see also Sect.~\ref{templx}), often leaving very 
little EM at modest temperatures and correspondingly weak
spectral lines of C, N, and O \citep{kaastra96}.  
For example, \citet{haisch94a} found only
highly-ionized spectral lines in the EUV spectrum of the intermediately active
solar analog $\chi^1$ Ori (Fe\,{\sc xvi}  and higher) whereas the EUV spectrum of
Procyon and $\alpha$ Cen is dominated by lower ionization stages, corresponding to
a DEM peaking at 1--2 MK, similar to the full-disk, non-flaring solar DEM
\citep{mewe95, drake95b, drake97}.

Double peaks are often found for active stars. They then reveal 
a minimum in the range of 10--20~MK \citep{lemen89, mewe96, guedel97a, huenemoerder03}. 
Filling in this trough does not seem to lead to consistent solutions (e.g., \citealt{lemen89, sanz01}),
although  bias could be introduced by  inaccurate atomic physics. 
On the other hand, the two peaks may have a physical foundation.  Continual flaring has been proposed \citep{guedel97c},
or separate families of magnetic loops dominated by two different temperature regimes \citep{sciortino99}. 
In less active stars, the hotter peak disappears, leaving a marked 
single EMD maximum just below 10~MK \citep{dupree93, brickhouse98, sanz01, sanz02}.

EMDs are often steeper on the low-$T$ side than  
single, constant-cross section loop models (e.g., the static loop models by \citealt{rosner78}) predict, 
and this is particularly true for the more active stars (\citealt{pasquini89, 
schmitt90a, dempsey93b, laming96, laming99, drake00, sanz02, scelsi04, telleschi04}). For example, the 
cooler branches of the stellar EMDs in Fig.~\ref{demsolarstellar} follow approximately $Q \propto T^3$.  
One remedy may be loops with an expanding cross section from the base to the apex, as computed by
\citet{vesecky79}. In that case, 
there is comparatively more hot plasma, namely the plasma located around the loop apex,
than cooler plasma. The EMD and the DEM would consequently steepen. Spectral fits using individual  
loops require, in specific cases, expansion  factors $\Gamma$, i.e., the ratio between cross sectional areas at 
the apex and at the footpoints, of between 2 and 50. Still, this model
may fail for some sources \citep{schrijver89b, ottmann93, schrijver02}.
Alternatively, the steep low-$T$ EMD slopes may be further evidence for continual flaring 
\citep{guedel03a}; Eq.~(\ref{demflare}) predicts slopes between 1 and 4, similar to what 
is often found in magnetically active stars (see Sect.~\ref{obsstochflare}).
 
An extremely steep (slope of $3-5$) EMD has consistently been derived for Capella 
\citep{brickhouse95, audard01b, behar01, mewe01, argiroffi03}. Here, the EMD shows a sharp 
peak around 6~MK that dominates the overall spectrum. These EM results also agree well with 
previous analyses based on EUV spectroscopy \citep{brickhouse00}. 
 
\subsection{Coronal temperature-activity relations}\label{templx}

Early low-resolution spectra
from {\it HEAO 1}, {\it Einstein}, {\it EXOSAT},  and  {\it ROSAT} implied the persistent presence of 
considerable amounts of plasma at unexpectedly high temperatures, $T > 10$~MK (e.g., \citealt{walter78b, swank81}),
particularly in extremely active stars. 
This is now borne out by EMDs  of many active stars that show significant contributions  
up to 30--40~MK, if not higher. 

\begin{figure} 
\centerline{\resizebox{0.85\textwidth}{!}{\includegraphics{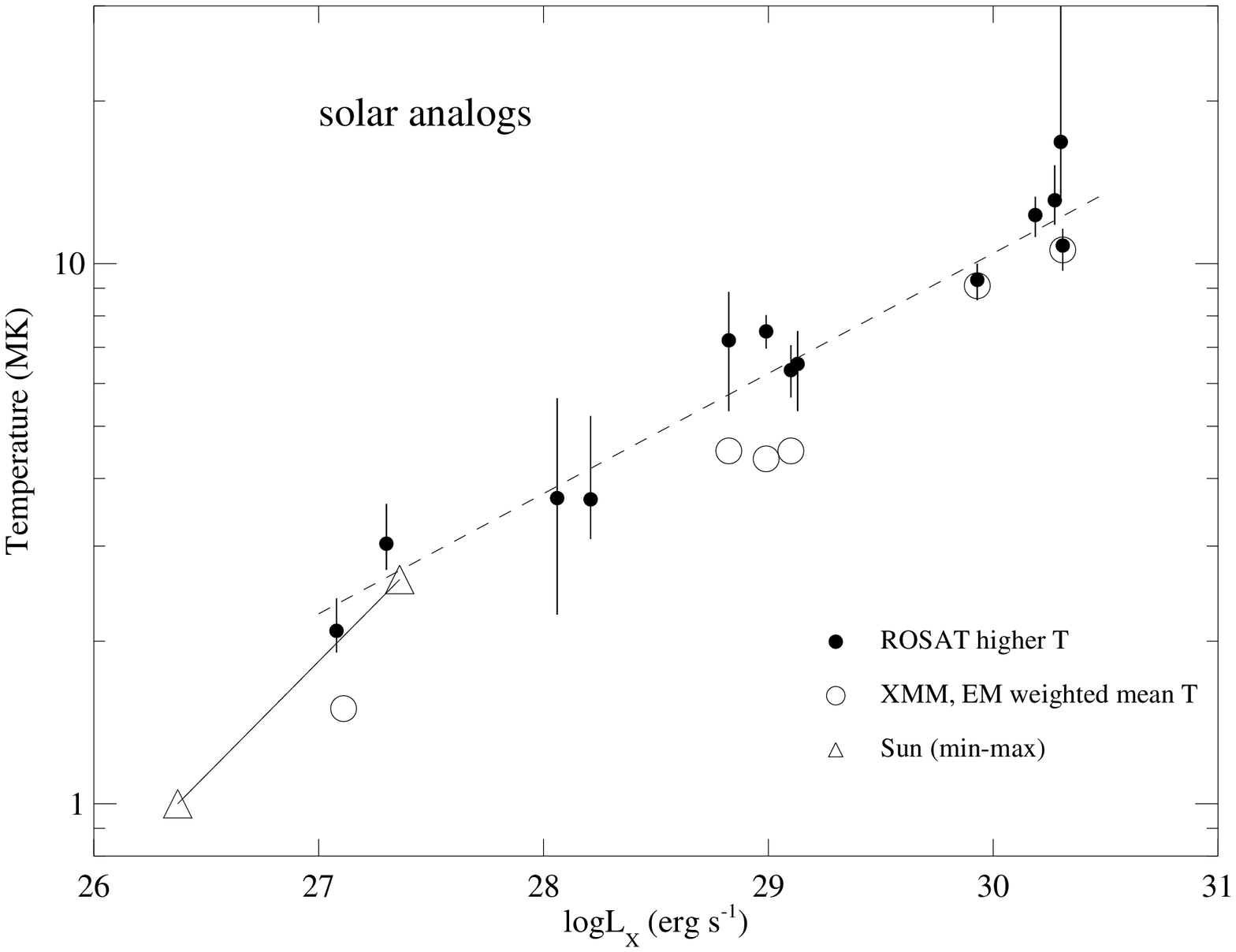}}}
\caption{Coronal temperature vs. X-ray luminosity for solar analogs. The filled circles are
from the sample of \citet{guedel97a} and \citet{guedel98} and refer to the hotter component in 2-$T$ fits
to {\it ROSAT} data. The solar points (triangles), covering the range from sunspot minimum to
maximum, were taken from \citet{peres00}. The open circles refer to EM-weighted averages
of log\,$T$; these values were  derived from full reconstructed EMDs  
\citep{telleschi04} and, for $\alpha$ Cen (at $L_X \approx 10^{27}$~erg~s$^{-1}$), 
from multi-$T$ fits \citep{raassen03a}. In both of the latter cases, {\it XMM-Newton} data
were used.} \label{temperaturelx}
\end{figure}

One of the surprising findings from stellar X-ray surveys is a relatively tight
correlation between the characteristic coronal temperature   
and the normalized coronal luminosity $L_X/L_{\rm bol}$:
{\it Stars at higher activity levels support hotter coronae} 
\citep{vaiana83, schrijver84, stern86, schmitt90a, dempsey93b, maggio94, gagne95a, schmitt95, huensch96b, 
guedel97a, preibisch97a, schmitt97, singh99}.  Instead of explicit temperature measurements, some authors used spectral 
hardness as a proxy for $T$, and the surface flux $F_X$ or the EM per unit area can be
used instead of $L_X/L_{\rm bol}$, but the conclusions remain the same. The example of solar analogs is shown
together with the Sun itself during its activity maximum and minimum in Fig.~\ref{temperaturelx}. Here,  
\begin{eqnarray}\label{TLx}
L_X \propto T^{4.5\pm 0.3} \\
EM  \propto T^{5.4\pm 0.6} 
\end{eqnarray}
where $L_X$ denotes  the total X-ray luminosity, but EM and $T$ refer to the ``hotter'' component in standard
2-$T$ fits to {\it ROSAT} data. Such relations extend further into the pre-main sequence domain where 
exceedingly hot coronae with temperatures up to  $\approx$100~MK are found \citep{imanishi01a}.

The cause of this relation is not clear. Three classes of models 
might apply: Phenomenologically, as the activity on a star increases, the corona becomes
progressively more dominated by hotter and denser features, for example active regions
as opposed to quiet areas or coronal holes. Consequently, the average stellar X-ray spectrum
indicates more hot plasma (\citealt{schrijver84, maggio94, guedel97a, preibisch97a, 
orlando00, peres00}; see Sect.~\ref{constituents}).

Increased magnetic activity also leads to more numerous interactions between adjacent magnetic field structures. 
The heating efficiency thus increases. In particular, we expect a higher rate of large flares. Such models naturally 
produce the temperature-activity correlation: The increased flare rate produces higher X-ray luminosity
because chromospheric evaporation produces more EM; at the same time,
the plasma is heated to higher temperatures in larger flares (as we will discuss in Sect.~\ref{flaretemp}). 
This mechanism was  simulated and discussed by \citet{guedel97a}, \citet{guedel97c}, and \citet{audard00} 
(see also Sect.~\ref{stochasticflares}).

\citet{jordan87} and \citet{jordan91} studied an EM-$T$ relation based on arguments of
a minimum energy loss configuration of the corona, assuming a fixed ratio between radiative losses
and the coronal conductive loss. They suggested a relation including the stellar gravity $g$ of the form
\begin{equation}\label{jordan}
EM \propto T^3g
\end{equation}
which fits quite well to a sample of observations with $T$ taken from single-$T$ fits to stellar coronal data.
The authors suggested that this relation holds because coronal heating directly relates to the production
rate of magnetic fields, and the magnetic pressure is assumed to scale with the thermal coronal pressure.
Equation~(\ref{jordan}) then follows directly.

\section{Electron densities in stellar coronae}\label{coronaldensities}

With the advent of high-resolution spectroscopy in the EUV range (by {\it EUVE} and {\it Chandra}) and
in X-rays (by {\it XMM-Newton} and {\it Chandra}), spectroscopic tools have become available to 
measure electron densities $n_e$ in coronae.
Coronal electron densities are important because they control radiative losses from the 
coronal plasma; observationally, they can in principle also be used in conjunction with EMs to derive approximate 
coronal source volumes. The spectroscopic derivation of coronal densities is subtle, however.  Two principal  methods  
are available.

\subsection{Densities from Fe line ratios}\label{fedensities}

The emissivities of many transitions of Fe ions in the EUV range 
are sensitive to densities  in the range of interest to  coronal research 
\citep{mewe85, mewe95, schmitt94a, schmitt96c, mathioudakis99}.
\citet{brickhouse95} have provided extensive tabulations of the relevant emissivities together 
with a critical review of their overall reliability. The different
density dependencies of different lines of the same Fe ion then also make their 
{\it line-flux ratios}, which (apart from blends) are easy to measure, useful 
diagnostics for the electron density. 

A review of the literature (see also \citealt{bowyer00}) shows a rather unexpected
segregation of coronal densities into two realms at different temperatures. The cool coronal plasma in 
inactive stars is typically found at low densities of order $10^9$~cm$^{-3}$ -- $10^{10}$~cm$^{-3}$.
In active stars, the cooler components may show elevated densities, but extreme values up to $>10^{13}$~cm$^{-3}$ 
have been reported for the hotter plasma component. A few notable examples follow.

For the inactive $\alpha$ Cen, \citet{mewe95} derived $n_e = 2\times 10^8$~cm$^{-3}$ -- $2\times 10^9$~cm$^{-3}$, 
in reasonable agreement with a measurement from {\it EUVE} \citep{drake97} and with typical solar coronal densities
\citep{landi98}. For the similarly inactive Procyon, 
\citet{schmitt94a, schmitt96c} and \citet{schrijver95} found somewhat higher values of approximately
$3\times 10^9$~cm$^{-3}$ -- $4\times 10^9$~cm$^{-3}$ from lines of Fe\,{\sc x--xiv}, with an allowed range
from $10^9$~cm$^{-3}$ to $10^{10}$~cm$^{-3}$.  

This picture does not  change much for intermediately active stars such as $\epsilon$ Eri
($n_e \approx 3\times 10^9$~cm$^{-3}$, \citealt{laming96, schmitt96b}), or
$\xi$ Boo A ($n_e \ga 10^{10}$~cm$^{-3}$, \citealt{laming99}). In general, the 
more active stars tend to show  somewhat higher densities.

An appreciable change comes with higher activity levels when we consider the hotter plasma. Some extremely 
high densities have been reported from line ratios of highly ionized Fe, for example:
Capella ($n_e = [0.04-1.5]\times 10^{13}$~cm$^{-3}$ from Fe\,{\sc xix-xxii}; \citealt{dupree93, schrijver95}),  
UX Ari ($n_e = [4.5\pm 2]\times 10^{12}$~cm$^{-3}$ from Fe\,{\sc xxi}; \citealt{guedel99}, also \citealt{sanz02}), 
AU Mic ($n_e = [2-5]\times 10^{12}$~cm$^{-3}$ from Fe\,{\sc xxi-xxii}; \citealt{schrijver95}), 
$\sigma$ Gem ($n_e  \approx 10^{12}$~cm$^{-3}$ from Fe\,{\sc xxi-xxii}; \citealt{schrijver95}, also \citealt{sanz02}), 
$\xi$ UMa B ($n_e = 5\times 10^{12}$~cm$^{-3}$ from Fe\,{\sc xxi-xxii}; \citealt{schrijver95}), 
44i Boo ($n_e > 10^{13}$~cm$^{-3}$ from Fe\,{\sc xix}; \citealt{brickhouse98}), 
HR~1099 ($n_e = [1.6\pm 0.4]\times 10^{12}$~cm$^{-3}$ from Fe\,{\sc xix, xxi, xxii}; \citealt{sanz02}; somewhat higher values in 
                                                                                     \citealt{osten04}), 
II Peg ($n_e = [2.5-20]\times 10^{12}$~cm$^{-3}$ from Fe\,{\sc xxi} and \,{\sc xxii}; \citealt{sanz02}), 
AB Dor ($n_e = [2-20] \times 10^{12}$~cm$^{-3}$ from Fe\,{\sc xx-xxii}; \citealt{sanz02}), 
$\beta$ Cet ($n_e = [2.5-16] \times 10^{11}$~cm$^{-3}$ from Fe\,{\sc xxi}; \citealt{sanz02}), 
and several short-period active binaries
($n_e \approx [3-10]\times 10^{12}$~cm$^{-3}$ from Fe\,{\sc xxi}, \citealt{osten02}).
However, most of these densities are only slightly above the low-density limit, and upper limits 
have equally been reported (for $\sigma^2$ CrB, $n_e \la 10^{12}$~cm$^{-3}$ from Fe\,{\sc xxi}, 
\citealt{osten00, osten03}; for HD 35850, $n_e \le [4-50]\times 10^{11}$~cm$^{-3}$ from Fe\,{\sc xxi}, 
\citealt{mathioudakis99, gagne99}).  

\citet{mewe01} used several line ratios of Fe\,{\sc xx, xxi}, and {\sc xxii} in {\it Chandra} data
of Capella, reporting $n_e \la (2-5)\times 
10^{12}$~cm$^{-3}$, in mild contradiction with {\it EUVE} reports cited above. More stringent upper 
limits of log\,$n_e = 11.52$ were obtained for Algol from Fe\,{\sc xxi} EUV line ratios \citep{ness02b}.
From a detailed consideration of Fe\,{\sc xxi}  line ratios in the {\it Chandra} HETG spectrum of 
Capella, \citet{phillips01} even concluded that the density measured by the most reliable 
Fe\,{\sc xxi} line ratio, $f(\lambda 102.22)/f(\lambda 128.74)$, is compatible with the low-density limit of 
this diagnostic (i.e., $n_e \la 10^{12}$~cm$^{-3}$); in fact, these authors re-visited 
previous measurements of the same ratio and suggested that they {\it all} represent the low-density 
limit for Capella. Likewise, \citet{ayres01a} find contradictory results from different density 
indicators in the EUV spectrum of $\beta$ Cet, and suggest low densities. Finally, \citet{ness04}
measured various Fe line ratios in a large sample of coronal stars. None of the stars showed
high densities from all line ratios, and all values were again close to
the low-density limit; moreover, a given line flux ratio appears to be identical for all considered stars,
within the uncertainties. Because it is unlikely that all coronae reveal the same densities, a more
natural assumption is that all measurements represent the low-density limit.

The observational situation is clearly unsatisfactory at the time of  writing. It is worrisome that
most measurements referring to the hotter plasma straddle the low-density limit of the respective ion but
tend to be systematically different for ionization stages that have similar formation temperatures.
At face value,  it is perhaps little surprising that the densities  do not come out even
higher, but this circumstance makes the measurements extremely vulnerable to systematic but
unrecognized inaccuracies in the atomic physics tabulations, and to unrecognized blends in some of
the lines. Slight shifts then have a dramatic effect on the implied densities, as can be nicely seen in the 
analysis presented by \citet{phillips01}.  The resolution of these contradictions requires
a careful reconsideration of atomic physics issues.

\subsection{Line ratios of He-like ions}\label{denshe}

\begin{figure} 
\hbox{
\resizebox{0.45\textwidth}{!}{\includegraphics{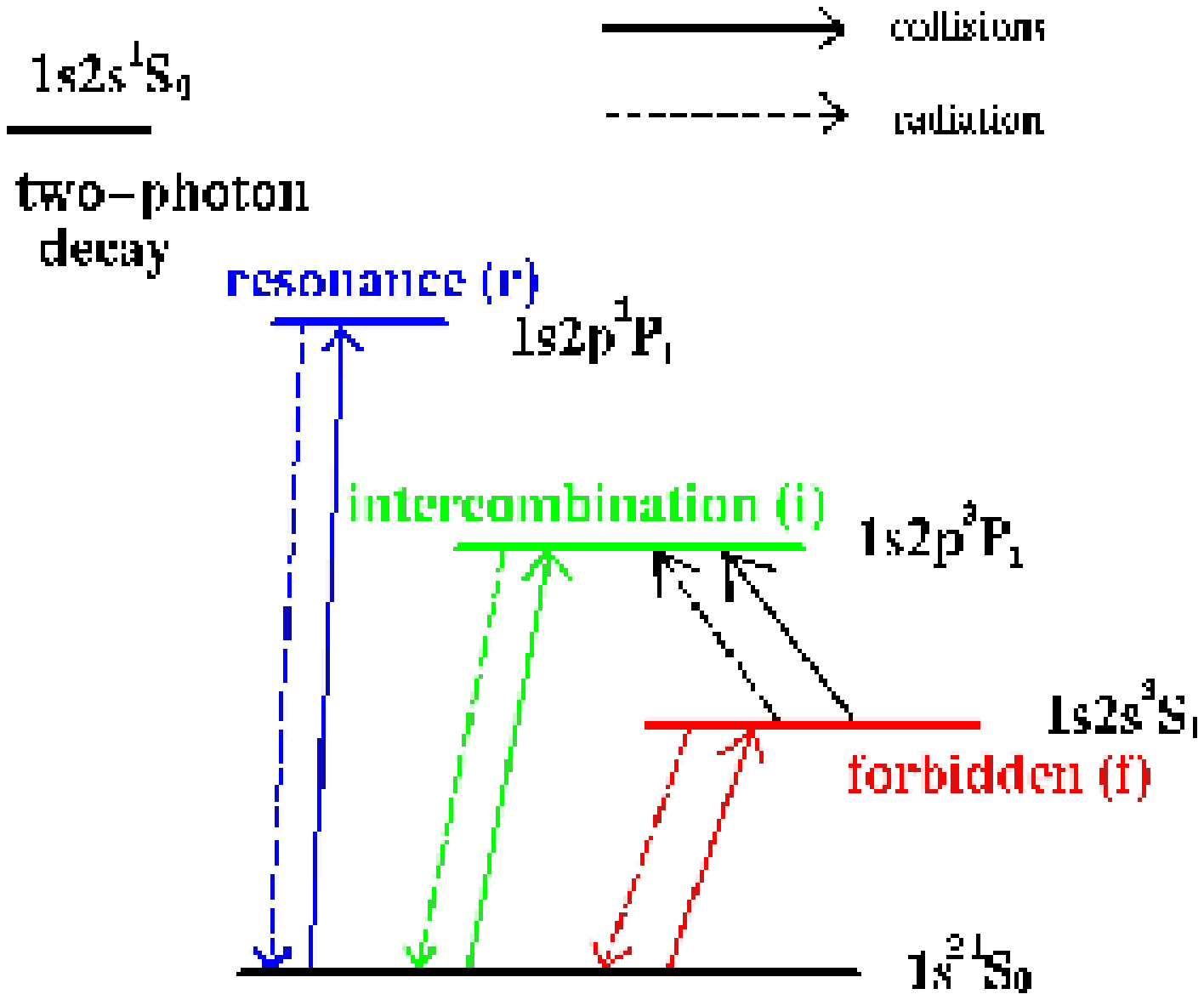}}
\resizebox{0.55\textwidth}{!}{\includegraphics{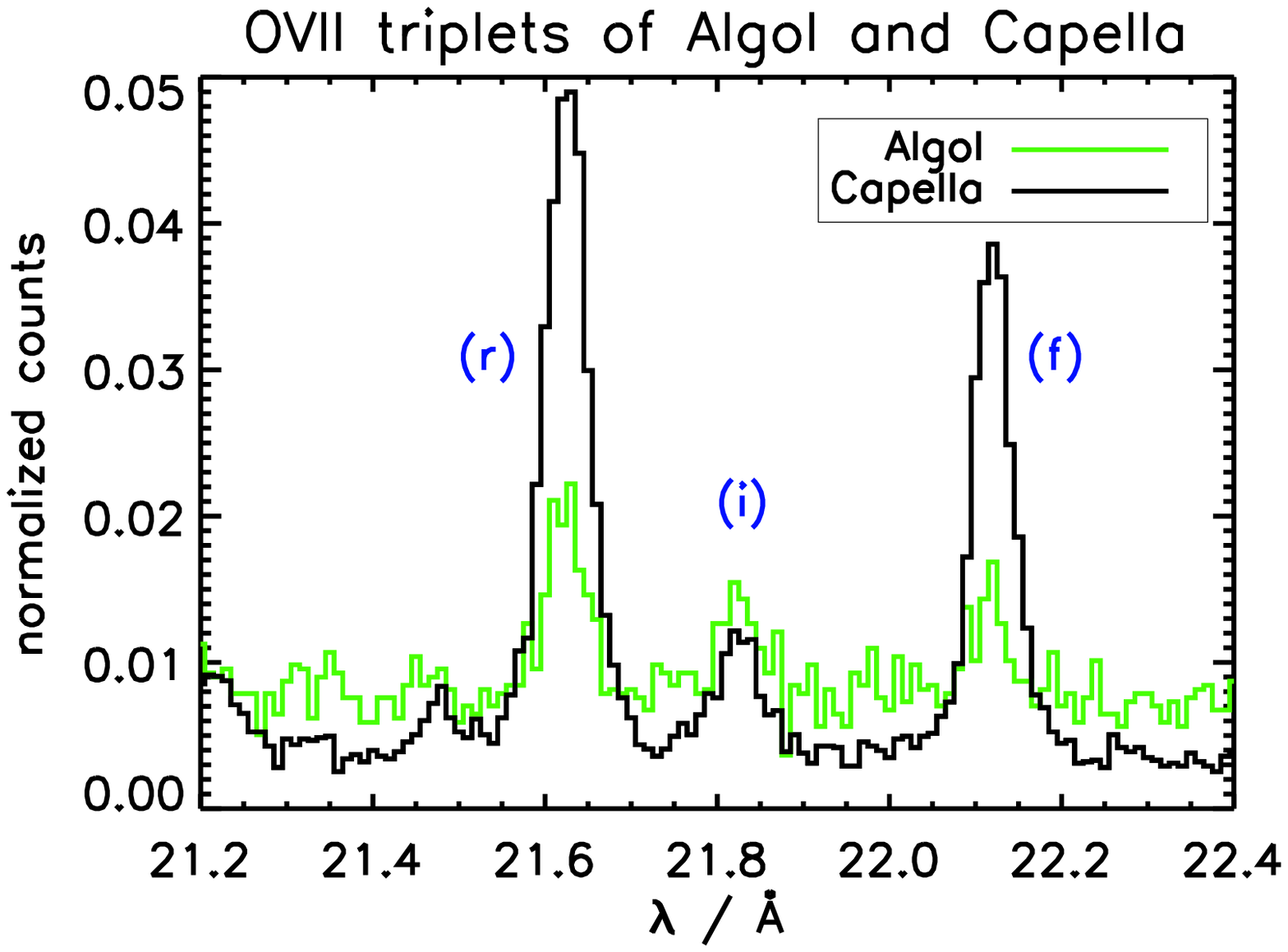}}
}
\vskip 0.4truecm\caption{{\it Left:} Term diagram for transitions in He-like triplets. The resonance, intercombination,
       and forbidden transitions are marked. The transition from $^3S_1$ to $^3P_1$ re-distributes electrons from the
       upper level of the forbidden transition to the upper level of the intercombination transition, thus making 
       the $f/i$ line-flux ratio density sensitive. In the presence of a strong UV field, however, the same 
       transition can be induced by radiation as well.
 {\it Right:} He-like triplet of O\,{\sc vii} for Capella (black) and Algol (green).
       The resonance (r), intercombination (i), and forbidden (f) lines are marked.
       The $f/i$ flux ratio of Algol is suppressed probably due to the strong UV radiation field
       of the primary B star (data from {\it Chandra}; both figures courtesy of J.-U. Ness).}\label{oxygentriplet}
\end{figure}

The He-like triplets of C\,{\sc v}, N\,{\sc vi}, O\,{\sc vii}, Ne\,{\sc ix}, Mg\,{\sc xi}, and Si\,{\sc xiii}  provide another interesting
density diagnostic for stellar coronae. Two examples are shown in Fig.~\ref{oxygentriplet}. The spectra show,
in order of increasing wavelength, the resonance, the intercombination, and the forbidden line of the O\,{\sc vii} triplet.
The ratio between the fluxes in the forbidden line and the intercombination line is sensitive to density
\citep{gabriel69} for the following reason: if the electron collision rate is sufficiently high, ions in the upper 
level of the forbidden transition, $1s2s\ ^3S_1$, do not return to the ground level, $1s^2\ ^1S_0$, instead the ions
are collisionally excited to the upper levels of the intercombination transitions, $1s2p\ ^3P_{1,2}$, from where 
they decay radiatively to the ground state (see Fig.~\ref{oxygentriplet} for a term diagram).
They thus enhance the flux  in the intercombination line
and weaken the flux in the forbidden line. The measured ratio $R = f/i$ of the forbidden to the intercombination 
line flux can be written as
\begin{equation}
R = {R_0 \over 1 + n_e/N_c} = {f\over i}
\end{equation}
where $R_0$ is the limiting flux ratio at low densities and $N_c$ is the critical
density at which $R$ drops to $R_0/2$. For C\,{\sc v} and N\,{\sc vi}, the photospheric radiation field needs
to be considered as well because it enhances the $^3S_1 - {^3}P_{1,2}$ transitions; the same
applies to higher-$Z$ triplets if a hotter star illuminates the X-ray source (see \citealt{ness01}). 
The tabulated parameters $R_0$ and $N_c$ are slightly dependent on the electron temperature in
the emitting source; this average temperature can conveniently  be confined by the temperature-sensitive  
$G$ ratio of the same lines, $G = (i+f)/r$ 
(here, $r$ is the flux in the resonance line $1s^2\ ^1S_0 - 1s2p\ ^1P_1$). A recent 
comprehensive tabulation is given in \citet{porquet01}; Table~\ref{lineratios} contains
relevant parameters for the case of a plasma that is at the maximum formation temperature of the
respective ion. A systematic problem with He-like triplets 
is that the critical density $N_c$ increases with the formation temperature of 
the ion, i.e., higher-$Z$ ions measure only high densities at high $T$, while
the lower-density analysis based on C\,{\sc v}, N\,{\sc vi}, O\,{\sc vii}, and Ne\,{\sc ix} is 
applicable only to cool plasma.

Stellar coronal He-like triplets have become popular with {\it Chandra} and {\it XMM-Newt\-on}.
For C\,{\sc v}, N\,{\sc vi}, and O\,{\sc vii}, early reports indicated densities either around or below the low density limit,
viz. $n_e \approx 10^9-10^{10}$~cm$^{-3}$ for Capella 
\citep{brinkman00, canizares00, audard01b, mewe01, ness01, phillips01}, 
$\alpha$ Cen \citep{raassen03a}, Procyon \citep{ness01, raassen02}, HR~1099 \citep{audard01a}, 
and II Peg \citep{huenemoerder01}. Note that this sample covers an appreciable range of activity. (A conflicting, higher-density 
measurement, $n_e \approx [2-3]\times 10^{10}$~cm$^{-3}$, was given by \citealt{ayres01b} for HR1099 and Capella). 
A low-density limit (log\,$n_e < 10.2$) is found for Capella also from the
Ne\,{\sc ix} triplet \citep{ness03b}. First unequivocal reports on significant, higher densities measured in O\,{\sc vii} came for 
very active main-sequence stars such as AB Dor \citep{guedel01b} and YY Gem \citep{guedel01a}, indicating
$n_e$ of several times $10^{10}$~cm$^{-3}$. The trend for higher densities in more active main-sequence stars is
consistently found across various spectral types  \citep{ness02a, raassen03b, besselaar03}, whereas active binaries 
may reveal either high or low  densities (\citealt{ness02b, huenemoerder01, huenemoerder03}), and low-activity stars
generally show low densities \citep{ness02a}. The most recent, comprehensive compilation of these trends can be 
found in \citet{ness04} who surveyed  O\,{\sc vii} and  Ne\,{\sc ix} triplets of a sample of 42 stellar systems across all levels 
of magnetic activity, and in \citet{testa04} who studied a sample of 22 stars with {\it Chandra}.

\begin{table}
\caption{Density-sensitive He-like triplets$^a$}
\label{lineratios}     
\begin{tabular}{llllll}
\hline\noalign{\smallskip}
Ion       &  $\lambda(r,i,f)$ (\AA)&   $R_0$          & $N_c$              & log\,$n_e$ range$^b$  &   $T$ range$^c$ (MK) \\
\noalign{\smallskip}\hline\noalign{\smallskip}
C\,{\sc v}       & 40.28/40.71/41.46      &  11.4            & $6\times 10^8$     &  7.7--10         &   0.5--2      \\
N\,{\sc vi}      & 28.79/29.07/29.53      &   5.3            & $5.3\times 10^9$   &  8.7--10.7       &   0.7--3      \\
O\,{\sc vii}     & 21.60/21.80/22.10      &   3.74           & $3.5\times 10^{10}$&  9.5--11.5       &   1.0--4.0    \\ 
Ne\,{\sc ix}     & 13.45/13.55/13.70      &   3.08           & $8.3\times 10^{11}$&  11.0--13.0      &   2.0--8.0    \\ 
Mg\,{\sc xi}     & 9.17/9.23/9.31         &   2.66$^d$       & $1.0\times 10^{13}$&  12.0--14.0      &   3.3--13     \\
Si\,{\sc xiii}   & 6.65/6.68/6.74         &   2.33$^d$       & $8.6\times 10^{13}$&  13.0--15.0      &   5.0--20     \\
\noalign{\smallskip}\hline
\multicolumn{6}{l}{$^a$data derived from \citet{porquet01} at maximum formation temperature of ion} \\
\multicolumn{6}{l}{$^b$range where $R$ is within approximately [0.1,0.9] times $R_0$ } \\
\multicolumn{6}{l}{$^c$range of 0.5--2 times maximum formation temperature of ion.} \\
\multicolumn{6}{l}{$^d$for  measurement with {\it Chandra} HETGS-MEG spectral resolution} \\
\end{tabular}
\end{table}

As for higher-$Z$ He-like triplets, reports become quite ambiguous, echoing both the results and the problems 
encountered in the analysis of Fe lines. \citet{mewe01} found
$n_e = 3\times 10^{12}$~cm$^{-3}$ -- $3\times 10^{13}$~cm$^{-3}$ in Capella from Mg\,{\sc xi} and Si\,{\sc xiii}  as measured by the {\it Chandra} LETGS. 
These high values agree with {\it EUVE} measurements (e.g., \citealt{dupree93}), but they contradict simultaneous 
{\it Chandra} measurements obtained from Fe\,{\sc xx-xxii} \citep{mewe01}. 
\citet{osten03} derived densities from He-like triplets,  Fe\,{\sc xxi} and Fe\,{\sc xxii} line ratios over a 
temperature range of $\approx 1-15$~MK. They found a sharply increasing trend: densities from
lines formed below 6~MK point at a modest electron density of a few times $10^{10}$~cm$^{-3}$, while those
formed above indicate densities exceeding $10^{11}$~cm$^{-3}$, possibly reaching up to a few times $10^{12}$~cm$^{-3}$.
Somewhat perplexingly, though, the  Si\,{\sc xiii}  triplet that is formed at similar  temperatures as 
Mg\,{\sc xi} suggests $n_e < 10^{11}$~cm$^{-3}$, and discrepancies of up to an order of magnitude become evident depending 
on the adopted formation temperature of the respective ion. The trend for an excessively high density implied by Mg can also be 
seen in the  analysis of Capella by \citet{audard01b} and \citet{argiroffi03}. 

Clearly, a careful reconsideration of line blends is in order. \citet{testa04} have measured Mg\,{\sc xi} densities in a large
stellar sample after modeling blends from Ne and Fe, still finding densities up to a few times $10^{12}$~cm$^{-3}$ but
not reaching beyond $10^{13}$~cm$^{-3}$. All measurements from Si\,{\sc xiii} imply an upper limit $\approx 10^{13}$~cm$^{-3}$,
casting some doubt on such densities derived from EUV Fe lines (see above). A trend similar to results from O\,{\sc vii} is
found again, namely that more active stars tend to reveal higher overall densities.

\begin{figure} 
\centerline{\resizebox{0.8\textwidth}{!}{\rotatebox{90}{\includegraphics{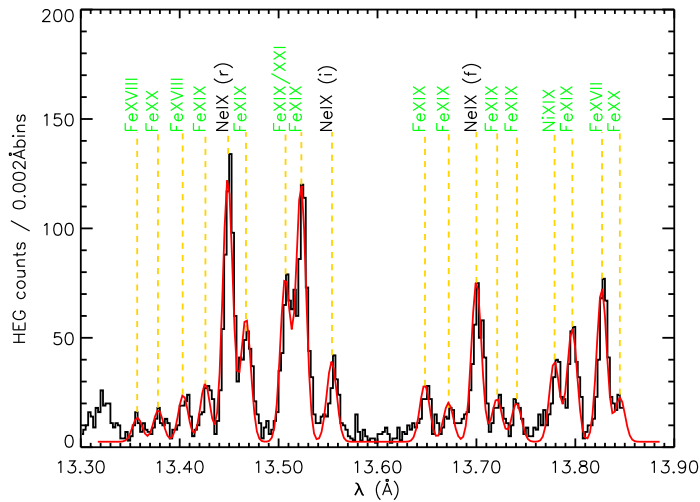}}}}
\caption{The spectral region around the Ne\,{\sc ix} triplet, showing a large number of Fe lines, some of them
   potentially blending the Ne lines if the resolving power is smaller than shown here
     (data from {\it Chandra} HETGS; the smooth red line shows a fit based on Gaussian line components; 
     figure courtesy of J.-U. Ness).}\label{neix}
\end{figure}

In the case of Ne, the problematic situation with regard to line blends is illustrated in Fig.~\ref{neix}.
If the density trend described above is real, however, then coronal loop pressures should vary by 3--4 
orders of magnitude.  This obviously requires different magnetic loop systems for the different pressure 
regimes, with a tendency that hotter plasma occupies progressively smaller volumes \citep{osten03, argiroffi03}.

Contrasting results have been reported, however. \citet{canizares00}, \citet{ayres01b}, and \citet{phillips01} found, from the 
{\it Chandra} HETG spectrum  of Capella, densities at, or below the low-density limit for Ne, Mg, and Si. A similar
result applies to II Peg \citep{huenemoerder01}.

A summary of the present status of coronal density measurements necessarily remains tentative.
Densities measured from Mg\,{\sc xi} and Si\,{\sc xiii}  may  differ greatly: despite their similar 
formation temperatures, Mg often results in very high densities, possibly induced by blends (see \citealt{testa04}); 
there are also discrepancies between densities derived from lines of He-like ions 
and from Fe\,{\sc xxi} and Fe\,{\sc xxii}  line ratios,  again for similar temperature ranges; and 
there is, lastly, disagreement between various authors who have used data from
different instruments. The agreement is better for the cooler plasma components measured with 
C\,{\sc v}, N\,{\sc vi}, and O\,{\sc vii}.
There, inactive stars generally show $n_e < 10^{10}$~cm$^{-3}$, whereas densities of active stars may
reach several times $10^{10}$~cm$^{-3}$, values that are in fact in good agreement with measurements
based on eclipses or rotational modulation (Sect.~\ref{coronalstructure}).

\subsection{Spectroscopic density measurements for inhomogeneous coronae}

Density measurements as discussed above have often been treated as physical parameters of an emitting 
source. However, because X-ray coronae are inhomogeneous, spectroscopic density measurements should 
not be taken at face value but should be further interpreted based on {\it statistical} models and
distributions of coronal features. In spectroradiometry parlance, the observed {\it irradiance}, also loosely called 
``flux'', from the stellar corona derives from the sum of the {\it radiances} over the entire visible coronal volume. 
Let us assume, for the sake  of argument, an isothermal 
plasma around the maximum-formation temperature, $T$. Let us assume, further, that the emitting volumes, $V$, are
distributed in electron density as a power-law, $dV/dn_e \propto n_e^{-\beta}$. We expect $\beta$ to be positive, 
i.e., low-density plasma occupies a large volume and high-density plasma is concentrated 
in small volumes. The densities inferred from the irradiance line ratios $R_{\mathrm{obs}}$ of He-like ions 
are then biased toward the highest densities occurring in reasonably large volumes because of the $n_e^2$ 
dependence of the luminosity.

\begin{figure} 
\centerline{\resizebox{0.7\textwidth}{!}{\includegraphics{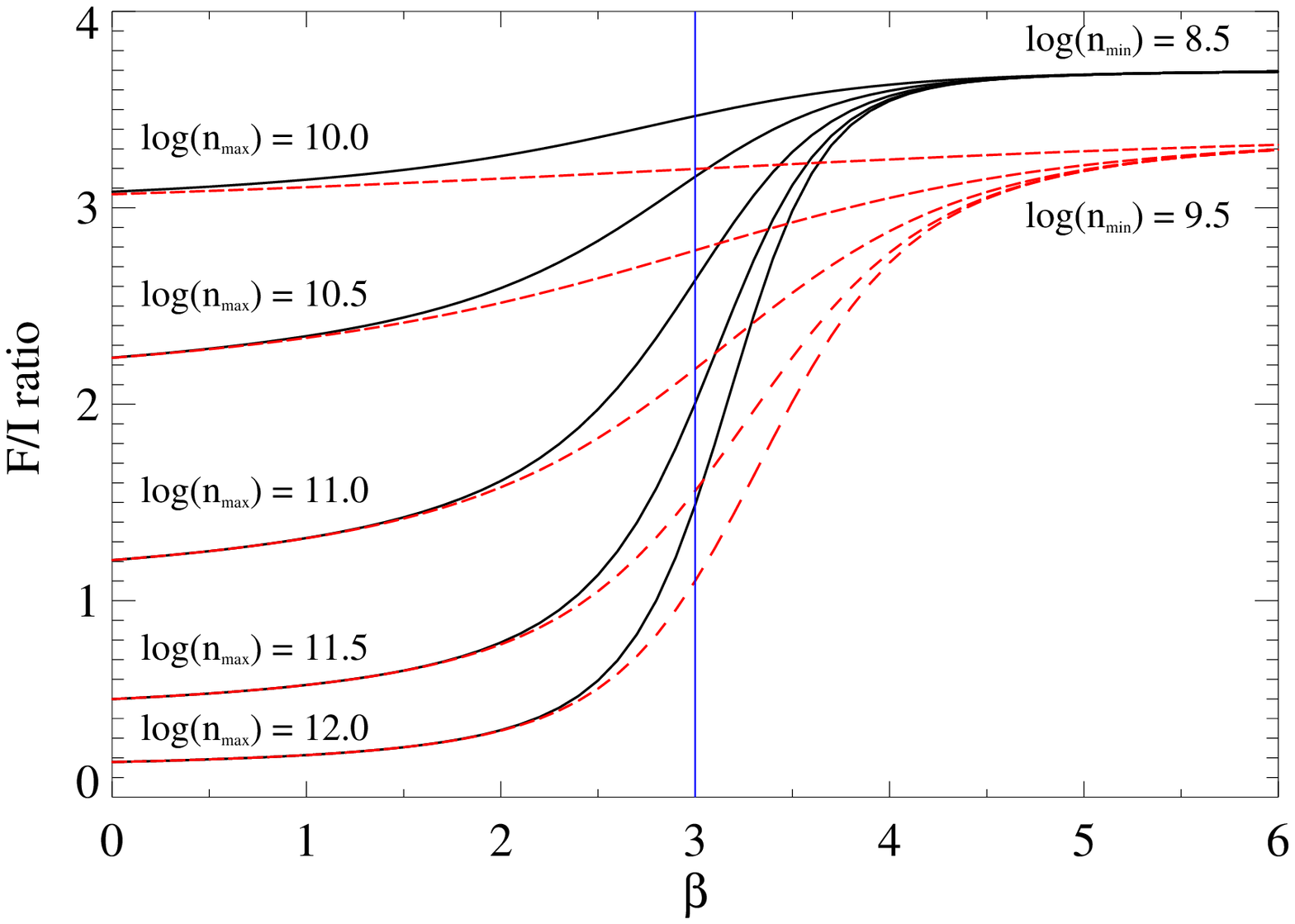}}}
\centerline{\resizebox{0.7\textwidth}{!}{\includegraphics{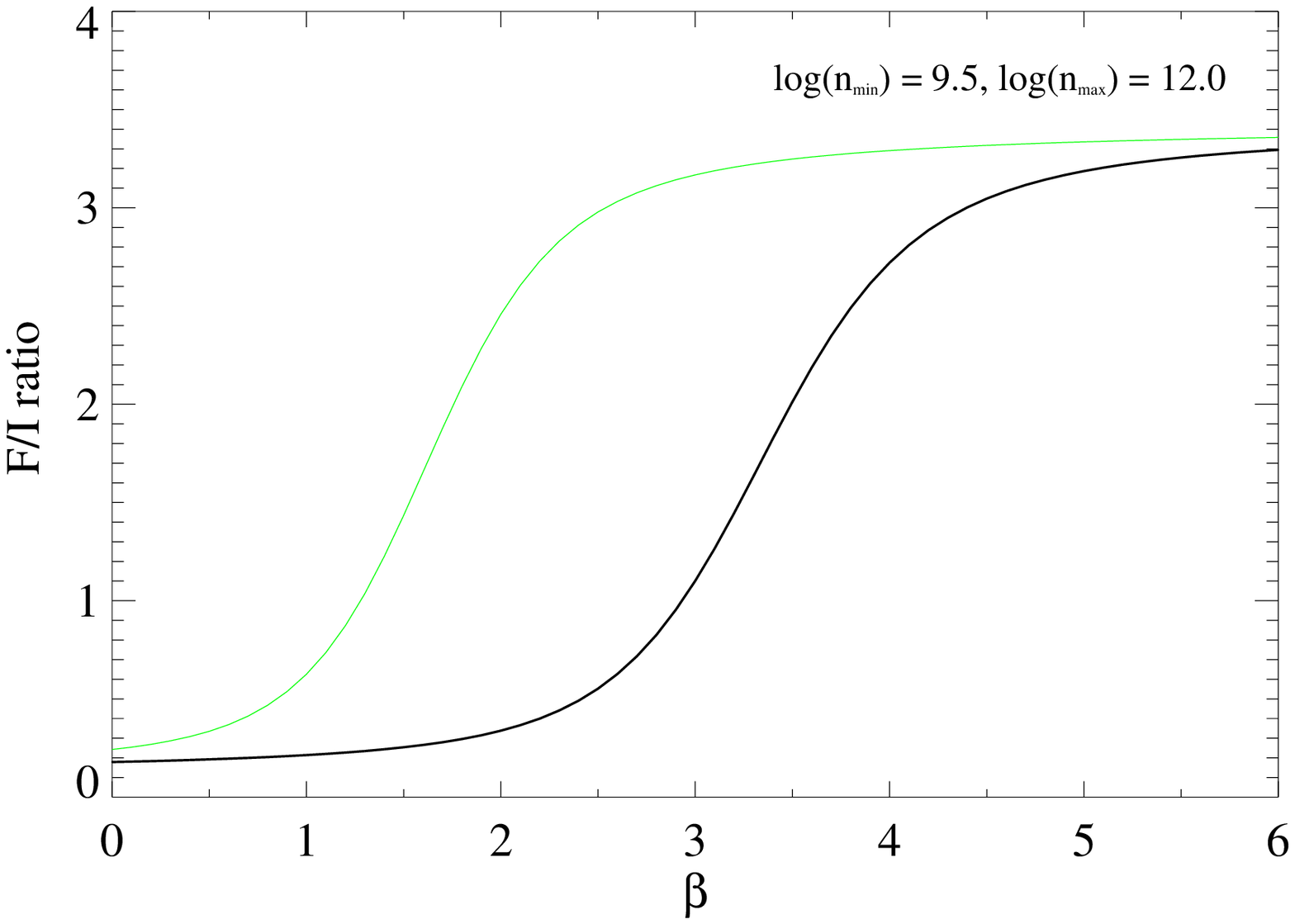}}}
\caption{{\it Top:} Calculated $R_{\mathrm{obs}} = F/I$ irradiance ratios for the He-like 
O\,{\sc vii} line triplet originating in an inhomogeneous corona with a radiance distribution 
resulting from a power-law volume-density distribution (see text for details). The $R$
ratio is plotted as a function of the power-law index $\beta$.
The different curves are for different ranges of densities at which
plasma is assumed to exist; the upper cutoff of
the adopted range is given by the labels at left, while labels at right
give the lower cutoff. The vertical line marks the transition from
the low-$R$ to the high-$R$ regime at $\beta = 3$. - {\it Bottom:} 
Example illustrating the $R_{\mathrm{obs}} = F/I$ irradiance ratio as a function of
$\beta$ (black line), compared with the $R$ ratio that would correspond to
an unweighted average of the densities of all volume elements considered (green). 
Atomic data from \citet{porquet01} were  used.
}\label{densinhom}
\end{figure}

The {\it observed} forbidden and intercombination line irradiances $F$ and $I$ are the 
contributions of the radiances $f$ and $i$, respectively, integrated over the distribution $dV/dn_e$ and corrected
for the stellar distance. The 
resulting observed $R_{\mathrm{obs}} = F/I$ ratios are plotted in 
Fig.~\ref{densinhom}a as a function of $\beta$ for reasonable ranges of coronal 
densities $[n_{e,{\rm min}}, n_{e, {\rm max}}]$ considered for the power-law volume distribution. 
The transition from low to high $R$ occurs around $\beta = 3$, which is a direct consequence 
of the $n_e^2$ factor under the integral sign.
A slight rearrangement of the distribution of active regions may thus
dramatically change $R_{\mathrm{obs}}$. Further, again due to the $n_e^2$ dependence
of the line flux, the ``inferred density'' is {\it not} an average over
the coronal volume. In Fig.~\ref{densinhom}b, the spectroscopically measured $R$ ratios 
from this distribution are compared with  the ratio that the linearly  {\it averaged} density itself would 
produce (for one example). It illustrates that the inferred densities are considerably biased to higher 
values, that is, smaller $R$ ratios, for a given volume distribution (or $\beta$).

In  a {\it global} coronal picture like that above, then, the $R_{\mathrm{obs}}$ 
values do not describe ``densities'' but the {\it steepness of the density distribution},
and a straightforward interpretation of densities and volumes from the singular
spectroscopic values can be misleading. The line ratio instead contains interesting information
on the distribution of coronal densities. More active MS
stars thus appear to maintain flatter density distributions, which could
be a consequence of the finite coronal volume: since the latter
must be shared by low- and high-density plasmas,
the more efficient heating in more active stars (e.g., the increased production 
of chromospheric evaporation)  produces a larger amount of high-density
volume, at the cost of the residual low-density volume, thus flattening 
the distribution and decreasing $R$. Alternatively, as Fig.~\ref{densinhom}a illustrates,
the cutoffs of the density distribution could be shifted to higher values for more active
stars while the slope $\beta$ remains similar. The interpretation of coronal density 
measurements thus naturally connects to coronal structure, which is the subject of the following section.
 
\section{The structure of stellar coronae}\label{coronalstructure}

The magnetic structure of stellar coronae is one of the central topics in our
research discipline. The extent and predominant locations of magnetic structures
currently hold the key to our understanding of the internal magnetic dynamo. For example,
compact or extended coronae may argue for or against the presence of a distributed dynamo.
All X-ray inferences of coronal structure in stars other than the Sun are so 
far indirect. This section describes various methods to infer 
structure in stellar coronae at X-ray wavelengths, and reviews the results thus obtained.

\subsection{Loop models}

\begin{figure} 
\centerline{
\hbox{
\resizebox{0.5\textwidth}{!}{\includegraphics{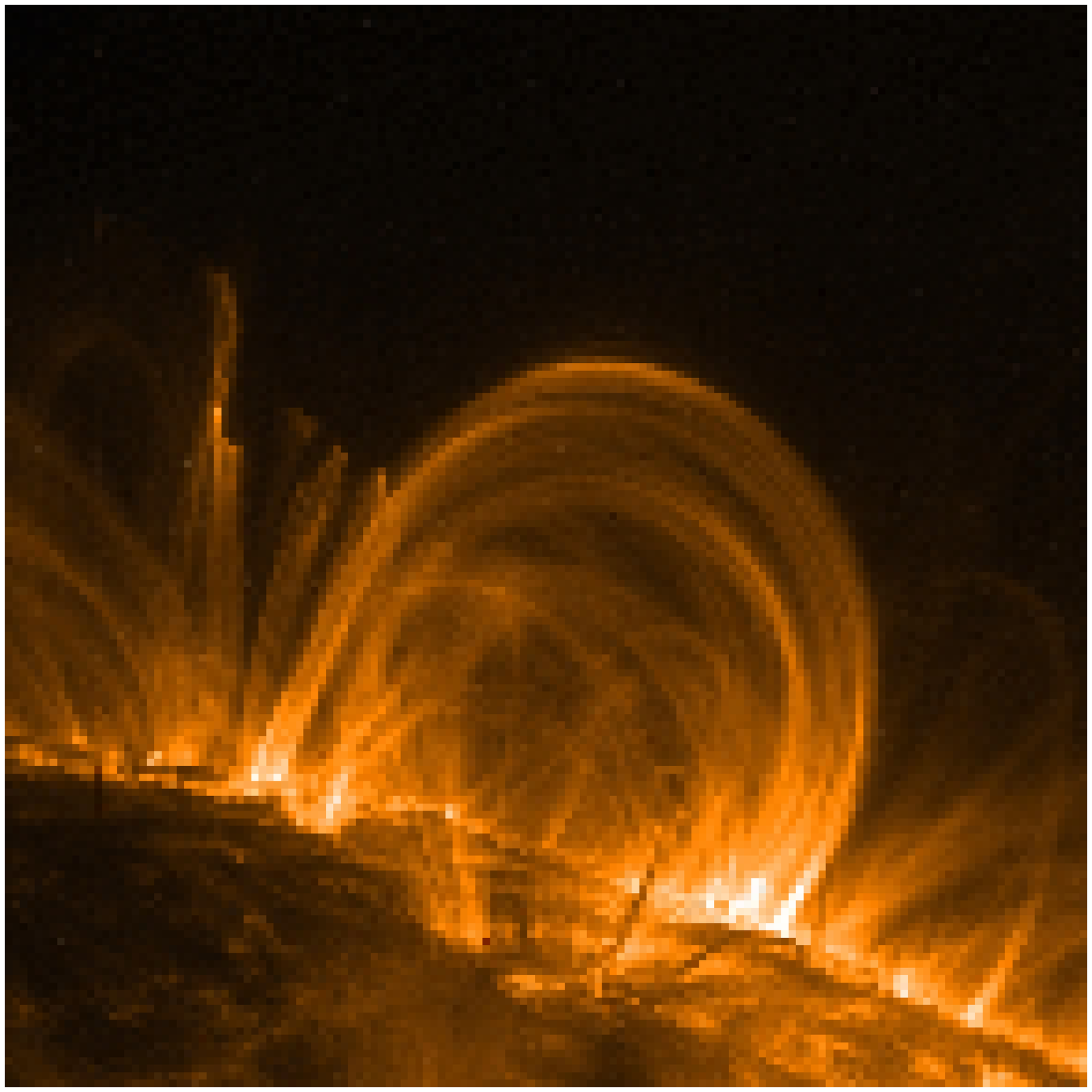}}
\resizebox{0.5\textwidth}{!}{\includegraphics{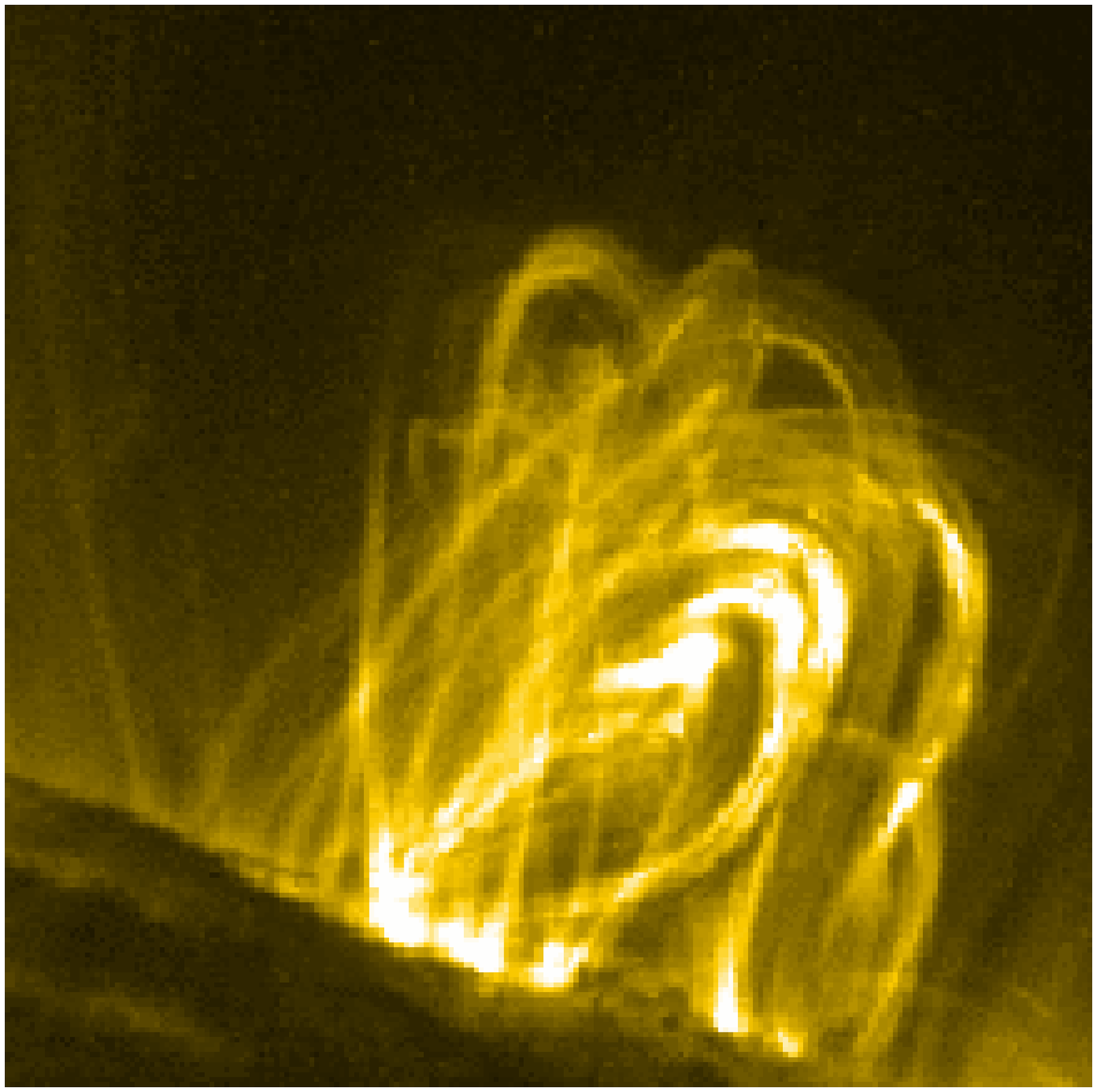}}
}
}
\caption{{\it Left:} Example of a solar coronal loop system observed by {\it TRACE}. {\it Right:}
Flaring loop system (observation by {\it TRACE} at 171\AA). Although these images show the emission from
relatively cool coronal plasma, they illustrate the possible complexity of magnetic fields.}\label{traceloop}
\end{figure}

Closed magnetic loops are the fundamental ``building blocks'' of the solar corona. When interpreting stellar
coronae of any kind, we assume that this concept applies as well, although caution is
in order. Even in the solar case, loops come in a wide variety of shapes and sizes (Fig.~\ref{traceloop}) and appear to imply heating 
mechanisms and heating locations that are poorly understood - see, for example, \citet{aschwanden00a}. Nevertheless, 
simplified loop models offer an important starting point for coronal structure studies and possibly for coronal heating diagnostics.
A short summary of some elementary properties follows.

Under certain simplifying assumptions, loop scaling laws can be derived. These have been widely applied to stellar coronae.
\citet{rosner78} (= RTV) have modeled hydrostatic loops with
constant pressure (i.e., the loop height is smaller than the pressure scale height). They also assumed absence of gravity, 
constant cross section, and uniform heating, and found two scaling laws relating the loop semi-length $L$ (in cm),
the volumetric heating rate $\epsilon$ (in erg~cm$^{-3}$s$^{-1}$), the electron pressure $p$ (in dynes~cm$^{-2}$), 
and the loop apex temperature  $T_{\rm a}$ (in K),
\begin{equation}\label{RTV}
T_{\rm a} = 1400(pL)^{1/3}; \quad\quad \epsilon = 9.8\times 10^4p^{7/6}L^{-5/6}. 
\end{equation}
\citet{serio81} extended these scaling laws to loops exceeding the pressure scale height $s_p$, whereby, however,
the limiting height at which the loops grow unstable is $(2-3)s_p$:
\begin{equation}
T_{\rm a} = 1400(pL)^{1/3}e^{-0.04L(2/s_H + 1/s_p)}; \quad\quad \epsilon = 10^5p^{7/6}L^{-5/6}e^{0.5L(1/s_H - 1/s_p)}.
\end{equation}
Here, $s_H$ is the heat deposition scale height. For loops with 
an area expansion factor  $\Gamma > 1$, \citet{vesecky79} (=  VAU) found numerical solutions that approximately follow the scaling 
laws \citep{schrijver89b} 
\begin{equation}
T_{\rm a} \approx 1400\Gamma^{-0.1}(pL)^{1/3}; \quad\quad T_{\rm a} = 60\Gamma^{-0.1}L^{4/7}\epsilon^{2/7}.
\end{equation}
Accurate analytical approximations to hydrostatic-loop solutions have been given by \citet{aschwanden02}
for uniform and non-uniform heating, including loops with expanding cross sections and loops heated near
their footpoints. I note in passing that the hydrostatic equations allow for a second solution for cool loops
($T < 10^5$~K) with essentially vanishing temperature gradients and small heights
(on the Sun: $< 5000$~km, the pressure scale height at $10^5$~K). These are transition 
region loops that can be observed in the UV region \citep{martens82, antiochos86b}. 

There are serious disagreements between some solar-loop observations and the RTV formalism so 
long as simplified quasi-static heating laws are assumed, the loops being more isothermal than
predicted by the models. There is, however, only limited understanding 
of possible remedies, such as heating that is strongly concentrated at the loop footpoints,
or dynamical processes in the loops (see, for example, a summary of this debate in
\citealt{schrijver02}).

Unstable solutions of large coronal loops with temperature inversions at the loop apex were 
numerically studied by \citet{collier88a}. If such loops are anchored on rapidly rotating 
stars, they may, in addition, become unstable under the influence of centrifugal forces.  
Once the latter exceed gravity, the pressure and the electron density grow outwards along 
the magnetic field, enhance the radiative loss rate and lead to a temperature inversion, 
which grows unstable. Rapid thermal cooling at the distant loop apex may then lead to 
condensations of prominence-like, magnetically  trapped and centrifugally supported, 
synchronously rotating cold gas, for which there is indeed  evidence  at distances of 
$\approx 3R_*$ around the rapid rotator AB Dor \citep{collier88a}. Further numerical 
studies, including various assumptions on the base pressure, surface magnetic field, 
and base conductive flux were presented by  \citet{unruh97}.

\subsection{Coronal structure from loop models}\label{loopmodels}

\subsubsection{Loop parameters} 
When we interpret stellar coronal spectra, we assume, to first order, that some physical
loop parameters map on our measured quantities, such as temperature and EM (and possibly
density), in a straightforward way. In the simplest approach, we assume that the
observed luminosity $L_X$ is produced  by
an ensemble of identical coronal loops with characteristic half-length $L$, surface filling
factor $f$, and an apex temperature $T$ used for the entire loop; then, on using Eq.~(\ref{RTV}) 
and identifying $L_X =  \epsilon V$, we obtain
\begin{equation}
L \approx 6\times 10^{16} \left({R_*\over R_{\odot}}\right)^2 {f\over L_X}T^{3.5}\quad {\rm [cm]}.
\end{equation}
This relation can only hold if $L$ is smaller than the pressure scale height. Based on this
expression, the luminous, hot plasma component in magnetically active stars seems to
invariably require either very large, moderate-pressure loops with a large filling factor,
or solar-sized high-pressure compact loops with a very small ($<1$\%) filling factor 
\citep{giampapa85, stern86, schrijver89b, giampapa96, guedel97a, preibisch97a, sciortino99}. 

\citet{schrijver84} modeled $T$ and EM of a sample of
coronal sources based on RTV loop models and found the following trends: 
i) Inactive MS  stars such as the Sun are covered to a large
fraction with large-scale, cool (2 MK) loops of modest  size ($0.1R_*$). ii) Moderately active dwarfs
are dominated by very compact, high-density, hot ($\approx$20~MK) loops that require large heating 
rates (up to 20 times more than for solar compact active region loops).
iii) The most active stars may additionally form rather extended loops with heights
similar to $R_*$.

\subsubsection{Loop-structure models}\label{loopstructure}
While the above interpretational work identifies spect\-ral-fit parameters such as $T$
or EMs with parameters  of theoretical loop models, a physically more appealing
approach involves full hydrostatic models whose calculated emission spectra are directly
fitted to the observations. While physically more realistic than multi-isothermal models,
the approach has its own limitations because it relies on a host of ad hoc parameters such as the 
location and distribution of heating sources within a loop, the loop geometry, the type of thermal 
conduction law inside the loop and, in particular, the unknown statistical distribution of 
all the loop parameters in a coronal ensemble. Fitting spectra calculated by use of full
hydrostatic models permits, however, to constrain possible combinations of these physical
loop parameters.

\citet{giampapa85} presented this type of numerical RTV-type loop models and,
assuming that one type of loop dominates, fitted them 
both to stellar X-ray and  transition-region UV fluxes. 
The modest success of the  fitting suggested  to them 
that their assumption was questionable and that various  structures do coexist 
in stellar atmospheres. \citet{stern86} fitted numerical loop models, 
characterized by the loop apex temperature $T_{\rm a}$, the loop semi-length $L$, and the expansion factor $\Gamma$, 
to spectra from the {\it Einstein} Imaging Proportional Counter (IPC), and the solutions were again constrained by 
UV observations. The successful fit results indicated magnetic  loops of modest size 
($L < 10^{10}$~cm) with modest filling factors ($f < 10-20$\%),  but with extreme, 
flare-like pressures ($p \ga 400$~dynes~cm$^{-2}$).   

\citet{giampapa96} extended loop studies to M dwarfs observed by {\it ROSAT}.
 They found that the low-$T$ component at
$\approx 1-2$~MK requires loops of small length ($L \ll R_*$) but high pressure
($p>p_{\odot}$), whereas the high-$T$ component at $\approx 5-10$~MK must be confined by rather long
loops ($L \approx R_*$) and  high base pressures ($p\approx 20$~dynes cm$^{-2} \gg
p_{\odot}$) with filling factors of order 0.1. These latter solutions however violate the applicability
of the RTV scaling law, since the loop height exceeds the pressure scale-height. \citet{giampapa96}
therefore speculated that this component is, in fact, related to multiple, very compact flaring regions with
a small filling factor, while the cooler, compact component relates to non-flaring active regions. 

The loop model approach has been extensively developed and 
further discussed in a series of papers by the Palermo group \citep{ciaravella96, ciaravella97, maggio96, maggio97,
ventura98}. The most notable results are: 
\begin{description}
\item [1.] To obtain a successful spectral fit to low-resolution data of a
single loop, one usually requires at least two isothermal plasma components. {\it The often-found
two spectral components from fits to low-resolution spectra must not, in general, be identified with
two loop families} \citep{ciaravella97}. 

\item [2.] Applications to observations often do, however, require
two loop families in any case, but for reasons more involved than the presence of two dominant thermal components.
In such cases, one finds relatively cool loops ($T = 1.5-5$~MK) with modest to high pressures ($p = 2-100$~dynes~cm$^{-2}$)
and hot ($T = 10-30$~MK), extreme-pressure loops ($p = 10^2-10^4$~dynes~cm$^{-2}$). The latter are - once again -
reminiscent of flaring loops with a very small surface filling factor \citep{maggio97, ventura98}. 

\item [3.] Low-activity
stellar coronae, on the other hand, may be sufficiently well described by  a single, dominant type of loop. This is the case for 
Procyon on which short, cool, low-pressure loops should occupy  10\% of the surface in this picture.
\end{description}
\citet{ottmann93b} found that the coronal  structure on AR Lac can be interpreted with essentially one
class of RTV loops with an apex temperature of 38~MK, although a better match may
be based on variable cross-section VAU loops, an approach followed by \citet{ottmann93},
who found loops with a half-length $L \approx 3\times 10^{11}$~cm and a filling factor
$f \le 10$\%.

\citet{vdoord97} applied analytic loop models including non-zero conductive flux at the loop footpoints
and variable expansion factors $\Gamma$ to {\it EUVE} spectra. The spectral inversion is problematic
because a given spectrum can be modeled by various sets of physical parameters, such as various combinations 
between loop expansion factors and loop base conductive fluxes. The range of solutions, however,
consistently requires loop expansion factors of 2--5, and in several stars, at least two loop families 
with different apex temperatures. 

\citet{sciortino99} similarly applied tabulated loop models of 
steady-state coronal (RTV) loops to interpret medium-resolution X-ray spectra of M dwarfs. 
They again found that at least two classes of magnetic loops, one with an apex temperature of $\approx 10$~MK 
and one with several tens of MK, are required. Although the solutions allow for a large range of 
base pressures and surface filling factors, both the cooler and the hotter loops are found to be 
quite compact, with lengths smaller than 0.1$R_*$ and filling factors of $10^{-4}$ to $10^{-3}$. 
A similar study was presented by \citet{griffiths99} who computed comprehensive static energy-balance
loop models to interpret EMDs of RS CVn binaries.  The double-peak structure in
the EMD again required two loop families with apex temperatures of 8 and 22~MK, respectively, 
both of modest size.

Finally, \citet{favata00b} concluded, from modeling of several X-ray flares on AD Leo
(Sect.~\ref{contheat}), that all magnetic loops involved in the flaring corona are of similar,
relatively compact dimension (0.3$R_*$). If these are the typical structures that form
the overall corona, then their filling factor is no larger than about 6\% despite the rather high
X-ray luminosity of this star. The loop pressure must consequently be large 
(70~dyne~cm$^{-2}$). 

\subsubsection{Conclusions and limitations} 
The most essential conclusion from these exercises is perhaps  that,
within the framework of these simplistic models, the loop heating rate required for
magnetically active stars may exceed values for typical solar loops 
by  orders of magnitude, pointing toward some enhanced
heating process reminiscent of the energy deposition in flares. 
The compactness of the hot loops and the consequent high pressures also set these coronal
structures clearly apart from any non-flaring solar coronal features.

The inferred geometric size  has important implications for dynamo theories. 
The apparent predominance of compact, localized sources 
suggests a predominance of local, small-scale magnetic fields. Such fields are expected from distributed dynamos
that have been postulated for fully convective stars. A possible conclusion for deeply convective
M dwarfs is that the solar-type $\alpha\omega$ dynamo does not operate, in agreement
with the absence of a convective boundary inside the star once it  becomes fully convective \citep{giampapa96}.
This conclusion is, however, not entirely valid in the light of other magnetic field measurements that
have revealed global components, e.g., by use of radio interferometry (Sect.~\ref{extendedcompact}). It is also
important to note that open solar magnetic fields carry little specific EM, and
that large-scale hot loops cannot be dense because the plasma cannot be magnetically confined in that case.
Little X-ray emission is therefore expected from such features, even though {\it they may well exist
as magnetic structures.}

While explicit loop models provide an appealing  basis for a physical 
interpretation of coronal structures, the limitations of our diagnostics should be kept in mind: 
\begin{description}
\item [1.] due to the 
degeneracy of solutions in the product $pL$ (see Eq.~\ref{RTV}), multiple, largely 
differing solutions are usually compatible with observations of finite quality and
of low energy resolution (e.g., \citealt{schmitt85b} for Procyon, \citealt{vdoord97} for further sources). 

\item [2.] the use of models based on one loop or one family of identical loops  (in $L$, $T_{\rm a}$
and hence $p$, $f$, $\Gamma$, and the heating profile along 
the loop) may  offer a number of degrees of freedom that is sufficiently large to describe a given 
spectrum satisfactorily,  yet the model is unlikely to describe any real corona. 
\end{description}
In this sense, like in the case of multi-isothermal models or in ``minimum flux'' coronal models,
loop  models parameterize the real situation in a manner that is not straightforward and 
that requires additional constraints from additional sources of  information.
I also refer to the extensive, critical discussion on this point by  \citet{jordan86}.	 

\subsection{Coronal structure from densities and opacities}\label{densityopacity}

Spectroscopically measured densities provide, in conjunction with EMs, important
estimates of emitting volumes. For example,
\citet{ness01} use the RTV scaling laws together with measured coronal temperatures
and electron densities inferred from He-like triplets of C, N, and O of Procyon.
They concluded that the dominant coronal loops on this star are  solar-like and low-lying
($L \la 10^9$~cm). For the more active Capella, \citet{mewe01} found very similar 
structures covering a few tens of percent of the surface, but additionally they inferred the presence of a 
hotter plasma at apparently very high densities (Sect.~\ref{fedensities}). This latter plasma
would have to be confined to within very  compact loops ($L \la 5\times 10^7$~cm) that cover an 
extremely small area  on the star ($f \approx 10^{-6} - 10^{-4}$).
In general, for increasing temperature, progressively higher pressures and progressively smaller
volumes are determined (\citealt{osten03, argiroffi03}; Sect.~\ref{denshe}). The confinement of such exceedingly high
densities in compact sources, with a size of a few 1000~km, would then require coronal magnetic field
strengths of order 1 kG \citep{brickhouse98}. In that case, the typical magnetic dissipation time is
only a few seconds for $n_e \approx 10^{13}$~cm$^{-3}$ if the energy is derived from the same magnetic fields,
suggesting that the small, bright loops light up only briefly. In other words, the stellar corona would
be made up of numerous ephemeral loop sources that  cannot be treated as being in a quasi-static equilibrium
\citep{vdoord97}.

The debate as to how 
real these densities are, continues, as we mentioned in Sect.~\ref{coronaldensities}. 
\citet{ness02b}, again using standard coronal-loop models, derived moderately compact loop sizes 
($L \approx 10^9-5\times 10^{10}$~cm) for Algol, the uncertainty being related to uncertain density measurements. 
\citet{testa04} and \citet{ness04} used density information from O\,{\sc vii},  Ne\,{\sc ix}, Mg\,{\sc xi},
and computed a rough stellar surface  filling factor $f$ of static loops. They found that $f$ remained 
at a few  percent for cool, O\,{\sc vii} emitting material for the most active stars. 
For less active stars, \citet{ness04} reported similar filling factors, whereas \citet{testa04} found
them to decrease with decreasing activity. For the hotter,
Mg\,{\sc xi} emitting material, the surface coverage rapidly increases in more active stars.
A possible interpretation involves a relatively  cool, inactive base corona that  remains 
unaltered while hotter and denser loops are added as one moves
to progressively more active stars - a larger rate of magnetic interaction between adjacent
active regions could then  be assumed to  heat the plasma, for example by flares.

The lack of measured optical depths $\tau$ due to resonance scattering in stellar coronae (see Sect.~\ref{opticaldepth}) 
can also be exploited to set limits to coronal size scales. \citet{mewe01} and \citet{phillips01} used formal upper limits on
$\tau$ derived from Fe\,{\sc xvii} line ratios of Capella together with densities derived from
Fe\,{\sc xx-xxii} line ratios (with the caveats mentioned above) to
estimate that the characteristic size of a single emitting region is in the range of
$(1-3)\times 10^8$~cm, i.e., very compact in comparison with Capella's radius. This procedure, however, 
is not  valid for more complicated geometries: if active regions are distributed across the
stellar surface, then optical depths may  cancel 
out in the observations, even if individual larger structures may be involved with non-zero 
optical depth \citep{mewe01}.

\subsection{Coronal constituents: Emission-measure interpretation}\label{constituents}

The average X-ray surface flux $F_X$ is a direct tracer for the type of structures that
can possibly cover the stellar surface.  Because we generally have no information on stellar coronal
inhomogeneities, the average stellar $F_X$ can obviously not be compared directly with $F_X$ values
of solar coronal structures. We can nevertheless derive quite meaningful constraints.
For example, the {\it average} $F_X$ on Proxima Centauri 
is only about one fifth of corresponding values for solar active regions while
it  exceeds solar quiet region values by one order of magnitude. Assuming, then,
that the emission is concentrated in equivalents of solar
active regions, their surface filling factor would be  as much as 20\%, compared
to 0.01--1\% for the Sun at the 1~dyne~cm$^{-2}$ level \citep{haisch80}. At the low
end of magnetic activity, there seem to be no stars
with an X-ray surface flux below $F_X \approx 10^4$~erg~cm$^{-2}$~s$^{-1}$ \citep{schmitt97}.
This flux coincides with the surface flux of solar coronal holes, suggesting 
that the least active MS stars are fully covered by coronal holes. 
Similar results were reported by \citet{huensch96b} for giants. 

If coronal holes, inactive regions, active regions, bright points,
small and large flares  are characteristically  different in their thermal structure and their surface 
flux, then one may  interpret a full-disk EMD as a linear 
superposition of these various {\it building blocks}.  
We may  start from the Sun by attempting to understand how 
various coronal features contribute to the integrated X-ray light \citep{ayres96, orlando00, orlando01, peres00}. 
Full-disk solar EMDs from {\it Yohkoh} images reveal broad distributions with steep slopes below the 
temperature peak and a gradual decline up to $10^7$~K. The interesting aspect is that the EMDs 
shift to higher temperatures both from activity minimum (peak at $\approx 1$~MK)
to maximum (peak at $\approx 2$~MK) and from X-ray faint to X-ray bright features (see also Fig.~\ref{demsolarstellar}). 

On the stellar side, we see similar shifts from a hotter to a cooler EMD on {\it evolutionary time scales}
as a star ages and becomes less active \citep{guedel97a, guedel97c}. For example, EMDs of intermediately active
stars closely resemble the solar cycle-maximum EMD (Fig.~\ref{demsolarstellar}), which indicates a 
nearly full surface coverage with active regions \citep{drake00}. From this point of view,
the Sun's magnetic cycle mimics an interval
of stellar activity during its magnetic cycle \citep{ayres96}. Moving to very active stars, large-scale structures 
between active regions and post-flare loops seem to become the dominating coronal components rather
than normal
active regions and bright points.  The increased luminosity is then a consequence of 
increased filling factors, increased loop base pressures, and higher temperatures
\citep{ayres96, guedel97a}. \citet{orlando01} showed that in a similar manner the increasing X-ray spectral 
hardness  from solar minimum to maximum on the one hand and from low-activity to high-activity solar analogs 
on the other hand can be explained by an increasing proportion of dense and bright coronal features, namely active
regions and in particular hot cores of active regions. 

However, if the Sun were entirely covered with active regions, the X-ray luminosity 
would amount to only $\approx (2-3)\times 10^{29}$~erg~s$^{-1}$ 
\citep{vaiana78, wood94}  with $L_X/L_{\rm bol} \approx 10^{-4}$ \citep{vilhu84}, 
short of $L_X$ of the most active early G-type analogs of the Sun by one order of 
magnitude. A similar problem is evident for extremely active, rapidly
rotating  FK Comae-type giants whose full disk $F_X$ exceeds non-flaring solar active region 
fluxes by up to an order of magnitude \citep{gondoin02, gondoin03a}. In those stars, the cooler
($<10$~MK) plasma component alone already fills the complete surface under the usual assumptions.

The DEMs of such active stars produce excessive emission around 10--20~MK 
\citep{guedel97a}, which incidentally is the typical range of solar flare temperatures. 
This led to the suggestion that the high-$T$ DEM is in fact due to
the superposition of a multitude of superimposed  but temporally unresolved flares (Sect.~\ref{stochasticflares}).
If added to a low-$T$ ``quiescent'' solar DEM, the time-integrated DEM of 
solar flares indeed produces a characteristic bump around 10--20 MK that compares
favorably with stellar DEMs \citep{guedel97a}. This happens because the flare EM decreases rapidly
as the temperature  decays, leaving a trace on the DEM only at relatively high temperatures,
in agreement with Eq.~(\ref{demflareint}) for time-integrated flares.
If a full distribution of flares contributes, including small flares with lower temperature,
then the entire DEM could be formed by the continually heating and cooling 
plasma in flares \citep{guedel97c}. The predicted steep low-$T$ slope (up to $\approx 4$)
of a  stochastic-flare DEM compares very favorably with observations of active stars
\citep{guedel03a}.

The reason for an increased fraction of hotter features in more active stars (Fig.~\ref{temperaturelx})
may be found in the coronal structure itself. Toward more active stars,
magnetic fields interact progressively more frequently due to their denser packing. In this view, an increased
heating rate in particular in the form of flares reflects the enhanced dynamo operation 
in rapidly rotating stars. Since flares enhance the electron density along with the temperature, 
one finds a predominance of hotter structures in more active stars (\citealt{guedel97a}; cf. also
the discussion in Sect.~\ref{templx}). 
Along these lines, \citet{phillips01} compared high-resolution spectra of Capella and solar flares,
finding a surprising overall agreement; similarly, \citet{guedel04} compared
the average X-ray spectrum of a large stellar flare with the spectrum of low-level emission
of the very active dMe binary YY Gem. The agreement between the spectral features, or equivalently, the
EMD, is compelling. The hypothesis of ongoing flaring in active stars that we have now
 repeatedly invoked  will be further discussed in
Sect.~\ref{stochasticflares}.

\subsection{Coronal imaging: Overview}

As the development of solar coronal physics has amply shown, {\it coronal imaging} will be an 
indispensable tool for studying the structure and heating of stellar coronae in detail. 
At present, coronal structure recognition based on direct 
or indirect imaging methods is still highly biased by observational constraints and by
the location of physical processes that heat plasma or accelerate particles.
Stellar coronal structure resolved  at radio wavelengths refers to extended (typically low-density)
closed magnetic fields into which high-energy electrons have been injected. X-ray emitting
plasma traces magnetic fields that have been loaded with dense plasma.
{\it But both types of structure, as well as coronal holes that are weak both
in X-rays and at radio wavelengths, refer to the 
underlying global distribution of surface magnetic fields.} 
Presently, only radio structures can be directly imaged by means of 
radio Very Long Baseline Interferometry (VLBI; for a review see \citealt{guedel02d}),
whereas thermal coronal sources can be mapped indirectly by using X-ray eclipses,
rotational modulation, or Doppler information as discussed below. 

With these limitations in mind, I now briefly summarize various image 
reconstruction methods for eclipses and rotational modulation (Sect.~\ref{activeregions}-\ref{clean}) and 
related results 
(Sect.~\ref{imageeclipse}-\ref{eclipseflare}). Sect.~\ref{doppler} and \ref{zeeman} address
alternative  methods for structure modeling. 

\begin{figure} 
\centerline{\resizebox{0.75\textwidth}{!}{\includegraphics{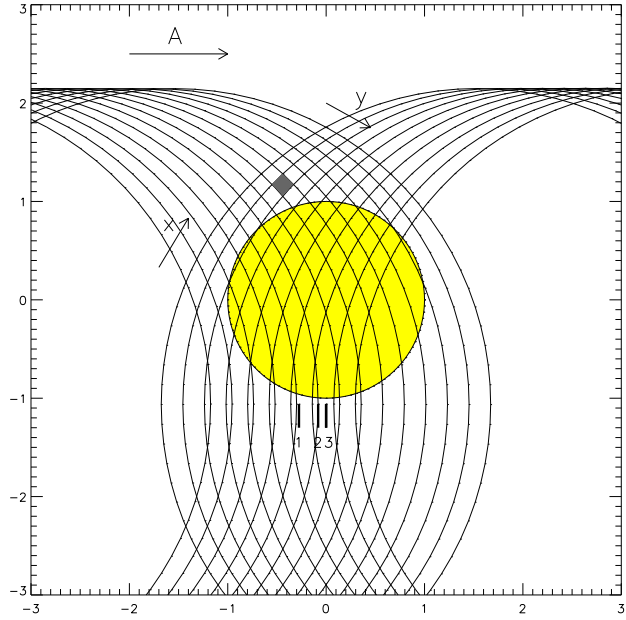}} }
\caption{Sketch showing the  geometry of an eclipsing binary (in this case, $\alpha$ CrB, figure
from \citealt{guedel03b}). The large circles
illustrate the limbs of the eclipsing star that moves from left to right in front of the eclipsed star 
(shown in yellow). The limbs projected at different times during ingress and egress
define a distorted 2-D  array $(x,y)$ of pixels (an example of a pixel is shown in gray).}\label{eclipsegeometry}
\end{figure}

In its 
general form,  the ``image'' to be reconstructed consists of volume elements at coordinates 
$(x,y,z)$  with {\it optically thin} fluxes $f(x,y,z)$ assumed to be constant in time. 
In the special case of negligible  stellar rotation during the observation, the 
problem can be reduced to a 2-D projection onto the plane of the sky, at the cost
of  positional information along the line of sight (Fig.~\ref{eclipsegeometry}). In general, 
one thus seeks the geometric brightness distribution $f(x,y,z) = f_{ijk}$ ($i,j,k$ being the
discrete pixel number indices) from a binned, observed light curve 
$F_s = F(t_s)$ that undergoes a modulation due to an eclipse or due to rotation.

\subsection{Active-region modeling}\label{activeregions}

In the most basic approach, the emitting X-ray or radio corona can be modeled by
making use of a small 
number of simple, elementary building blocks that are essentially described by their size, 
their brightness, and their location. This approach is the 3-D equivalent
to standard surface spot modeling. 
Preferred building block shapes are radially directed, uniformly bright, optically thin,
radially truncated spherical cones with their apexes at the stellar center. Free parameters 
are their opening angles, their heights above the stellar surface, their radiances, and 
their central latitudes and longitudes. These parameters are then varied until the 
model  fits the observed light curve. 
For extensive applications to eclipsing binaries, see, for example, \citet{white90},
\citet{ottmann93}, and \citet{culhane90}.

A minimum solution was presented by \citet{guedel95c} for a rotationally modulated star.
If a rotationally modulated feature is {\it invisible} during a phase interval $\varphi$ of
the stellar rotation, then all sources contributing to this feature must be confined
to within a maximum volume, $V_{\rm max}$, given by
\begin{equation}\label{minimumvolume}
{V_{\rm max}\over R_*^3} = {\psi\over 3} - {(2\pi - \varphi)(1+{\rm sin}^2i) \over 6{\rm sin}i}
                 +{2{\rm cot}(\chi/2)\over 3{\rm tan}i}
\end{equation}
where ${\rm tan}\chi = {\rm tan}(\varphi/2){\rm cos}i$, 
      ${\rm sin}(\psi/2) = {\rm sin}(\varphi/2){\rm sin}i$ with $0 \le \psi/2\le \pi/2$,
      and $\chi$ and $\varphi/2$ lie in the same quadrant ($i$ is the stellar inclination,
$0 \le i \le \pi/2$). Together with the modulated fraction of the luminosity, lower limits to average 
electron densities in the modulated region follow directly.

\subsection{Maximum-entropy image reconstruction}\label{mem}
  
We follow the outline given in \citet{guedel03b} to discuss maximum entropy methods (MEM) for
image reconstruction. They are applicable both to rotationally modulated light curves
and to eclipse observations. The standard MEM  selects among all images $f_{ijk}$ (defined in units of counts per volume element)
that are  compatible with the observation, the one that minimizes the Kullback contrast 
(``relative entropy'')
\begin{equation}\label{K}
K = \sum_{i,j,k} f_{ijk} \ln \frac{f_{ijk}}{f^a_{ijk}}
\end{equation}
with respect to an a priori image $f^a_{ijk}$, which is usually unity inside the allowed area or volume
and vanishes where no brightness is admitted. Minimizing $K$ thus introduces the least possible information 
while being compatible with the observation. The contrast $K$ is minimum if $f_{ijk}$ is proportional to 
$f_{ijk}^a$ and thus flat inside the field of view, and it is maximum if the whole flux is 
concentrated in a single pixel $(i,j,k)$. The compatibility with the observed count light curve is measured by
$\chi^2$,
\begin{equation}
\chi^2 = \sum_s \frac{(F^*_s-F_s)^2}{F^*_s}
\label{chi2}
\end{equation}
where $F_s$ and $F^*_s$ are, respectively, the observed number of counts and the number of counts predicted from 
$f_{ijk}$ and the eclipse geometry. Poisson statistics usually requires more than 15 counts per bin. 
Finally, normalization is enforced by means of the constraint
\begin{equation}
N = \frac{ \left( {\displaystyle f^{\rm tot} - \sum_{ijk} f_{ijk}}\right)^2  }{f^{\rm tot}}
\label{N}
\end{equation}
where $f^{\rm tot}$ is the sum of all fluxes in the model. 

The final algorithm minimizes the cost function
\begin{equation}
C = \chi^2 + \xi K + \eta N\, .
\label{F}
\end{equation}
The trade-off between the compatibility with the observation, normalization, and unbiasedness
is determined by the Lagrange multipliers $\xi$ and $\eta$ such that the reduced $\chi^2$ 
is $\la 1$, and normalization holds within a few percent. For applications of this method and 
variants thereof, see \citet{white90} and \citet{guedel03b}.

\subsection{Lucy/Withbroe image reconstruction}\label{lucy}

This method (after \citealt{lucy74} and \citealt{withbroe75}, the latter author discussing
an application to spectral line flux inversion) was extensively used by 
\citet{siarkowski92}, \citet{siarkowski96}, \citet{pres95}, \citet{guedel01a}, \citet{guedel03b} and
\citet{schmitt03} to image X-ray  coronae of  eclipsing binaries; it is applicable to pure rotational modulation as well.  
The method is formally related to maximum likelihood methods although the iteration and its convergence  
are methodologically different \citep{schmitt96f}. The algorithm iteratively adjusts fluxes in a 
given set of volume elements based on the mismatch between the model and the observed light curves 
in all time bins to which the volume element contributes.  At any given time $t_s$ during the eclipse, 
the observed  flux $F_s$ is the sum of the fluxes $f_{ijk}$ from all pixels that are unocculted:
\begin{equation}\label{withbroe1}
F(t_s) = {\displaystyle \sum_{i,j,k}} f_{ijk}\mathbf{m}_s(i,j,k)
\end{equation}
where $\mathbf{m}_s$ is the ``occultation matrix'' for the time $t_s$:  it puts, for any given time
$t_s$, a weight of unity to all visible pixels and zero to all invisible pixels (and intermediate values 
for partially occulted pixels). Since $F_s$ is given, one  needs to solve Eq.~(\ref{withbroe1})
for the flux distribution, which is done iteratively as follows:
\begin{equation}\label{withbroe2}
f_{ijk}^{n+1}= 
    f_{ijk}^{n}\frac{{\displaystyle \sum_s \frac{F_o(t_s)}{F^n_m(t_s)}}\mathbf{m}_s(i,j,k)}
                       {{\displaystyle \sum_s \mathbf{m}_s(i,j,k)}}
\end{equation}
where $F_o(t_s)$ and $F^n_m(t_s)$ are, respectively, the observed flux and the model flux (or counts)
in the bin at time $t_s$, both for the iteration step $n$. Initially, 
a plausible, smooth distribution of flux is assumed, e.g., constant brightness, or
some $r^{-p}$ radial dependence. 

\subsection{Backprojection and Clean image reconstruction}\label{clean}
 
If rotation can be neglected during an eclipse, for example in long-period detached binaries, then the limb of the eclipsing star is 
projected at regular time intervals onto the plane of the sky and therefore onto a specific
part of the eclipsed corona, first during ingress, later during
egress. The two limb sets define a 2-D grid of distorted, curved pixels (Fig.~\ref{eclipsegeometry}). The brightness 
decrement during ingress or, respectively, the brightness increment during egress within a time step $[t_s, t_{s+1}]$ originates from 
within a region confined by the two respective limb projections at $t_s$ and $t_{s+1}$. Ingress and 
egress thus each define a 1-D image by backprojection from the light curve gradients onto the plane 
of the sky. The relevant reconstruction problem from multiple geometric projections is known in 
tomography. The limiting case of only two independent projections can be augmented by a CLEAN step, as follows. 
The pixel with the largest {\it sum} of projected fluxes  from ingress and egress is assumed to represent 
the location of a real source. A fraction, $g < 1$, of this source flux  is then subtracted 
from the two projections and saved on a clean map, and the process is iterated until all flux is 
transferred onto the latter. This method was described by \citet{guedel03b} who applied it to a total stellar eclipse
of $\alpha$ CrB, with consideration of different gain factors $g$ and post-fit flux redistributions to study multiple
solutions.

\subsection{Coronal structure inferred from eclipses}\label{imageeclipse}

I now review selected results from analyses of X-ray eclipses, and, in subsequent subsections,
of rotational modulation and Doppler measurements. Some important
parameters are summarized in  Table~\ref{structuretable}.

\begin{table}
\begin{minipage}{1.0\textwidth}
\caption{X-ray coronal structure inferred from eclipses and rotational modulation}\label{structuretable}
\label{tab:1}       
\begin{tabular}{lllllll}
\hline\noalign{\smallskip}
Star            & Spectrum        & Extended$^a$       & $n_e^b$     & Compact$^a$  &  $n_e^b$   & Reference$^c$  \\
                &                 & height ($R_*$) &               &  height ($R_*$) &       &            \\
\noalign{\smallskip}\hline\noalign{\smallskip}
AR Lac          &  G2~IV+K0~IV    &  $\approx 1$   &  0.29         &   0.01    & 4--6        & 1       \\
AR Lac          &  G2~IV+K0~IV    &  1.1--1.6	   &  0.2--0.8     &   0.06    & $>5$	     & 2  	\\
AR Lac          &  G2~IV+K0~IV    &  0.7--1.4	   &  0.3--0.8     & 0.03--0.06& 6--60	     & 3	\\
AR Lac          &  G2~IV+K0~IV    &  $\approx 1$   &  0.12--1.8    &	-      &	-    & 4   \\
Algol           &  B8~V+K2~IV     &   0.8          &	...	   &	-      &  -	     & 5        \\
Algol$^d$       &  B8~V+K2~IV     &   -            &  -		   &$\la 0.5$  & $\ga 9.4$   & 6         \\
Algol$^d$       &  B8~V+K2~IV 	  &   -            &  -		   &  0.1      & $\la  3$    & 7      \\
TY Pyx          &  G5~IV+G5~IV 	  &  $\approx$1--2 &  0.02--3       &   -       &    -       & 8   \\
XY UMa          &  G3~V+K4~V      &   -            &  -            & $\le 0.75$&    ...      & 9	   \\
VW Cep$^d$      &  K0~V+G5~V      &  0.84          &  5            &   -       &    -        & 10        \\
$\alpha$ CrB    &  A0~V+G5~V      &   -            &	-	   &$\la 0.2$  & $\la 3$     & 11        \\
$\alpha$ CrB    &  A0~V+G5~V      &   -            &	-	   &$\la 0.1$  & 0.1--3      & 12         \\
EK Dra          &  dG0e	          &   -            &	-	   &$\la 0.2$  & $\ga 4$     & 13         \\
YY Gem          & dM1e+dM1e       &   -            &	-	   & 0.25--1   &  0.3--3     & 14         \\
V773 Tau$^d$    &  K2~V+K5~V      &                &		   &$\approx 0.6$& $\ge 20$  & 15         \\
\noalign{\smallskip}\hline
\end{tabular}
\footnotetext{
NOTES: $^a$~Extended structures of order $R_*$, compact structures significantly smaller.  \\
$^b$~Electron density in $10^{10}$~cm$^{-3}$ for extended and compact structures, respectively.  \\
$^c$~References: 1 \citet{walter83b}; 2 \citet{white90}; 3 \citet{ottmann93b}; 4 \citet{siarkowski96}; 5 \citet{ottmann94};
                6 \citet{schmitt99}; 7 \citet{schmitt03}; 8 \citet{pres95}; 9 \citet{bedford90}; 10 \citet{choi98};
	       11 \citet{schmitt93c};  12 \citet{guedel03b}; 13 \citet{guedel95}; 14 \citet{guedel01a}; \citet{skinner97}.\\ 
$^d$~Refers to observation of eclipsed/modulated {\it flare}. \\
}
\end{minipage}
\end{table}

\subsubsection{Extent of eclipsed features} 
 
Some shallow X-ray eclipses in tidally interacting binary systems of the RS CVn, Algol, 
or BY Dra type have provided important information
on extended coronal structure. For example, Walter et al. (1983) concluded
that the coronae in the AR Lac binary components  are bi-modal in size, 
consisting of compact, high-pressure (i.e., 50--100 dynes cm$^{-2}$)
active regions with a scale height $<R_*$, while the subgiant K star is additionally surrounded by an extended (2.7$R_*$)
low-pressure corona. This view was supported by an analysis of {\it ROSAT} data by
\citet{ottmann93b}. In the analysis of \citet{white90}, the association of different
regions with the binary components remained ambiguous, and so did the coronal
heights, but the most likely arrangement again required at least one
compact region with $p > 100$ dynes cm$^{-2}$ and favored an additional {\it extended}, low-pressure
coronal feature with $p \approx 15$~dynes~cm$^{-2}$ and a scale height of $\approx R_*$. 
Further, a hot component pervading the entire binary system was implied from the 
absence of an eclipse in the hard ME detector on {\it EXOSAT}. \citet{culhane90} similarly 
observed a deep eclipse in TY Pyx in the softer band of {\it EXOSAT} but a clear absence thereof 
in the harder band, once more supporting a model including an  extended, hot component.  
From an offset of the eclipse relative to the  optical first contact in Algol, \citet{ottmann94} 
estimated the  height of the active K star corona to be $\approx 2.8R_{\odot}$.

\citet{white86} inferred, from the {\it absence} of X-ray dips or any modulation in the light curve of Algol, a 
minimum characteristic coronal scale height of 3$R_{\odot}$ (i.e., about 1$R_*$). Similar arguments were used by  
\citet{jeffries98a}  for the short period system XY UMa to infer  a corona that must 
be larger than $1R_{\odot}$ unless more compact structures sit at high latitudes. This latter possibility
should in fact be reconsidered for several observations that entirely lack modulation. Eclipses or rotational
modulation can be entirely absent if the active regions are concentrated toward one  of the polar regions.
This possibility has found quite some attention in recent stellar research.

\begin{figure}[t!] 
\centerline{\resizebox{0.72\textwidth}{!}{\includegraphics{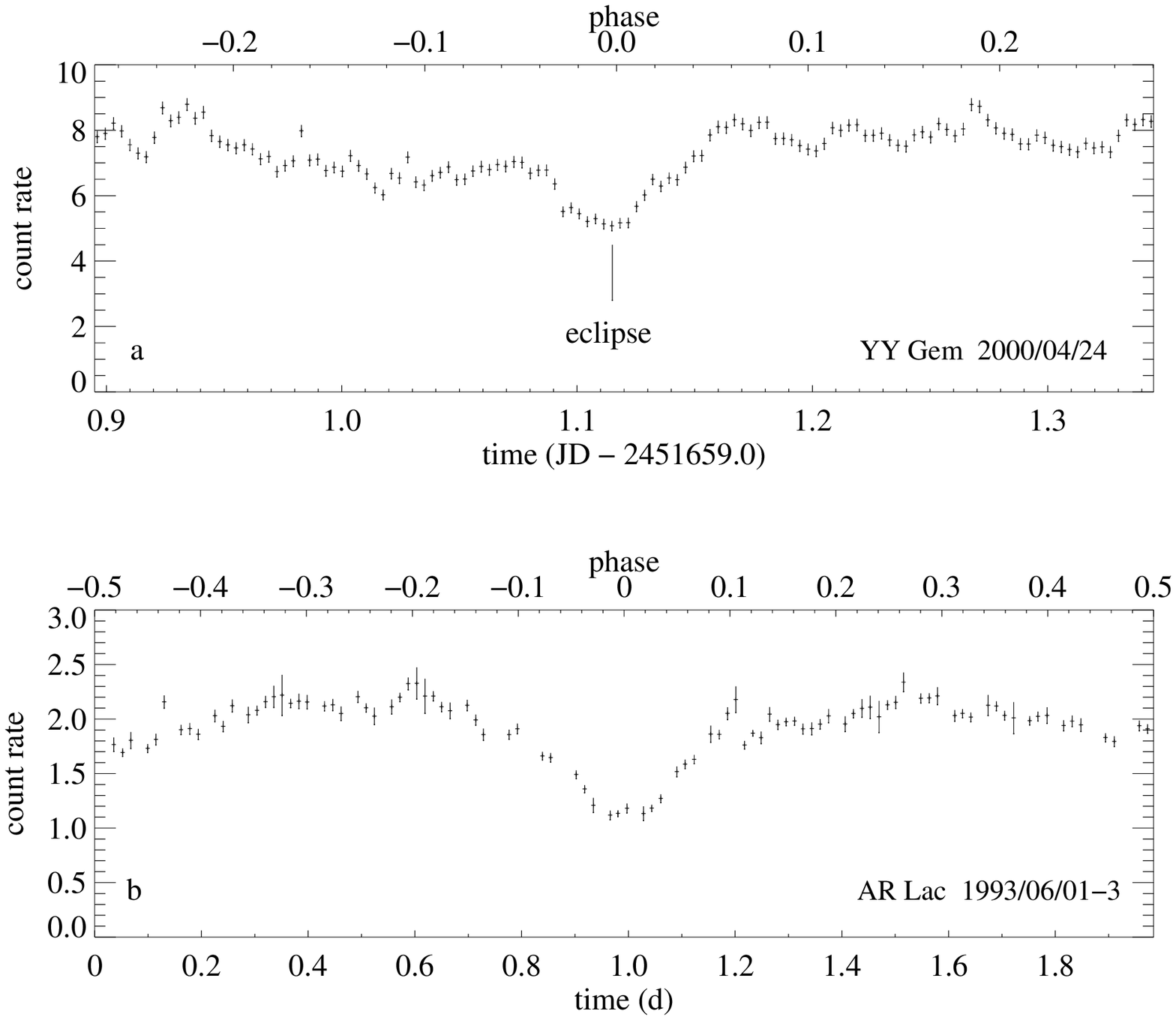}}}
\center{
\hbox{
\vbox{
\resizebox{0.61\textwidth}{!}{\includegraphics{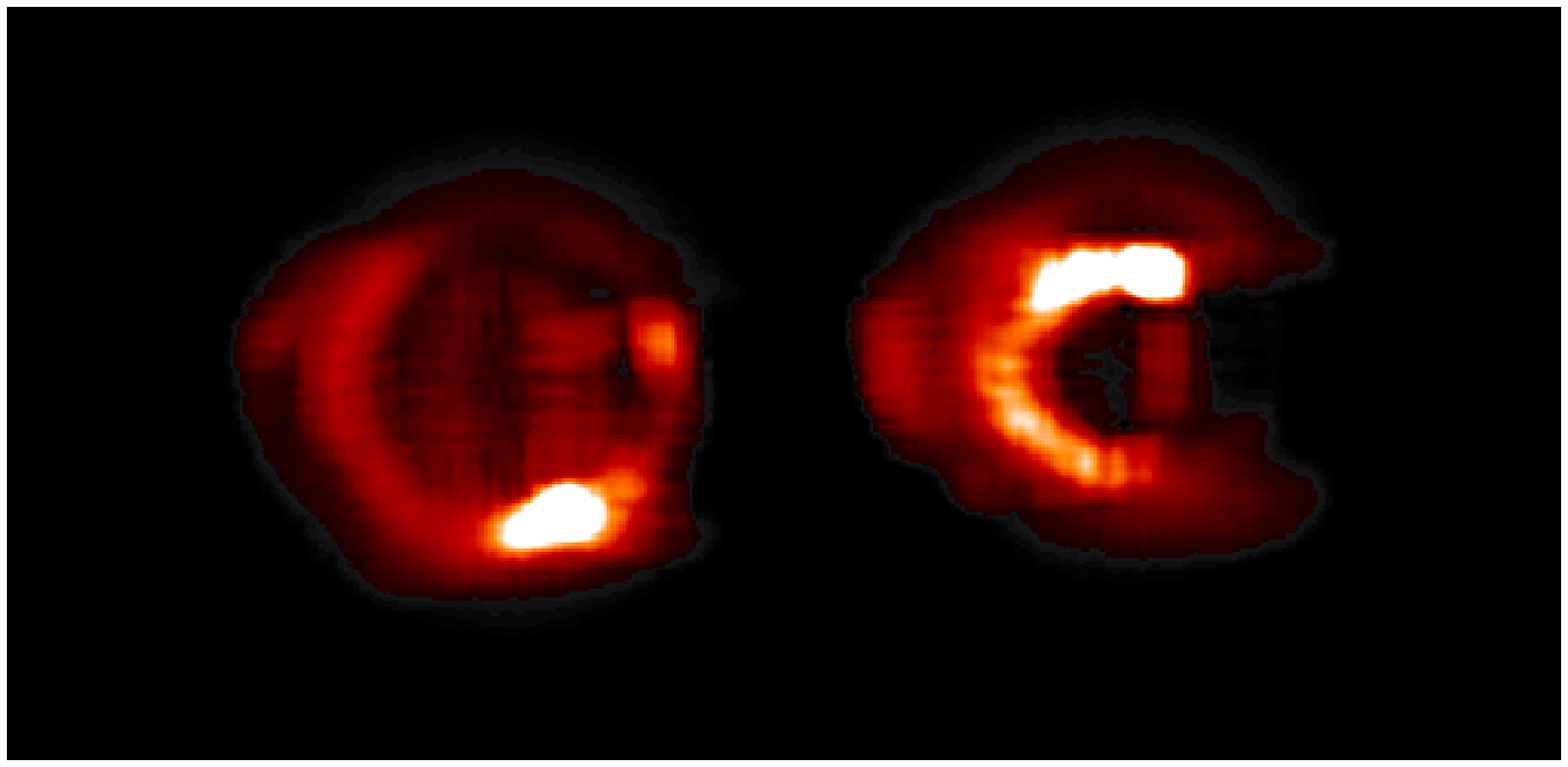}}\\
\resizebox{0.61\textwidth}{!}{\includegraphics{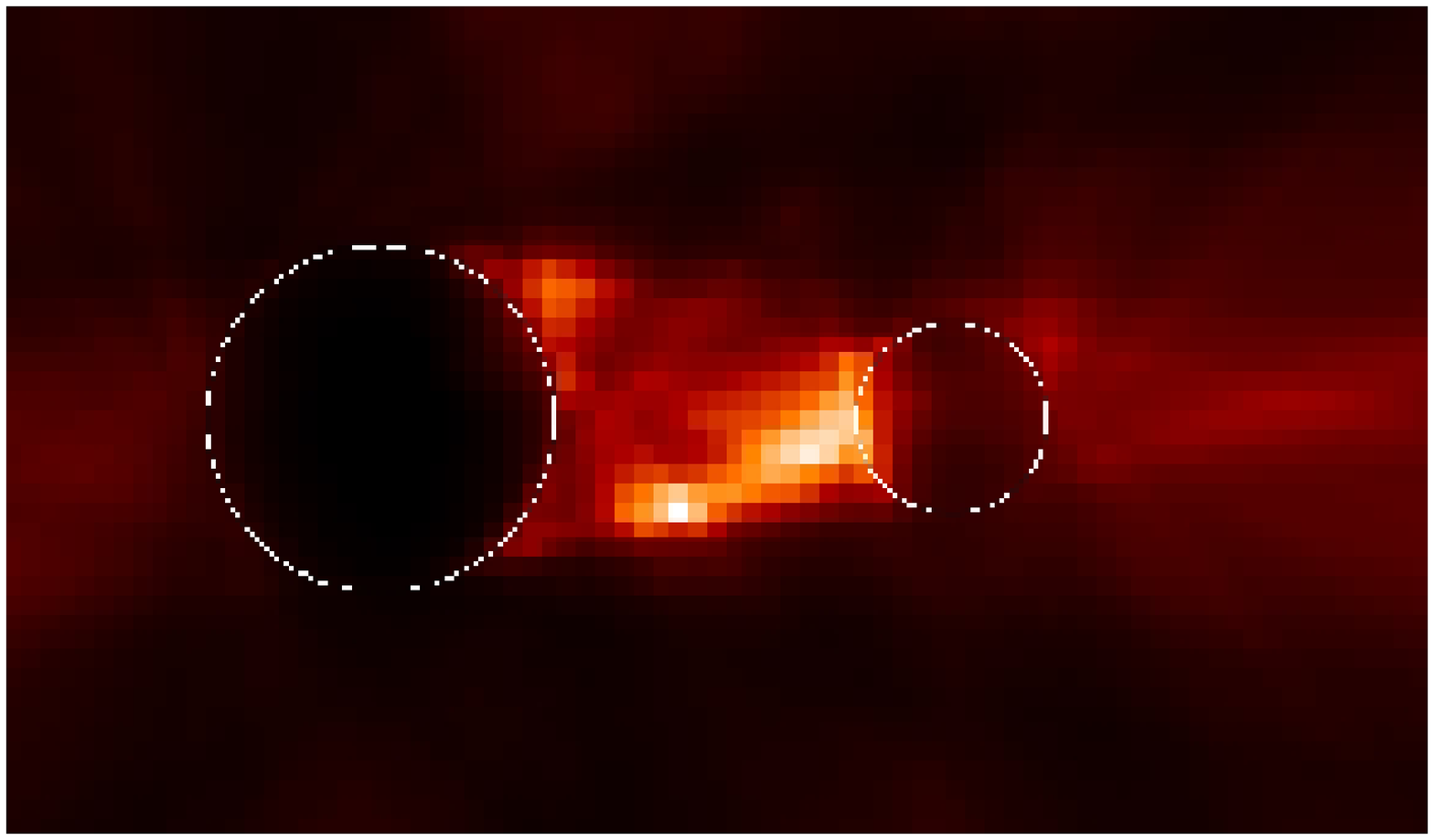}}
}
}
}
\vskip 0.2truecm\caption{Two examples of eclipses and the corresponding coronal image reconstructions. 
{\it From top to bottom:} Light curve of the YY Gem system (from \citealt{guedel01a}, observation with
{\it XMM-Newton} EPIC);
light curve of the AR Lac system (after \citealt{siarkowski96}, observation with {\it ASCA} SIS);
reconstructed image of the coronal structure of, respectively, YY Gem and AR Lac.
The latter figure shows a solution with intrabinary emission. 
(The light curve of AR Lac is phase-folded; the actual observation started around phase 0; data and image
for AR Lac courtesy of M. Siarkowski.)
}\label{eclipsefig}
\end{figure}

Detailed studies of light curves that cover complete binary orbits with {\it ASCA} raise, however, 
some questions on the reliability of the derived structure sizes. Unconstrained iterations
of the light curve inversion algorithm do converge to structures that are extended
on scales of $R_*$ (Fig~\ref{eclipsefig}); but constrained solutions exist that sufficiently represent the
light curves with  sources no larger than $0.3R_*$ \citep{siarkowski96}.
Nevertheless, detailed studies of the imaging reconstruction strategy
and the set-up of initial conditions led \citet{pres95} to conclude that
X-ray bright sources do indeed exist  far above the surfaces in the TY Pyx system, most
likely  located between the two components.  The latter configuration includes the possibility of 
magnetic fields connecting the two stars.
Interconnecting magnetic fields would draw implications for magnetic heating
through reconnection between intrabinary magnetic fields, as was
suggested by \citet{uchida84, uchida85} and was also proposed from radio observations
of the RS CVn-type  binary CF Tuc \citep{gunn97}, the Algol-type binary  V505 Sgr \citep{gunn99}, and the pre-cataclysmic binary 
V471 Tau \citep{lim96}. The X-ray evidence remains ambiguous at this time, and alternative X-ray methods
such as Doppler measurements \citep{ayres01b} have not added  support to this hypothesis (Sect.~\ref{doppler}).

\subsubsection{Structure and location of coronal features}

The eclipse light curves often require {\it asymmetric, inhomogeneous} coronae, with bright features sometimes 
found on the leading stellar hemispheres (e.g., \citealt{walter83b, ottmann93b, ottmann94}), but often also
on the hemispheres facing each other \citep{bedford90, culhane90, white90, siarkowski92, siarkowski96, pres95}.
This may, as mentioned above, have important implications for intrabinary magnetic fields.
For the dMe binary YY Gem, \citet{doyle90} reported a preferred occurrence of optical flares also
on the  two hemispheres facing each other, and this hypothesis has been supported by the timing
of X-ray flares \citep{haisch90b}. However, X-ray image
reconstruction from light curves does not require any emission significantly beyond $2R_*$ 
(\citealt{guedel01a}, Fig.~\ref{eclipsefig}): interconnecting magnetic fields are thus not supported in this case.
Similarly, a deep eclipse observed on XY UMa (in contrast to an observation reported by
\citealt{jeffries98a}) places the eclipsed material on the hemisphere of the 
primary that faces the companion, but judged from thermal loop models, the sources are suggested to be 
low-lying \citep{bedford90}. A concentration of activity on the inner hemispheres could, alternatively, 
be induced by tidal interactions and may therefore not require any interconnecting magnetic fields
\citep{culhane90}.

Eclipse modulation also confines the latitudes $b$ of the eclipsed material. For AR Lac, 
$b = 10^{\circ}-40^{\circ}$ (\citealt{walter83b}, and similarly in \citealt{white90}, 
and \citealt{ottmann93b}). \citet{bedford90} infer $-30^{\circ} \le b \le  +30^{\circ}$ from a deep eclipse 
on XY UMa. Most active regions on the dMe binary YY Gem (dM1e+dM1e) 
are concentrated around $\pm(30^{\circ}-50^{\circ})$ \citep{guedel01a}, in good agreement with Doppler imaging of 
surface active regions \citep{hatzes95}. The confinement of the inhomogeneities leads to 
the somewhat perplexing result that these most active stars reveal
``active'' X-ray filling factors of no more than 5--25\% despite their being in the
saturation regime (Sect.~\ref{rotation}; see \citealt{white90, ottmann93b}). 

\subsubsection{Thermal properties of coronal structures}

The presence of distinct compact and extended coronae may reflect the presence of different thermal
structures - in fact, the distinction may simply be due to different
scale heights if magnetic fields do not constrain the corona further. The average radial
density profile of YY Gem with a scale height of $\approx [0.1-0.4]R_*$ derived from eclipse reconstruction
is in good agreement with the pressure scale height of the one component of the  plasma that dominates
the X-ray spectrum \citep{guedel01a}. Compact sources are often inferred with size
scales comparable to solar active regions. Based on such arguments, the extended structures are
more likely to be associated with the hottest persistent plasma in active binaries
(\citealt{walter83b, white90, rodono99}, but see \citealt{singh96a} and 
\citealt{siarkowski96} for alternative views). Inferred pressures of up to $>$100~dynes~cm$^{-2}$ make these regions
appear like continuously flaring active regions \citep{walter83b, white90}.
Alternatively, they could contain low-density, slowly cooling gas ejected from large 
flares. This view would be consistent with radio VLBI observations. The latter have mapped
non-thermal flares that expand from compact cores to extended ($\gg 1R_*$) halos where they cool 
essentially by radiation (\citealt{guedel02d} and references therein). Conjecture about different 
classes of thermal sources is 
again not unequivocal, however. \citet{ottmann93b} and \citet{ottmann94} found equivalent 
behavior of soft and hard spectral components during eclipses in AR Lac and Algol, respectively, and argued in favor of a 
close spatial association of hot and cool coronal components regardless of the overall spatial extent. 
This issue is clearly unresolved.

\begin{figure}[t!] 
\centerline{
\hbox{
\resizebox{0.55\textwidth}{!}{\includegraphics{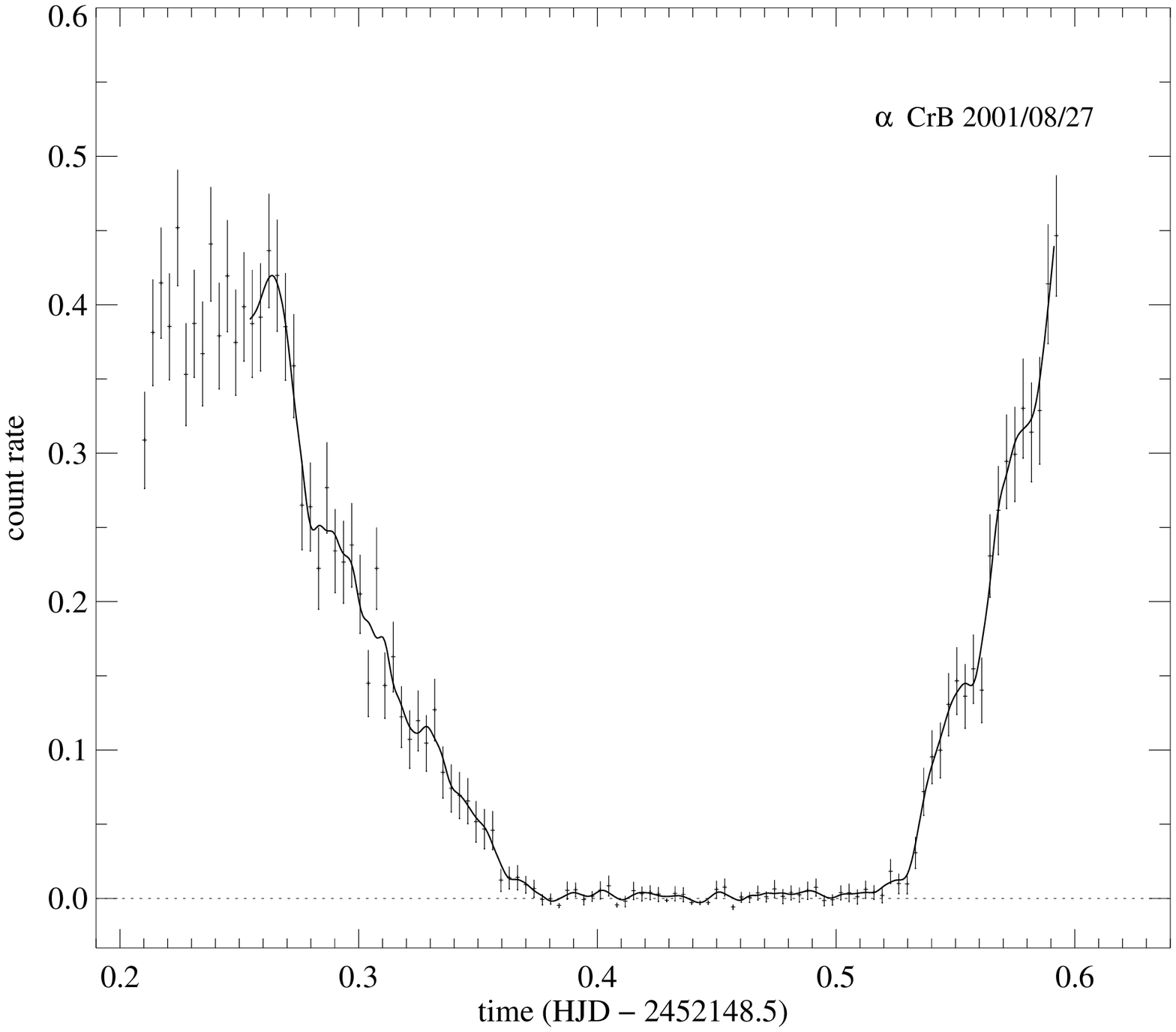}}
\resizebox{0.45\textwidth}{!}{\includegraphics{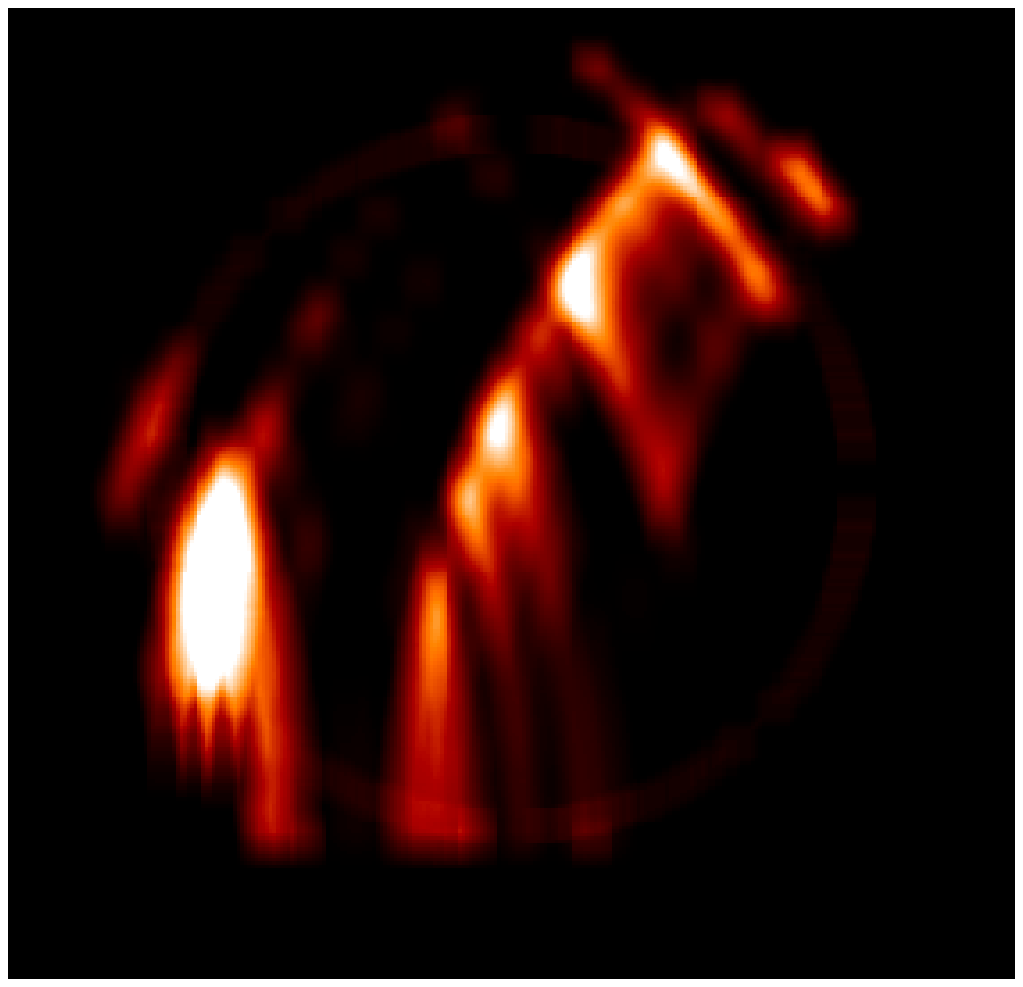}}
}
}
\vskip 0.2truecm\caption{Light curve and image reconstruction of the A+G binary
$\alpha$ CrB. The left panel shows the light curve from observations with {\it XMM-Newton}, 
the right panel illustrates the reconstructed X-ray brightness distribution  on the G star (after 
\citealt{guedel03b}).}\label{eclipsefig2}
\end{figure}

\subsection{X-ray coronal structure in other eclipsing binaries}

Among wide, non-interacting eclipsing stars, $\alpha$ CrB provides a particularly well-suited
example because its X-ray active, young solar analog (G5~V) is totally eclipsed every 17 days
by the optical primary, an A0~V star that is perfectly X-ray dark. Other parameters are
ideal as well, such as the non-central eclipse, the eclipse time-scale of a few hours, and the relatively
slow rotation period of the secondary. Eclipse observations obtained by {\it ROSAT} \citep{schmitt93c}
and by {\it XMM-Newton} \citep{guedel03b} were used to reconstruct projected 2-D images
of the X-ray structure. They consistently reveal patches of active regions across the face of 
the G star; not much material is found significantly beyond its limb (Fig.~\ref{eclipsefig2}). The structures tend to be 
of modest size ($\approx 5\times 10^9$~cm), with large, X-ray faint areas in between, although the
star's luminosity exceeds that of the active Sun by a factor of $\approx$30.  
These observations imply moderately high densities in the 
emitting active regions, and both studies mentioned above yielded average electron densities in the
brightest active regions of a few $10^{10}$~cm$^{-3}$.

X-ray light curves of eclipsing contact binary systems have shown sharp dips \citep{vilhu86, gondoin04a}  that could be
interpreted as being due to compact sources probably located in the ``neck'' region
that connects the two stars \citep{vilhu86} although \citet{gondoin04a} inferred quite an extended corona from
static loop models. \citet{brickhouse98}, in contrast, placed
a very compact source, with an extent of order $10^8$~cm, near the polar region of the primary, 
complemented by a more extended low-density corona that contributes most of the
X-ray light.

\subsection{Inferences from rotational modulation}\label{rotmod}

The Sun often shows rather pronounced rotational 
modulation in X-rays as few active regions rotate into and out of view. 
Observations of X-ray rotational modulation are exceptional  among stars, one of the main 
reasons being that the X-ray brightest rapid rotators are highly active; such stars are
probably covered with numerous active regions, and intense flaring may further   
veil low-amplitude modulations.  

\begin{figure} 
\centerline{
\hbox{
\resizebox{0.48\textwidth}{!}{\includegraphics{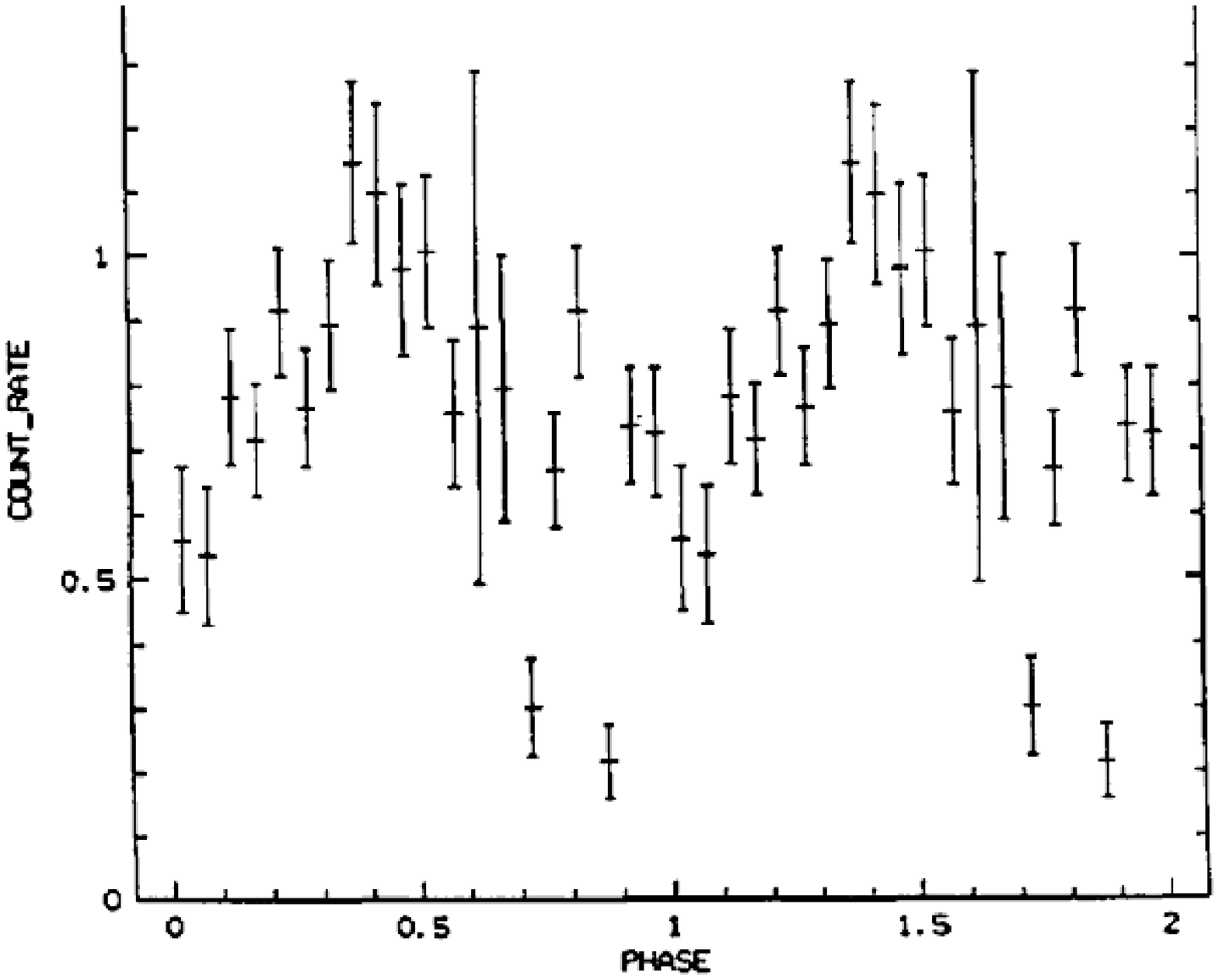}}
\hskip 0.5truecm\resizebox{0.52\textwidth}{!}{\includegraphics{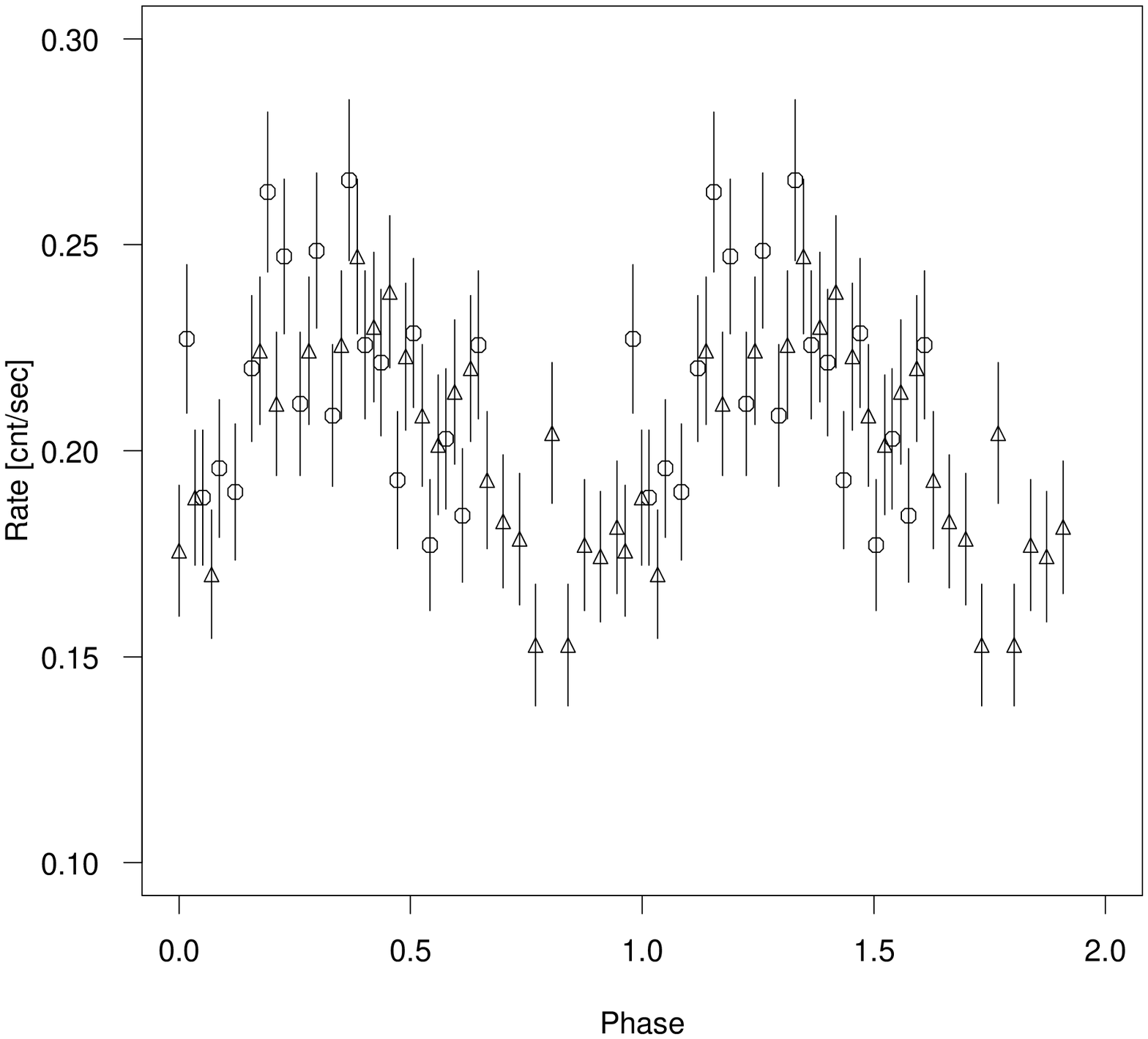}}
}
}
\caption{Two examples of X-ray rotational modulation in active solar analogs:
{\it a (left):} EK Dra \citep{guedel95}, and 
{\it b (right),} the supersaturated VXR45 (right figure courtesy of A. Marino, after \citealt{marino03a}). Both
light curves are phase-folded.}\label{rotmodstars}
\end{figure}

Among somewhat less active stars, the young solar analog EK Dra has shown rotational modulation 
both at X-ray and radio wavelengths \citep{guedel95}, and in X-rays it is predominantly the cooler material that 
shows this modulation. This argues against flares contributing to the signal. The depth and 
length of the modulation (Fig.~\ref{rotmodstars}a) constrains the X-ray coronal  height, and also the 
electron densities to $n_e > 4\times 10^{10}$~cm$^{-3}$, in agreement with spectroscopic measurements \citep{ness04}. 
This leads to the conclusion that much of the emitting material is concentrated in large ``active regions''. 
\citet{collier88b} reported a similar finding for a weak modulation in AB Dor: this  again 
suggested that the cooler loops are relatively compact. It is worthwhile mentioning, though, that 
X-ray rotational modulation has been
difficult to identify on this star (\citealt{whites96} and references therein; \citealt{vilhu93, maggio00, 
guedel01b}); a weak modulation has been reported by \citet{kuerster97}. Interestingly, radio observations of AB Dor
reveal two emission peaks per rotation that probably relate to preferred active longitudes \citep{lim92}. 

Because the X-ray luminosity in ``supersaturated stars'' (Sect.~\ref{rotation})
is also below the empirical maximum, rotational modulation would give important 
structural information on the state of such coronae. A deep modulation in VXR45 (Fig.~\ref{rotmodstars}b) 
suggests that extreme activity in these stars is again {\it not} due to complete coverage of the surface with active 
regions (\citealt{marino03a}).

Among  evolved stars,  the RS CVn binary HR~1099 (K1~IV + G5~V) 
has consistently shown X-ray and EUV rotational modulation, with an X-ray  maximum at 
phases when the larger K subgiant is in front \citep{agrawal88, drakej94, audard01a}.
Because the X-ray material is almost entirely located on the K star \citep{ayres01b}, the  
rotationally modulated material  can be located - in contrast to the binaries studied 
through eclipses - on the K star hemisphere 
that {\it faces away} from the companion (\citealt{audard01a}). A particularly clear example was presented by 
\citet{ottmann94} for Algol that showed a closely repeating pattern over three stellar orbits, 
testifying to the stability of the underlying coronal structure on time scales of 
several days. The strong modulation combined with a large inclination angle further 
suggests that most of the modulated material is located at moderate latitudes. \citet{schmitt96d} and
\citet{gunn97} found strong rotational modulation on  the RS CVn-type binary CF Tuc. Here, \citet{gunn97}  
speculated that the emitting corona is associated with the larger K subgiant and is facing toward
the smaller companion, thus again opening up a possibility for intrabinary magnetic fields.

Somewhat unexpectedly, even extremely active protostars appear to show signs
of coronal inhomogeneities.  \citet{kamata97} observed
sinusoidal variations in one such object and tentatively  interpreted it as the
signature of rapid ($P \approx 1$~d) rotation. If this interpretation is
correct, then once again we infer that these extremely
active stars  are not fully covered by coronal active regions.

\subsection{Rotationally modulated and eclipsed X-ray flares}\label{eclipseflare}

{\it Rotational modulation} of flares, or the absence thereof, contributes very valuable 
information on densities and the geometric size of flaring structures. 
\citet{skinner97} found compelling evidence for a rotationally  modulated flare on
the T Tau star V773 Tau. By making use of Eq.~(\ref{minimumvolume}), they inferred, independent of any
flare model, a minimum electron density of $2\times 10^{11}$~cm$^{-3}$ in the flaring region. This immediately 
implies that the decaying plasma is subject to continuous heating. Otherwise, the plasma would freely
cool on a time scale of $\approx 1.5$~hrs, an order of magnitude shorter than observed. Geometric
considerations then lead to a source of modest size at high latitudes. 
\citet{stelzer99} developed a ``rotating flare'' model that combines 
flare decay with self-eclipse of the flaring volume by the rotating star. They successfully fitted light curves of
four large stellar flares with slow rises and flat peaks. Alternative 
explanations for such anomalous light curves are possible, however.

If a long-lasting flare with time scales exceeding one orbit period shows no eclipse in its course,
then the flare either occurred near one of the polar regions, or it is geometrically 
large.  \citet{kuerster96} argued in favor of the latter possibility; they modeled a flare light-curve 
of CF Tuc. Nevertheless,  a partial eclipse may also have affected the decay of that flare. 
\citet{maggio00} suggested flaring loops located at latitudes higher
than 60$^{\circ}$ based on the absence of eclipse features during a large flare on 
AB Dor. This would allow for more modest flare sizes, a possibility that
is clearly supported by detailed flare modeling results (Sect.~\ref{flares}).

\begin{figure} 
\centerline{\hbox{\hskip 1truecm
\resizebox{0.7\textwidth}{!}{\includegraphics{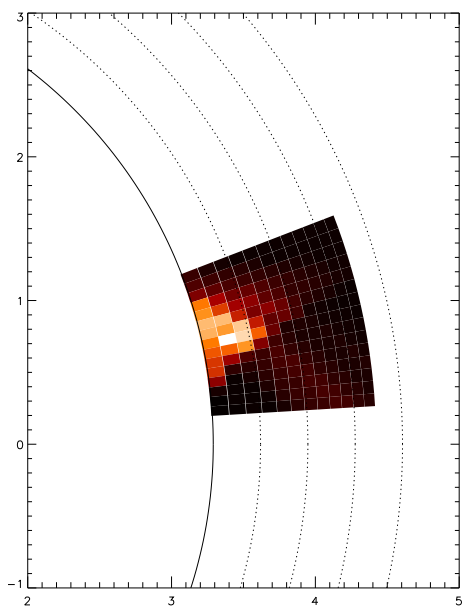}}
}}
\caption{Limb view of an X-ray flare on Algol, reconstructed from an 
eclipse light curve. Axis labels  are in units  of $R_{\odot}$. Dashed circles
give height in steps of 0.1$R_*$ (figure courtesy of J.-U. Ness, after \citealt{schmitt03}).}\label{algolflare}
\end{figure}

\citet{choi98} were the first to describe a full {\it eclipse} of an X-ray flare in progress in
a binary system, namely in the contact binary system VW Cep. During a narrow dip in the flare decay,
the X-ray flux returned  essentially to the pre-flare level. Geometric
considerations then placed the flare near one of the poles of the primary star, with a size
scale of order $5.5\times 10^{10}$~cm or somewhat smaller than the secondary star. 
The authors consequently inferred an electron density of $5\times 10^{10}$~cm$^{-3}$. 
A polar location was also advocated for a flare on Algol observed across an eclipse by 
\citet{schmitt99}. The  flare emission was again eclipsed completely, and judged from the known system geometry, 
the flare was located above one of the poles, with a maximum source height of no more than approximately 
0.5$R_*$, implying a minimum electron density of $9.4\times 10^{10}$~cm$^{-3}$ if the volume filling 
factor was unity. 
A more moderate flare was observed during an eclipse in the Algol system by \citet{schmitt03} (Fig.~\ref{algolflare}).
In this case, the image reconstruction required an equatorial location, with a compact flare source of
height $h \approx 0.1R_*$. Most of the  source volume exceeded densities of
$10^{11}$~cm$^{-3}$, with the highest values at $\approx 2\times 10^{11}$~cm$^{-3}$. Because
the {\it quiescent} flux level was attained throughout the flare eclipse, the authors
argued that its source, in turn, must be concentrated near the polar region with a modest filling
factor of $f < 0.1$ and  electron densities of $\approx 3\times 10^{10}$~cm$^{-3}$. 

\begin{figure}[t!] 
\centerline{
\resizebox{0.7\textwidth}{!}{\rotatebox{270}{\includegraphics{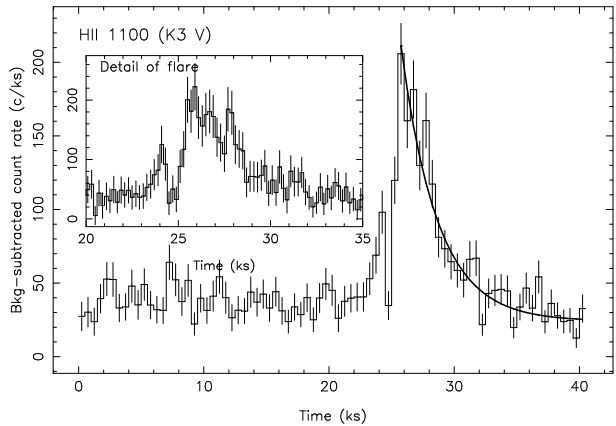}}}
}
\vskip 0.2truecm\caption{A flare on the Pleiades member HII 1100 that shows a short dip
early in the rise time, with the flux dropping back to the pre-flare level. One hypothesis
(that requires confirmation) refers to an eclipsing Jupiter-like planet (figure courtesy of K. Briggs, after 
\citealt{briggs03}, based on {\it XMM-Newton} observations).}\label{starplanet}
\end{figure}

Further candidates for eclipsed flares may 
be found in \citet{bedford90} for the short-period binary XY UMa as suggested by \citet{jeffries98a}, 
and in \citet{guedel01a} for the eclipsing M dwarf binary YY Gem. \citet{briggs03} reported
on an interesting eclipse-like feature in a flare in progress on an active Pleiades member
(Fig.~\ref{starplanet}). Again, the flux returned
precisely to pre-flare values for a short time, only to resume the increase of the flare light-curve 
after a brief interval. A somewhat speculative
but quite possible cause for the observed dip could also be a planet in transit. The existence of such a
planet  obviously needs confirmation. 
Some important
parameters related to the flares discussed in this section are summarized in Table~\ref{structuretable}.

\subsection{Inferences from Doppler measurements}\label{doppler}

\begin{figure} 
\centerline{\resizebox{0.85\textwidth}{!}{\includegraphics{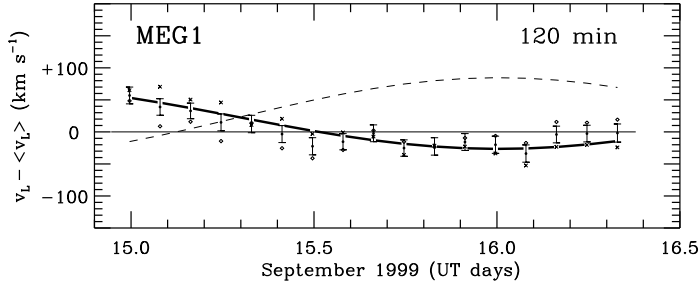}}}
\caption{Doppler motion in the HR~1099 system measured from line shifts in the {\it Chandra}
 HETGS/MEG spectrum. The predicted radial velocity curves are shown solid (K star) and dashed (G star).
 The measurements clearly locate the X-ray emission predominantly on the K star (figure courtesy  of T. Ayres, after
 \citealt{ayres01b}).}\label{hr1099doppler}
\end{figure}

Doppler information from X-ray spectral lines may open up new ways of imaging 
stellar coronae as they rotate, or as they orbit around the center of gravity in binaries.
First attempts are encouraging although the instrumental limitations are still severe.
\citet{ayres01b} found Doppler shifts with amplitudes of $\approx 50$~km~s$^{-1}$
in X-ray lines of HR~1099 (Fig.~\ref{hr1099doppler}). Amplitudes and phases clearly agree with the line-of-sight orbital 
velocity of the subgiant K star, thus locating the bulk of the X-ray emitting plasma  on this
star, rather than in the intrabinary region.
Periodic line {\it broadening} in YY Gem, on the other hand, suggests that both components are 
similarly X-ray luminous \citep{guedel01a}; this is expected, because this binary
consists of two almost identical  M dwarfs. \citet{huenemoerder03}  found Doppler motions in
AR Lac to be compatible with coronae on both companions if the plasma is close to the photospheric 
level. For the  contact binary 44i Boo, \citet{brickhouse01}  reported periodic line 
shifts corresponding to a total net velocity change over the full orbit of 180 km~s$^{-1}$. From the 
amplitudes and the phase of the rotational modulation \citep{brickhouse98}, 
they concluded that two dominant X-ray sources were present, one being very compact and the other
being extended, but both being located close to the stellar pole of the larger companion. 

A rather new technique employs {\it coronal} forbidden lines in the UV or optical range, making use of
spectral resolving powers that are still out of reach to  X-ray astronomy. \citet{maran94} and \citet{robinson96}
were the first to report the detection of the Fe\,{\sc xxi} $\lambda1354$ line in {\it HST} observations of
active stars. \citet{linsky98} presented a detailed analysis of this line in the Capella system.
\citet{ayres03b} summarized the current status and presented a survey of further possible
UV line candidates. In the UV range, the second promising candidate is the Fe\,{\sc xii}~$\lambda1349$ line
while few other transitions are sufficiently strong for detection. An analogous study  for the far-UV range was 
presented by \citet{redfield03}. In the UV range accessible from the ground, \citet{schmitt01} for the first time recorded the Fe\,{\sc xiii} 
$\lambda$3388.1 line formed at 1.6~MK in a spectrum of the dMe dwarf CN Leo. 

The collected results from these observations
are still modest - compared to the X-ray bibliography! Tentative results seem promising for further
investigation: 
\begin{description}
\item [1.] the lines observed so far are essentially only thermally and rotationally broadened, i.e., significant 
bulk Doppler shifts due to mass flows in flares have not (yet) been detected. 
\item [2.] several very active
stars appear to show some excess broadening; it could possibly be due to rotational velocities of extended coronal regions 
located high above the stellar surface. Line broadening may in this case  provide some important information 
on the overall coronal size \citep{ayres03b, redfield03}.
\end{description}

\subsection{Inferences from surface magnetic fields}\label{zeeman}
 
Information on coronal structure can also be derived indirectly from surface Zeeman-Doppler 
images as developed for and applied to the stellar case by \citet{jardine02a}, \citet{jardine02b}, and 
\citet{hussain02} and further references therein. Because Zeeman-Doppler images provide
the radial and azimuthal magnetic-field strengths as a function of position, one could
in principle derive the coronal magnetic field structure by 3-D extrapolation. This 
requires a number of assumptions, however. \citet{jardine02a} and \citet{jardine02b} studied the case of potential field
extrapolation, i.e., the coronal magnetic field follows $\mathbf{B} = -\nabla\Psi$, where
$\Psi$ is a function of the coordinates. Because the
field must be divergence free, one requires $\nabla^2\Psi = 0$. The solution involves associated
Legendre  functions in spherical coordinates, and the boundary conditions, namely the measured
magnetic field strengths on the surface, fix the coefficients. 

\begin{figure} 
\centerline{
\hbox{
\resizebox{0.5\textwidth}{!}{\includegraphics{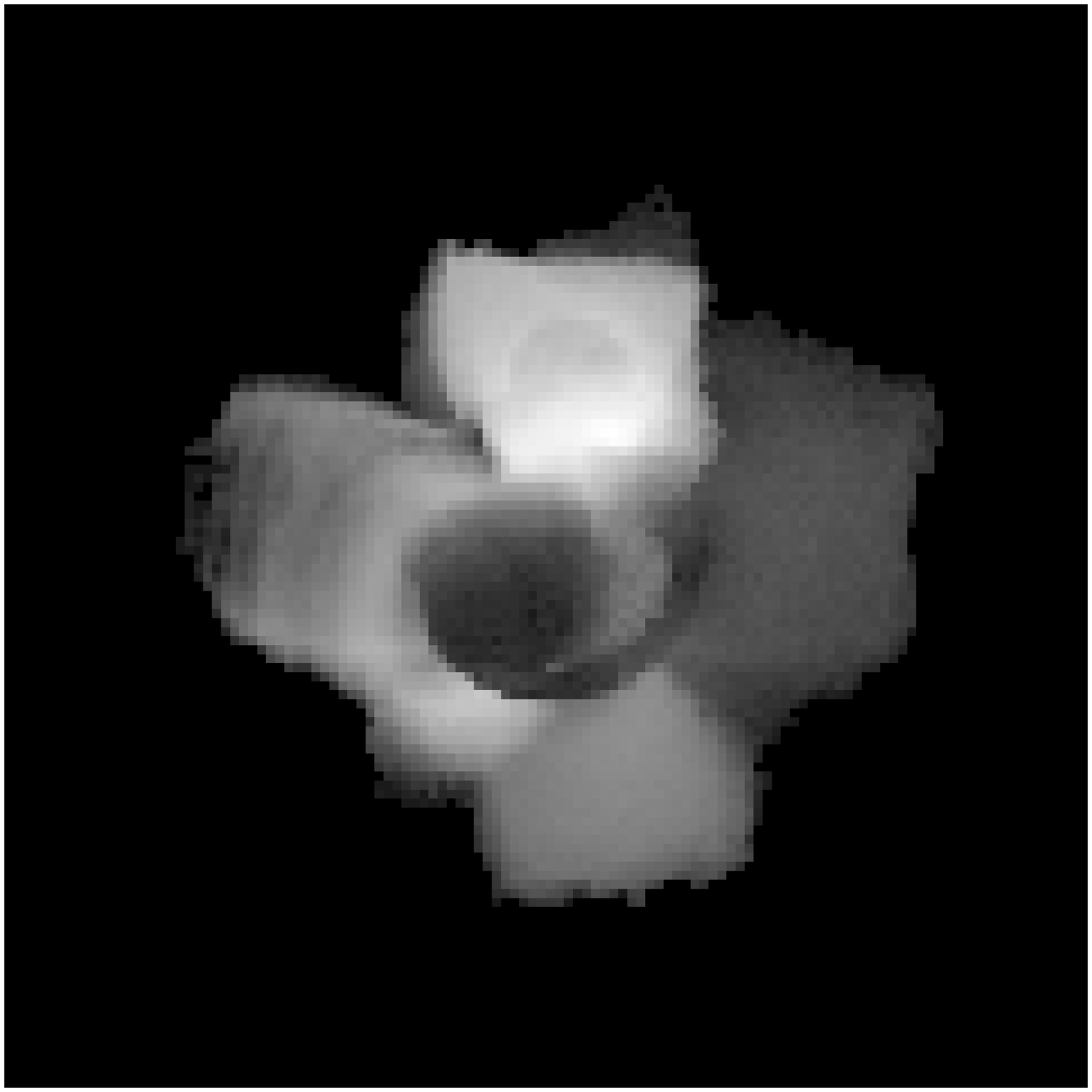}}
\resizebox{0.5\textwidth}{!}{\includegraphics{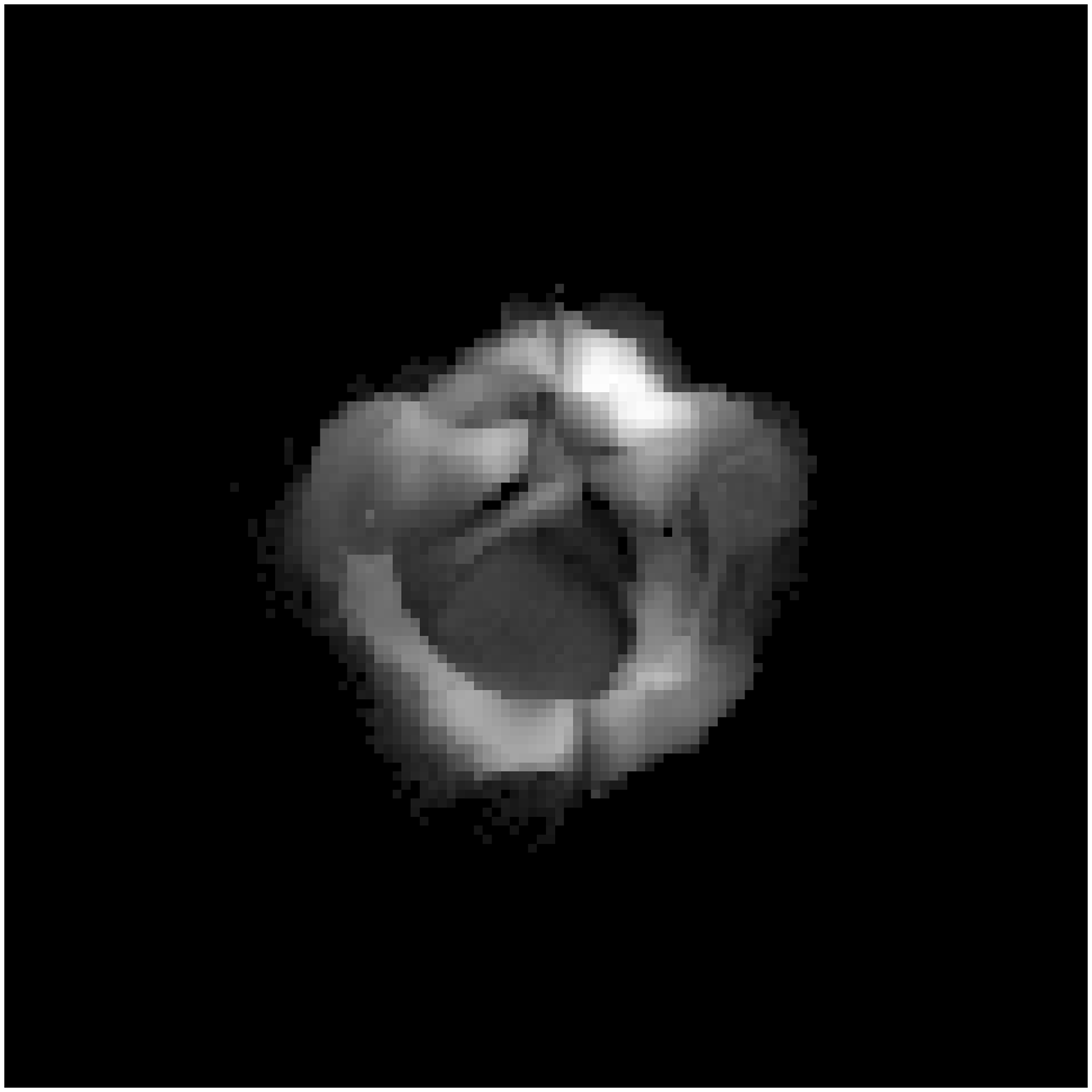}}
}
}
\caption{Reconstruction of the X-ray corona of AB Dor based on extrapolation from Zeeman Doppler
Imaging. Brightness encodes emission measure along the line of sight. The left panel  shows 
a solution for an EM-weighted density of $4\times 10^8$~cm$^{-3}$, the right panel for
$1.5\times 10^{10}$~cm$^{-3}$ (figure courtesy of M. Jardine, after  \citealt{jardine02b}).}\label{jardine}
\end{figure}

The model requires further 
parameters such as the base thermal pressure with respect to the local magnetic pressure, and
some cutoff of the corona at locations where the thermal pressure might open up the
coronal field lines. Various computed models (Fig.~\ref{jardine}) recover, at least qualitatively, the total EM,
the average density, and the low level of rotational modulation observed on very active stars 
such as AB Dor. The highly complex 
coronal structure, involving both very large magnetic features and more compact loops 
anchored predominantly at polar latitudes (as implied by the Doppler images), 
suppresses X-ray rotational modulation to a large extent. 

The modeling is  delicate
because i) part of the star cannot usually be Doppler imaged, ii) fine structures are not
recognized in the available Doppler images, in particular so in dark areas, and iii) the active
corona is not in a potential configuration. The first point was addressed by \citet{jardine02a} and \citet{jardine02b}
and the last by \citet{hussain02} who extended the models to include some form of 
currents in force-free fields. It is interesting that the models show various locations where the
gravity, centrifugal, and Lorentz forces are in equilibrium. These are the places where distant 
H$\alpha$ prominences may condense, for which there is indeed evidence 
in AB Dor out to distances of $5R_*$ \citep{collier88a, donati99}.

\subsection{Extended or compact coronae?}\label{extendedcompact}

As the previous discussions imply, we are confronted with mixed evidence for predominantly extended 
(source height $>R_*$) and  predominantly compact ($\ll R_*$) coronal structures or a 
mixture thereof. There does not seem to be unequivocal agreement on the type of structure
that generally prevails. Several trends can be recognized, however, as summarized below. 

{\it Compact coronal structure.} Steep (portions of the) ingress and egress
light curves or prominent rotational modulation unambiguously argue in favor of short 
scale lengths perpendicular to the line of sight (e.g., for the G star in AR Lac, see 
\citealt{walter83b, white90, ottmann93b}; or for $\alpha$ CrB, see \citealt{schmitt93c, guedel03b}). 
Common to all are relatively high inferred densities ($\approx 10^{10}$~cm$^{-3}$). The pressures of
such active regions may exceed pressures of non-flaring solar active regions by up to two orders 
of magnitude.  Spectroscopic observations of high densities and loop modeling
add further evidence for the presence of some rather compact sources 
(Sect.~\ref{coronaldensities} and \ref{loopmodels}). Flare modeling also provides modest sizes,
often of order 0.1--1$R_*$, for the involved magnetic loops (Sect.~\ref{flares}).

{\it Extended structure.} Here, the arguments are less direct and are usually
based on the absence of deep eclipses, or very shallow ingress and egress curves
(e.g., the K star in AR Lac, see \citealt{walter83b, white90}). Caution is in order in
cases where the sources may be located near one of the polar regions; in those cases, eclipses and
rotational modulation may also be absent regardless of the source size. Complementary
information is available from flare analysis (see Sect.~\ref{flares}) that in some cases 
does suggest quite large loops. The caveat here is that simple single-loop models may not
apply to such flares.
Clear evidence is available from radio interferometry that proves the
presence of large-scale, globally ordered magnetic fields (see references in \citealt{guedel02d}). 
The existence of prominent extended, closed magnetic fields on scales $>R_*$ is therefore 
{\it also} beyond doubt for several active stars.

The most likely answer to the question on coronal structure size is 
therefore an equivocal one: Coronal magnetic structures follow a size distribution from
very compact to extended ($\ga R_*$) with various characteristic densities,
temperatures, non-thermal electron densities, and surface locations. This
is no different from what we see on the Sun even though various features observed on stars
stretch the comparison perhaps rather too far for comfort: various structures may
predominate, depending on the magnetic activity level, on the depth of the convection zone,
or on binary characteristics.

\section{Stellar X-ray flares}\label{flares}

Flares arise as a consequence of a sudden energy release and relaxation process of the magnetic field in 
solar and stellar coronae. Present-day models assume that the energy is accumulated and
stored in non-potential magnetic fields prior to an instability that most likely implies
reconnection of neighboring antiparallel magnetic fields. The energy is brought
into the corona by turbulent footpoint motions that tangle the field lines at larger heights.
The explosive energy release becomes measurable across the electromagnetic spectrum
and, in the solar case, as high-energy particles in interplanetary space as well. For a review, I refer the
reader to \citet{haisch91a}. 

Flares are ubiquitous among coronal stars, 
with very few exceptions. Apart from the entire main sequence \citep{schmitt94c},
flares have been found among giants (\citealt{welty94} for FK Com) and hybrid stars \citep{kashyap94, huensch95}, and 
in clump giants both pre-He-flash \citep{ayres99} and post-He-flash (\citealt{ayres01a}, Fig.~\ref{flareexamples} below), 
partly with extremely long  time-scales of up to several days and with signs of continued activity.

Flares have prominently figured in solar studies, and it
is once again solar physics that has paved the way to the interpretation of stellar 
flares, even if not all features are fully understood yet. The complexity that flares reveal to
the solar astronomer is inaccessible in stellar flares, especially in the absence of
spatially resolved observations. Simplified concepts, perhaps tested for solar examples, must suffice. The
following sections summarize the ``stellar astronomer's way'' of looking at flares.

\subsection{General properties and classifications}\label{flareclass}

X-ray flares on the Sun come, roughly speaking, in two varieties, known as {\it compact} and {\it long-duration}
(also ``two-ribbon'') flares \citep{pallavicini77}. The former variant shows a simple structure,
usually consisting of one or a few individual loops that brighten up on time scales of minutes. They are 
of modest height and show high densities. The most likely mechanism leading to compact flares is an interaction
between neighboring loops. In contrast, the second type of flare shows decay time scales of up to 1--2
hours; their magnetic field structures are large ($10^4-10^5$~km) and the densities are low. Complex loop
arcades that are anchored in two H$\alpha$ ribbons are regularly involved. The most likely flare 
process relates to an opening up of magnetic fields (e.g., by a filament eruption) and subsequent relaxation 
by closing the ``open'' field lines. Related flare classifications have been made for hard X-rays as well.

Applications of the solar classification scheme to stars, in particular magnetically
active stars, should be treated with caution. The possibility of intrabinary magnetic
fields or significant tidal effects in close binary stars, extremely dense packing of
magnetic fields, polar magnetic fields, magnetic fields from a distributed dynamo, 
magnetospheres of global dimensions, and star-disk magnetic fields in pre-main sequence 
stars may lead to energy release configurations that are unknown on the Sun. The comprehensive
{\it EXOSAT} survey of MS stellar flares by \citet{pallavicini90a} testifies to this problem, with
some flares showing more rapid decays than rises, multiple peaks, abrupt drops etc.

Nevertheless, these authors did find evidence for a class of flares
with short rise times of order of minutes and decay times of order of a few tens of minutes, proposed to be analogs of
solar {\it compact flares}, and flares with long decay times exceeding one hour, reminiscent
of two-ribbon (2-R) flares on the Sun.
Flares with a rapid rise on time scales smaller than the dynamical time scale of filament
eruption are unlikely to be 2-R flares but are more suggestive of compact flares
\citep{vdoord88}, while longer rise times may indicate 2-R flares (e.g., \citealt{vdoord89}).
Typical e-folding decay times of large flares on active stars are found to be several kiloseconds 
\citep{gotthelf94, monsignori96, sciortino99, maggio00}. Decay times of 5--15~ks 
are not exceptional for young stars, e.g., in the Pleiades or in star-forming regions \citep{gagne95a, stelzer00a}.
In any case, the natural approach to understanding stellar coronal X-ray flares has been 
to extend solar concepts to stellar environments. 
I will now first discuss several flare models that have been repeatedly applied to observations in
the literature;  the subsequent sections will summarize a few notable results derived from 
stellar flare observations.

\subsection{General flare scenario}\label{flaremodel}

A ``standard picture'' has emerged from numerous solar flare studies, comprising roughly 
the following features. The flare reconnection region, located somewhere at large coronal heights, primarily 
accelerates electrons (and possibly ions) up to MeV energies.  The accelerated electrons precipitate
along the magnetic fields into the chromosphere where they heat  the cool plasma to coronal 
flare temperatures, thus ``evaporating'' part of the chromosphere into the corona. The high-energy electron
population emits radio gyrosynchrotron radiation and, upon impact, non-thermal hard X-ray (HXR) 
brems\-strah\-lung, and it generates optical continuum+line radiation. These emissions are well correlated on time
scales as short as seconds (e.g., \citealt{hudson92} for HXR and white light
flares; \citealt{kosugi88} for hard X-rays and gyrosynchrotron emission). The soft X-rays, in 
contrast, develop only as the closed magnetic fields are filled with plasma on
time scales of tens of seconds to minutes. Various elements of this scenario may vary from flare to flare.

From the X-ray point of view, the above model implies a characteristic evolution of flare EM and temperature.
As the initial energy release suddenly heats part of the chromospheric plasma, very high
temperatures are reached rapidly. As large amounts of plasma are streaming into
the corona, cooling starts while the luminosity is still increasing as a consequence
of increasing densities. The flare temperature thus peaks before the EM does, or analogously, harder emission
peaks before softer emission, a feature that is regularly
observed  in solar and stellar flares\footnote{This should  not be mistaken for the {\it Neupert Effect} 
(Sect.~\ref{neuperteffect});
the present effect is entirely due to plasma cooling while the Neupert effect involves a 
physically different non-thermal population of electrons. The different hard and soft light-curves
do not per se require non-thermal plasma, nor multiple components.}
 \citep{landini86, haisch87, vilhu88, doyle88a, vdoord88, 
vdoord89, haisch90b, stern92a, monsignori96, guedel99, maggio00, guedel04} and in numerical 
simulations of flares (e.g., \citealt{cheng91, reale04}).

\subsection{Cooling physics}

Flares cool through radiative, conductive, and possibly also
volume expansion processes. We define the flare decay phase as the episode when
the net energy loss by cooling exceeds the energy gain by heating, and
the total thermal energy of the flare plasma  decreases. The thermal energy decay
time scale $\tau_{\rm th}$ is defined as
\begin{equation}\label{decay}
\tau_{\rm th} = {E\over \dot{E}}
\end{equation}
where $E \approx 3n_e kT$  is the total thermal energy density
in the flaring plasma of electron density $n_e$ and temperature $T$,
and $\dot{E}$  is the volumetric cooling loss rate (in erg~cm$^{-3}$~s$^{-1}$). 
For conduction across temperature gradients in parallel magnetic fields, the
mean loss rate per unit volume is
\begin{equation}\label{condloss}
\dot{E_c} = {1\over L}\kappa_0 T^{5/2}{dT\over ds} \approx {4\over 7L^2}\kappa_0 T^{7/2} 
\end{equation}
where $s$ is the coordinate along the field lines, and the term $\kappa_0 T^{5/2}dT/ds$ 
is the conductive flux in the approximation of \citet{spitzer62}, to be evaluated near the loop 
footpoint where $T$ drops below $10^6$~K, with 
$\kappa_0 \approx 9\times 10^{-7}$~erg~cm$^{-1}$s$^{-1}$K$^{-7/2}$. 
Eqs.~(\ref{decay}, \ref{condloss}) define the {\it conductive time scale} $\tau_{\rm th} \equiv \tau_c$.
The second equation in (\ref{condloss}) should be used only as an approximation for non-radiating
loops with a constant cross section  down to the loss region
and with uniform heating (or for time-dependent cooling of a constant pressure loop without
heating; for the factor of 4/7, see \citealt{dowdy85, kopp93}).
We have used $L$ for the characteristic dimension of the source along the
magnetic field lines, for example the half-length of a magnetic loop. 
Strictly speaking, energy is not lost by conduction but is
redistributed within the source; however, we consider energy lost when it is conducted
to a region that is below X-ray emitting temperatures, e.g., the 
transition region/chromosphere at the magnetic loop footpoints.
For expressions relevant for loops with varying cross sections,
see \citet{vdoord89}.

Radiative losses are by bremsstrahlung (dominant for $T \ga 20$~MK), 2-photon continuum,  
bound-free, and line radiation.  We note that the plasma composition in terms
of element abundances can modify the cooling function $\Lambda(T)$, but the correction 
is of minor importance because stellar flares are usually rather hot. At relevant
temperatures, the dominant radiative losses are by bremsstrahlung, which is little
sensitive to modifications of the heavy-element abundances. The energy loss rate is
\begin{equation}
\dot{E_r} =  n_en_H\Lambda(T)
\label{radloss}
\end{equation}
 (or $n_e^2\Lambda^{\prime}(T)$ in an alternative definition, with $n_{\rm H} \approx
0.85n_e$ for cosmic abundances). For $T \ge 20$~MK, 
$\Lambda(T) = \Lambda_0T^{\gamma} \approx 10^{-24.66}T^{1/4}$~erg~cm$^3$~s$^{-1}$ (after \citealt{vdoord89} and \citealt{mewe85}).
Eqs.~(\ref{decay}, \ref{radloss}) define the {\it radiative time scale} $\tau_{\rm th} \equiv \tau_r$.

\subsection{Interpretation of the decay time}\label{decaytime}

Equations (\ref{decay}), (\ref{condloss}), and (\ref{radloss}) describe the decay of the thermal
energy, which in flare plasma is primarily due to the decay of  temperature
(with a time scale $\tau_T$) and density. In contrast, the observed light curve decays 
(with a time
scale $\tau_d$ for the {\it luminosity}) primarily due to the decreasing EM and, to a 
lesser extent, due to the decrease  of $\Lambda(T)$ with decreasing temperature above $\approx 15$~MK. 
From the energy equation, the thermal energy decay time scale $\tau_{\rm th}$ is found to be
\begin{equation}\label{decaylc}
{1\over \tau_{\rm th}} = \left(1-{\gamma\over 2}\right){1\over \tau_T} + {1\over 2\tau_d}
\end{equation}
where the right-hand side is usually known from the observations (see \citealt{vdoord88} for
a derivation). The decay time scale
of the EM then follows as $1/\tau_{\rm EM} = 1/\tau_d - \gamma/\tau_T$.
\citet{pan97} derived somewhat different coefficients in Eq.~(\ref{decaylc}) for the assumption of  
constant volume or constant mass, including the enthalpy flux.
In the absence of  measurements of $\tau_T$, it is often assumed that $\tau_{\rm th} = \tau_d$ 
although this is an inaccurate approximation. For  a freely cooling loop, $\tau_{\rm EM}
= \tau_T$ (Sect.~\ref{contheat}), and a better replacement is therefore 
$\tau_{\rm th} = 2(\gamma + 1)\tau_d/3$.

In Eq.~(\ref{decaylc}), $\tau_{\rm th}$ is usually set to be $\tau_r$ or $\tau_c$ or,
if both loss terms are significant, $(\tau_r^{-1} + \tau_c^{-1})^{-1}$, taken at the beginning
of the flare decay (note again that a simple identification of $\tau_r$ with $\tau_d$ is not accurate). 
If radiative losses dominate, the density immediately follows
from Eqs.~(\ref{decay}, \ref{radloss})
\begin{equation}\label{cooltime}
\tau_{\rm th} \approx {3kT\over n_e\Lambda(T)}
\end{equation}
and the characteristic size scale $\ell$ of the flaring plasma or the flare-loop
semi-length $L$ for a sample of $N$ identical loops follow from
\begin{equation}
{\rm EM} = n_en_H (\Gamma + 1)\pi\alpha^2 N L^3 \approx n^2\ell^3 
\end{equation}
where $\alpha$ is the aspect ratio (ratio between loop cross sectional diameter at the base  
and total length $2L$) and $\Gamma$ is the  loop expansion factor.
The loop height for the important case of dominant radiative losses follows 
to be \citep{white86, vdoord88} 
\begin{equation}
H = \left({8\over 9\pi^4}{\Lambda_0^2\over k^2}\right)^{1/3}\left({{\rm EM}\over T^{3/2}}\tau_r^2\right)^{1/3}
   \left(N\alpha^2\right)^{-1/3}(\Gamma+1)^{-1/3}.
\end{equation}
A lower limit to $H$ is found for  $\tau_r \approx \tau_c$ in the same treatment:
\begin{equation}
H_{\rm min} = {\Lambda_0 \over \kappa_0\pi^2}{{\rm EM}\over T^{3.25}}\left(N\alpha^2\right)^{-1}.
\end{equation}
$N$, $\alpha$, and $\Gamma$ are usually unknown and treated as free parameters within
reasonable bounds. Generally, a small $N$ is compatible with dominant radiative cooling.
   
This approach gives a first estimate for  the flare-loop size, but it provides only 
an upper limit to $\ell$ and a lower limit to $n_e$,  for the following reason. Equation
(\ref{cooltime}) assumes free cooling without a heating contribution. If heating continues
during the decay phase, then $\tau_{\rm obs} > \tau_{\rm th}$, hence the implied
$n_{e, {\rm obs}} < n_e$; in other words, the effective cooling function $\Lambda$ is reduced and, therefore,
the apparent $\ell_{\rm obs} > \ell$. The effect of continued heating will be discussed in 
Sect.~\ref{contheat} and \ref{2R}.

Decay time methods have been extensively used by, among others, \citet{haisch80}, \citet{vdoord88}, \citet{jeffries90},
\citet{doyle91}, \citet{doyle92b}, \citet{ottmann94b}, \citet{mewe97}, and \citet{osten00} 
for the interpretation of large flares. For small $N$, most authors found
loop heights of the order of a few $10^{10}$~cm and inferred densities of 
a few times $10^{11}$~cm$^{-3}$ (see Sect.~\ref{obsflares}).
\citet{pallavicini90a} inferred typical flare densities, volumes, and magnetic loop lengths 
for various strong flares on M dwarfs, concluding that $n_e$
tends to be higher than electron densities in solar compact flares, while the  volumes are more reminiscent
of solar two-ribbon (2-R) flares.

The simplest approach involving full coronal loop models assumes
cooling that is completely governed either by conduction or radiation. 
\citet{antiochos76, antiochos78} have treated a conductively-driven flaring loop, first with
static and then with evaporative cooling, i.e., respectively, without mass flow and with subsonic mass upflows (under
time-constant pressure). Under the assumption that radiation is negligible, Antiochos \& Sturrock
obtained as loop-apex temperatures 
\begin{equation}
T_{\rm stat}(t) = T_0\left(1 + {t\over \tau_{c,0}}\right)^{-2/5}, \quad\quad 
T_{\rm evap}(t) = T_0\left(1 + {t\over \tau_{c,0}}\right)^{-2/7} 
\end{equation}
Here, the relevant timescale is $\tau_{c,0} = 5p_0/(2\kappa_0T_0^{7/2})(L/1.6)^2$ \citep{cargill94}, and
the subscript 0 refers to values at the beginning of the cooling phase.
This formulation has been used by \citet{pan97} as part of an extended flare model involving variable 
mass or variable volume.
For the more likely case of dominant radiative cooling, \citet{antiochos80} and 
\citet{cargill94} give, for static cooling and cooling with
subsonic draining, respectively, a temperature at the position $s$ along the loop of 
\begin{equation}
T(s, t) = \left\{ 
   \begin{array}{ll}
     T_0(s)\left[1 - (1-\gamma)\displaystyle{{t\over \tau_{r,0}}}\right]^{1/(1-\gamma)}                                    & \mbox{\quad static} \\
     T_0(s)\left[1 - \displaystyle{{3\over 2}}\left(\displaystyle{{1\over 2}}-\gamma\right)\displaystyle{{t\over \tau_{r,0}}}\right]^{1/(1/2-\gamma)} & \mbox{\quad draining} 
   \end{array}
   \right. 
\end{equation}
where $\tau_{r,0}$ is the radiative loss time (Eq.~\ref{cooltime}) at the cooling onset.
 
A more general treatment was considered by \citet{pan95} for a flare on the active main-sequence binary
CC Eri, including both radiative and conductive losses. They gave a differential equation
for the development of the temperature as a function of the measured EM and $T$, which then serves
to determine $L$ in the conductively-dominated case, and the flaring volume in the radiative case.

\subsection{Quasi-static cooling loops}

\citet{vdoord89} derived the energy equation of a cooling magnetic loop
in such a way that it is formally identical to a static loop \citep{rosner78},
by introducing a slowly varying flare heating rate that balances the total
energy loss, and a possible constant heating rate during the flare decay. This specific
solution thus proceeds through a sequence of different (quasi-)static loops
with decreasing temperature.

The general treatment  involves continued heating that keeps the cooling loop 
at coronal temperatures. If this constant heating term is zero, one finds
for free quasi-static cooling
\begin{eqnarray}
T(t) &=& T_0(1 + t/3\tau_{r,0})^{-8/7}   \\
L_r(t) &=& L_{r,0}(1 + t/3\tau_{r,0})^{-4}  \\
n_e(t) &=& n_{e,0}(1 + t/3\tau_{r,0})^{-13/7}.  
\end{eqnarray}
 where $L_r$ is the total radiative loss rate, and $\tau_{r,0}$  is the radiative 
loss time scale (Eq.~\ref{cooltime})  at the beginning of the flare decay.

This prescription is equivalent to requiring a constant ratio  between
radiative and conductive loss times, i.e., in the approximation of $T \ga 20$~MK 
($\Lambda \propto T^{1/4}$)
\begin{equation}\label{decayratio}
{\tau_r \over \tau_c} = {\rm const}{T^{13/4}\over {\rm EM}} \approx 0.18.
\end{equation}
Accordingly, the applicability of the quasi-static cooling approach can be supported
or rejected based on the run of $T$ and EM during the decay phase. Note, however, 
that a constant ratio (Eq.~\ref{decayratio}) is not a sufficient condition to fully justify this 
approach.

The quasi-static cooling model also predicts a particular shape of the
flare DEM \citep{mewe97} - see Eq.~(\ref{demflareint}).
\citet{vdoord89}, \citet{stern92a}, \citet{ottmann96}, \citet{kuerster96}, \citet{mewe97} 
and many others have applied this model. The approach has recently been criticized, however, 
because the treatment of heating is not physically self-consistent \citep{reale02b}.

\subsection{Cooling loops with continued heating}\label{contheat}

The characteristics of the flare decay itself strongly depend on the amount of
ongoing heating. Several models include continued heating with some prescription
(e.g., the quasi-static cooling approach, or two-ribbon models).
Continued heating in several large flares with a rapid rise has 
been questioned for cases where the thermal plasma energy content
at flare peak was found to be approximately equal to the total radiative energy
during the complete flare. If that is the case, the flare 
energy has been deposited essentially before the flare peak
\citep{vdoord88, jeffries90, tagliaferri91}. Other flares, however, exhibit evidence 
of reheating or continued heating during the decay phase (e.g., \citealt{tsuboi98}).

\begin{figure} 
\centerline{\resizebox{0.75\textwidth}{!}{\rotatebox{270}{\includegraphics{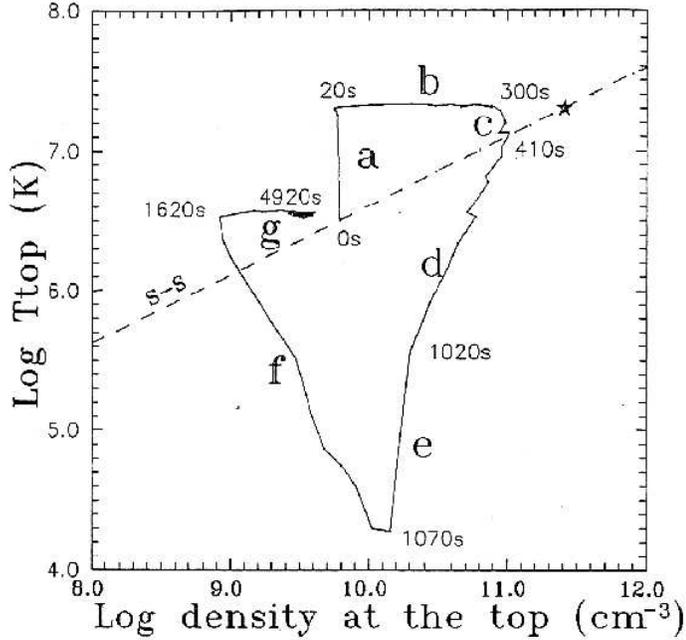}}}}
\caption{Density-temperature diagram of a hydrodynamically simulated flare.
           The flare loop starts from an equilibrium (S-S, steady-state loop according to
	   \citealt{rosner78}); (a) and (b) refer to the heating phase; at (c), the heating 
	   is abruptly turned off, after which the loop cools rapidly (d, e), and only slowly
	   recovers toward a new equilibrium solution (f, g) due to
	   constant background heating (from \citealt{jakimiec92}).}\label{jakimiec}
\end{figure}

Whether or not flaring loops indeed follow a quasi-static cooling path
is best studied on a density-temperature diagram (Fig.~\ref{jakimiec}). Usually, characteristic
values $T = T_{\rm a}$ and $n_e = n_{e, {\rm a}}$ as measured
at the loop apex are used as diagnostics. For a magnetic loop in hydrostatic
equilibrium, with constant cross section assumed, the RTV scaling law~(\ref{RTV})
requires stable solutions $(T, n_e)$ to be located 
where $T^2 \approx 7.6\times 10^{-7} n_e L$ (for $n_e = n_i$). On a diagram of 
log\,$T$ vs. log\,$n_e$, all solutions
are therefore located on a straight line with slope $\zeta = 0.5$. \citet{jakimiec92}
studied the paths of hydrodynamically simulated flares with different
heating histories. The initial rapid heating leads to a rapid increase of
$T$, inducing increased losses by conduction. As chromospheric evaporation 
grows, radiation helps to balance the heating energy input. 
The flare decay sets in once the heating rate drops. At this moment, 
depending on the amount of ongoing heating, the magnetic loop is too dense to 
be in equilibrium, and the radiative losses exceed the heating rate, resulting 
in a thermal instability. In the limit of no heating during the decay, that is, an abrupt 
turn-off of the heating at the flare peak, the slope of the path becomes
\begin{equation}
\zeta \equiv {d{\rm ln}~T\over d{\rm ln}~n_e} \equiv {\tau_n\over \tau_T} = 2  
\end{equation}
implying $T(t) \propto n_e^{\zeta}(t) =  n_e^2(t)$ (see \citealt{serio91} for further discussion).
Here, $\tau_T$ and $\tau_n$ are the e-folding decay times of the temperature and 
the electron density, respectively, under the assumption of exponential decay laws.
Only for a non-vanishing  heating rate 
does the loop slowly recover and eventually settle on a new equilibrium
locus (Fig.~\ref{jakimiec}). In contrast, if heating continues and is
very gradually  reduced, the loop decays along the static solutions ($\zeta = 0.5$). 
Observationally, this path is 
often followed by large solar flares \citep{jakimiec92}. Clearly, a  comprehensive 
description of stellar flares should thus consider continued heating throughout the flare.
However, the observables typically available to the stellar astronomer are  only 
the run of the X-ray luminosity (hence the EM) 
and of the characteristic temperature $T$. The volume and the 
density $n_e$ are unknown and need to be estimated from other parameters.

To this end, \citet{reale97} replaced $n_e$ by the observable $\sqrt{\rm EM}$
and thus assumed a constant flare volume, and further introduced the following generalization. 
In the freely cooling case after an abrupt heating turnoff,
the entropy per particle at the loop apex  decays on the thermodynamic decay time
\begin{equation}
\tau_{\rm td} = 3.7\times 10^{-4} {L\over T_0^{1/2}} \quad {\rm [s]}
\end{equation}
where $T_0$ is the flare temperature at the beginning of the decay (\citealt{serio91}; see
\citealt{reale93} for an extension to loops with $L$ up to a pressure scale height). 
For the  general case with continued heating, \citet{reale97} wrote
\begin{equation}\label{taulc}
\tau_{\rm LC} = 3.7\times 10^{-4} {L\over T_0^{1/2}}F(\zeta) \quad {\rm [s]}
 \end{equation}
where $F(\zeta) \equiv \tau_{\rm LC}/\tau_{\rm td}$ is a correction function 
depending on the heating decay time via $\zeta$, and $\tau_{\rm LC}$ is
the observed X-ray light curve decay timescale. $F(\zeta)$ is therefore to
be numerically calibrated for each X-ray telescope.

The important point is that $F(\zeta)$ has been empirically found from 
solar observations and numerical simulations to be a hyperbolic or exponential
function with three parameters that can be determined for a given instrument.
With known $F$, Eq.~(\ref{taulc}) can be solved for $L$. This scheme thus offers
i) an indirect method to study flaring loop geometries ($L$), ii) a way of  
determining the rate and decay time scale of continued heating via $F(\zeta)$
and $\tau_{\rm td}$, and iii) implications for the density decay time via 
$\tau_n = \zeta\tau_T$. Conditions of applicability include $\zeta \ge 0.3$ and  a
resulting loop length $L$ of less than one pressure scale height \citep{reale97}.

Loop sizes derived from this method agree with direct observations
on the Sun. Solar observations  of moderate flares  with the Solar Maximum Mission {\it SMM}
implied  $\zeta$ between $\sim$0.5 and 2 \citep{sylwester93}, although \citet{reale97} found
a predominance of values around $\sim 0.3 - 0.7$, i.e., flares with substantial sustained heating.

This method has been applied extensively in the interpretation of stellar flares. 
Modeling of large flares yields reasonable
flare loop sizes that are often smaller than those inferred from other methods. 
Loop sizes comparable with sizes of solar active regions have been found.  But 
such loops may also be of order of one stellar radius for M dwarfs such
as Proxima Centauri \citep{reale88, reale04}. The flaring region
may thus comprise a significant fraction of the stellar corona on such stars.
Giant flares with reliable measurements of temperatures are particularly well suited 
for an application of this method. Examples have been presented for EV Lac (a flare with a $>$100 fold 
increase in count rate in the {\it ASCA} detectors, reaching temperatures of $~70$~MK; 
\citealt{favata00a}, Fig.~\ref{flareexamples}), AB Dor \citep{maggio00}, and Proxima Centauri
\citep{reale04}. The same approach was also used for pre-main sequence stars by \citet{favata01}, including
comparisons with other methods.

The moderate loop sizes resulting from this method have important implications
for coronal structure (see Sect.\ref{coronalstructure}). The magnetic loops related to several flares observed
on AD Leo, for example, all seem to be of fairly similar, modest size (half-length $L \approx 0.3R_*$).
They are therefore likely to represent active region magnetic fields, although under this
assumption a quite small filling factor of 6\% is obtained (cf. \citealt{favata00b} for details).

Applications have also often shown that considerable heating
rates are present during the decay phase of large stellar flares (e.g., \citealt{favata00a, 
favata00b}). Values of $\zeta$ as small as 0.5 are frequently
found; this corresponds to a decay that is $\approx$ 3--5 times slower  
than predicted from free cooling on the thermodynamic time scale.

\subsection{Two-Ribbon flare models}\label{2R}
 
\begin{figure} 
\centerline{\resizebox{1\textwidth}{!}{\includegraphics{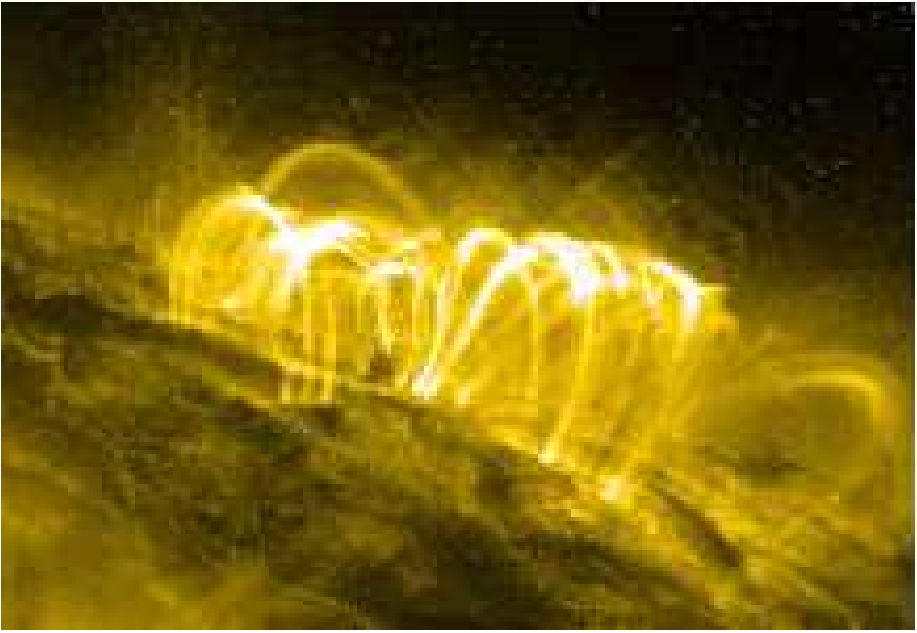}}}
\caption{{\it Trace} image of a flaring magnetic loop arcade.}\label{2R_Trace}
\end{figure}

An approach that is entirely based on continuous heating (as opposed to cooling) 
was developed for the two-ribbon (2-R) class of solar flares. An example of this flare type is 
shown in Fig.~\ref{2R_Trace}.
The 2-R flare model devised initially by \citet{kopp84}
is a parameterized magnetic-energy release model.
The time development of the flare light-curve is completely determined
by the amount of energy available in non-potential magnetic fields, and by
the rate of energy release as a function of time and geometry as the fields
reconnect and relax to a potential-field configuration.
Plasma cooling is not included in the original model; it is assumed
that a portion of the total energy is radiated into the observed X-ray band,
while the remaining energy will be lost by other mechanisms.  
An extension that includes approximations to radiative and conductive losses 
was described in \citet{guedel99} and \citet{guedel04}. 2-R flares are well 
established for the Sun (Sect.~\ref{flareclass}, Fig.~\ref{2R_Trace}); they often lead to large,   
long-duration flares that may be accompanied by mass ejections.  
 
The magnetic fields are, for convenience, described along meridional
planes on the star by Legendre polynomials $P_n$ of order $n$, up to the height of the
neutral point; above this level, the field is directed radially, that is, the
field lines are ``open''.  
As time proceeds, field lines nearest to the 
neutral line move inward at coronal levels and reconnect at progressively
larger heights above the neutral line. The reconnection point thus moves upward
as the flare proceeds, leaving closed magnetic-loop systems underneath. 
One loop arcade thus corresponds to one N-S aligned lobe between two zeros of $P_n$ in 
latitude, axisymmetrically continued over some longitude in E--W direction.
The propagation of the neutral point in height, $y(t)$, with a 
time constant $t_0$, 
is prescribed by ($y$ in units of $R_*$, measured from the star's center)
\begin{equation}\label{yt}
y(t) = 1 + {H_{\rm m}\over R_*}\left(1- e^{-t/t_0}\right)
\end{equation}
\begin{equation}\label{ht}
H(t) \equiv [y(t) - 1]R_*
\end{equation}
and the total energy release of the reconnecting arcade per radian in longitude is
equal to the magnetic energy lost by reconnection,
\begin{equation}
{{\rm d}E\over {\rm d}y} = {1\over 8\pi}2n(n+1)(2n+1)^2R_*^3B^2I_{12}(n)
        { y^{2n}(y^{2n+1} - 1)\over [n+(n+1)y^{2n+1}]^3} 
\end{equation}
\begin{equation}\label{dedt}
{dE\over dt} ={dE\over dy}{dy\over dt} 
\end{equation}
\citep{poletto88}. In Eq.~(\ref{yt}), $H_{\rm m}$ is the maximum height of 
the neutral point for $t\rightarrow \infty$; typically, $H_{\rm m}$  is assumed 
to be equal to the latitudinal extent of the loops, i.e.,
\begin{equation}\label{height}
H_{\rm m} \approx {\pi\over n+1/2}R_*
\end{equation}
for $n > 2$ and $H_{\rm m} = (\pi/2)R_*$ for $n = 2$.
Here, $B$ is the  surface magnetic field strength at the axis of symmetry,
and $R_*$ is the stellar radius. Finally, $I_{12}(n)$ corresponds to 
$\int [P_n({\rm cos}\theta)]^2d({\rm cos}\theta)$ evaluated between the
latitudinal borders  of the lobe (zeros of ${\rm d}P_n/{\rm d}\theta$), 
and $\theta$ is the  co-latitude. 

The free parameters are $B$ and the efficiency of the energy-to-radiation 
conversion, $q$, both of which determine the normalization of the light curve;
the time scale of the reconnection process, $t_0$, and the polynomial degree $n$ determine the 
duration of the flare; and the geometry of the flare is fixed  by 
$n$ and therefore the asymptotic height $H_{\rm m}$ of the reconnection point.
The largest realistic 2-R flare model is based on the Legendre polynomial of degree $n = 2$; 
the loop arcade then stretches out between the equator and the stellar poles. Usually,
solutions can be found for many larger $n$ as well. However, because a larger $n$ requires 
larger surface magnetic field strengths, a natural limit is set to $n$ within
the framework of this model. Once the model solution has been established, further 
parameters, in particular the electron density $n_e$, can be inferred.

The Kopp \& Poletto model is applicable 
after the initial flare trigger mechanism has terminated, although 
\citet{pneuman82} suggested   that
 reconnection may start in the earliest phase of loop
structure development. 

2-R models have been proposed for interpretation of stellar flares
on phenomenological grounds such as high luminosities, long
decay time scales, white-light transients, or X-ray absorption possibly by transits of  prominences,
by \citet{haisch81}, \citet{haisch83}, \citet{poletto88}, \citet{tagliaferri91}, \citet{doyle92b}, 
\citet{franciosini01}, \citet{guedel99}, \citet{guedel04}, to mention a few.  \citet{poletto88} emphasized 
the point that large solar  flares predominantly belong to this class rather than to compact, single-loop events. 
Many authors found overall sizes of order $R_*$, i.e., they obtained small $n$. 

Comparisons of 2-R model predictions with explicitly measured heating rates during large
flares indicate an acceptable  match from the rise to the early decay phase \citep{vdoord89} although the observed later
decay episodes are often much slower than any model parameter allows \citep{osten00, kuerster96, guedel04}.
Part of the explanation may be additional heating within the cooling loops.

\citet{katsova99} studied a
giant long-duration flare on the dMe star AU Mic. The observed decay time of this flare exceeded any plausible
decay time scale based on pure radiative cooling by an order of magnitude, suggestive of a 2-R flare. 
Although an
alternative model based on a very large, expanding CME-like plasma cloud at low densities 
has been presented \citep{cully94}, the spectroscopically measured high densities of the 
hot plasma argue for a coronal source undergoing significant post-eruptive heating. 
A very slowly evolving flare observed on the RS CVn-type binary HR~5110, with a decay time of several
days, suggested
large source scales but showed clear departures from any model fit; the authors
proposed that the flare was in fact occurring in intrabinary magnetic fields
\citep{graffagnino95}.

\subsection{Hydrodynamic models}\label{hydrodyn}

Full hydrodynamic calculations aim at solving the equations of mass,
momentum, and energy conservation comprehensively and under
quite general conditions, although certain simplifying initial conditions,
boundary conditions, and plasma physical approximations must be made.
The equations treat the plasma as a thermodynamic fluid, an approximation 
that is sufficiently good except for the critical footpoint 
region of magnetic loops where the mean free path of the electrons exceeds
the  temperature and pressure scale heights. In this region, non-local effects
such as saturation of the conductive flux due to the free-streaming limit to
the heat transport must be included. Treatment of time-dependent ionization
and recombination is also required  at chromospheric levels. The latter
define boundary conditions to the coronal simulations, therefore extensive chromospheric 
models must be included  (see, e.g., \citealt{reale88}). 
Radiation can usually be approximated sufficiently well in terms of a cooling 
function integrating all continuum, free-bound, and line losses at a given 
temperature in ionization equilibrium. The presence of
non-thermal, accelerated electron populations is often not treated, and 
the magnetic fields are assumed to be rigid, with the only purpose
of  confining the plasma. Heating terms are adjusted empirically, with a distribution
both in space and time. 

Accepting these approximations, state-of-the art simulations have given
deep insights into the flare process and the nature of spatially unresolved 
stellar X-ray flares. The methods have been comprehensively tested with,
and applied to, solar flares \citep{pallavicini83, reale95}. For a description
of various codes and numerical solar flare simulations, see \citet{nagai80}, 
\citet{peres82}, \citet{pallavicini83}, \citet{peres87}, and \citet{mariska89}. For
stellar applications, I refer the reader to \citet{cheng91}, \citet{katsova97}, \citet{reale02}, and \citet{reale04}.

For stellar flare simulations, there are two major unknowns. First, the energy input
can be in the form of high-energy particle beams or of direct heating, 
leading to different preferred heat source locations such as the footpoints or the loop 
apex. Fortunately, conduction equalizes effects due to spatially non-uniform
heating rapidly enough so that the choice is of limited relevance 
\citep{peres87, reale88}.  Second,  the basic magnetic configuration of the flares 
is unknown a priori, including initial conditions in the pre-flare plasma
such as pressure or density. Pure hydrodynamic models assume single flaring loops (e.g., 
\citealt{peres82}), or complexes of independent loops (e.g., \citealt{reale04}),  
while magnetohydrodynamic (MHD) codes  allow for an extension to interacting  loops or loop arcades 
(see below). By varying parameters, comparison of the simulations with
observed parameters such as the light curve and the temperature development eventually confine magnetic 
loop geometries, magnetic fields, heating durations, electron densities, and some of 
the initial conditions.

A parameter study appropriate for flares on dMe stars was performed by \citet{cheng91}. They pointed out
that multiple solutions exist for given light curves, but that all solutions fulfill an
overall linear relation between flare peak EM and total energy released in the flare, with good
agreement between the simulations and  the observations. This relation emerges because
larger energy release produces more extensive chromospheric evaporation. The trend saturates, however,
once the energy input becomes too large, because increased radiative losses suppress further
conductive evaporation. This, then,  suggests that {\it extremely large flares cannot be produced in
moderately-sized magnetic loops}. The simulations further showed plasma evaporative upflows in the early
episodes that reach velocities of 500--2000~km~s$^{-1}$. Such velocities can in principle be
measured spectroscopically. 

\citet{reale88, reale04} studied two giant flares on Proxima Centauri using
hydrodynamic approaches. Both flares have been suggested to be close analogs
of gradual, very large solar events, with the important difference that 
their geometric sizes ($5\times 10^9$~cm -- $10^{10}$~cm), while similar to large-scale
solar events, comprise a significant fraction of the stellar corona. For the second
flare, \citet{reale04} showed  explicitly that  multiple
flaring loops are required to describe the light curve. This led them to suggest that the 
actual geometry should be a loop arcade in analogy to the 2-R model, of which a 
few dominant loops with similar heights were modeled like isolated flaring loops. 
The same flare modeled with the 2-R approach (Sect.~\ref{2R}) indeed leads to very similar
magnetic loop heights \citep{guedel04}. Large flares
on late-type M dwarfs therefore constitute large-scale disturbances of the 
corona possibly inducing global effects, an assertion that has been supported  
by interferometric radio observations \citep{benz98}.

\subsection{Magnetohydrodynamic models}\label{mhd}

\citet{shibata99} and \citet{shibata02} discussed novel flare scaling laws based on their earlier
MHD work on X-point reconnection in large flares. They reported a simple scaling relation between
flare peak temperature $T$, the loop magnetic field strength $B$, the pre-flare loop electron density $n_0$, 
and the loop semi-length $L$ under the condition of dominant conductive cooling (appropriate for the early phase of a flare),
\begin{equation}\label{shibata1}
T \approx 1.8\times 10^4 B^{6/7}n_0^{-1/7}L^{2/7}~{\rm [K]}.
\end{equation}
 The law follows from the balance between conduction cooling 
($\propto T^{7/2}/L^2$, Eq.~\ref{condloss}) and magnetic reconnection heating ($\propto B^3/L$).
Assuming loop filling through chromospheric evaporation and balance between
thermal and magnetic pressure in the loop, two further ``pressure-balance scaling laws'' follow:
\begin{eqnarray}\label{shibata2}
{\rm EM}&\approx& 3\times 10^{-17} B^{-5}n_0^{3/2}T^{17/2}~{\rm [cm^{-3}]}\\
{\rm EM}&\approx& 2\times 10^8     L^{5/3}n_0^{2/3}T^{8/3}~{\rm [cm^{-3}]}.
\end{eqnarray}
An alternative scaling law applies if the density development in the initial flare
phase is assumed to follow balance between evaporation enthalpy-flux and conduction flux, although the
observational support is weaker,
\begin{equation}\label{shibata3}
{\rm EM}\approx 1\times 10^{-5} B^{-3}n_0^{1/2}T^{15/2}~{\rm [cm^{-3}]}.
\end{equation}
And third, a steady solution is found for which the radiative losses balance conductive
losses. This scaling law applies to a steady loop,
\begin{equation} 
{\rm EM} \approx \left\{ 
   \begin{array}{ll}\label{demflaremhd}
       10^{13} T^4L~\quad{\rm [cm^{-3}]}.            & \mbox{\quad for\quad  $T < 10^7$~K } \\
       10^{20} T^3L~\quad{\rm [cm^{-3}]}.            & \mbox{\quad for\quad  $T > 10^7$~K } 
   \end{array} 
   \right. 
\end{equation} 
and is equivalent to the RTV scaling law (Eq.~\ref{RTV}).

The advantage of these scaling laws is that they make use exclusively of the flare-peak
parameters $T$, EM, $B$ (and the pre-flare density $n_0$) and do not require knowledge of
the time evolution of these parameters. The models have been applied to flares on
protostars and T Tau stars (see Sect.~\ref{flaretemp}).

\subsection{Summary of methods}

Despite their considerable sophistication, stellar flare models remain crude approximations so
long as we have little a priori knowledge of the magnetic field topology. Solar flares 
reveal complexities that go far beyond any of the standard models described above. Nevertheless, 
stellar flare scenarios have been useful tools to roughly assess characteristic
flare sizes, densities, and heating rates. Several models have been applied to solar flares
as well, which has tested their  reliability.

In some cases, simple light curve decay analysis (Eq.~\ref{cooltime}) or the quasi-static cooling
model result in excessively large loop semi-lengths $L$ \citep{schmitt94c, favata99, favata00a}. Alternative methods 
such as the  heating-decay model (Sect. \ref{contheat}) may give more moderate values.
On the other hand, seemingly different methods may also result in overall agreement for
the magnetic structure size. For example, \citet{endl97} compared the 2-R approach with
the quasi-static cooling formalism for a large flare on an RS CVn binary and found similar 
heights of the flaring structures ($\approx 1R_*$). \citet{guedel04} and \citet{reale04}
compared the 2-R model, full hydrodynamic simulations and the  heating-decay model, again 
finding good overall agreement (loop sizes of order $1R_*$ on Proxima Centauri). \citet{covino01}
compared loop lengths obtained for several recently-modeled large stellar flares based
on a simple decay-time formalism (Eq.~\ref{cooltime}) and on the heating-decay model;
the agreement was once more rather good. The authors argued that neglecting
heating during the decay increases the model length in the decay-time formalism, but at the same time one ignores
conductive cooling.  The density is thus overestimated, which decreases the model loop size again. 
In  this sense, approaches such as the heating-decay model or the analytic
2-R model are preferred not necessarily (only) because of their predictive power but because of their
physically founded basis and hence reliability, and their support from direct solar observations. 
Full hydrodynamic or MHD simulations of course provide the closest description of the actual
processes, but a realistic simulation requires a careful choice of a number of unknown parameters.
 
One of the main results that have come from extensive modeling of stellar X-ray flares 
is that extremely large stellar flares require large volumes
under all realistic assumptions for the flare density. This is because, first, the energy derives 
from the non-potential portion of the magnetic fields that are probably no stronger than a few 
100~G in the corona; and second, small-loop models require higher pressure to produce the observed 
luminosity, hence requiring excessively strong magnetic fields. This is in line with the findings by  
\citet{cheng91} from hydrodynamic simulations.  This is not to say that magnetic loops must be
of enormous length - a number of interpretation methods suggest the contrary even for very 
large flares. But at this point, the concept of single-loop
approaches  becomes questionable particularly as large flares on the Sun often involve 
very complex arrangements of magnetic structures: the large volumes  do not necessarily involve large
heights but {\it large surface area} (see example in \citealt{reale04}).
 
Alternative models are available in the stellar literature, although 
they have mostly been applied to singular cases. Most notably, I mention the coronal mass ejection
model by \citet{cully94} that was applied to a giant flare on AU Mic observed by 
{\it EUVE} \citep{cully93}.
\citet{fisher90} derived equations for the evolution of a constant-cross section
flaring magnetic loop with uniform but time-varying pressure and volumetric
heating rate (for time scales that are long compared to the sound transit time, i.e., assuming quasi-hydrostatic
equilibrium). Although their model is not quasi-static by requirement,
 it can be well approximated by  ``equivalent static loops'' having the
same length and the same column depth as the evolving loop. A major advantage
of this formulation is that it includes the evaporation phase of the
flare, i.e., essentially the flare rise phase. 
The quality of the model was described in detail by \citet{fisher90}, together with
applications to a solar and an optical stellar flare. Another application was described by 
\citet{hawley95} for an EUV flare on AD Leo. Table~\ref{flaretable}  reports the refined 
results presented by \citet{cully97} for a typical
low-abundance corona. A further  model including evaporative cooling was 
presented by \citet{pan97} in an application to a strong flare on an M dwarf.

There are a number of further, ``unconventional'' models, e.g., star-disk magnetic flares in pre-main
sequence stars or intrabinary flares - see, among others, \citet{skinner97} and \citet{montmerle00}.

\subsection{Observations of stellar X-ray flares}\label{obsflares}

\begin{figure} 
\centerline{\resizebox{0.85\textwidth}{!}{\includegraphics{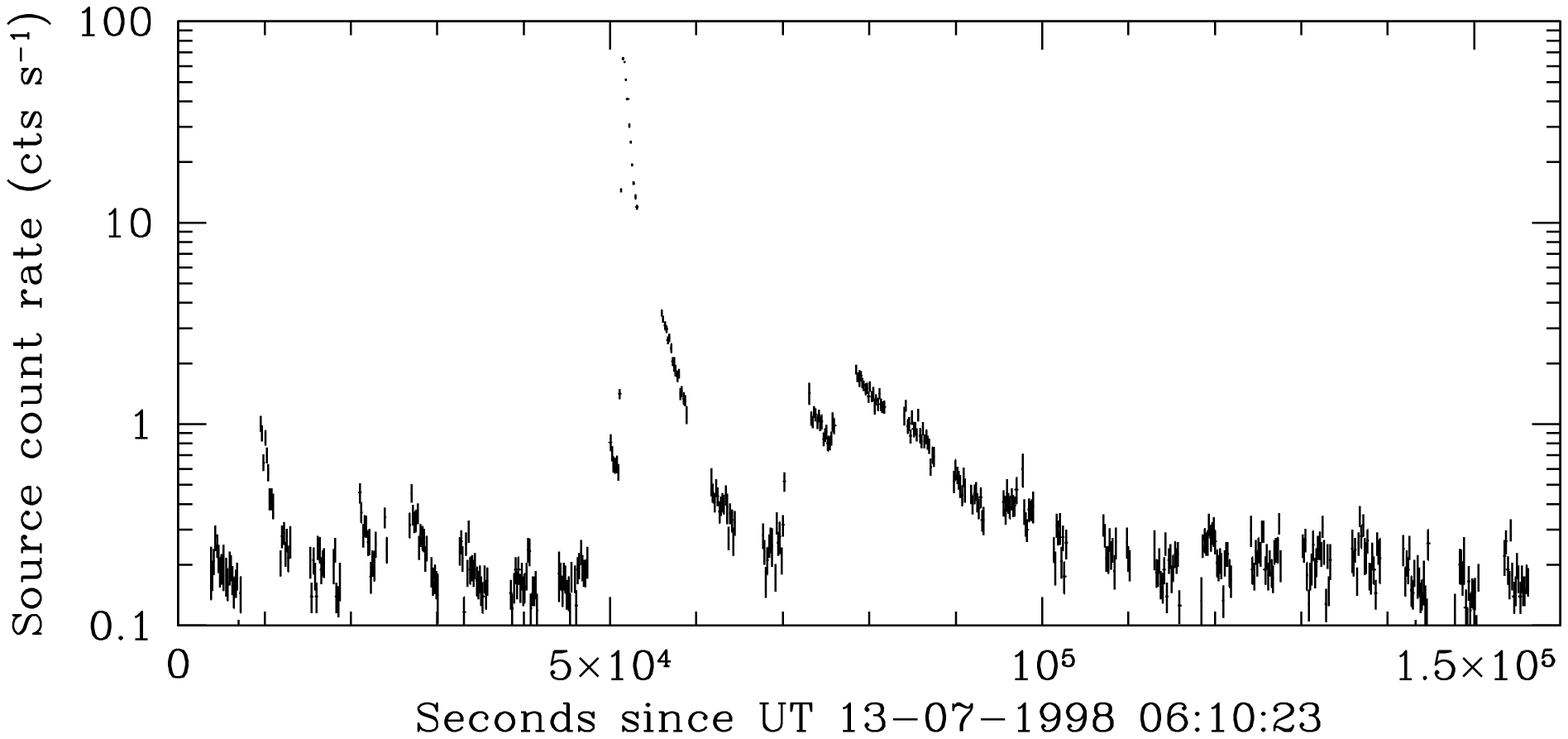}}}
\centerline{\resizebox{0.85\textwidth}{!}{\includegraphics{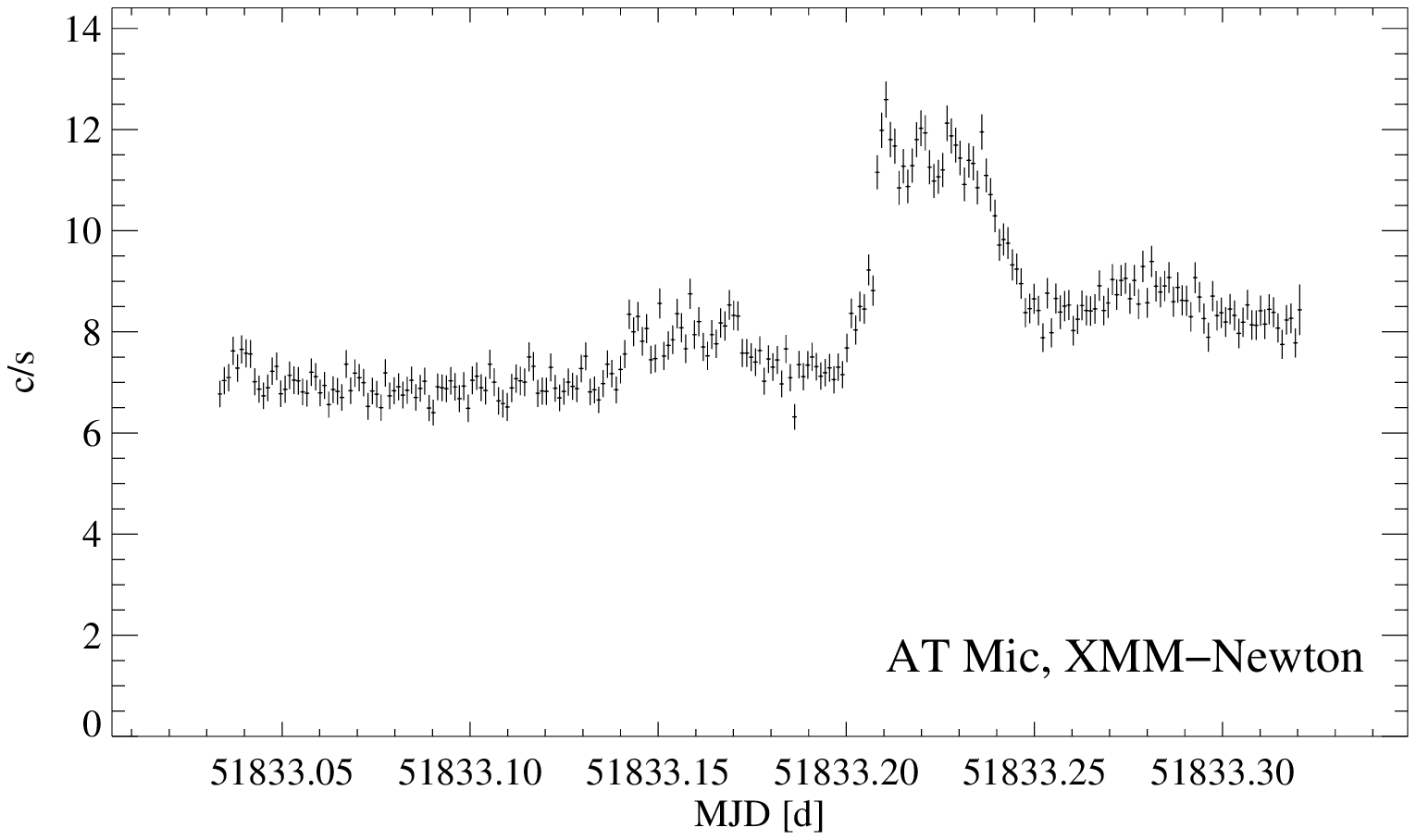}}}
\hbox{\resizebox{0.51\textwidth}{!}{\rotatebox{90}{\includegraphics{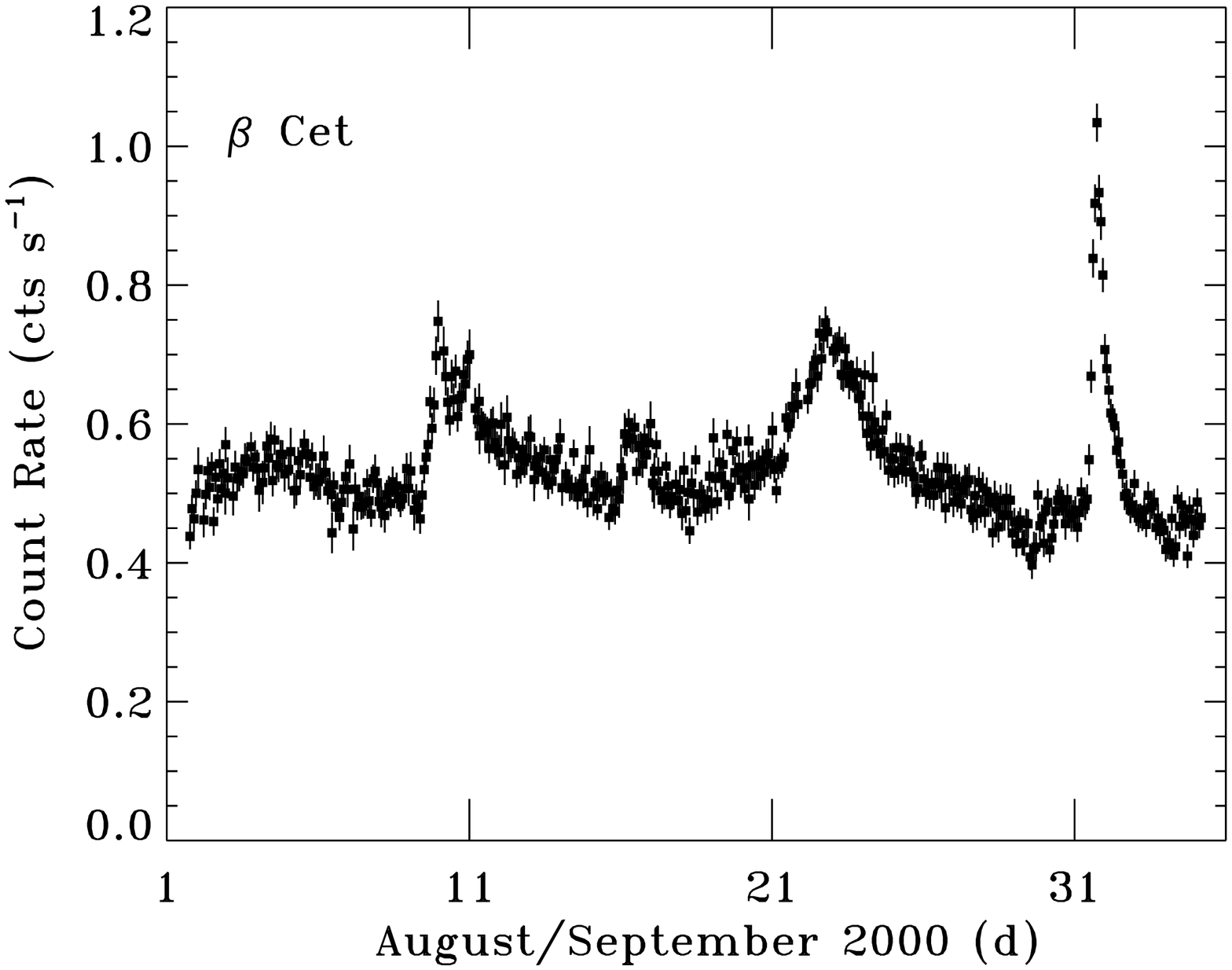}}}
      \resizebox{0.49\textwidth}{!}{\rotatebox{90}{\includegraphics{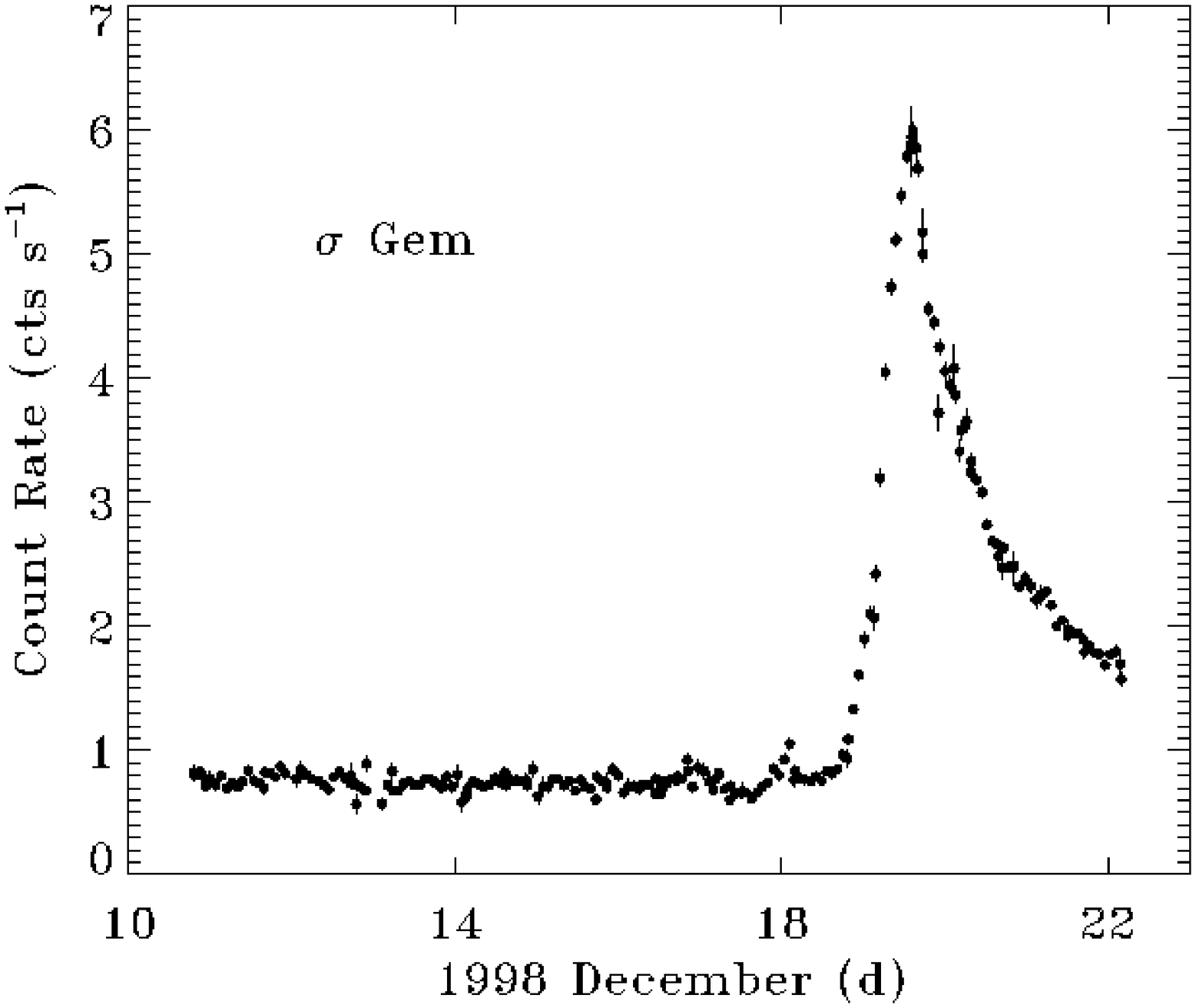}}}
}
\caption{Four largely different examples of stellar flares. {\it Top}: A very large
flare on EV Lac, showing a rapid rise and a long double-exponential decay (figure courtesy of F. Favata,
after \citealt{favata00a}, observations with {\it ASCA} GIS.) -
{\it Middle:} Modest, flat-topped flare on AT Mic (from \citealt{raassen03b}, observations
with {\it XMM-Newton} EPIC).    
{\it Bottom left:} Sequence of very slowly decaying flares on the giant $\beta$ Cet 
(figure courtesy of R. Osten, after \citealt{ayres01a}, observations
with {\it EUVE}).
{\it Bottom right:} Rapid rise and very slow decay of a flare on the RS CVn binary $\sigma$ Gem 
(figure courtesy of R. Osten, observations
with {\it EUVE}).
}\label{flareexamples}
\end{figure}

Table~\ref{flaretable} summarizes properties of a list of stellar X-ray events that have been modeled with one of
the above approaches in the literature (I have added a few selected published flares for which modeling is not available
but for which good temperature and EM measurements have been given). Further flares on pre-main
sequence stars have been discussed by \citet{imanishi03}. Some flares have
reported durations of several days (\citealt{graffagnino95, kuerster96}; see also \citealt{ayres99, ayres01a}), 
some show X-ray luminosities up to $10^{33}$~erg~s$^{-1}$  (e.g., \citealt{preibisch95, tsuboi98}), and some show
temperatures in excess of 100~MK \citep{tsuboi98, guedel99, favata99}. A number of extreme flares were already reported 
from the pre-{\it Einstein} era, with RS CVn binaries recognized as their origin \citep{pye83}. 
Unusual shapes of flares have frequently been reported, such as flares with slow rise and rapid decay
\citep{haisch87}, flares with secondary peaks or separate reheating events during the primary 
decay \citep{white86, katsova99, guedel04},
flares with double-exponential decays (\citealt{cully93, osten99, favata00a, reale04}, see Fig.~\ref{flareexamples}), 
flat-topped flare light-curves  (\citealt{agrawal86b, osten99, raassen03b}, see Fig.~\ref{flareexamples}),  
flares with very long rise times \citep{tagliaferri91}, or flares in which the temperature stays essentially 
constant during the decay phase \citep{graffagnino95}.

\begin{table}
\rotatebox{90}{
\begin{minipage}{1.0\textheight}
\caption{Flare interpretation: Summary of results}
\label{flaretable}       
\scriptsize
\begin{tabular}{lllllllllll}
\hline\noalign{\smallskip}
Star$^a$        & &   log\,$L_X^b$     & log\,$E_X^c$      & log\,EM$^d$ &$T_{\rm max}^e$ & $\tau^f$     & model$^g$  & log\,$n_{e, {\rm max}}^h$   & log\,(scale)$^i$   & Reference \\
                & & (erg~s$^{-1}$) & (erg)         & (cm$^{-3}$)  &($10^6$~K)    & (s)        &            & (cm$^{-3}$)             & (cm) &   \\
\noalign{\smallskip}\hline\noalign{\smallskip}
Prox Cen        & E &  28.08	   &  31.32	    & 51.0 & 20 	 &  1750       &  H	      &  10.70-11.11 & 9.65	& \citet{reale88} \\
   "            & E &  28.30  	   &		    & 51.0 & 27 	 &	       &  2R          &  12.00	     & 9.15  	& \citet{poletto88, haisch83}  \\
   "            & E &    	   &  31.54         &	   &		 &  1300       &  D	      &  11.18       & 9.78	& \citet{byrne89}  \\		 
Prox Cen        & E &  27.87	   & (30.95)	    & 51.1 & 17 	 &  1200       &  D	      &  11.26	     & 9.30	& \citet{haisch80, haisch81} \\  
Prox Cen        & X &  28.59	   &  32.18	    & 51.3 &	 25	 &1400-4400    &   2R         & 11.48-11.70  &  10.23	& \citet{guedel04} \\
  "             & X &		   &		    &	   &		 &	       &   H	      & 12.30	     &$\approx 9.80$&  \citet{reale04} \\
EQ Peg          & E &  30.48	   &		    & 53.3 & 44 	 &  2400       &  2R          &  12.78	     & 9.61  	& \citet{poletto88} \\
EQ Peg          & Ex&  29.30 	   & (32.56)	    & 51.7 &  26	 &  1800       & D	      &  11.30	     &  10.11  	& \citet{haisch87}  \\
EQ Peg          & Ex&  30.2   	   &  34.64         & 53.3 &  42  	 &  2400       & ...   	      & ...          &  ...	&  \citet{pallavicini90a}\\   
EV Lac          & H &  29.48  	   & (32.59)	    & 53.3 & (20)	 &$\la 1300$   & D	      &  11.60-12.00 & 9.70	& \citet{ambruster84} \\     
EV Lac          & R &		   &		    & 52.2 & ...         & 38400       & Q	      &  10.48       & 11.58 	& \citet{schmitt94c} \\
EV Lac          & A &  31.18	   &  34.18	    & 53.9 & 73  	 & 1800        & Q	      &  11.78	     & 10.50  	& \citet{favata00a} \\
   "            & A &		   &		    &	   &		 &	       & C            &  11.30-12.30 &  9.92  	& \citet{favata00a} \\
AD Leo          & A &  28.95	   &  32.11	    & 52.1 &  20	 & 570         & C            &  $>$11.70    &  $<$9.80	& \citet{favata00b} \\
AD Leo          & A &  29.40  	   &  32.91	    & 52.7 &  25	 & 1180        & C            &  12.53	     &  9.42 	& \citet{favata00b} \\
AD Leo          & A &  28.85	   &  32.20	    & 52.4 &  38	 & 3500        & C            &  12.04	     &  9.65	& \citet{favata00b} \\
AD Leo          & R &  29.26  	   &  32.08	    & 52.3 &  35	 & 680         & C            &  12.30       &  9.41	& \citet{reale98,favata00b} \\
AD Leo          & R &  29.28  	   &  32.40	    & 51.9 &  12	 & 8040        & C            &  $>$11.18    &  $<$10.03& \citet{favata00b}  \\ 
AD Leo          & E &  29.51  	   &  33.40	    & 52.0 &  12	 & 5400        & C            &  11.48	     &  9.92  	& \citet{favata00b} \\
AD Leo          & EU&		   &		    & 52.3 &  13	 &	       & FH	      &  10.60	     &  9.68 	& \citet{cully97}  \\
AD Leo          & EU&		   &		    & 52.3 &  $<$32	 &	       & FH	      &  11.64	     &  9.54 	& \citet{cully97}  \\
AT Mic          & Ex&  29.80  	   &  33.10    	    & 52.7 &  39    	 &  1800       & ...   	      & ...          &  ...	&  \citet{pallavicini90a}\\   
AU Mic          & EU&  (30.92)	   &  35.46	    & 53.0 & ...	 &   35000     &	      &  10.70	     &10.54-10.85& \citet{katsova99}	     \\
Gl 644          & Ex&  29.91 	   &  33.20	    & 52.5 &	  50	 &  950        &  D	      & 11.90	     &  9.90	&  \citet{doyle88a} \\  	  
YZ CMi          & E &  28.90 	   &  31.56	    & 51.3 &	  20	 &  600        &  D	      & 11.15	     &  10.00	&  \citet{kahler82}\\
YY Gem          & X &  29.65   	   &  33.63    	    & 52.1 &   40   	 &  960        &  C   	      &  11.8         & 9.10	&  \citet{stelzer02}\\   
YY Gem          & Ex&  30.67   	   &  34.0     	    & 53.3 &   64   	 &  3900       & ...   	      &  ...         &  ...	&  \citet{pallavicini90a}\\   
Castor          & Ex&  30.90 	   &  33.70	    & 53.7 &  48	 & 300-600     & D	      &  12.54	     &  9.71	& \citet{pallavicini90b}    \\
AB Dor          & B &  32.00	   & 35.78-35.90    & 54.7 & 109	 &  3840       &  C           & 11.73-12.20  &  10.34  	& \citet{maggio00}	     \\
AB Dor          & B &  32.00	   & 35.78-35.90    & 54.6 & 110	 &  3420       &  C           & 11.49-11.95  &  10.49  	& \citet{maggio00}	     \\
AB Dor          & X &($\approx 30.4$)&  34.00	    & 52.7 &	33	 &  3850       & C            & 11.41	     &  10.20	&  \citet{guedel01b} \\ 
CC Eri          & R &  29.48	   &  33.46	    & 52.6 & 28 	 &  (9500)     &   2R         & 10.0-11.1    &9.95-10.10&  \citet{pan95}\\
    "           & R &		   &		    &	   &		 &             & D	      & 11.04-11.1   & ...  	&  \citet{pan95} \\	    
EQ1839.6+8002   & G &  31.23	   &  34.00	    & 54.0 & 100	 &  629        &  D	      & 12.41	     & 9.23-9.92& \citet{pan97} \\   
   "            & G &		   &		    &	   &		 &	       & D	      & 11.60-12.23  &9.50-10.10& \citet{pan97}\\
Gl 355          & R &  31.04	   &  34.95	    & 54.1 &  52	 &  10100      &  C           & 11.30	     &  10.72  	&  \citet{covino01} \\
  "             & R &		   &		    &	   &		 &	       &     D        & 11.11	     & 10.80  	&  \citet{covino01} \\
\noalign{\smallskip}\hline
\end{tabular}
\end{minipage}
}
\end{table}

\setcounter{table}{3}
\begin{table}
\rotatebox{90}{
\begin{minipage}{1.0\textheight}
\caption{Flare interpretation (continued)}
\scriptsize
\begin{tabular}{lllllllllll}
\hline\noalign{\smallskip}
Star$^a$        & &   log\,$L_X^b$     & log\,$E_X^c$      & log\,EM$^d$ &$T_{\rm max}^e$ & $\tau^f$     & model$^g$  & log\,$n_{e, {\rm max}}^h$   & log\,(scale)$^i$   & Reference \\
                & & (erg~s$^{-1}$) & (erg)         & (cm$^{-3}$)  &($10^6$~K)    & (s)        &            & (cm$^{-3}$)             & (cm) &   \\
\noalign{\smallskip}\hline\noalign{\smallskip}
$\pi^1$ UMa     & Ex&  30.00 	   &  33.30	    & 52.9 & 30 	 &  1000       &  D	      &  11.87	     &   9.86 	& \citet{landini86} \\
H~II 1516       & R &  31.20	   &  (34.90)	    & 54.3 & 25 	 & 5000        &   Q	      & 11.11	     & $<10.98$ &  \citet{gagne95a}  \\
H~II 1032       & X &  30.69	   &  (33.99)	    & 53.5 & 23.5	 & 2000        &   C          & 11.70	     &  10.06  	&  \citet{briggs03} \\  				       
H~II 1100       & X &  30.21	   &  (33.67)	    & 53.1 & 15.4	 & 2900        &   C          & 11.70	     &  9.88    &  \citet{briggs03}\\
$\sigma^2$ CrB  & Ex&  30.97 	   &  34.38	    & 53.8 &  95	 &  1700       &  D	      &  11.78	     &  10.17  	& \citet{vdoord88}	     \\
$\sigma^2$ CrB  & E &  30.78 	   &  34.30	    & 53.5 &  25	 &$\approx$2000&  D	      & 11.95	     &  10.04 	& \citet{agrawal86b} \\
HR~1099         & X &  ...   	   & ...      	    & 54.2 &  36    	 & ...         &  ...  	      & ...          &  ...	&  \citet{audard01a}\\   
AR Lac          & R &  32.30	   &  37.00	    & 55.0 &  88	 &  11000      &  Q	      &  11.23	     &  10.90 	& \citet{ottmann94b} \\
II Peg          & Ex&  31.04  	   &$\ga$35.20      & 53.9 & 44 	 &  40000      &	    D &$\la$10.4     &$\ga 11$ 	& \citet{tagliaferri91} \\
II Peg          & A &  30.41 	   &  34.43	    & 53.3 &	  36	 &  10000      &	   Q  &  10.90	     &  10.90	& \citet{mewe97} \\
II Peg          & G &($\approx 32$)&  35.26	    & 54.5 &  65	 &  2081       &  D	      &  11.88	     &  10.46  	& \citet{doyle91, doyle92b}		\\
UX Ari          & G &  32.30	   &  $>37.00$	    & 54.9 &  85	 &	       &  D	      &  9.3-10.3    &  ... 	& \citet{tsuru89}      \\
UX Ari          & G &  31.78	   &  35.48         & 54.1 &  105	 &	       &  D	      &  11.23	     &  ... 	& \citet{tsuru89}  \\
UX Ari          & A &  32.15	   & (36.30)	    & 54.8 & 120	 &             &   2R	      & 11.0-12.0    & 11.0-11.7& \citet{guedel99}  \\    
UX Ari          & B &  32.00	   &$\ga 36.70$     & 54.9 &$\approx 100$&$\approx 47000$& 2R         & 10.78-11.00  &11.28-11.48& \citet{franciosini01} \\
HU Vir          & R &  32.15	   &  36.89         & 55.2 &$\approx 60$ &  67600      & 2R           & ...	     & 11.3-11.6 &  \citet{endl97}    \\  
  "             & R &		   &		    &	   &		 &	       &    Q	      & 10.32        &  11.72	& \citet{endl97} \\
XY UMa          & Ex&  30.85 	   &  33.70	    & 53.3-.8 &    ...   &$\approx$1000&     D        & 11.48-12.0   & 10.4-11.1& \citet{jeffries90}\\ 
Algol           & R &  32.20	   &  36.85         & 55.0 &$\approx 100$ &30400        &   Q	      &$\approx 11.7$&  11.22	& \citet{ottmann96} \\  
Algol           & Ex&  31.20	   &  35.00	    & 54.0 &	78	 & 4350        &  Q	      & 12.00        &  10.71	& \citet{vdoord89}\\
  "             & Ex&  31.15	   &  35.00	    & 54.0 &   60	 & 7000        &  D	      & 11.48	     & 10.3-10.7& \citet{white86}  \\
Algol           & G &  31.00	   &  (35.34)	    & 54.1 &	67	 & 22000       &  Q	      & 10.70	     &$\approx 11.28$&  \citet{stern92a}\\
Algol           & B &  32.48	   &  37.15         & 55.1 & 140	 &  64000      &   Q	      & 10.52	     &$\approx 11.84$& \citet{favata99}\\
  "             & B &		   &		    &	   &		 & 49600       & C            & 10.3-10.48   &  11.72	&  \citet{favata99}\\
CF Tuc          & R &  31.85 	   &  37.15         & 55.0 & 45 	 &$\approx$80000&    Q	      & $\ge$9.95    &  $>11.50$&  \citet{kuerster96}\\   
   "            & R &		   &		    &	   &		 &	       &  2R          & $\le$11.6    &  ... 	& \citet{kuerster96}\\
$\beta$ Boo     & R &  29.48  	   &  32.23 	    & 52.0 &$\ga  17$& 540        & D	      & 11.92	     &  9.62  	& \citet{huensch95}\\
HD 27130        & E & $\ga 31$     & $\ga 34.48$    & 53.5-54.0  &  50   & 2500        &    D	      & 11.60	     &  10.2   & \citet{stern83}\\
HD283572        & R &  31.58	   &  35.08	    & 54.6 & 30-75	 &  12000      &   C          & 11.45-11.92  &  10.5 	& \citet{favata01} \\	    
V773 Tau        & A &  32.23	   & $>37.00$	    & 55.2 & 42 	 & 45000       & RM           & $\ge 11.3$   &$\le 11.18$& \citet{skinner97}\\	     
V773 Tau        & A &  33.00	   &  37.00	    & 55.7 &	  115    & 8200        &   D	      & 11.48	     &  11.10  	& \citet{tsuboi98}\\
  "             & A &		   &		    &	   &		 &	       & C            & 11.72-12.20  &  10.68 	& \citet{favata01} \\	  
LkH$\alpha$92   & R &  32.70       &  36.60         & 55.7 & 43 	 & 7800        &   D	      & 11.18	     &$\ge 10.9$& \citet{preibisch93}\\
    "           & R &		   &		    &	   &		 & 5600        & C            & 11.68-12.15  &  10.36	& \citet{favata01}   \\     
P1724           & R &  33.26       &  37.70         &(56.3)&$\approx 30$ &32000        &   D	      & 10.48        & ...  	& \citet{preibisch95}\\
YLW15           & A &  32.30	    &  36.78         & 54.8 & 65	  & 31000      &  Q	      & 10.70	      & 11.49	 &  \citet{tsuboi00}\\         
  "             & A &		   &		    & 55.2 &		 &	       & C            & 10.11-10.58  & 11.47 	&  \citet{favata01}\\	      
\noalign{\smallskip}\hline
\end{tabular}
\end{minipage}
}
\end{table}

\setcounter{table}{3}
\begin{table}
\rotatebox{90}{
\begin{minipage}{1.0\textheight}
\caption{Flare interpretation (continued)}
\scriptsize
\begin{tabular}{lllllllllll}
\hline\noalign{\smallskip}
Star$^a$        & &   log\,$L_X^b$     & log\,$E_X^c$      & log\,EM$^d$ &$T_{\rm max}^e$ & $\tau^f$     & model$^g$  & log\,$n_{e, {\rm max}}^h$   & log\,(scale)$^i$   & Reference \\
                & & (erg~s$^{-1}$) & (erg)         & (cm$^{-3}$)  &($10^6$~K)    & (s)        &            & (cm$^{-3}$)             & (cm) &   \\
\noalign{\smallskip}\hline\noalign{\smallskip}
YLW 16A         & A &  32.2  	   &       	    & 54.5 &  137  	 &  ...        & ...   	      & ...          &  ...	&  \citet{imanishi01a}\\   
EL29            & A &  (31.70)	   &  35.60         & 54.5 & 45 	 & 8000        & D	      & 11.30	     & 11.40 	&  \citet{kamata97}\\	      
R CrA           & A &  (31.00)	   &  35.30         & 53.6 & 70 	 & 20000       &  D	      & 11.00	     & ...  	&  \citet{tsuboi00}\\	      
SR 24           & A &  (33.00)	   &  36.78         & 55.5 & 80 	 & 6000        & D	      & 11.70	     & ...  	&  \citet{tsuboi00}\\	      
ROXs31          & A &  (33.26)	   &  37.11         & 55.8 & 70 	 & 7000        & D	      & 11.48	     & ...  	&  \citet{tsuboi00}\\		 
ROXA 2          & A &  31.40	   &  35.08	    & 54.0 & 80 	 & 6100        & D	      & 11.51	     & 11.10 	&  \citet{kamata97}\\		 
SSV 63          & A &  32.54	   & (36.60)        & 55.3 & 75 	 & 12000       &  D	      & 11.00	     & 11.00 	&  \citet{ozawa99}\\		 
ROX 20          & E &  32.00	   & $>35.90$	    & 54.8-55.1&  23	 & 6300        &  D	      & 10.48-10.78  & 11.00 	&  \citet{montmerle83}\\      
MWC~297         & A & $\ga$32.69   & $\ga$37.43     & 55.5 &  78         & 56400       &   D          &    10.64     & 11.4     &  \citet{hamaguchi00} \\
Sun, compact$^j$ & - &  26-27	   & 29-31  	    & 47-49& 10-30	 & 1000        & ...	      & 11-12        &  8-9     & \citet{landini86} \\
Sun, 2-R$^j$     & - &  27-28	   &  32	    & 49-50& 10-30 	 & 10000       &   ...        & 10-11	     &  10      & \citet{landini86} \\
\noalign{\smallskip}\hline
\end{tabular}
\footnotetext{\scriptsize 
NOTES: Values in parentheses are derived from parameters explicitly given by the authors.
If no object name is given, observation is the same as next above.
Note that spectral energy ranges may vary.\\
$^a$ Code after star name gives observing satellite; A = {\it ASCA},  B = {\it BeppoSAX}, 
        E = {\it Einstein}, EU = {\it EUVE}, Ex = {\it EXOSAT}, G = {\it Ginga},  H = {\it HEAO 1}, R = {\it ROSAT}, X = {\it XMM-Newton}.\\
$^b$ Logarithm of peak X-ray luminosity.\\
$^c$ Logarithm of total X-radiated energy.\\
$^d$ Logarithm of peak emission measure.\\
$^e$ Maximum temperature.\\
$^f$ E-folding decay time of luminosity.\\
$^g$ Model interpretation: 
       D = estimates from flare decay;
       Q = quasi-static cooling; 
       C = heating/cooling model after \citet{reale97}; 
       2R = two-ribbon model; 
       H = hydrodynamic simulation;
       RM = rotational modulation; 
       FH = evaporation model after \citet{fisher90}.\\
$^h$ Logarithm of peak  electron density.\\
$^i$ Loop lengths and semi-lengths as given by some authors were converted to (semi-circular) loop heights. \\
$^j$ For solar characteristics, see also \citet{montmerle83}, \citet{stern83}, \citet{landini86}, \citet{white86}, 
                                                \citet{vdoord88}  \citet{jeffries90}, \citet{pallavicini90a}, \citet{huensch95},
                                             and further references therein.}
\end{minipage}
}
\end{table}

\subsection{Flare temperatures}\label{flaretemp}

\begin{figure} 
\centerline{\resizebox{0.85\textwidth}{!}{\includegraphics{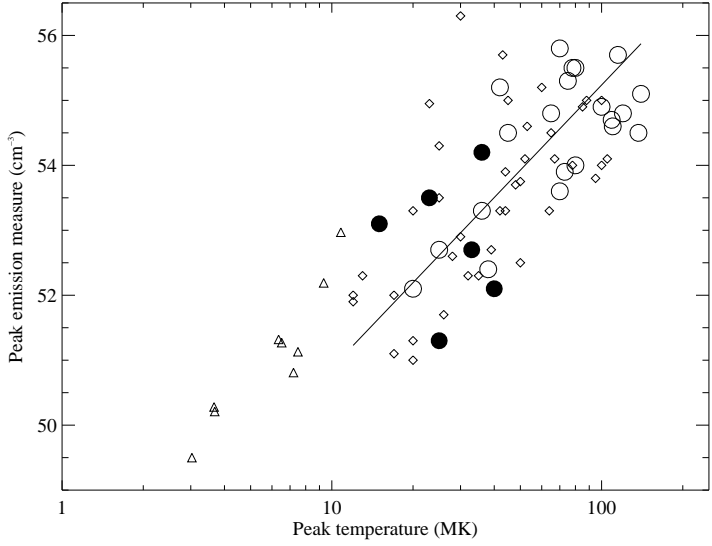}}}
\caption{Peak temperatures and EMs of the flares listed in Table~\ref{flaretable}. Key to the symbols:
Filled circles: {\it XMM-Newton} observations. Open circles: {\it ASCA}  or {\it BeppoSAX} observations.
Small diamonds: observations from other satellites. The solid line shows a regression fit (Eq.~\ref{EM_T}).
Triangles  represent non-flaring parameters of 
the G star sample from \citet{guedel97a} and \citet{guedel98}, referring to the hotter plasma component in 2-$T$ 
spectral fits to {\it ROSAT} data.}\label{tem_flare}
\end{figure}

When the flare energy release  evaporates plasma into the corona, heating and cooling
effects compete simultaneously, depending on the density and temperature profiles in a given flare. It is
therefore quite surprising to find a broad correlation between peak temperature $T_p$ and peak emission 
measure EM$_p$, as illustrated in Fig.~\ref{tem_flare} for the sample reported in Table~\ref{flaretable}.\footnote{If ranges 
are given in the table, the geometric mean of the minimum and maximum was taken; upper or lower limits were treated as measured values.}
A regression fit gives (for 66 entries)

\begin{figure} 
\centerline{\resizebox{0.85\textwidth}{!}{\includegraphics{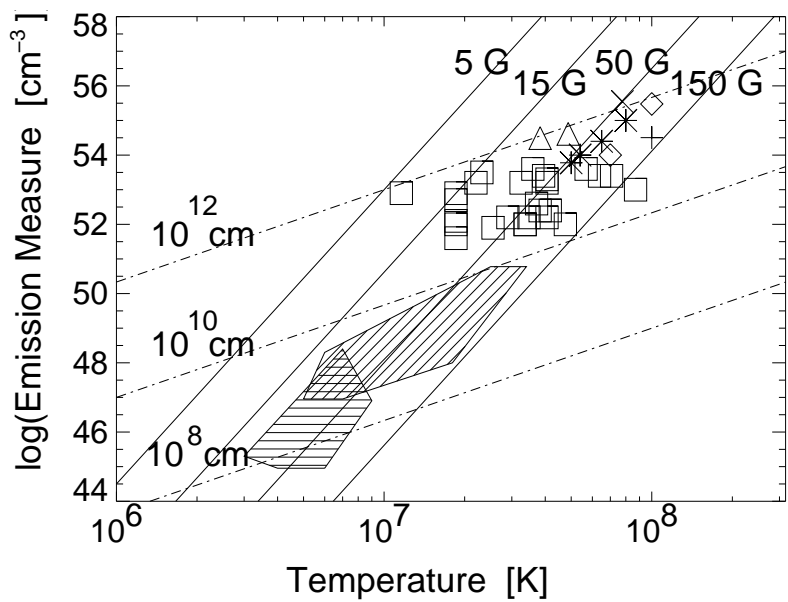}}}
\caption{Theoretical EM-$T$ relations based on the reconnection model by
Shibata \& Yokoyama, showing lines of constant loop length $L$ and
lines of constant magnetic field strength $B$. Hatched areas are loci reported for
solar flares, and other symbols refer to individual stellar flares in star-forming
regions (figure courtesy of K. Shibata and T. Yokoyama, after \citealt{shibata02}).}\label{shibata}
\end{figure}

\begin{equation}\label{EM_T}
EM_p \approx T_p^{4.30\pm 0.35}.
\end{equation}
The correlation overall indicates that {\it larger flares are hotter.} 
A similar relation was reported previously for solar flares \citep{feldman95}, with a suggestion
that it extends to selected large stellar flares. Although the trend shown in Fig.~\ref{tem_flare} does not smoothly 
connect to the  Feldman et al. solar-flare relation (theirs being lower in EM or hotter for a given EM, see also Fig.~\ref{shibata}), 
their stellar loci agree with ours. Various selection effects or data analysis biases related to limited S/N ratios and detector
energy resolution may prohibit an accurate comparison
of the two relations. Also, \citet{feldman95} measured the temperature at the time when the EM reached its peak, whereas
Table~\ref{flaretable} reports the maximum temperature that often occurs slightly before the EM peak. Consideration of
this effect could only increase the disagreement.  A handful of stellar flares for which the  temperature was
determined precisely at the EM peak yield  the same  slope as given in Eq.~(\ref{EM_T}).

On the other hand, the correlation is reminiscent of  the $T-L_X$ correlation for the ``non-flaring'' coronal stars in 
Fig.~\ref{temperaturelx} at cooler temperatures. This same sample is plotted as triangles in Fig.~\ref{tem_flare}, again 
only for the hotter plasma component (data from \citealt{guedel97a}). The stars follow the same slope as the flares,
albeit at cooler temperatures, and for a given temperature, the EM is higher. This trend may suggest that flares systematically 
contribute to the hot plasma component, although we have not temporally averaged the flare temperature 
and EM for this
simple comparison. For this model, we would require that the coronal emission of more luminous stars is dominated by 
larger and hotter flares. If flares are distributed in energy as described in Sect.~\ref{energydistribution}, then
a larger number of flares will generate both a higher luminosity and a shift to higher average 
temperatures indeed.

\citet{shibata99} and \citet{shibata02} interpreted the EM-$T$ relation as presented by \citet{feldman95} 
based on their MHD flare scaling laws (\ref{shibata2}). The observed loci of the flares require 
loop magnetic field strengths similar to solar flare values ($B \approx 10-150$~G) but the
loop lengths must increase toward larger flares. This is seen in Fig.~\ref{shibata} where lines of
constant $L$ and $B$ are plotted for this flare model. The same applies to the flares in Fig.~\ref{tem_flare};
typical loop lengths would then be $L \approx 10^{11}$~cm.

\subsection{Flare densities}

The density of a flaring plasma is of fundamental importance because
it determines the time scales of radiation and of several plasma-physical instabilities.
For spatially unresolved observations, densities can be inferred either indirectly from a
flare decay analysis (see Sect.~\ref{decaytime}--\ref{2R}) or directly by measuring density-sensitive line
ratios as described in Sect.~\ref{coronaldensities}.

Good examples from {\it solar} studies are relatively rare. \citet{mckenzie80} presented O\,{\sc vii} 
He-like triplets (Sect.~\ref{denshe}) during a large solar flare and found
$f/i$ ratios around unity close to the flare peak, implying densities
of up to $2\times 10^{11}$~cm$^{-3}$. Much shorter flares were discussed
by \citet{doschek81}; in those cases, the densities reached peaks around $(10-20)\times
10^{11}$~cm$^{-3}$ as measured from O\,{\sc vii}, but this occurred during the flare rise, 
while electron densities of a few times $10^{11}$~cm$^{-3}$ were derived during the flare peak. 
The estimated masses and volumes, on the other hand, steadily
increased. For the more relevant hot temperatures, line ratio diagnostics based on
Fe\,{\sc xxi}, Fe\,{\sc xxii}, or (He-like) Fe\,{\sc xxv}  have been employed. \citet{doschek81} and references therein reported 
densities derived from Fe\,{\sc xxv} ($T > 10$~MK) that are similar to those measured from 
O\,{\sc vii}. \citet{phillips96} inferred very high densities of $10^{13}$~cm$^{-3}$ from
Fe\,{\sc xxii}  about one minute after the Ca\,{\sc xix} flare peak, and $n_e \approx (2-3)\times 10^{12}$~cm$^{-3}$
from  Fe\,{\sc xxi} five minutes later. \citet{landi03} recently used various density
diagnostics for a modest solar limb flare. From Fe\,{\sc xxi} lines, they derived
densities of up to $3\times 10^{12}$~cm$^{-3}$, but there are conflicting
measurements for lower ionization stages that reveal much lower densities, 
comparable to pre-flare values (see also further references in their paper). If the density values 
at $\approx 10^7$~K are real, then  pressure equilibrium cannot be assumed for flaring loops; 
a possible explanation involves spatially separate volumes for the O\,{\sc vii} and the Fe\,{\sc xxi--xxv} 
emitting plasmas.

\begin{figure} 
\hbox{\resizebox{1.\textwidth}{!}{\includegraphics{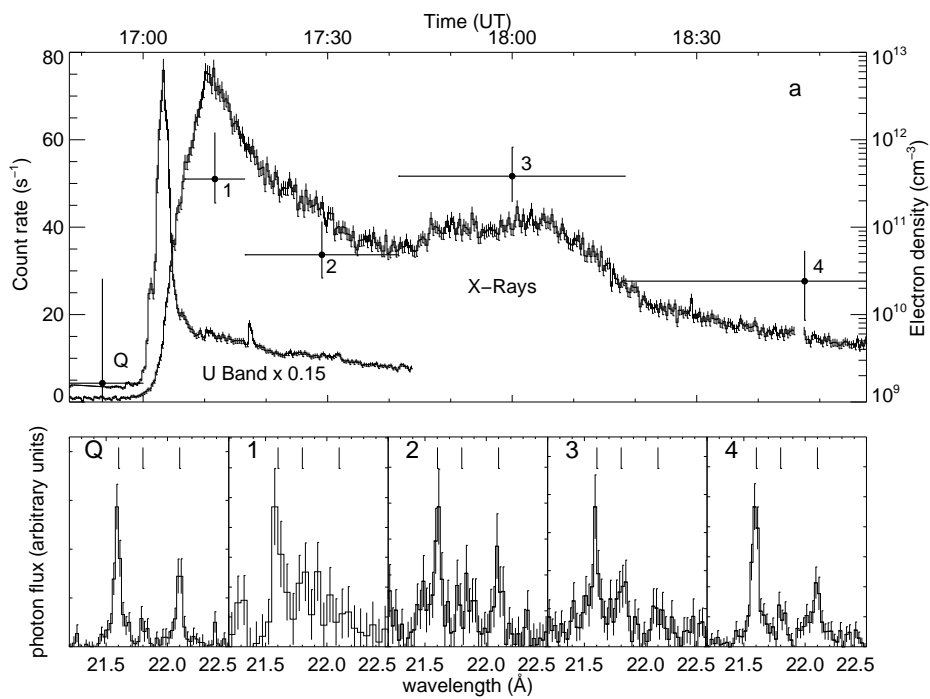}}}
\caption{Flare on Proxima Centauri, observed with {\it XMM-Newton}. The top panel shows
the X-ray light curve and the much shorter  U band flare (around 17 UT). The bottom panel
shows the O\,{\sc vii} He-like triplets observed during various time intervals of the flare. 
The locations of the $r, i$, and $f$ lines are marked by vertical lines. 
The resulting electron densities are given in the top panel by the crosses, where the
horizontal arm lengths indicate the time intervals over which the data were integrated, 
and the right axis gives the logarithmic scale 
(after \citealt{guedel02a, guedel04}).}\label{proxcenflare}
\end{figure}

{\it Stellar} flare density measurements are in a rather infant state as they require 
high signal-to-noise ratios over the short time of a flare, and good contrast against the
steady stellar emission.
A large flare on Proxima Centauri provided first evidence for significant
density variations as derived from the He-like O\,{\sc vii} triplet (and more tentatively,
from Ne\,{\sc ix}; \citealt{guedel02a, guedel04}, Fig.~\ref{proxcenflare}). The densities rapidly increased from
$n_e < 10^{10}$~cm$^{-3}$ to $\approx 4\times 10^{11}$~cm$^{-3}$ at  flare peak,
then again rapidly decayed to $\approx 2\times 10^{10}$~cm$^{-3}$, to increase again
during a secondary peak, followed by a gradual decay. The instantaneous mass involved in the cool,
O\,{\sc vii} emitting source was estimated at $\approx 10^{15}$~g, resulting in similar (instantaneous) potential and 
thermal energies in the cool plasma, both of which are much smaller than the total radiated 
X-ray energy. It is probable that the cool plasma
is continuously replenished by the large amount of material that is initially heated to higher 
temperatures and subsequently cools to O\,{\sc vii} forming temperatures and below.
The measured densities agree well with estimates from hydrodynamic simulations \citep{reale04}.
A marginal signature of a density increase was also recorded in O\,{\sc vii}  during a modest flare on
YY Gem \citep{stelzer02}, and in Mg\,{\sc xi} during a flare on $\sigma^2$ CrB although the 
density stayed similarly high outside the flare \citep{osten03}. Further marginal
indications for increased densities during flares were reported for AD Leo \citep{besselaar03}, 
AT Mic \citep{raassen03b}, and AU Mic \citep{magee03}.

Measurements from Fe line ratios in the EUV have been hampered by the apparently 
rather large  {\it quiescent} densities that already reach values expected for flares 
(see Sect.~\ref{fedensities}). They therefore generally 
show little  evidence for density {\it increases} during flares \citep{osten99}.
\citet{monsignori96} inferred $n_e$ up to $1.5\times 10^{13}$~cm$^{-3}$ from from the Fe\,{\sc xxi} $\lambda 142.2/\lambda 128.7$
flux ratio during a giant flare on AU Mic although the detection is again marginal. 
\citet{sanz01} and \citet{sanz02} found slightly higher densities during flares on $\lambda$ And, HR 1099,
and $\sigma$ Gem based on Fe\,{\sc xxi}   compared to quiescence (log\,$n_e \approx 12.9$ vs. 12.1 in $\lambda$ And), 
but again all values are extremely high.

\subsection{Correlation with UV and optical flares}\label{corropt}

Optical continuum flares (``white light flares''), often
observed in the Johnson U band, are tracers of the impulsive phase of the flare.
The emission is presumably due
to enhanced continuum emission after chromospheric heating, possibly following
electron impact (e.g.,  \citealt{hawley95}). In the  chromospheric evaporation scenario, U band
bursts are expected to occur during the rise phase of soft X-ray flares (Sect.~\ref{neuperteffect}).
Early examples were reported by \citet{kahler82}, \citet{dejager86}, and \citet{dejager89} for flares on 
YZ CMi, BY Dra, and UV Cet, respectively.  The total optical energy output is approximately 0.1--1 times the X-ray output, 
a range that has been confirmed by many other observations, including solar data (\citealt{kahler82} and 
references therein). The flare amplitudes and radiated energies appear to be correlated in the X-ray and
UV ranges \citep{mitra04}.

Occasionally, very good correlations are found between soft X-ray and H$\alpha$ flares. 
The total energy emitted in H$\alpha$ amounts to approximately 5\% of the 
soft X-ray losses \citep{doyle88a} but may, in exceptional cases, reach order unity \citep{kahler82}.
The physical causes of this correlation were discussed by \citet{butler93};
direct photoionization of chromospheric material by coronal 
soft X-rays is problematic because the H$\alpha$ ribbons seen
on the Sun are narrowly confined to the magnetic footpoints, and the
correlation is not necessarily detailed in time. A more likely explanation
involves the same electrons that also induce the continuum white light
flare in the chromosphere, as discussed above.

In some cases, however, optical and X-ray flares may be uncorrelated.
\citet{haisch81} described an X-ray flare without an accompanying signal in the optical or the UV.
Vice versa, \citet{doyle86} and \citet{doyle88b} 
reported H$\gamma$ and U band flares on YZ CMi with no indication of a simultaneous response in soft X-rays. 
They suggested that the heating occurred in low-lying loops in the transition region that did not reach coronal levels;
alternatively, absorption of the X-rays by cold overlying material is a possibility, or heating by proton
beams that are more efficient at inhibiting chromospheric evaporation.

\subsection{Correlation with radio flares}

Similar to the case of U band flares, we expect that non-thermal radio events, produced by accelerated particles, 
precede X-ray flares.
Reports on such correlations have been rather mixed. \citet{kundu88} described  a poor correlation
between X-ray events and radio flares observed during an {\it EXOSAT}-VLA joint survey program of dMe stars,
although there may be contemporaneous flaring in the two wavelength regions (see also \citealt{kahler82}).
Since this program was carried out at relatively long radio wavelengths (6~cm and 20~cm), the observed radio bursts were 
probably produced by a coherent
emission process that requires relatively few electrons in an unstable energy distribution. 
We cannot expect a  one-to-one correspondence in time for those cases. 
A special diagnostic case of correlated behavior can be observed for gyrosynchrotron radio flares at
higher radio frequencies - see below.

\subsection{The ``Neupert Effect''}\label{neuperteffect}

Radio gyrosynchrotron, hard X-ray, and optical emissions 
are induced on time scales of the electron propagation (seconds), and therefore essentially
develop proportionally to the influx of high-energy particles if long-term trapping does not
occur.  On the other hand, the cooling time of a thermal plasma in an extended coronal loop is governed by 
radiation and conduction with typical time scales of several minutes  to hours. 
The X-ray radiation therefore develops roughly proportionally to 
the accumulating thermal coronal energy\footnote{We ignore the detailed evolution of the flare
temperature and the density and thus the EM; the evolution of $T$ and $n_e$ may even be coupled,
see Sect.~\ref{contheat}; strictly speaking, in Eq.~\ref{neupert} one should refer to the thermal energy in the hot plasma rather than
to the X-ray luminosity which may not be proportional (see \citealt{guedel96}).}. To first order thus, for the radio (R), optical
(O), and hard X-ray (HXR) luminosities,
\begin{equation}\label{neupert}
L_\mathrm{R, O, HXR}(t) \propto {d\over dt}L_{\rm X}(t),
\end{equation}
a relation that has first been formulated for solar radio and X-ray flares \citep{neupert68} and that has become known
as the ``Neupert Effect''. It is a good diagnostic for the chromospheric evaporation process and has been well
observed on the Sun in most impulsive and many gradual flares \citep{dennis93}. The search
for stellar equivalents has been more challenging.

\citet{hawley95} observed the dMe star AD Leo in the
EUV (as a proxy for X-rays) and in the optical during extremely long 
gradual flares. Despite the long time scales and considerable time gaps in the EUV
observations, the presence of the Neupert relation (Eq.~\ref{neupert}) was demonstrated (in its integrated form).

\citet{guedel96} observed the similar dMe  binary UV Cet during several shorter
flares in X-rays  and at radio wavelengths.
A Neupert dependence between the light curves was clearly evident and was found to be very similar
to the behavior of some gradual solar flares. Furthermore, the  ratio between the energy losses in the two 
energy bands was derived to be similar to the corresponding luminosity ratio ``in quiescence''.
Similar results apply to RS CVn binaries: \citet{osten04} reported the Neupert effect in X-ray, EUV, and
radio observations of HR~1099, again with radiative X-ray/radio energy ratios that are close to
quiescent conditions. 

An example linking X-rays with white-light emission was presented by \citet{guedel02a}
in observations obtained with {\it XMM-Newton}. The flare is illustrated in 
Fig.~\ref{proxcenflare}. Equation (\ref{neupert}) is closely followed by the two light curves during the early part of  
the flare. The same temporal behavior was  identified in a sequence of small flares during
the same observation, suggesting that a considerable fraction of the low-level radiation 
is induced by temporally overlapping episodes of chromospheric evaporation. 

Although the relative timing provides important support for the evaporation model, 
the absolute energy content in the fast electrons must also be sufficient to 
evaporate and heat the observed plasma. Given the uncertain nature of the optical
white-light flares, the energy content of the high-energy particles 
is difficult to assess. The situation is somewhat better
at radio wavelengths although the spectral modeling of the non-thermal electron
population is usually rather incomplete and order-of-magnitude. During a pair of 
gradual soft X-ray flares observed on the RS CVn binary $\sigma$ Gem \citep{guedel02b}, 
Eq.~(\ref{neupert}) 
was again followed very closely, suggesting that electrons are capable of heating 
plasma over extended periods. The total injected electron energy was found to equal or possibly largely
exceed the associated X-ray losses albeit with large margins of uncertainty. Essentially 
{\it all} of the  released energy could therefore initially be contained in the fast electrons.

Similar timing between {\it radio} and X-ray flare events is seen in
previously published light curves, although the Neupert effect was not discussed.
Notable examples include flares described in  
\citet{vilhu88}, \citet{stern92b},  \citet{brown98}, and \citet{ayres01b}. If optical
emission is taken as a proxy for the radio emission, further examples can be found in
\citet{doyle88b}, \citet{kahler82}, \citet{dejager86}, and \citet{dejager89}.
The  Neupert effect is observed neither in each solar 
flare (50\% of solar  gradual flares show a different behavior; \citealt{dennis93}),
nor in each stellar flare. Stellar counter-examples include an impulsive optical flare 
with following gradual radio emission \citep{vdoord96}, gyrosynchrotron emission that 
peaks after the soft X-rays \citep{osten00}, an X-ray depression during strong radio flaring
\citep{guedel98}, or the absence of any X-ray response during radio flares \citep{fox94,  
franciosini99}.

\subsection{Non-thermal hard X-rays?}

In solar
flares, non-thermal hard X-rays begin to dominate the spectrum beyond 15--20~keV. Typically,
these spectral components are power laws since the electron distributions in energy
are power laws \citep{brown71}. An ultimate
test of the flare evaporation scenario in large flares on magnetically active stars
would consist in  the detection of non-thermal
hard X-ray components during the soft X-ray flare rise. Observations with 
{\it Ginga} up to $\approx$ 20~keV initially seemed to suggest the presence of such emission in
quiescence \citep{doyle92a} although \citet{doyle92b} showed that an unrealistically large
number of electrons, and therefore an unrealistically large rate of energy release, 
would be involved. A continuous emission measure
distribution with a tail up to very high temperatures, $T \ga 10^8$~K, can
explain the data self-consistently, in agreement with previous arguments given by \citet{tsuru89}.

\begin{figure} 
\centerline{\resizebox{0.85\textwidth}{!}{\rotatebox{270}{\includegraphics{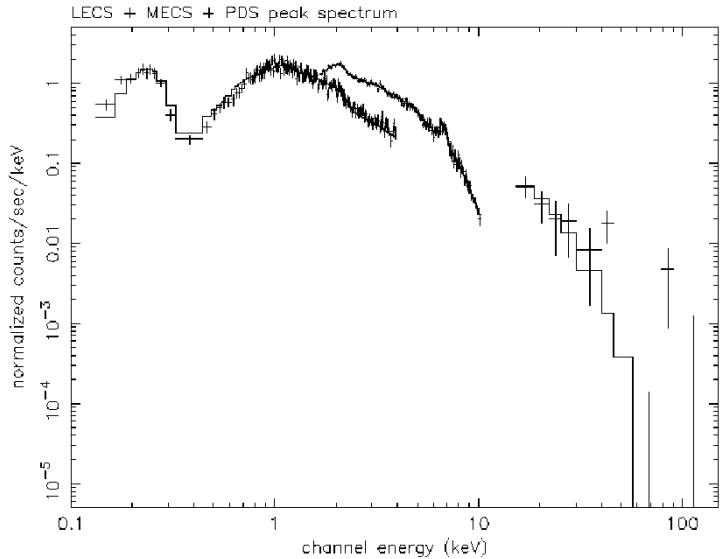}}}}
\caption{X-ray spectra taken during a large flare on UX Ari by three detectors on board {\it BeppoSAX}. The histogram
represents a fit based on thermal plasma components; it describes the data 
acceptably well up to about 40~keV (figure courtesy of E. Franciosini, after \citealt{franciosini01}).}\label{hardx}
\end{figure}

Up to the present day, no compelling evidence  has been reported for non-thermal X-rays from stellar coronae.
The Phoswich Detector System (PDS) instrument on board {\it BeppoSAX} was sensitive enough to detect photons
up to 50--100~keV during large stellar flares, but all recorded spectra could
be modeled sufficiently well with thermal plasma components. Examples
include flares on UX Ari \citep{franciosini01},
AR Lac \citep{rodono99}, Algol \citep{favata99}, and AB Dor \citep{pallavicini01}. 
The case of UX Ari is shown in  Fig.~\ref{hardx}. Although the
PDS signal was strongest just before the early peak phase of the flare when non-thermal
contributions are indeed expected, this is also the phase when the hottest plasma
is formed, so that the light curves make no strong argument in favor of non-thermal emission either.
The same applies to observations of giant flares on AB Dor \citep{pallavicini01}.

An interesting suggestion for non-thermal contributions was put forward by \citet{vilhu93}:
the small equivalent width of the Fe K line at 6.7~keV during flares on AB Dor could be due
to a continuum level that is enhanced by an additional non-thermal (power-law) component.
The principal uncertainty is the location of a lower cut-off in the electron distribution and
consequently the turnover in the non-thermal spectral contribution. Model calculations predict that this 
view is tenable only if the magnetic fields are weaker than 50~G \citep{vilhu93}.
 
The elusive non-thermal X-ray components are a classic case for the absence of evidence not
giving any evidence of absence: to the contrary, the very efficient production of non-thermal
{\it radio} emission in many of these active coronae both during flares and during (putative)
``quiescence''  is clear proof that large numbers of high-energy electrons are present in 
active coronae. We need more sensitive detectors to trace their radiative signatures, which will
hold unparalleled information on the primary energy release in stellar coronae.

\section{The statistics of flares}\label{stochasticflares}

The study of coronal structure confronts us with several problems that are difficult
to explain by scaling of solar coronal structure: i) Characteristic 
coronal temperatures increase with increasing magnetic activity (Sect.~\ref{templx}). 
ii) Characteristic coronal densities are typically higher in active than in inactive stars  
(Sect.~\ref{coronaldensities}), and pressures in hot loops can be exceedingly
high  (Sect.~\ref{loopstructure}).
iii) The maximum stellar X-ray luminosities exceed the levels expected from complete 
coverage of the surface with solar-like active regions by up to an order
of magnitude (Sect.~\ref{constituents}). iv) Radio observations reveal a persistent population 
of non-thermal high-energy electrons in magnetically active stars even if
the lifetime of such a population should only be tens of minutes 
to about an hour  under ideal trapping conditions in coronal loops \citep{guedel02d} and 
perhaps much less due to efficient scattering of electrons into the chromosphere
\citep{kundu87}. Several of these features 
are reminiscent of flaring, as are some structural elements in stellar coronae. If flares
are important for any of the above stellar coronal properties indeed, then we must consider
the effects of frequent flares that may be  unresolved in our observations but that may make up
part, if not all, of the ``quiescent'' emission. 

\subsection{Correlations between quiescent and flare emissions}\label{flarequicorr}

In 1985, three papers \citep{doyle85, skumanich85, whitehouse85} reported an unexpected, linear
correlation between the time-averaged  power from optical flares and the low-level, 
``quiescent'' X-ray luminosity. 
This correlation could  suggest that the mechanism that produces the optical flares also heats the
plasma, that is, the quasi-steady, slowly varying X-ray emission may be the product of {\it stochastic
flaring.} \citet{pearce92} showed that over the entire solar magnetic cycle, the monthly average
soft X-ray luminosity scales in detail and linearly with the rate of detected H$\alpha$ flares,
again supporting a picture in which a continuous distribution of flares at least contributes to the overall
coronal heating. This is echoed in the observation that stellar ``quiescent'' luminosity correlates approximately
linearly with the rate of X-ray flares (above some lower energy threshold; \citealt{audard00}, 
Fig.~\ref{flaredistrib}a below).

\citet{guedel93} discussed a global relation between non-thermal
radio luminosities of active stars and their ``quiescent'' X-ray luminosities. Since the short
lifetime of MeV electrons in coronal magnetic fields implies frequent acceleration,
a possible explanation again involves stochastic flares: the flare-accelerated
electrons could themselves act as the heating agents via chromospheric evaporation.
The smoking gun came with the observation that the total radio and X-ray outputs
of solar flares follow the same correlation \citep{benz94}. 
Similarly, \citet{haisch90a} found that
the ratio between energy losses in coronal X-rays and in the chromospheric Mg\,{\sc ii} lines
is the same in flares and in quiescence. This also applies if UV-filter observations
are used instead of Mg\,{\sc ii} fluxes \citep{mitra04}.
Finally, \citet{mathioudakis90} reported a tight correlation
between $L_X$ and H$\gamma$ luminosity that it the same for flares and for ``quiescence''. 
These observations point to an intimate relation between
flares and the overall coronal emission.

\subsection{Short-term coronal X-ray  variability}\label{corvar}

Further suggestive evidence for a connection between steady emission and coronal flares has come from
the study of light curves. A strong correlation 
between H$\gamma$ flare flux and simultaneous low-level X-ray flux in dMe stars suggests
that a large number of flare-like events are always present 
\citep{butler86}. Continuous low-level variability due to flares has been frequently reported  
for active M dwarfs in particular, but also for earlier-type dwarfs 
\citep{pollock91, vilhu93, kuerster97, mathioudakis99, gagne99}. Evidence  has  also 
been reported for giants, including stars close to saturation (\citealt{haisch94b, ayres01a}; Fig.~\ref{flareexamples}) 
and hybrid stars \citep{kashyap94}. \citet{montmerle83} estimated from light curves that 50\% of the observed
X-ray emission in young stars in the $\rho$ Oph star-forming region is due to relatively strong flares.
\citet{maggio00} found low-level variability in the very active AB Dor at a level 
of 20--25\% which they suggested to be due to ongoing low-level flaring (see similar conclusions in \citealt{stern92b}).
When variability is studied for different plasma components,
it is the hotter plasma that predominantly varies, while the cooler component is steady \citep{giampapa96}. 
An obvious suggestion is therefore that the high-temperature coronal component in active stars is the 
result of ongoing flaring.

\subsection{Stochastic variability - what is ``quiescent emission''?}\label{stochvariability}

The problem has been attacked in several dedicated statistical studies. 
While \citet{ambruster87} found significant continuous variability on time scales of several minutes in an {\it Einstein} 
sample of M dwarfs, \citet{collura88} concluded, from a similar investigation based on {\it EXOSAT} data,
 that the low-level episodes are truly ``quiescent''. \citet{pallavicini90a} found 
that approximately half of all investigated {\it EXOSAT} light curves show some residual variability, while the
other half are constant. However, such statements can only be made within 
the limitations of the available sensitivity that also limits the ability to resolve fluctuations in time. \citet{schmitt88} 
and \citet{mcgale95} discussed and simulated in detail to what extent flares can be statistically detected 
in light curves as a  function of the quiescent count rate, the relative flare amplitude, and the decay time.
The essence is that a statistically significant number of counts must be collected during the
finite time of a flare in order to define sufficient contrast against the ``quiescent'' background - otherwise
the light curve is deemed constant.

Higher sensitivity became available with {\it ROSAT}, and the picture indeed began to change.
Almost all {\it ROSAT} X-ray light curves of M stars are statistically variable on short ($\la$ 1 day)
time scales \citep{marino00}; this applies to a lesser extent to F-K dwarfs \citep{marino03b}. 
More specifically, the light curve
luminosity distribution evaluated over time scales of hours to days is very similar to the equivalent
distribution derived for solar flares, which  suggests that the overall stellar light curves of dM
stars are variable in the same way as a statistical sample of solar flares \citep{marino00}.
Such variability may thus  dominate flux level differences in snapshot observations taken several 
months apart \citep{kashyap99}.

\begin{figure} 
\centerline{\resizebox{0.90\textwidth}{!}{\includegraphics{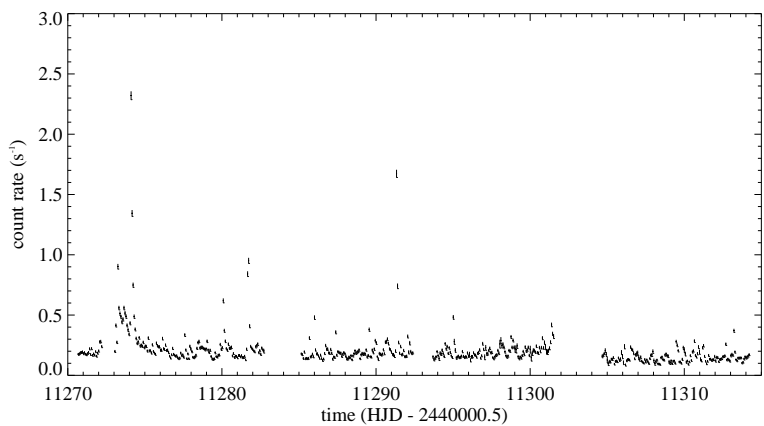}}}
\caption{A long light curve of the dMe star AD Leo, obtained by the DS instrument on {\it EUVE}. Most of the discernible
variability is due to flares (after \citealt{guedel03a}).}\label{adleoeuve}
\end{figure}

The {\it EUVE} satellite, while not being very sensitive, secured many observations from long monitoring
programs that lasted up to 44 days. Some of these light curves reveal an astonishing level of continuous
variability in main-sequence stars (\citealt{audard00, guedel03a}; Fig.~\ref{adleoeuve}), in RS CVn binaries 
(\citealt{osten99, sanz02}; Fig.~\ref{flareexamples}), and in giants (\citealt{ayres01a}; Fig.~\ref{flareexamples}).
Some of those data were used to investigate statistical properties of the flare energy distribution 
(see Sect.~\ref{energydistribution}).

\begin{figure} 
\centerline{\resizebox{0.95\textwidth}{!}{\includegraphics{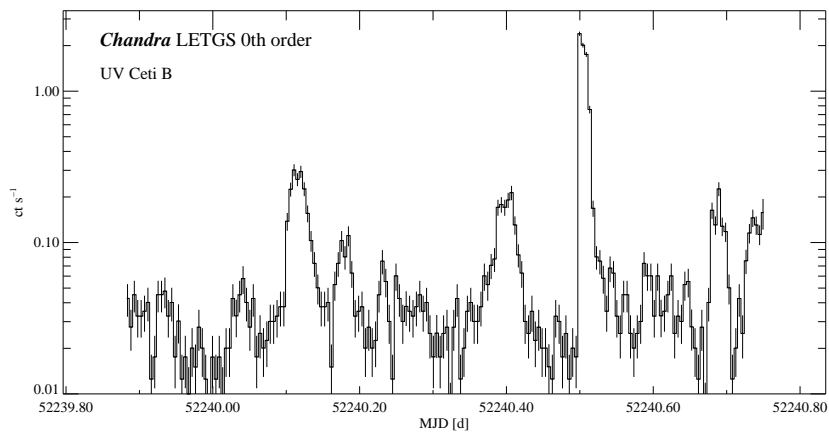}}}
\caption{Light curve of UV Ceti B, observed with the {\it Chandra} LETGS/HRC during about
1 day. Note the logarithmic flux axis (figure courtesy of M. Audard, after \citealt{audard03b}).}\label{uvcetlight}
\end{figure}

With the advent of the much increased sensitivity offered by {\it XMM-Newton} and {\it Chandra},
weaker flares were uncovered, and they occur - expectedly  - at higher rates.
\citet{guedel02a, guedel04} presented sensitive X-ray light curves of Proxima Centauri  
in which no time intervals longer than a few tens of minutes could be described
as constant within the given sensitivity limit. Incidentally,
this star was deemed constant outside obvious flares in the much less sensitive 
{\it EXOSAT} study by \citet{pallavicini90a}. Frequent faint, flare-like X-ray fluctuations  
in this observation were often accompanied - in fact slightly preceded - by  U band bursts, the latter 
being a signature of the initial bombardment of the chromosphere by high-energy electrons 
as discussed in Sect.~\ref{neuperteffect}. \citet{audard03b} estimated that no more than 30\%, and probably 
much less, of the long-term average X-ray emission of UV Cet can be attributed to any sort of steady emission, even
outside obvious, large flares. On the contrary, almost the entire 
light curve is resolved into frequent, stochastically occurring flares of various amplitudes (Fig.~\ref{uvcetlight}).
Many further observations  from the new observatories reveal almost continual flaring 
\citep{stelzer02, stelzer03, besselaar03, raassen03b}.  For active binaries, \citet{osten02} found that the flux 
distributions of the low-level emission significantly deviate from the Poisson distributions expected from a constant source,
once again pointing to continual variability.

\subsection{The solar analogy}

If variability is found in stellar light curves, its characteristic time scale is typically at least 
3--5 minutes and often longer than 10 minutes \citep{ambruster87, pallavicini90a, guedel02a}. 
This motivated several authors to interpret low-level variability as being due to slow reconfigurations of active
regions and emerging flux rather than due to stochastic flaring. The latter was expected to 
reveal itself in the form of short-term fluctuations, recalling the concept of ``microflaring'' in the solar
corona.

There is, however, a widespread misconception that should briefly be discussed. A popular opinion has
it that larger flares last longer, and that microvariability in stars
should therefore express itself in short-term flickering. This view is not entirely correct. 
Statistical studies of solar flares usually do not find clear evidence for a dependence
between flare duration and flare amplitude over quite wide a range in energy \citep{pearce88, 
feldman97, shimizu95}.  The distributions are  dominated
by the scatter in the duration, with durations ranging from 1 to 20 minutes even in
the domain of quite small solar events. \citet{aschwanden00b}  investigated scaling laws from 
solar nanoflares to large flares, covering 9 orders of magnitude in energy. They reported
that the radiative and conductive time scales do not depend on the flare
size. \citet{guedel03a} inspected the brightest 
flares in a long-duration EUV observation of AD Leo, again finding no
trend.  The latter authors
also studied the sample of selected large flares from {\it different} stars reported
by \citet{pallavicini90a}; although a weak trend  was found if four orders of
magnitude in energy were included ($\tau \propto E^{0.25}$), the scatter again
dominated, and it is unclear what form of selection bias was introduced given
that the sample consists only of well-detected bright flares (for the same reason, I refrain 
from performing statistics with the flares in Table~\ref{flaretable}; both the decay times
and the total energies are subject to selection bias). 

\begin{figure} 
\centerline{\resizebox{0.95\textwidth}{!}{\includegraphics{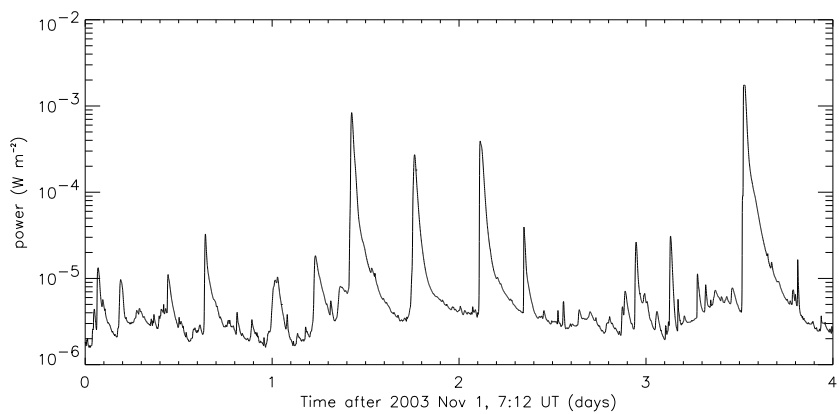}}}
\caption{{\it GOES} full-disk solar X-ray light curve, observed in the 1.5--12~keV band in
        November 2003. The abscissa gives time after 2003 November 1, 7:12 UT in days.}\label{goeslight}
\end{figure}

In fact, the claim that light curves are strictly bi-modal, separating detected flares from
truly quiescent episodes, is tantamount to requiring that the peak flux of each flare must exceed a certain
fraction of the ``quiescent'' emission level;  since this level varies over at least 6 orders
of magnitude in cool stars, the most luminous stars would be bound to produce only flares that
exceed the largest solar events by many orders of magnitude, biased such that they are detected by available
detectors. This view is not supported by solar observations: the solar flare rate increases steeply toward lower radiative
energies, with no evidence (yet) for a lower threshold (e.g., \citealt{krucker98}). Figure \ref{goeslight}
shows an example of a {\it GOES} light curve in the 1.5-12~keV range, purposely selected during an extremely
active period in November 2003. While  the {\it GOES} band is harder than typical bands
used for stellar observations, it more clearly reveals the level of the underlying variability (a typical
detector used for stellar observations would see much less contrast).  If the solar
analogy  has any merit in interpreting stellar coronal X-rays, then  
low-level emission in stars that do show flares {\it cannot} be truly quiescent, that is, constant or slowly varying exclusively due
to long-term evolution of active regions, or due to rotational modulation.
A measure of {\it flare rates} is therefore not meaningful unless it refers to flares above a 
given luminosity or energy threshold.
This is - emphatically - not to say that steady emission is absent. However, once we accept the solar analogy
as a working principle, the question is not so much about the presence of large numbers of flares, but to what
extent they contribute to the overall X-ray emission from coronae.

\subsection{The flare-energy distribution}\label{energydistribution}

The suggestion that stochastically occurring flares may be largely responsible for
coronal heating is known as the ``microflare'' or ``nanoflare'' hypothesis in solar 
physics \citep{parker88}. Observationally, it is supported by evidence  for the presence of
numerous  small-scale flare events occurring in the solar corona 
at any time (e.g., \citealt{lin84}). Their distribution in energy is  a 
power law,
\begin{equation}\label{powerlaw}
\frac{dN}{dE} = k E^{-\alpha} 
\end{equation}
where $dN$ is the number of flares per unit time with a total 
energy in the  interval [$E,E+dE$]. If $\alpha\ge 2$, then the energy integration (for a given 
time interval) diverges for $E_{\rm min} \rightarrow 0$, that is, by 
extrapolating the power law to sufficiently small flare energies, {\it any} 
energy release power can be attained. This is not the case for $\alpha <2$.
Solar studies have repeatedly resulted in $\alpha$ values of $1.6-1.8$ for ordinary
solar flares \citep{crosby93}, but some recent studies of low-level  flaring
suggest $\alpha = 2.0 - 2.6$ \citep{krucker98, parnell00}.

\begin{table}[b!] 
\caption{Stellar radiative flare-energy distributions} 
\label{flarestat}       
\begin{tabular}{lllll} 
\hline\noalign{\smallskip}
Star sample	       & Photon energies      & log\,(Flare             & $\alpha$	 & References \\
         	       & [keV]                & energies)$^a$         &          	 &            \\
\noalign{\smallskip}\hline\noalign{\smallskip}
M dwarfs	       &  0.05--2   	      & $30.6-33.2$ & 1.52$\pm 0.08$ & \citet{collura88} \\    
M dwarfs	       &  0.05--2   	      & $30.5-34.0$ & 1.7$\pm 0.1$   & \citet{pallavicini90a} \\
RS CVn binaries        & EUV		      & $32.9-34.6$ & 1.6	     & \citet{osten99} \\
Two G dwarfs	       & EUV		      & $33.5-34.8$ & 2.0--2.2       & \citet{audard99} \\
F-M dwarfs	       & EUV		      & $30.6-35.0$ & 1.8--2.3       & \citet{audard00} \\
Three M dwarfs         & EUV		      & $29.0-33.7$ & 2.2--2.7       & \citet{kashyap02}\\
AD Leo  	       & EUV\&0.1--10         & $31.1-33.7$ & 2.0--2.5       & \citet{guedel03a}  \\
AD Leo  	       & EUV                  & $31.1-33.7$ & $2.3\pm 0.1$   & \citet{arzner04} \\
\noalign{\smallskip}\hline
\multicolumn{5}{l}{$^a$Total flare-radiated X-ray energies used for the analysis (in ergs).}
\end{tabular}
\end{table}

Relevant stellar studies have been rare (see Table~\ref{flarestat}). Early investigations lumped several stars
together to produce meaningful statistics. A set of M dwarf flares 
observed with {\it EXOSAT} resulted in $\alpha \approx 1.5$ \citep{collura88}, and similarly,
the comprehensive study by \citet{pallavicini90a} of an {\it EXOSAT} survey of M dwarfs
implied $\alpha \approx 1.7$. \cite{osten99} used {\it EUVE} data to perform a similar investigation of 
flares on RS CVn-type binaries, and again by lumping stellar samples together they 
inferred $\alpha \approx 1.6$.

However, several biases may affect statistical flare studies, all related to the 
ill-determined problem of flare identification in stellar observations.
First, detecting flares in light curves is a problem of contrast. Poisson counting
statistics increases the detection threshold for flares on top of 
a higher continuous
emission level. The low-energy end of the distribution of detected flares is
therefore ill-defined and is underrepresented. 
Further, large flares can inhibit the detection of the more numerous weak
flares during an appreciable fraction of the observing time. And lastly, the
detection threshold also depends on stellar distance; by lumping stars together,
the low-energy end of the distribution becomes invariably too shallow.

\begin{figure} 
\centerline{
\hbox{\hskip 0.6truecm
\resizebox{0.54\textwidth}{!}{\includegraphics{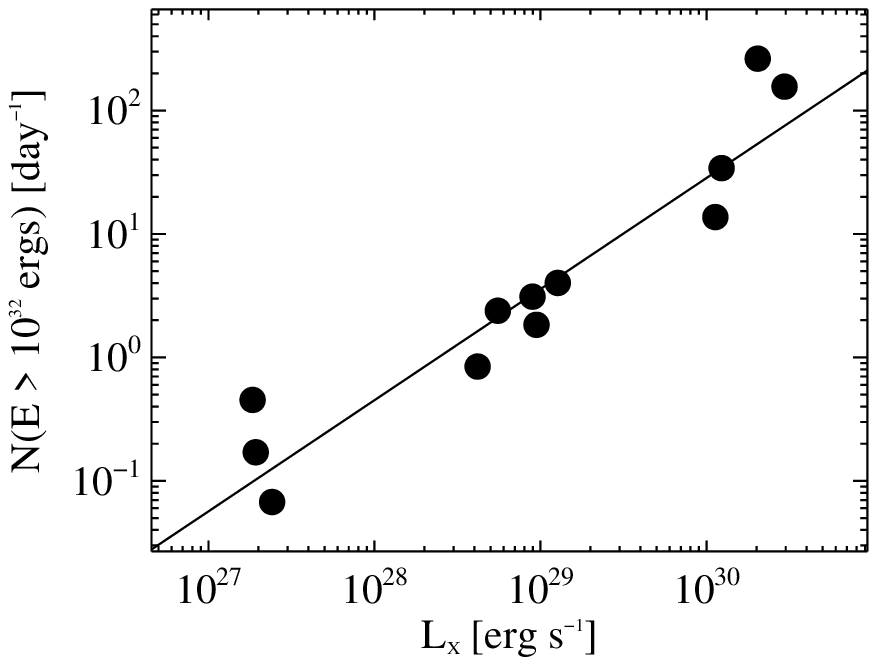}}
\hskip 0.4truecm\resizebox{0.49\textwidth}{!}{\includegraphics{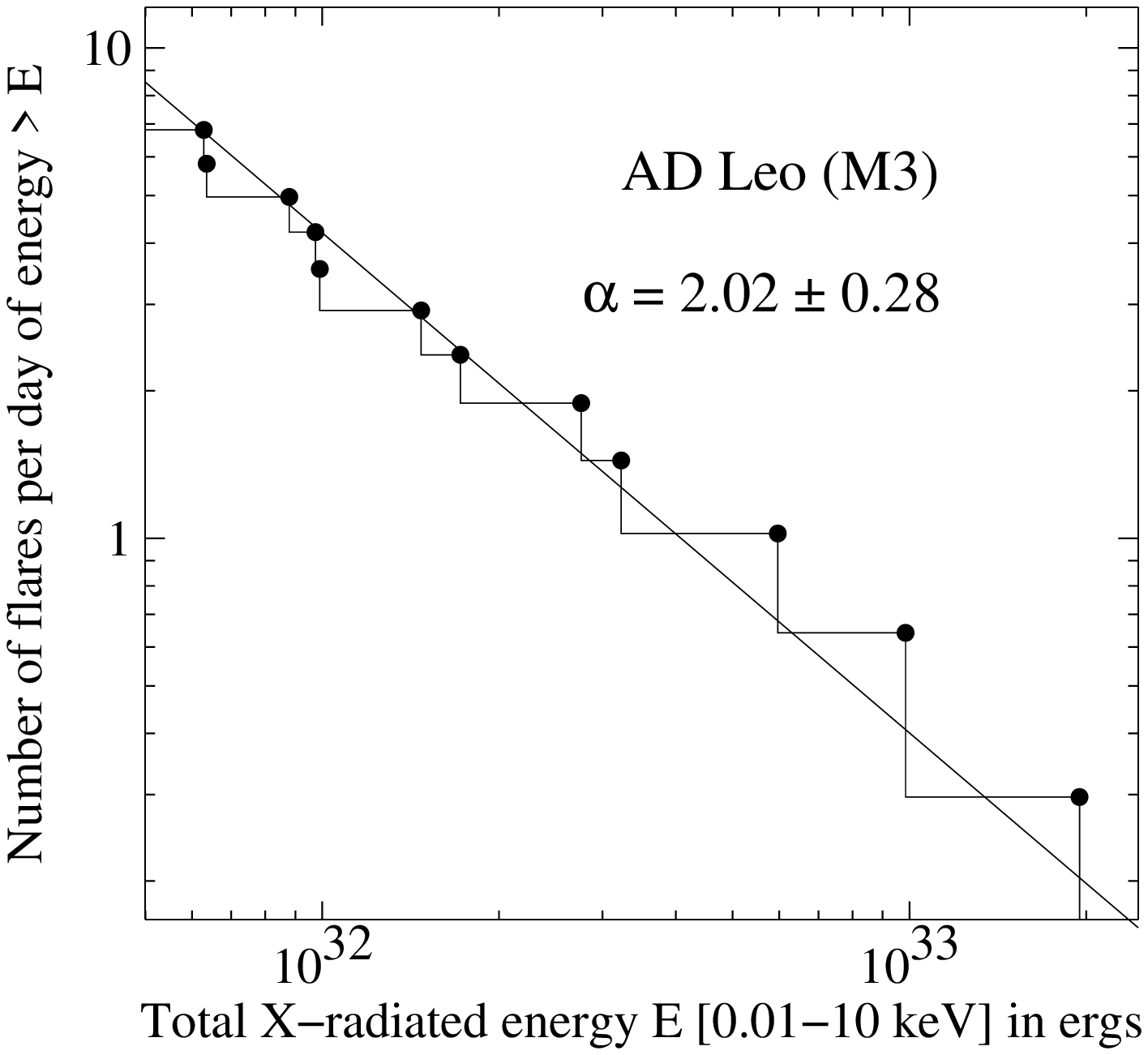}}
}
}
\vskip 0.5truecm
\caption{{\it Left:} The rate of flares above a threshold of $10^{32}$~erg in total
radiated X-ray energy is plotted against the low-level luminosity for several stars, together
with a regression fit.
{\it Right:} Flare energy distribution for AD Leo,
using a flare identification algorithm for an observation with {\it EUVE} (both figures courtesy of M. 
Audard, after \citealt{audard00}).}\label{flaredistrib}
\end{figure}

To avoid biases of this kind,
\citet{audard99} and \citet{audard00}  applied a flare search 
algorithm to {\it EUVE} light curves of individual active main-sequence stars, taking 
into account flare superpositions and
various binning to recognize weak flares, and performing the analysis on individual light curves.
Their results indicate a predominance of relatively steep power laws including 
$\alpha \ge 2$ (an example is shown in Fig.~\ref{flaredistrib}b).

Full forward modeling of a superposition of stochastic flares was applied to EUV and
X-ray light curves by \citet{kashyap02} and \citet{guedel03a} based on Monte Carlo simulations, 
and by \citet{arzner04} based on an analytical formulation.\footnote{Note that the distribution of measured fluxes
does {\it not} describe the flare amplitude distribution. The problem of inverting the former to
obtain the latter was analytically solved by \citet{arzner04}.}
The results of these investigations are in full agreement, converging to $\alpha \approx 2.0 - 2.5$
for M dwarfs (Table~\ref{flarestat}). If the power-law flare energy distribution extends by about 1--2 orders of 
magnitude below the actual detection limit in the light curves, then the {\it entire} emission could be explained
by stochastic flares. The coronal heating process in magnetically active stars would - in this extreme limit - be one solely due to 
{\it time-dependent} heating by flares, or, in other words, the X-ray corona would be an entirely hydrodynamic
phenomenon rather than an ensemble of hydrostatic loops.

\subsection{Observables of stochastic flaring}\label{obsstochflare}

In order for stochastic flaring to be an acceptable coronal heating  mechanism, it should 
explain a number of X-ray observables. Flares develop characteristically in 
EM and $T$ (fast rise to peak, slow decay). Therefore, the superposition of a statistical 
ensemble of flares produces a characteristic time-averaged DEM. For simple flare decay laws,
the resulting DEM is analytically given by Eq.~(\ref{demflare}) that fits observed, steeply 
rising DEMs excellently \citep{guedel03a}.
An extension of the \citet{kopp93} formalism can be used numerically to find approximations of a DEM that
results by time-averaging the EM$(T,t)$ of a population of cooling, stochastic flares drawn from
a distribution with a prescribed 
$\alpha$; these DEMs show two characteristic maxima, the cooler one being induced by the large rate of 
small flares, and the hotter one being due to the few large flares with high temperature and EM, somewhat similar
to double-peaked DEMs that were derived from observations; in this interpretation, the higher rate
of larger flares in more active stars shifts the EM to higher temperatures, as
has been found in evolutionary sequences of solar analogs \citep{guedel97a, guedel97c, skinner98}.

\begin{figure} 
\centerline{
\vbox{\hbox{
\resizebox{0.5\textwidth}{!}{\rotatebox{90}{\includegraphics{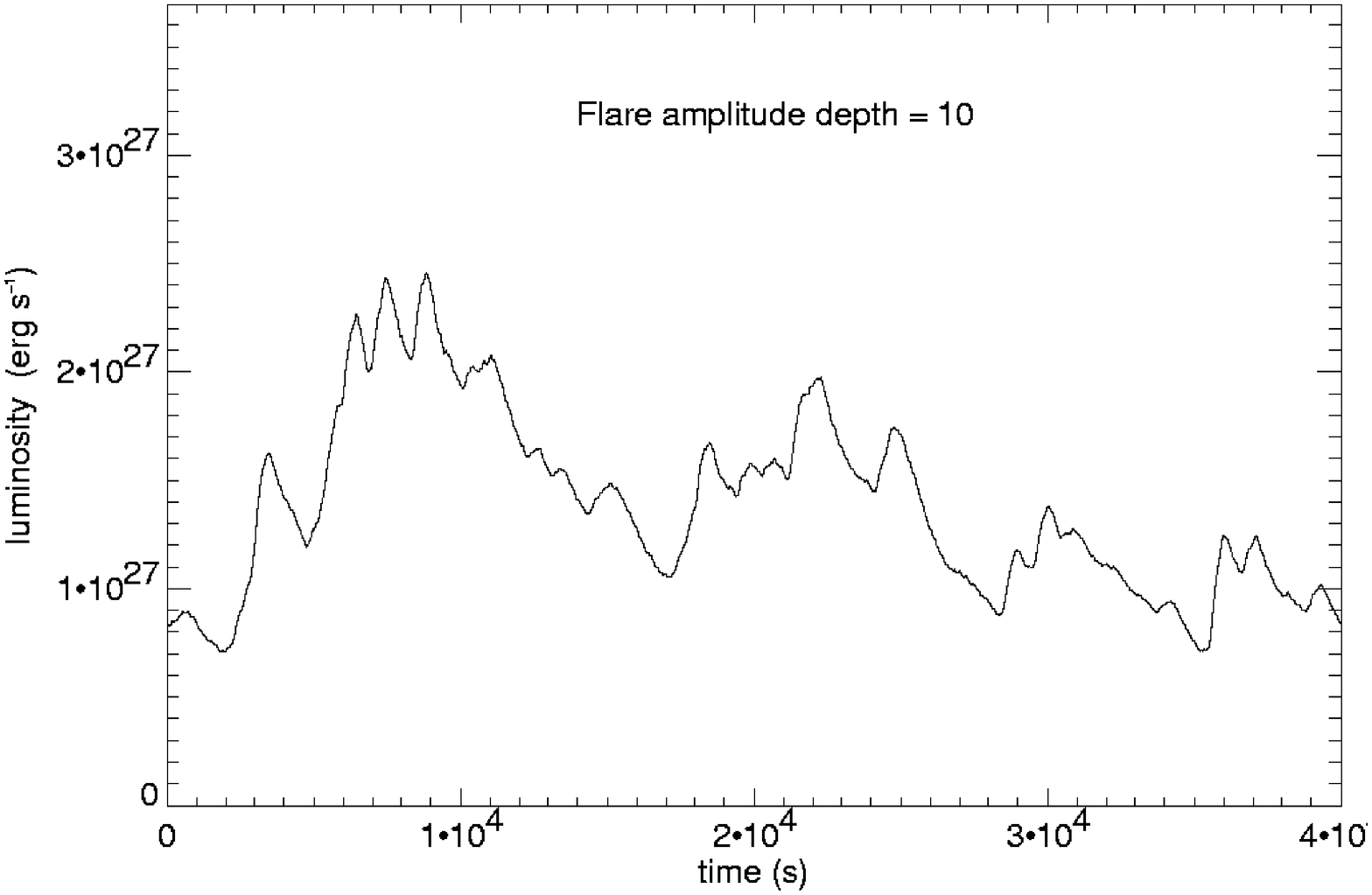}}}
\resizebox{0.5\textwidth}{!}{\rotatebox{90}{\includegraphics{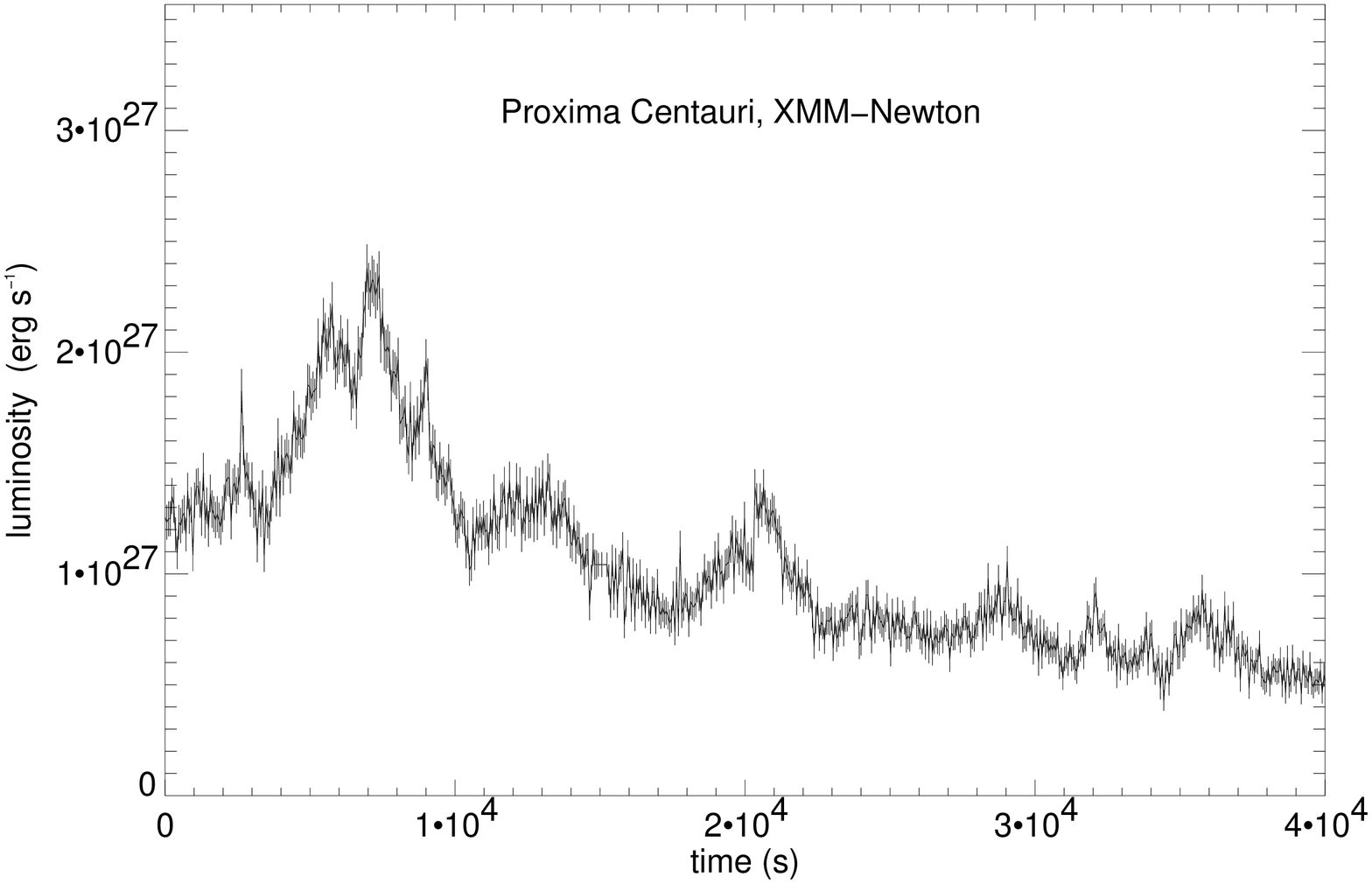}}}
}
\hbox{
\resizebox{0.5\textwidth}{!}{\rotatebox{90}{\includegraphics{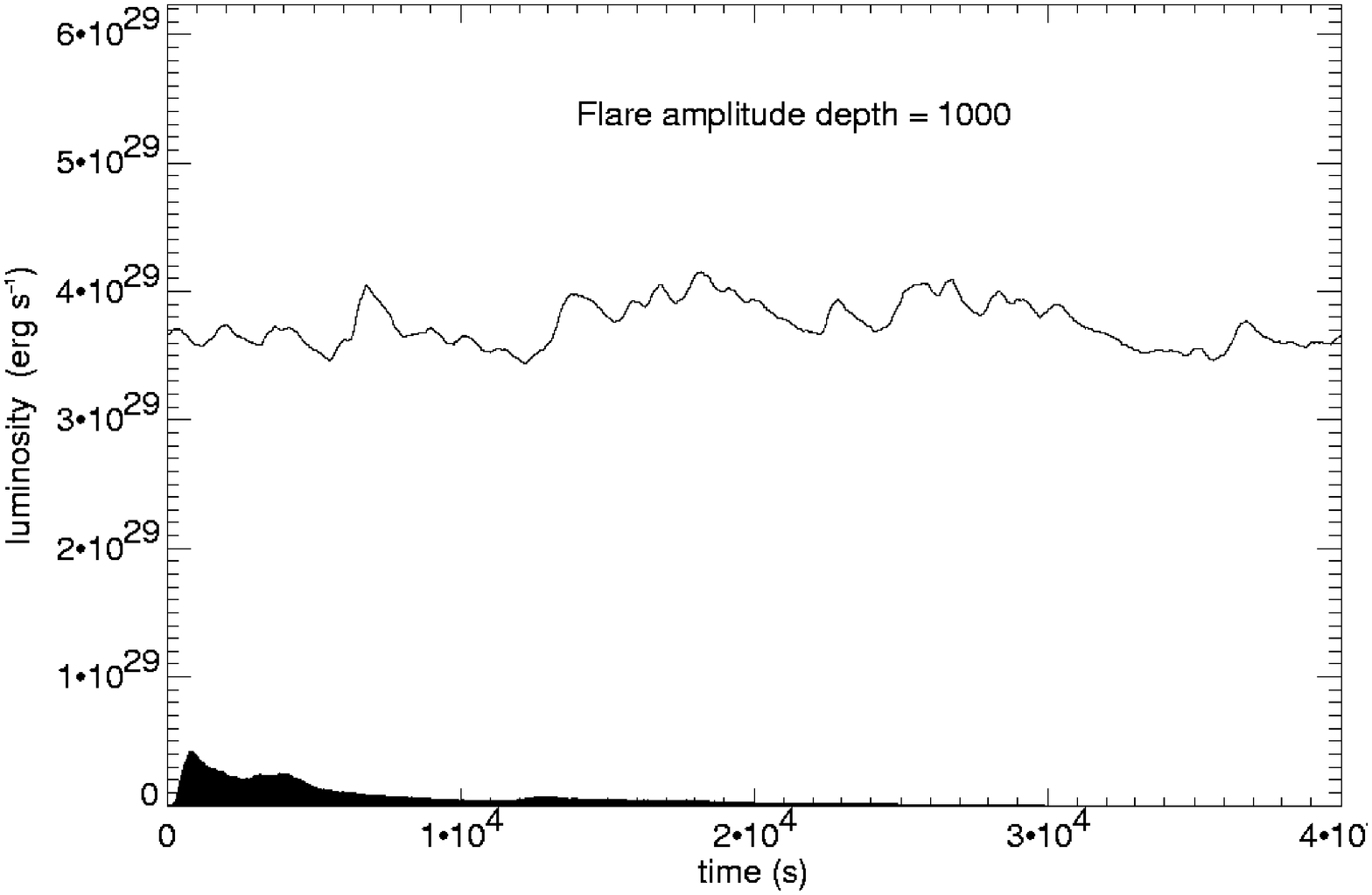}}}
\resizebox{0.5\textwidth}{!}{\rotatebox{90}{\includegraphics{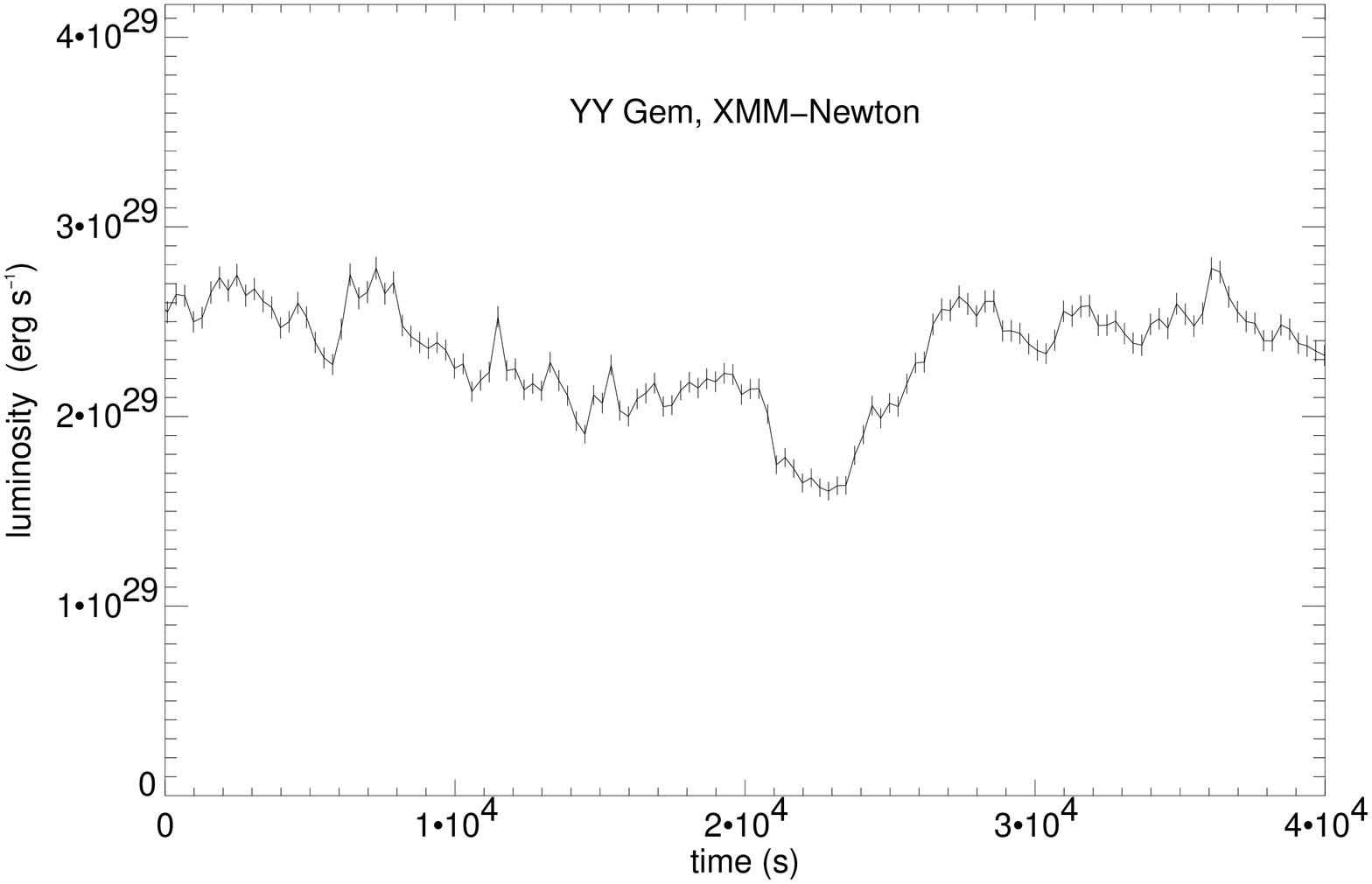}}}
}
}
}
\caption{X-ray light curves. {\it Left panels:} Synthetic light curves from superimposed
flares; largest to smallest amplitude = 10 (upper) and 1000 (lower plot). No truly quiescent emission
has been added. A
giant flare on Proxima Centauri \citep{guedel02a} has been used as a shape template, 
shown filled in the lower plot for the  maximum amplitude
contributing to the light curve. {\it Right panels:} Observed X-ray light curves  for comparison. Proxima
Centauri (upper plot, \citealt{guedel04}) and YY Gem (lower plot, including eclipse; after 
\citealt{guedel01a}).}\label{flaresuperpos}
\end{figure}

As for the {\it light curves}, the superposition of stochastically occurring flares produces 
an apparently steady baseline emission level that is constantly present and that may be mistaken
for a truly quiescent component \citep{kopp93}, in particular for large $\alpha$ 
values in Eq.~(\ref{powerlaw}). Figure \ref{flaresuperpos} shows simulations of superimposed flares 
drawn from a power-law distribution with $\alpha = 2.2$, compared with observations. No truly
steady emission has been added. The large flare
observed on Proxima Centauri \citep{guedel04} was used as a shape template from which all 
flares in the simulation were scaled.
The first example uses flare energies spread over a factor of ten only, the second uses three
orders of magnitude. In the latter case, the light curve has smoothed out to an extent that it is
dominated by ``quiescent'' emission. This effect is stronger for higher $\alpha$. The individual peaks are merely the peaks of the most energetic
flares  in the ensemble.  For small (dM) stars, the available stellar area may in fact constrain the 
energy range of stochastic flares because flaring active regions may cover a significant
fraction of the surface, which limits the number
of simultaneous flares (see upper example in Fig.~\ref{flaresuperpos}, e.g., Proxima Centauri) and thus makes 
lower-mass stars statistically more variable.

Finally, spectroscopic {\it density} measurements of a time-integrated, stochastically flaring corona 
should  yield values equivalent to the densities derived from time-integrating the
spectrum of a large flare. There is suggestive evidence in support of this. While large stellar X-ray flares
achieve peak electron densities of several times $10^{11}$~cm$^{-3}$,
the time-integrated X-ray spectrum of the Proxima Centauri flare described by 
\citet{guedel02a, guedel04} (Fig.~\ref{proxcenflare}) yields a characteristic density of log\,$n_e \approx 10.5 \pm 0.25$
derived from O\,{\sc vii} triplet, which compares favorably with  densities in 
magnetically active MS stars during low-level emission (\citealt{ness01, ness02a}, Sect.~\ref{denshe}).

\section{X-ray absorption features and prominences}

X-ray spectra are sensitive to photoelectric absorption by cooler foreground gas. The absorption column 
is a powerful diagnostic for the amount of cool circumstellar gas although little can be said about
its distribution along the line of sight. X-ray attenuation by the interstellar
medium is generally weak for field stars within a few 100~pc, but it becomes very prominent for deeply 
embedded stars in star formation regions, or stars that are surrounded by thick accretion disks. 

However, anomalous absorption is sometimes also recorded in more solar-like, nearby stars, in particular
during large flares. The observed column densities may
vary by typically a factor of two on short time scales. Examples were presented by
\citet{haisch83} for Proxima Cen, \citet{ottmann96} for  Algol, and by \citet{ottmann94b} for  AR Lac.
Much larger absorption column densities have occasionally been measured, up to $4\times 10^{22}$~cm$^{-2}$ 
in a flare on V733 Tau \citep{tsuboi98} and $3\times 10^{21}$~cm$^{-2}$ in the course of a large flare
on Algol \citep{favata99}. The cause of this anomalous absorption is not clear. From the solar analogy,
it is possible that prominences related with the flare region pass in front of the
X-ray source and temporarily shadow part of it. Coronal mass ejections sometimes accompany solar
flares. If the material cools sufficiently rapidly by radiation and expansion, it may also
attenuate the X-rays. The detection of ejected mass is of some relevance in binaries. In systems like Algol,
the mass may flow onto the binary companion or form a temporary accretion disk 
around it \citep{stern92a}. This could  explain why excess absorption has been found also during low-level emission 
episodes in several  Algol-type binaries \citep{singh95}. In rapidly rotating single stars, increased amounts of
hydrogen may condense out of large cooling loops that grow unstable near their apexes, a suggestion that has found 
direct support in optical observations of AB Dor  (e.g., \citealt{collier88a, donati99}).

\section{Resonance scattering and the optical depth of stellar coronae}\label{opticaldepth}

When studies of the initial spectra from the {\it EUVE} satellite encountered unexpectedly
low line-to-continuum ratios (e.g., \citealt{mewe95}), line suppression by optical depth effects due to
resonant scattering in stellar coronae surfaced as one possible explanation \citep{schrijver94}. 
Although we nowadays attribute these anomalies mostly to abundance anomalies and  to
an excess ``pseudo''-continuum from weak lines, a search for non-zero optical 
depths in coronae is worthwhile because it may be used as another tool to study coronal structure. 
Non-negligible optical depths would also affect our interpretation of stellar coronal X-ray spectra for
which we usually assume that the corona is entirely optically thin. 

Resonant scattering requires optical depths
in the line centers of $\tau \ga 1$. The latter is essentially proportional
to $n_e\ell/T^{1/2}$ \citep{mewe95}. For static coronal loops, this implies
$\tau \propto T^{3/2}$ (\citealt{schrijver94}; e.g., along a loop or for a
sample of nested loops in a coronal volume). Numerical values for $\tau$ are given
in \citet{schrijver94} and \citet{mewe95} for several EUV lines. If the optical depth
in a line is significant in a coronal environment, then the absorbed photon will be
re-emitted (``scattered'') prior to collisional de-excitation. 
Continuum photons, on the other hand, are much less likely to be scattered. The effects
of resonant line scattering into and out of the line of sight, however,
cancel in  a homogeneous source, but the stellar surface that absorbs down-going photons breaks 
the symmetry: The line-to-continuum ratio  is reduced if the emitting volume is
smaller than the scattering volume and lies closer to the stellar surface \citep{schrijver94,
mewe95}. This situation is fulfilled in a corona since the scattering efficiency decreases only
with $n_e$ while the emissivity decreases more rapidly, $\propto n_e^2$. \citet{schrijver94} 
predicted suppression of the line fluxes by up to a factor of two in the most extreme cases. 

A number of applications to {\it EUVE} spectra from stars at different activity levels have been 
given by \citet{schrijver95}. The authors concluded that significant optical depths may be present in 
particular in inactive stars such as $\alpha$ Cen and Procyon. The scattering layer would most 
probably be an extended hot envelope or a stellar wind. This interpretation was subsequently 
challenged, however, by \citet{schmitt96a} in a detailed study of the EUV continuum and a comparison of 
flux levels in the {\it ROSAT} spectral range. They attributed the anomalously low line-to-continuum ratios
to an excess EUV  continuum that builds up from the superposition of many weak lines that 
are not tabulated in present-day codes.

In the X-ray range, the usually  strong Ne-like Fe\,{\sc xvii} lines  provide excellent diagnostics.
The  Fe\,{\sc xvii} $\lambda$15.01 line is often chosen for its large oscillator strength. Its flux is compared
with the fluxes of Fe\,{\sc xvii}   lines with  low-oscillator strengths  such as those at 15.26~\AA\ and at 16.78~\AA. 
There is considerable uncertainty in the atomic physics, however, that has equally affected 
interpretation of {\it solar} data. While \citet{schmelz97} and \citet{saba99}
found evidence for optical depth effects in the Fe\,{\sc xvii} $\lambda$15.01
line, recent laboratory measurements  of the corresponding flux ratios differ significantly from previous 
theoretical  calculations \citep{browng98, brown01, laming00}. New calculations have recently
been  presented by, for example, \citet{doron02}.

{\it Stellar} coronal optical depths have become accessible in the X-ray range with {\it Chandra} and
{\it XMM-Newton}. No significant optical depth effects were found for Fe\,{\sc xvii}  $\lambda$15.01
for a restricted initial sample of stars such as Capella and Procyon \citep{brinkman00, ness01, mewe01, phillips01}.
\citet{ness01} calculated upper limits to optical depths based on measured EM and electron
densities, assuming homogeneous sources. They found no significant optical depths
although none of these stars is very active. \citet{huenemoerder01}, \citet{ness02b} 
and \citet{audard03a} 
extended the  Fe\,{\sc xvii} diagnostics to the high end of magnetic activity, viz. II Peg, Algol and a sample of RS CVn
binaries, respectively, but again reported no evidence for optical depth effects. The problem was
comprehensively studied by \citet{ness03a} in a survey of 26 stellar coronae observed with
{\it XMM-Newton} and {\it Chandra} across all levels of activity. They again used the Fe\,{\sc xvii}  $\lambda 15.27/\lambda 15.01$
and $\lambda 16.78/\lambda 15.01$ diagnostics as well as the ratios between the resonance and the forbidden
lines in He-like line triplets of O\,{\sc vii} and Ne\,{\sc ix}. Many line ratios
are at variance with solar measurements  and with calculated predictions, but the  latter themselves
are uncertain (see above). The interesting point is, however, that
the flux ratios are similar for all stars except for those with the very coolest coronae for which line blends are suspected to bias
the flux measurements. \citet{ness03a} concluded that optical depth effects are absent on {\it all}
stars at least in the relevant temperature regime, rather than requiring non-zero but identical
optical depth in such variety of stellar coronae. These conclusions also extend to
line ratios in Ly$\alpha$ series (e.g., Ly$\alpha$:Ly$\beta$ for O\,{\sc viii}, Ne\,{\sc x}, or 
Si\,{\sc xiv}, \citealt{huenemoerder01, osten03}).

Related effects were previously considered, but also questioned, for the Fe K complex at 
6.7--7~keV in intermediate-resolution observations \citep{tsuru89, stern92a, singh96a}.
Lastly, \citet{guedel04} measured Fe\,{\sc xvii} flux ratios during a
strong flare on Proxima Centauri but again found neither discrepant values nor a time evolution that
would contradict an optically thin assumption. 

To conclude, then, it appears that no X-ray optical depth 
effects have unambiguously been detected in any stellar corona investigated so far, notwithstanding
the large range of geometries, temperatures, and densities likely to be involved in stars
across the spectrum of activity, including large flares. Further checks of individual cases
or of large flares remain worthwhile, however, given the potential diagnostic power of resonant scattering 
effects.\footnote{{\it Note added in proof:} A recent report by Testa et al. (2004, ApJ, 609, L79) indicates
evidence of resonance scattering in Ly$\alpha$/Ly$\beta$ line-flux ratios of O\,{\sc viii} and Ne\,{\sc x} in
the RS CVn binaries II Peg and IM Peg. The inferred path lengths vary between $2\times 10^{-4}R_*$ and 
$4\times 10^{-2}R_*$.}

\section{The elemental composition of stellar coronae}\label{composition}

It is quite well established that the solar corona and the solar wind show an elemental composition 
at variance with the composition of the solar photosphere. Whatever the reason for the discrepancy, 
our interest in understanding element abundances in stellar coronae is twofold:
Observationally, because they shape the X-ray spectra from which we derive basic coronal parameters;
and physically, because abundance anomalies reflect diffusion processes and element fractionation mechanisms 
in the chromosphere and transition region, and possibly the physics of mass transport into the corona.

\subsection{Solar coronal abundances: A brief summary}\label{solarabun}

Measurements of the composition of the solar corona have revealed what is commonly known as the 
``First Ionization Potential (FIP) Effect'': Essentially, elements with a  FIP above $\approx 10$~eV
(e.g., C, N, O, Ne, Ar) show photospheric abundance ratios with respect to hydrogen, while elements with a smaller 
FIP (e.g., Si, Mg, Ca, Fe) are overabundant by a factor of a few. This picture was comprehensively summarized
in the extensive work by \citet{meyer85a, meyer85b} who showed that the same
FIP effect is also present in the solar wind, in solar energetic particles, and - quite surprisingly - in
cosmic rays.\footnote{I note in passing that Meyer normalized the abundances such that the low-FIP element
abundances were photospheric and the high-FIP abundances depleted, whereas present-day wisdom has the 
high-FIP elements at photospheric levels, and the low-FIP elements {\it enriched} in the corona 
\citep{feldman92}.} The latter finding immediately suggested that the seeds of cosmic rays
may perhaps be ejected by active stellar coronae that are subject to a similar, solar-like abundance anomaly.
An extension of the currently accepted picture was given by \citet{feldman92} who discussed
abundance anomalies in various solar features, pointing out that the degree of the FIP anomaly
varies from feature to feature. 

I refer the interested stellar reader to the extensive solar literature on 
the solar FIP effect, and in particular to a review of solar FIP models
by \citet{henoux95}, overviews presented by \citet{jordan98} and in 
the papers by \citet{drake95b} and \citet{laming95}. I only briefly summarize a few principal points.
Current thinking is that a fractionation process, probably involving electric and/or magnetic fields
or pressure gradients, occurs at chromospheric levels where low-FIP elements are predominantly
ionized and high-FIP elements are predominantly neutral.
Ions and neutrals are then affected differently by electric and magnetic fields. A successful
model for the FIP effect will eventually have to explain why and how low-FIP elements are 
transported into the corona at an enhanced rate. 

The FIP effect is most
pronounced in relatively evolved solar coronal features such as old loops, but also in 
magnetically open regions. In contrast, young, compact active regions and newly emerged
structures reveal photospheric composition. The latter mixture is also most evident in flares, which suggests
that new material is brought up from photospheric/chromospheric layers that has not been
subject to fractionation.  Notable exceptions exist, such as Ca-rich flares \citep{sylwester84},
and Ne-rich flares \citep{schmelz93}.

In the light of the diverse FIP anomalies in the solar corona, stellar observations should
obviously be compared with Sun-as-a-star data. To this end, \citet{laming95} have studied
the FIP anomaly by making use of full-disk solar spectra. While they confirmed the presence of an overall
coronal enrichment of low-FIP elements by factors of about 3--4, they somewhat surprisingly reported
an absence of a FIP bias at subcoronal temperatures ($< 1$~MK). They suggested that the strength
of the FIP effect is in fact a function of altitude, with the lower-temperature emission
being dominated by the supergranulation network that shows photospheric abundances.

\subsection{Stellar coronal abundances: The pre-{\it XMM-Newton/Chandra} view}

Early X-ray observatories typically lacked the spectral resolution required to
make solid statements about coronal abundances.
Some indications of possible  depletions of Fe were reported during very strong flares, as will
be summarized in Sect.~\ref{flareabun}. Also, some early medium-resolution spectra apparently required
``anomalous abundances'' for a successful model fit \citep{walter78b, swank81}. Despite these initial
developments, element abundances became a non-issue as the available detectors simply did not
permit their unambiguous determination. It was customary to adopt a coronal composition  equal to
the solar photospheric mixture. The basis for this assumption is that most stars in the solar neighborhood 
belong to Population I, and for these stars we expect a near-solar composition, even if the
Sun is a somewhat metal-rich star. On the other  hand, the known FIP bias of the solar corona should
be more of a concern when interpreting stellar spectra, but the large variations in the solar
coronal composition would have made any ``standard coronal abundance'' tabulation quite arbitrary. As we now know, adopting such
a coronal standard would have been useless.

When {\it ASCA} introduced routine medium-resolution X-ray spectroscopy based on
CCD technology with a resolving power of $R \approx 10-30$, and {\it EUVE} allowed for well-resolved EUV 
line spectroscopy with $R \approx 300$, some classes of magnetically active stars 
started to show perplexing abundance features that were  neither comparable with those of the solar corona
nor compatible with any pattern expected from the  photospheric composition (which
is often  not well determined either).

The only way to reconcile thermal models with the observed CCD
spectra was to introduce depleted abundances in particular 
of Mg, Si, and Fe but also of other elements \citep{white94, white96, antunes94, gotthelf94, 
drakes94a, drakes96, singh95, singh96a, mewe96, kaastra96}. Abundances 
of Fe as low as 10--30\% of the solar photospheric value were  regularly reported. 
Little in the way of systematic trends was present for the various other elements \citep{drakes96, kaastra96}. 
Some observations with {\it ASCA} indicated the presence 
of a ``relative'' FIP effect because the high-FIP elements were more depleted than the 
low-FIP elements while all abundances were subsolar, but the evidence was marginal and was not followed by
Fe (\citealt{drakes94a, tagliaferri97}; \citealt{guedel97a} for Mg).
Generally, metal depletion was found to be strongest  in the most
active stars \citep{singh95, singh99, drakes96}. 

{\it EUVE} confirmed the considerable metal depletion for magnetically active stars
based on unexpectedly low Fe line-to-continuum ratios 
\citep{stern95a}. Extreme cases such as CF Tuc showed almost no Fe lines, requiring Fe 
abundances as low as 10\% solar \citep{schmitt96d}. Further examples
of significant metal depletion were reported by \citet{rucinski95} and \citet{mewe97}.

{\it EUVE} spectra, however, also revealed a solar-like FIP-related bias, but this was found
exclusively for weakly or intermediately active stars such as $\alpha$ Cen \citep{drake97}, $\epsilon$ Eri 
\citep{laming96} and $\xi$ Boo A \citep{laming99, drake01b}.
\citet{mewe98a} found further indications for a FIP effect in $\alpha$ Cen also from {\it ASCA}
observations, but no magnetically active stars were ever reported with a similar coronal composition.
Cosmic rays thus no longer appeared to be related to active stars as far as the abundance mix 
was concerned \citep{drake95b}.

A peculiarity was found for the inactive Procyon, namely identical 
coronal and photospheric compositions \citep{drake95a, drake95b}. A possible cause of
this anomaly among anomalies is that the supergranulation network,
assumed to be of photospheric composition as in the solar case, reaches to higher temperatures.

As discussed  in Sect.~\ref{demmeth}, the X-ray spectroscopic abundance determination is strong\-ly
tangled with the derivation of the DEM, and the caveats and debates described there apply here.
The problem is particularly serious in low-resolution
X-ray spectra, as discussed by \citet{singh99} and \citet{favata97a}.
Despite a decent capability of CCD detectors to recover temperatures and EMs, some caveats apply to the abundance 
determination in particular if the considered spectral ranges are too restricted. The state of the field 
remained unsatisfactory, and the lack of systematics made much of the physical interpretation quite ambiguous.

\subsection{Stellar coronal abundances: New developments with {\it XMM-Newton} and {\it Chandra}}\label{abunnew}

\begin{figure} 
\centerline{\resizebox{0.85\textwidth}{!}{\rotatebox{270}{\includegraphics{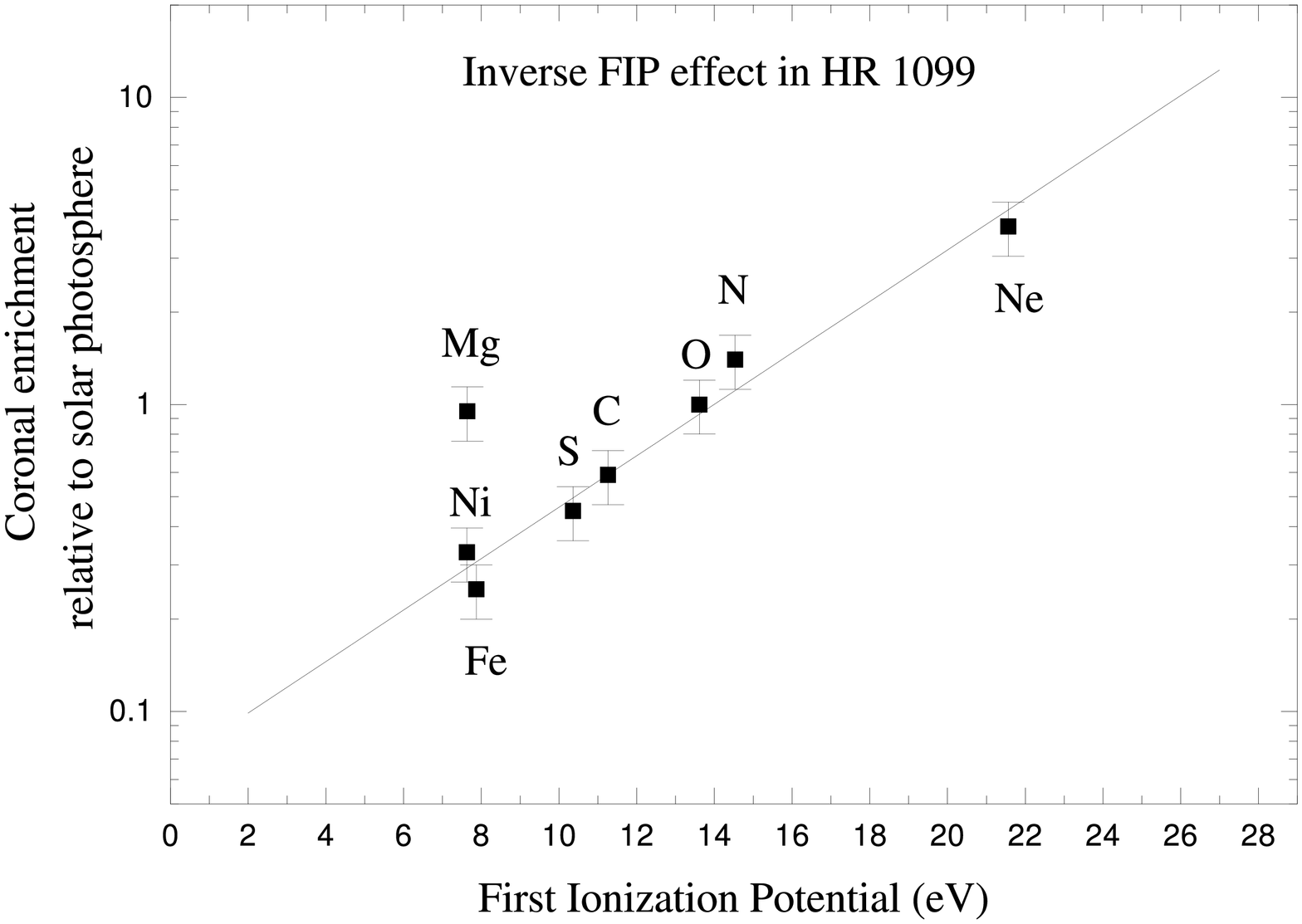}}}}
\caption{Inverse FIP effect in the corona of  HR~1099. The coronal element abundance ratios with respect to
oxygen and normalized to the solar photospheric ratios are plotted as a function of the
FIP of the respective element (after \citealt{brinkman01}).}\label{ifip}
\end{figure}

At least some clarification came  with the advent of 
high-resolution X-ray grating spectroscopy. Early observations of HR~1099 and AB Dor with the
{\it XMM-Newton} Reflection Grating Spectrometer uncovered a new, systematic FIP-related
bias in magnetically active stars: in contrast to the solar case, {\it low-FIP abundances are 
systematically depleted with respect to high-FIP elements} (\citealt{brinkman01, guedel01b, audard01a}, Fig.~\ref{ifip}),
 a trend that has been coined the ``inverse FIP effect'' (IFIP).
As a consequence of this anomaly, the ratio between the abundances of Ne (highest FIP) and Fe (low FIP)
is unusually large, of order 10, compared to solar photospheric conditions. These trends 
have been widely confirmed for many active stars and from the various gratings available on 
{\it XMM-Newton} or {\it Chandra}  (e.g., \citealt{drake01, huenemoerder01, huenemoerder03, raassen03b, 
besselaar03}, to name a few). With respect to the  
hydrogen abundance, most elements in active stars remain, however, depleted \citep{guedel01a, 
guedel01b, audard01a}, and this agrees with  the overall findings reported previously from
low-resolution spectroscopy. 
Strong Ne enhancements can be seen, in retrospect, also in many low-resolution
data discussed in the earlier literature, and the IFIP trend has now also been traced into the 
pre-main sequence domain  by \citet{imanishi02}.

\begin{figure} 
\hbox{\hskip -0.1truecm
\resizebox{0.5\textwidth}{!}{\includegraphics{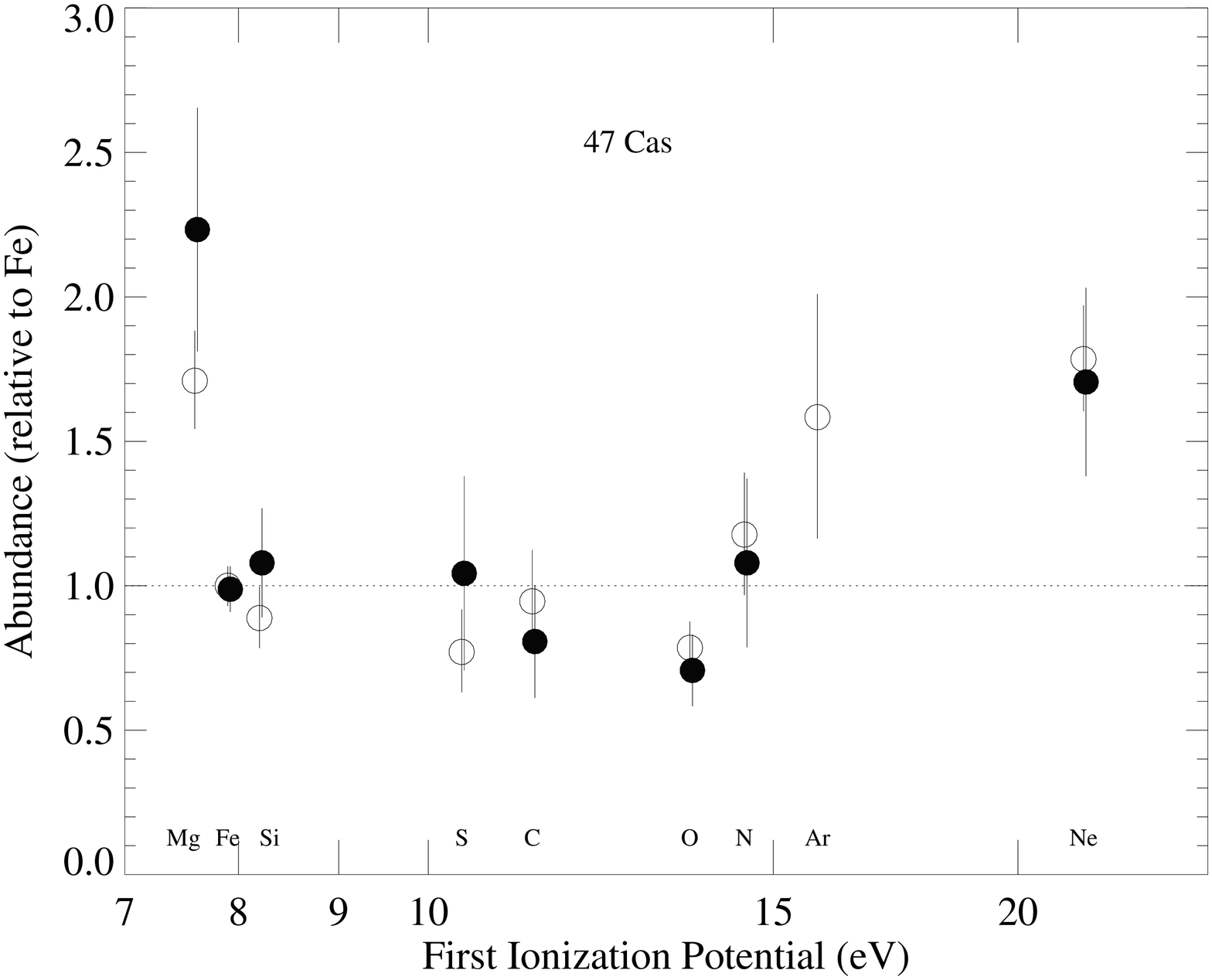}}
\resizebox{0.5\textwidth}{!}{\includegraphics{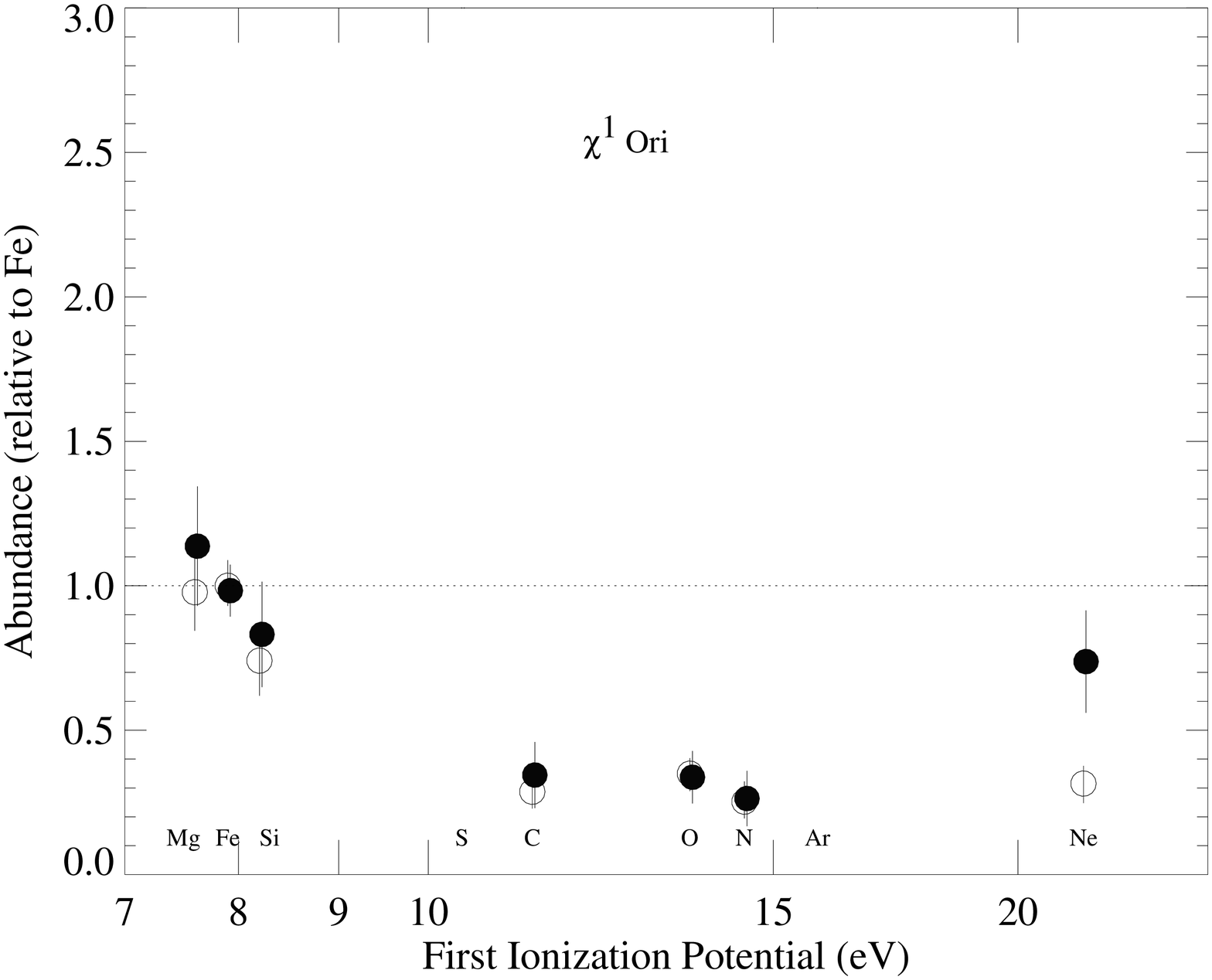}}
}
\caption{Coronal abundance determination for solar analogs. {\it Left:} 47 Cas, a very active near-ZAMS
star; {\it right:} $\chi^1$ Ori, an intermediately active solar analog. Abundances are given
relative to Fe, and refer to solar photospheric abundances as given by \citet{anders89} and \citet{grevesse99}.
Filled circles refer to determinations that used selected line fluxes of Fe for the DEM reconstruction;
open circles show values found from a full spectral fit (figures courtesy of A. Telleschi, after \citealt{telleschi04}).
 }\label{abun_solar}
\end{figure}

When stellar spectra covering a wide range of magnetic activity are compared, only highly active stars 
show the presence of an IFIP pattern. In intermediately active stars, flat abundance distributions are recovered
\citep{audard03a}. The abundances  revert to a normal, solar-type FIP anomaly for
stars at activity levels of  log\,$L_X/L_{\rm bol} \la -4$ (\citealt{guedel02c, telleschi04}, Fig.~\ref{abun_solar}). Whenever the IFIP
pattern is present, all abundances appear to be sub-solar, but the Fe/H abundance
ratio gradually rises with decreasing coronal activity. The transition from an IFIP
to a solar-like FIP abundance pattern and from very low Fe abundances to mild depletion seems
to coincide with i) the transition from coronae with a prominent hot ($T \ga 10$~MK)
component to cooler coronae, and ii) with the transition from prominent non-thermal
radio emission to the absence thereof \citep{guedel02c}.

\begin{table}
\caption{Element abundances$^a$ from high-resolution spectroscopy}
\label{abuntable}       
\begin{tabular}{llllllll}
\hline\noalign{\smallskip}
Star            & I$^b$ &{$\bar{T}$}& Fe           & Ne/Fe & O/Fe &  Mg/Fe  &        Reference	           \\
\noalign{\smallskip}\hline\noalign{\smallskip}
Procyon         & L & 1.4  &	0.66		   & 1.5   & 1.0  & 1.1	    & \citet{raassen02}		\\
Procyon         & L & 1.45 &	0.98		   & 1.08  & 0.37 & 1.66    & \citet{sanz04}    	\\
Procyon         & R & 1.8  &	1.1		   & 1.04  & 0.68 & ...	    & \citet{raassen02}		\\
$\alpha$ Cen A  & L & 1.5  & 	1.36		   & 0.37  & 0.3  & 1.01    & \citet{raassen03a}	\\
$\alpha$ Cen B  & L & 1.8  & 	1.43		   & 0.38  & 0.23 & 1.12    & \citet{raassen03a}	\\
Prox Cen        & R & 3.7  &	0.51		   & 1.6   & 0.6  & 2.1	    & \citet{guedel04}  	\\
$\epsilon$ Eri  & L & 4.0  &	0.74		   & 1.35  & 0.53 & 0.95    & \citet{sanz04}     	\\
$\chi^1$ Ori    & R & 4.4  &	0.87		   & 0.73  & 0.33 & 1.12    & \citet{telleschi04}$^c$  	\\
$\kappa^1$ Cet  & R & 4.5  &	1.18		   & 0.95  & 0.39 & 1.94    & \citet{telleschi04}$^c$  	\\
$\pi^1$ UMa     & R & 4.5  &	0.81		   & 0.62  & 0.32 & 1.24    & \citet{telleschi04}$^c$  	\\
AD Leo          & R & 5.8  &	0.34		   & 2.5   & 1.21 & 1.13    & \citet{besselaar03}       \\
AD Leo          & L & 6.1  &	0.39		   & 3.43  & 1.64 & 0.6	    & \citet{besselaar03} 	\\
Capella         & R & 5.0  &    0.92               & 0.64  & 0.32 & 1.22    & \citet{audard03a}         \\  
Capella         & R & 6.5  &	0.84		   & 0.50  & 0.50 & 1.08    & \citet{audard01b} 	\\
Capella         & L & 6.5  &	1.0		   & 0.5   & 0.48 & 0.91    & \citet{argiroffi03}       \\
YY Gem          & R & 7.6  &	0.21		   & 3.62  & 1.42 & 0.81    & \citet{guedel01a} 	\\
$\sigma^2$ CrB  & H & 9    &	0.46		   & 1.40  & 0.55 & 0.99    & \citet{osten03}	        \\
EK Dra          & R & 9.1  &	0.72		   & 1.01  & 0.51 & 1.54    & \citet{telleschi04}$^c$  	\\
AT Mic          & R & 9.2  &	0.30		   & 4.8   & 3.2  & 1.4	    & \citet{raassen03b}	\\
47 Cas          & R & 10.6 &	0.50		   & 1.68  & 0.70 & 2.21    & \citet{telleschi04}$^c$ 	\\
AB Dor          & R & 10.0 &	0.40         	   & 4.8   & 2.23 & 0.95    & \citet{sanz03}		\\
AB Dor          & R & 11.4 &	0.33		   & 3.04  & 1.22 & 0.83    & \citet{guedel01b} 	\\
V851~Cen        & R & 11   &	0.56		   & 5.5   & 1.76 & 1.6	    & \citet{sanz04}    	\\
$\lambda$ And   & H & 11   &	0.37		   & 2.23  & 1.35 & 1.56    & \citet{sanz04}	        \\
$\lambda$ And   & R & 13.2 &    0.2                & 5.3   & 1.75 & 2.95    & \citet{audard03a}         \\  
VY Ari          & R & 11.3 &    0.18               & 7.0   & 2.2  & 1.83    & \citet{audard03a}         \\  
Algol           & L & 12   &    0.25               & 2.61  & 0.99 & 1.37    & \citet{schmitt04}  \\
HR~1099         & H & 13   &	0.30		   & 10    & 3.0  & 2.5	    & \citet{drake01}		\\
HR~1099         & R & 14   &    ...                & 15.6  & 3.9  & 3.7     & \citet{brinkman01}        \\
HR~1099         & R & 14.4 &	0.22		   & 4.2   & 1.55 & 0.45    & \citet{audard01a} 	\\
HR~1099         & R & 14.8 &    0.20               & 6.6   & 2.75 & 0.9     & \citet{audard03a}         \\  
AR Lac          & H & 15   &	0.74		   & 2.16  & 0.81 & 0.95    & \citet{huenemoerder03}	\\
UX Ari          & R & 15.1 &    0.14               & 13.4  & 4.0  & 2.21    & \citet{audard03a}         \\  
II Peg          & H & 16   &	0.15		   & 14.9  & 7.4  & 2.7	    & \citet{huenemoerder01}	\\
\noalign{\smallskip}\hline
\multicolumn{8}{l}{$^a$All abundance relative to solar photospheric values: \citet{anders89} except for Fe:}\\ 
\multicolumn{8}{l}{\citet{grevesse99}} \\
\multicolumn{8}{l}{$^b$Instrument: R = {\it XMM-Newton} RGS; H = {\it Chandra} HETGS; L = {\it Chandra} LETGS}\\
\multicolumn{8}{l}{$^c$Based on their method 2 using the SPEX database}\\
\end{tabular}\end{table}

\begin{figure} 
\centerline{
\vbox{\hbox{
\resizebox{0.5\textwidth}{!}{\includegraphics{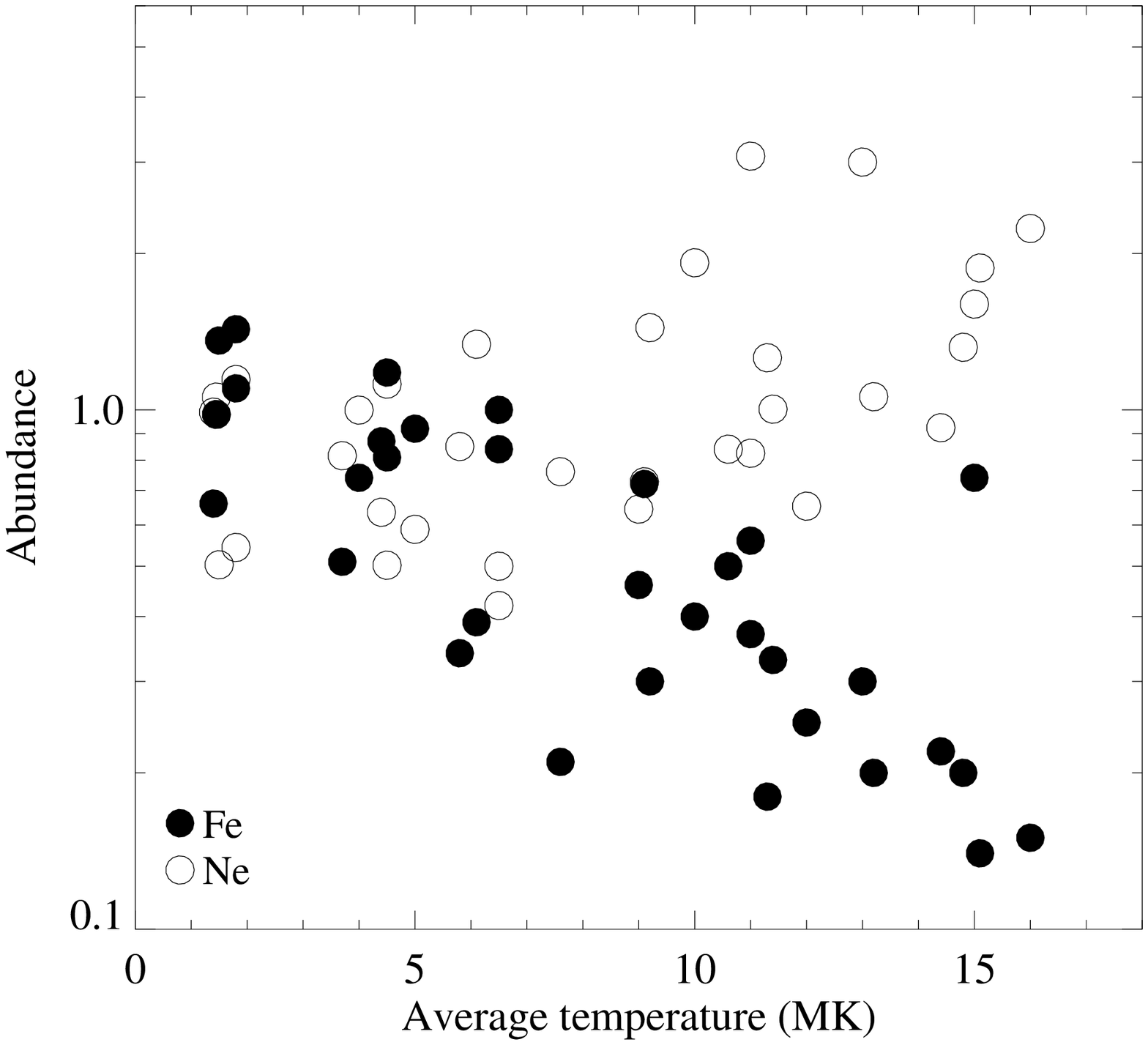}}
\resizebox{0.5\textwidth}{!}{\includegraphics{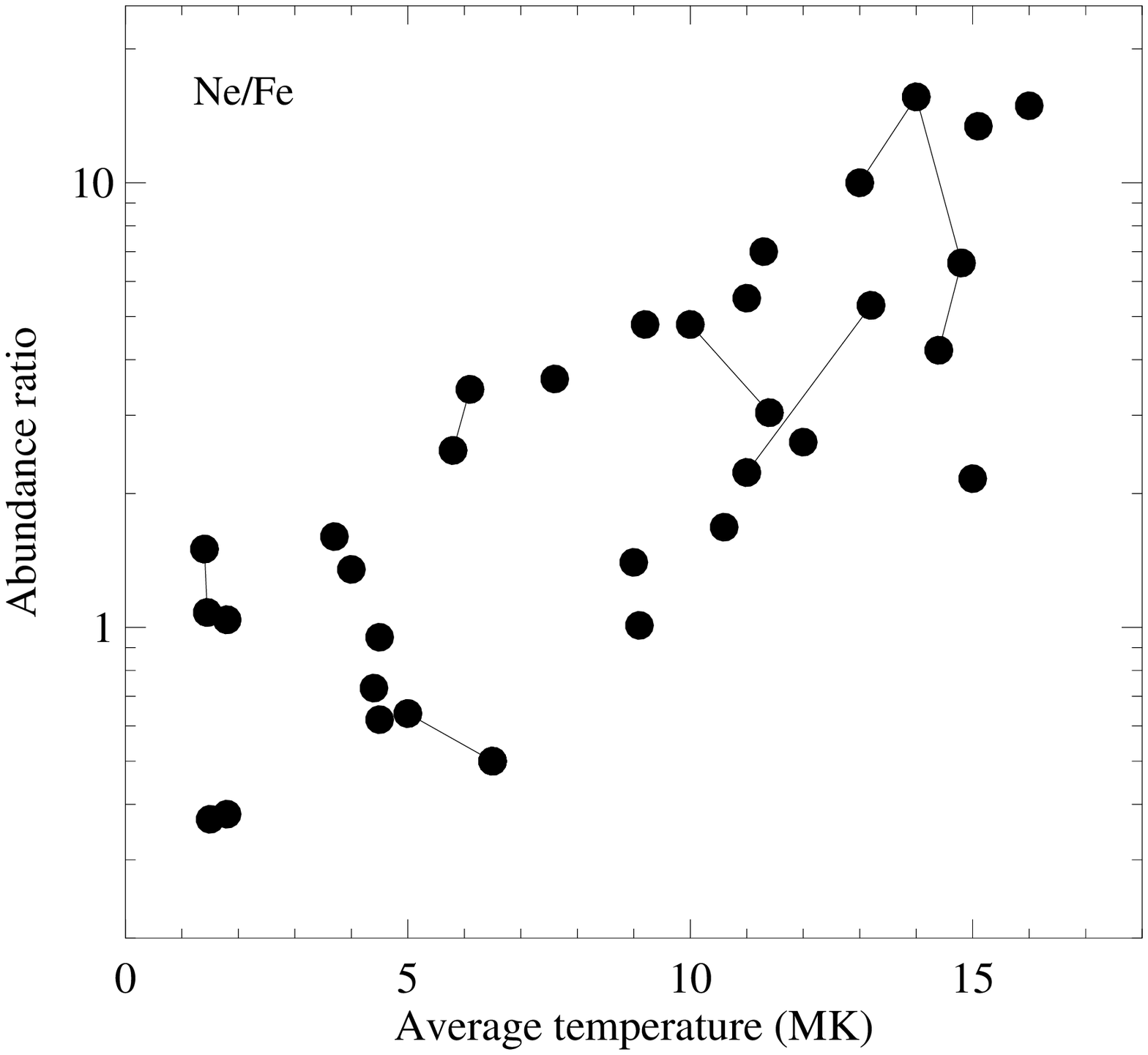}}
} 
\hbox{
\resizebox{0.5\textwidth}{!}{\includegraphics{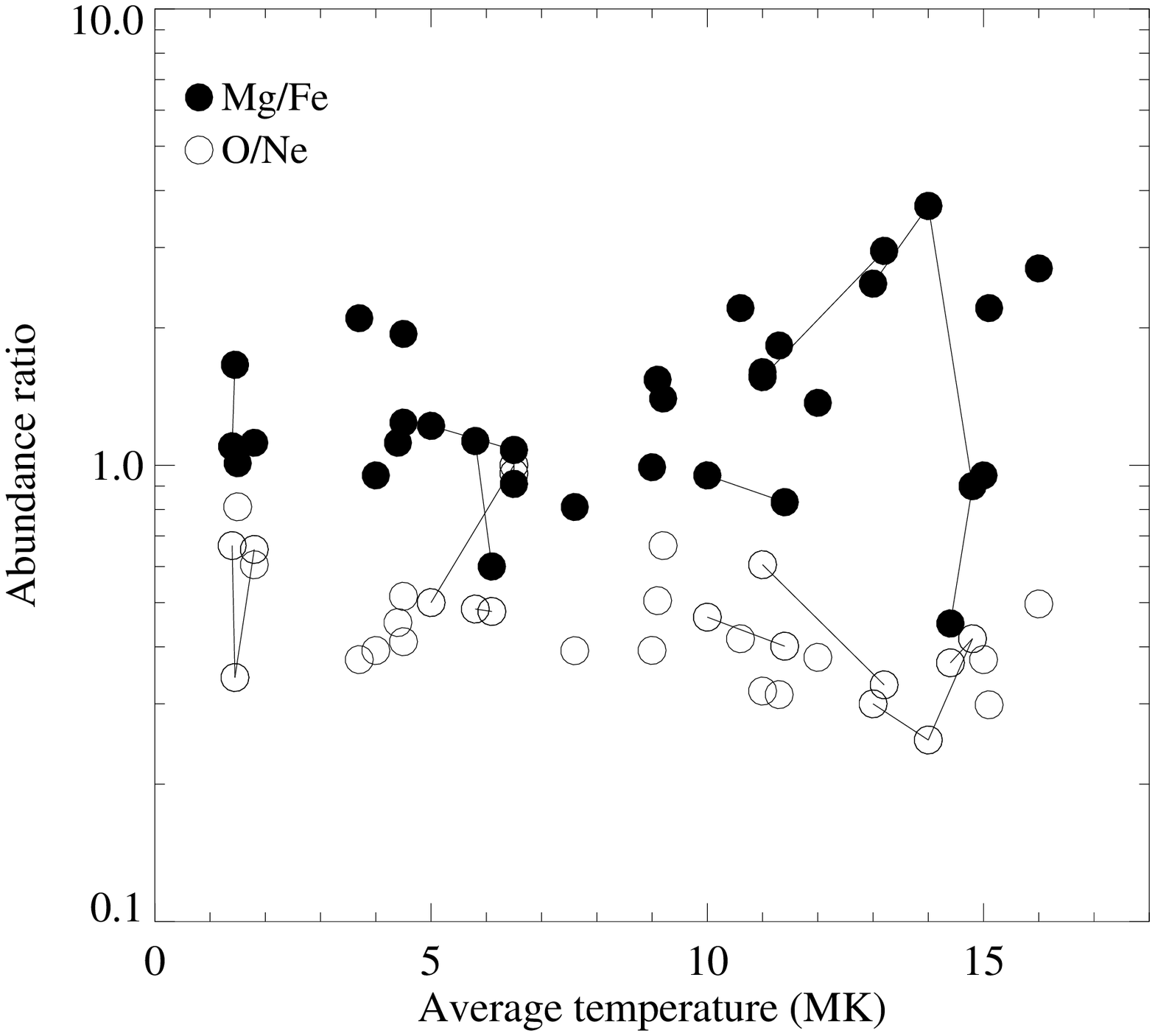}}
\resizebox{0.5\textwidth}{!}{\includegraphics{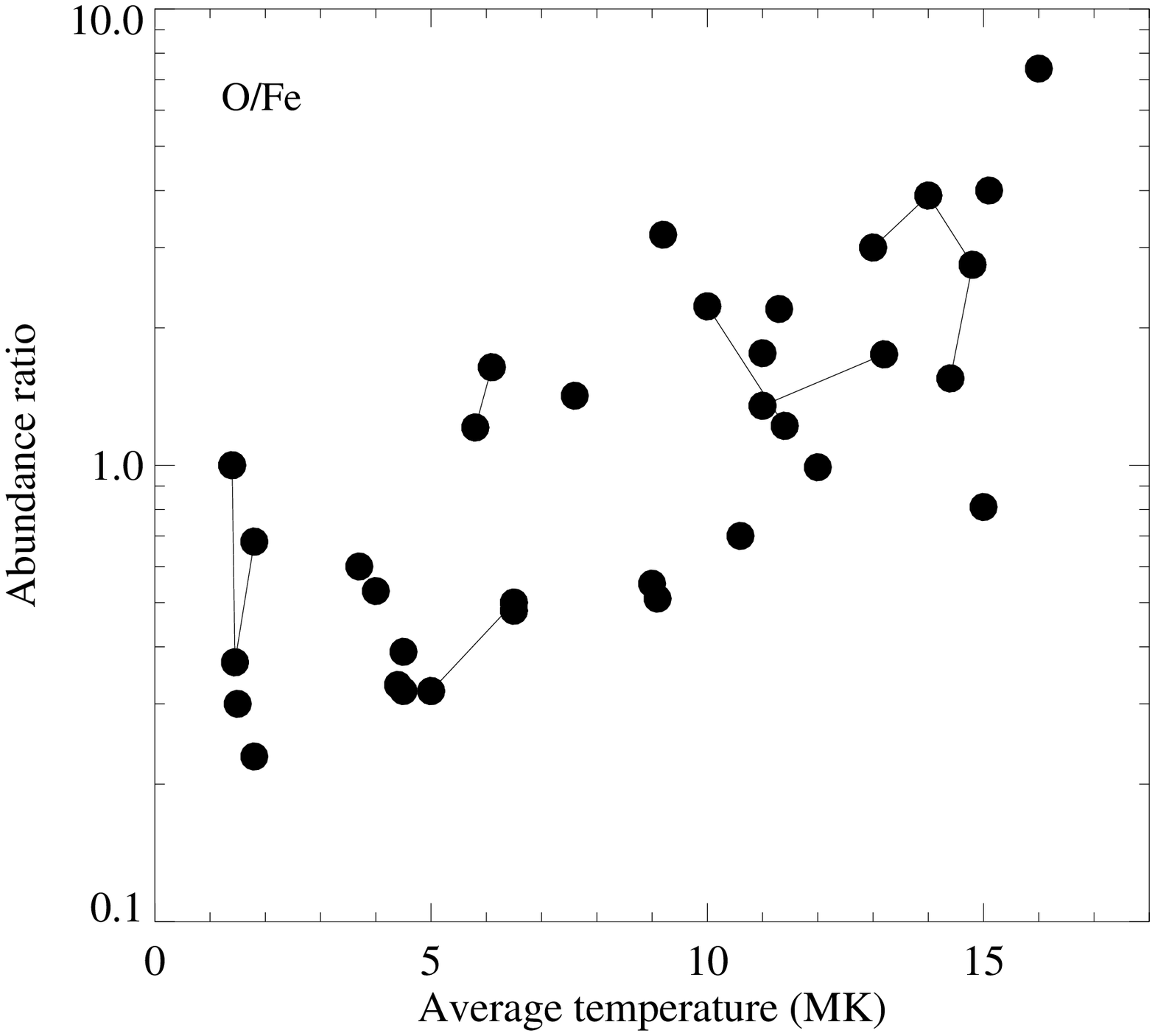}}
}
}
}
\caption{Abundances as a function of the average coronal temperature from high-resolution spectroscopy
with {\it XMM-Newton} and {\it Chandra}. {\it Top left:} Abundances of Fe (filled circles) and Ne (open circles),
relative to solar photospheric values for the sample reported in Table~\ref{abuntable}. {\it Top right:} Similar, but the ratio between the Fe and
Ne abundance is shown. {\it Lower left:} Similar, for the Mg/Fe (filled) and O/Ne ratios (open circles).
{\it Lower right:} Similar, for the O/Fe ratio. Lines connect points referring to the same star analyzed by different authors.}\label{abunplot}
\end{figure}

In order to illustrate the above trends more comprehensively, Table~\ref{abuntable} summarizes a few 
important parameters from recent abundance determinations based on high-resolut\-ion spectroscopy.
The table gives absolute Fe abundances, ratios of the high-FIP elements O and Ne with respect to
Fe, and the ratios between the two low-FIP element abundances of Mg and Fe and between
the two high-FIP element abundances of Ne and O. Direct comparison of reported
abundances should generally be treated with caution because various solar photospheric standards have
been adopted. As far as possible, I transformed the abundances to refer to the solar values of \citet{anders89} except
for Fe, for which I adopted the value given by \citet{grevesse99}. No attempt has been made to quote error
estimates. Errors are extremely difficult to assess and include systematics from calibration
problems and from the inversion 
method, and most importantly uncertainties in the atomic parameter tabulations. It is unlikely that
any measurement represents its ``true''  value within better than 20\%.
The average coronal temperature was estimated either from the logarithmic EMD, EM(log\,$T$),
or was calculated as the EM-weighted mean of log\,$T$ in the case of numerically listed EMDs 
or results from multi-$T$ fits.
Figure \ref{abunplot} shows the relevant trends. Clearly, the low-FIP elements (Mg, Fe) vary in concert, and so
do the high-FIP elements (O, Ne). But the absolute Fe abundance significantly drops with increasing 
activity, and  the Ne/Fe ratio sharply increases as a consequence. The trend for O/Fe is very similar, with 
ratios that are lower by typically a factor of two. The transition from the FIP to the IFIP pattern  
for O/Fe occurs at average temperatures of about 7--10~MK.\footnote{The trends are independent of the spectral inversion
method used to determine the abundances and the EMD; 17 spectra were fitted as a whole, while 17 spectra were analyzed
with various iterative inversion methods using extracted line fluxes; both subsamples show identical trends.}

\subsection{Systematic uncertainties}

The abundance results presently available from high-resolution spectroscopy refer to data from different
detectors, and various methods have been used to analyze the spectra. Some systematic deviations
are found when we compare the reported abundance values (see Table~\ref{abuntable}). This comes as no surprise if we recall
the discussion in Sect.~\ref{demmeth} on the various difficulties inherent in the spectral inversion problem. 
Nevertheless, the situation is not as bad as it might appear, in particular once we review the actual needs for a 
physical model interpretation. I discuss examples from the recent literature, including some results from
medium-resolution observations.

Capella has been a favorite target for case studies given its very high signal-to-noise
spectra available from various satellites. Despite some discrepancies, the abundances
derived from {\it ASCA}, {\it EUVE} \citep{brickhouse00}, {\it BeppoSAX} \citep{favata97c}, 
{\it XMM-Newton} \citep{audard01b}, and {\it Chandra} \citep{mewe01} agree satisfactorily: Fe and Mg
are  close to solar photospheric values (see Sect.~\ref{photosphericabun} below).

Several analyses have been presented for the RS CVn binary HR~1099 (for example by \citealt{brinkman01, audard01a, drake01, audard03a}).
Although systematic differences are found between the derived abundances, 
different atomic data\-bases (APEC, SPEX, HULLAC) and different instruments or instrument combinations
(e.g., RGS+EPIC MOS on {\it XMM-Newton}, or HETGS on {\it Chandra}) were used during various stages of coronal activity (and at various stages 
of early instrument calibration efforts), making any detailed comparison suspect. For example, \citet{audard03a} 
showed conflicting results from two analyses that differ exclusively by their using different atomic databases. 
The overall trends, however, show gratifying agreement, also for several further magnetically active binaries.  All results 
indicate low Fe (0.2--0.3 times solar photospheric for HR~1099), a high Ne/Fe abundance ratio ($\approx 10$), a high Mg/Fe 
abundance ratio ($\approx 2$ in several active stars), and an overall trend for increasing abundances with increasing FIP. 
The same conclusions were reported by \citet{schmitt04} for an analysis of the Algol 
spectrum based on different reconstruction methods, the main uncertainty being due to large systematic errors 
in the atomic physics parameters. 

\citet{telleschi04} systematically studied element abundances and EMDs for solar analogs
at various activity levels, applying approaches that involve either spectral fits of sections of the spectra, or
iterative reconstructions based on selected emission line fluxes.  The two iteration methods yielded nearly
identical abundances (Fig.~\ref{abun_solar}). A more fundamental problem was recognized in an unbiased determination of line 
fluxes in the presence of line blends, and in the use of different spectroscopic databases.

Expectedly, there is less agreement between results from high-resolution data and  analyses that make
use of low- or 
medium-resolution observations (from, e.g., {\it ASCA} or {\it BeppoSAX}), but several trends are noteworthy. First, 
high Ne/Fe abundance ratios are in fact found in a number of
reports based on {\it ASCA} observations, even if the agreement is not always convincing.
The following ratios have again been converted to
the \citet{anders89} and  \citet{grevesse99} solar photospheric abundances (see also Table~\ref{abuntable}):
For II Peg, Ne/Fe = 4.5 ({\it ASCA}, \citealt{mewe97}) vs.  15 ({\it Chandra}, \citealt{huenemoerder01}); 
for AR Lac,  3.4 ({\it ASCA}, \citealt{white94}) vs. 2.2 ({\it Chandra}, \citealt{huenemoerder03});
for $\lambda$ And, 3.4 ({\it ASCA}, \citealt{ortolani97}) vs.  4.0--5.3 ({\it XMM-Newton}, \citealt{audard03a}) and      
                                                   2.2 ({\it Chandra}, \citealt{sanz04});      
for HR~1099, 5.9 ({\it ASCA}, \citealt{osten04}) vs.  4.2--10 ({\it Chandra}, \citealt{audard01a, audard03a, drake01});      
for Algol, 1.4--2.3 ({\it ASCA}, \citealt{antunes94}) vs.  2.6 ({\it Chandra}, \citealt{schmitt04}).      
Similar trends apply to the Mg/Fe ratios:
For II Peg, Mg/Fe = 1.2 ({\it ASCA}, \citealt{mewe97}) vs.  2.7 ({\it Chandra}, \citealt{huenemoerder01}); 
for AR Lac,    1.0--1.4 ({\it ASCA}, \citealt{white94, kaastra96}) vs. 0.95 ({\it Chandra}, \citealt{huenemoerder03});    
for $\lambda$ And, 1.2--2.2 ({\it ASCA}, \citealt{ortolani97}) vs.  2.5--3.0 ({\it XMM-Newton}, \citealt{audard03a}) and 
                                                   1.6 ({\it Chandra}, \citealt{sanz04});      
for HR~1099, 1.6 ({\it ASCA}, \citealt{osten04}) vs. 0.45--2.5 ({\it Chandra}, \citealt{audard01a, audard03a, drake01});      
for Algol, 0.8--1.2 ({\it ASCA}, \citealt{antunes94}) vs.  1.4 ({\it Chandra}, \citealt{schmitt04}).     
The discrepancies are more severe and systematic for several other elements given the strong blending in
low-resolution data (see, e.g., discussion in \citealt{huenemoerder03}). 

In general, abundance ratios are 
more stable than absolute abundances that require a good measurement of the continuum level. This, in turn,
requires an accurate reconstruction of the DEM.
But what are we to make out of the residual uncertainties and discrepancies in abundance ratios 
from the interpretation of high-resolution data?  It is perhaps important to recall the situation
in the solar corona. There, abundances change spatially, and also in time as a given coronal 
structure ages \citep{feldman92, jordan98}. As a further consequence, abundances are likely 
to vary with temperature across the corona. Second, the DEM represents a complex mixture of quiet
plasma, evolving active regions, bright spots, and flares at many luminosity levels. 
A ``best-fit'' parameter derived from a
spectrum therefore represents a sample mean related to a statistical distribution of
the respective physical parameter. Whatever error ranges are reported, they
inevitably refer to this sample mean whereas the actual spread across the physically
distinct sources in a stellar corona remains presently unknown. A sound physical interpretation
will  have to address abundance ratios in specific coronal sources. In this context, 
excessive accuracy may well be meaningless for observational studies even
if atomic physics problems were absent - the essence for further physical
interpretation are trends such as those shown in Fig.~\ref{abunplot}.

\subsection{Coronal and photospheric abundances}\label{photosphericabun}

All coronal plasma ultimately derives from the respective stellar photosphere.
Strictly speaking, therefore, we should define abundance ratios with respect to the underlying 
photospheric abundances. Unfortunately, these are all too often poorly known, or
unknown altogether. 

Photospheric metallicities, $Z_{\rm phot}$, have been reported for a number of RS CVn binaries (see, among others, \citealt{randich93, randich94};
the latter authors, however, cautioned against the  strict use of their measurements as true metallicity indicators).
Unexpectedly, at least for population I stars, many of these systems turn out to be rather metal poor. 
If the coronal abundances are compared with these photospheric metallicities,  the putative coronal underabundances
may disappear entirely. This seems to be the case for Capella \citep{favata97c, brickhouse00}.
Similarly, \citet{white94} reported comparable
coronal and photospheric abundances for the RS CVn binary AR Lac, and \citet{favata97b}
found the coronal metallicity of VY Ari (0.4 times solar) to be in the midst of measured 
photospheric metallicities of RS CVn binaries as a class. \citet{ortolani97} measured,
from low-resolution {\it ROSAT} data, a
coronal metallicity, $Z_{\rm cor}$, in $\lambda$ And that is  nearly 
consistent with the stellar photospheric level, although {\it ASCA} data indicated lower 
metallicity, and several discrepant sets of photospheric abundances have been published (see \citealt{audard03a}). 
The trend agrees with abundance {\it ratios} being the same in
the corona and in the photosphere of this star \citep{audard03a}. The RS CVn binary CF Tuc has revealed particularly low 
coronal metal abundances of 0.1 times solar photospheric values \citep{schmitt96d, kuerster96}, 
but again this finding is accompanied by low measured photospheric abundances \citep{randich93}. 
The same appears to apply to the RS CVn/BY Dra binaries TY Pyx ($Z_{\rm cor} \approx 0.5-0.7$,
$Z_{\rm phot} \approx 0.63$, \citealt{franciosini03, randich93}) and HD 9770 ($Z_{\rm cor} \approx 0.5-0.6$,
$Z_{\rm phot} \approx 0.30-0.35$, \citealt{tagliaferri99}). 
Although \citet{drakes94a} found that the spectrum  of the giant $\beta$ Cet cannot be fitted even 
when assuming the known photospheric abundances, \citet{maggio98} reported  rough agreement between
photospheric (90\% confidence ranges for Si: 1.4--3.2, Fe: 0.7--2.5 times solar) and coronal
abundances (Si: 0.9 [90\%: 0.3--2.4], Fe: 0.7 [90\%: 0.5--2] times solar), albeit with 
large uncertainties.

But new complications have surfaced. \citet{ottmann98}  critically reviewed previous photospheric abundance determinations and
revisited the problem using a sophisticated  spectroscopic approach to derive
all relevant stellar parameters self-consistently. They challenged reports of very low photospheric 
metallicities in otherwise normal stars, finding at best mild underabundances (e.g., 40--60\% solar Mg, Si, and
Fe values for II Peg and $\lambda$ And), and metal abundances very close to solar for
young, nearby solar analogs ($\kappa^1$ Cet and $\pi^1$ UMa). An application of the technique to a solar
spectrum returned the correct solar values. These more solar-like values seem closer to what should be expected
for nearby, young or intermediately old population I stars \citep{feltzing01}. The apparent photospheric
underabundances in active stars may be feigned by chromospheric filling-in of the relevant lines
and due to photometric bias from large dark-spot areas, an explanation that   
Randich et al. also put forward to explain their significantly differing abundances for the two components 
in some active binaries. In the light of these reports, several authors returned to recover significantly 
depleted coronal abundances also relative to the respective  photospheres.  

An illustrative example is HR~1099. 
Recent coronal abundance determinations for this star converge to subsolar values in particular
for Fe (between 0.2--0.3, \citealt{audard01a, audard03a, drake01}), and 
these values  seem to superficially agree with an Fe abundance of  $\approx 0.25$ measured for the photosphere 
\citep{randich94}. On the other hand, \citet{strassmeier00} and \citet{savanov91} reported photospheric Fe abundances of,
respectively, 0.6--0.8 and 1.0 times solar, which would be in agreement with values expected from the 
age of this system
\citep{drake03a}, thus arguing in favor of a real depletion of the coronal Fe. Similarly,
\citet{covino00} found $Z_{\rm cor} \approx 0.2$ for II Peg (\citealt{mewe97} even give Fe/H $\approx 0.1$), 
whereas the photosphere is 3 times richer in metals ($Z_{\rm phot} = 0.6$ according to \citealt{ottmann98}).
And \citet{huenemoerder01} reported a coronal Fe abundance four times below  photospheric for
II Peg, and  various abundances are found at 60\% of the photospheric values in AR Lac 
\citep{huenemoerder03}.

Particularly interesting test samples are nearby, young  solar analogs for which
reliable photospheric abundances have been reported, not too surprisingly being consistent with
solar photospheric values (e.g., Mg, Si, and Fe given by \citealt{ottmann98} for $\kappa^1$ Cet and
$\pi^1$ UMa; several elements [Al, Ca, Fe, Ni] reported by \citealt{vilhu87b} for AB Dor; further values 
listed by \citealt{cayrel01} and references therein
also for $\chi^1$ Ori; with several individual elements being close to solar, large discrepancies
appear  rather unlikely for other elements). A corresponding coronal X-ray study
based on {\it XMM-Newton} spectroscopy, however, indicated significant overall metal depletion on the one 
hand, where the Fe depletion is stronger for higher activity levels, and  a relative FIP bias on the
other hand; the latter changes from ``normal'' to ``inverse'' with increasing activity level.  Both 
trends  disagree with photospheric patterns \citep{guedel02c, telleschi04}. 

Other important test cases are stars that reveal
strong, undisputed {\it photospheric} deviations from the solar abundance pattern.
``Super-metal rich'' stars are stars with measured photospheric Fe abundances [Fe/H]$\ge 0.2$
(logarithmic, relative to the solar photosphere).
\citet{maggio99} observed two extreme cases with [Fe/H]$ = 0.25$ (30 Ari B) and
[Fe/H]$ =0.305$ ($\eta$ Boo). Surprisingly, not only were the relative coronal abundances 
[Mg/Fe], [Si/Fe], and [O/Fe] found to be close  to the abundances of otherwise similar field stars 
with near-solar composition, but the absolute coronal [Fe/H] abundance was also derived to
be near-solar. The authors suspected a number of artefacts related to low-resolution
CCD spectroscopy, however.
At the opposite end, namely in metal-poor Population II stars, the simple lack of 
metals available in the stellar material should reflect strongly in the coronal
emission. A striking case was presented by \citet{fleming96}: The  binary HD~89499 
with [Fe/H] = --2.1 shows X-ray spectra that are essentially 
line-free, that is, they are dominated by bremsstrahlung. Because the emissivity of such 
plasma is much lower than the emissivity of a plasma with solar composition,
the material can efficiently be heated to higher temperatures, and 
indeed all plasma is detected at $T \approx 25$~MK, with no significant amounts of cooler material.

There are nevertheless important examples of stars that show no indication of
coronal metal depletion also in the light of new photospheric abundance measurements. 
\citet{drake01b} found a slight enhancement
of coronal vs. photospheric abundances in the intermediately active $\xi$ Boo A, in agreement with
a measured solar-like FIP effect. A trend for smaller metal deficiencies toward less active stars
has been noted earlier \citep{singh95}, and the least active stars such as Procyon \citep{drake95b, raassen02} 
and $\alpha$ Cen generally show solar-like metallicities, with the additional possibility of a solar-type 
FIP effect \citep{drake97, mewe98a, mewe98b, raassen03a}.  
The very active late-F star HD 35850 with measured solar photospheric 
Fe abundance also requires near-solar or only slightly subsolar abundances
\citep{gagne99} although there is disagreement with the analysis given by 
\citet{tagliaferri97}. A few further examples with near-photospheric composition were discussed
by \citet{sanz04}.

The above discussion amply illustrates the inconclusive and unsatisfactory present observational
status of the field.  The easy access to coronal abundance diagnostics makes further, 
comprehensive photospheric abundance  determinations very desirable.

\subsection{Flare metal abundances}\label{flareabun}

Abundances may change in particular in flares because the evaporation process  brings new photospheric or
chromospheric material into the 
corona. Peculiar effects have also  been noted in some solar flares, as described in Sect.~\ref{solarabun}.

There was some early evidence for enhanced or depleted Fe abundances in large stellar flares,
mostly on the basis of the line-to-continuum ratio for the Fe K$\alpha$ line at 6.7~keV.
\citet{stern92a} found an equivalent width
of only 25\% of the expected value in a flare on Algol, possibly indicating a corresponding 
Fe underabundance. A similar effect was noted by \citet{tsuru89}. In both cases, 
suppression due to resonance scattering was discussed as an alternative explanation.
\citet{doyle92a} reported a very low Fe abundance (33\% solar) during a large flare on II Peg. 
\citet{ottmann96} noticed an increase of the coronal metallicity from 0.2 to 0.8$Z_{\odot}$
during a giant flare on Algol observed by {\it ROSAT} in the 0.1--2.4~keV range, although the low 
spectral resolution makes such measurements problematic, for example if the dominant plasma found in simple
spectral fits is far from the relevant maximum line formation temperatures.  
\citet{white86}, on the other hand, concluded for a flare on Algol that the Fe abundance was within
20\% of the solar photospheric value.  Further reports of
unusual equivalent widths of the Fe K$\alpha$ line have been reported by \citet{doyle91}, \citet{tsuru92}, and
\citet{vilhu93}.

Higher-resolution broad-band spectra permitted measurements of other {\it individual} elements and
more complicated thermal structures. \citet{guedel99}, \citet{osten00} and \citet{audard01a} 
investigated the temporal evolution of several abundances
during  large, gradual flares on RS CVn binaries. They all found increases of low-FIP element abundances
(e.g., of Fe, Si, Mg) during the flare maximum, and a rapid decay back to pre-flare values during
the later phases of the flare. In contrast, a large flare on Proxima Centauri revealed no FIP-related abundance
evolution although all abundances appeared to be elevated in proportion during a narrow interval
around the flare peak \citep{guedel04}. Similarly, \citet{osten03} found an increase of the 
Fe abundance during a flare on $\sigma^2$ CrB by a factor of two, while the other elements 
increased accordingly, with no systematic trend related to the FIP.

Another, albeit less detailed, approach involves the monitoring of the ``global'' metallicity
$Z_{\rm cor}$, i.e., abundances
are assumed a priori to vary in proportion. The metallicity   is then predominantly driven by Fe, 
in particular if the 
Fe K$\alpha$ complex at 6.7~keV is accessible.  \citet{mewe97}, \citet{tsuboi98}, \citet{favata99}, 
\citet{favata00a}, \citet{guedel01b}, \citet{covino01}, and \citet{osten02}  noted peak enhancements of $Z_{\rm cor}$ by factors 
of about 3--4 during flares on II Peg, V773~Tau, Algol, EV Lac, AB Dor, Gl~355, and EI Eri, respectively. 
In contrast, \citet{maggio00} and \citet{franciosini01} reported the absence of a significant 
metallicity enhancement in flares on AB Dor and UX Ari, respectively, although the upper limits were still 
consistent with enrichment factors of $\approx 3$. 

Despite some variations of the theme, it seems to be certain  that flares can change the elemental
composition of the plasma, and the trend is toward increasing metallicity in flares on active stars
that otherwise show metal-depleted coronae. Some ideas for models will be discussed in the next section.

\subsection{Theoretical models for abundance anomalies}

It has proven particularly difficult to interpret the abundance anomalies in stellar coronae.
This should  come as no surprise given that the situation is not all that different for
the solar corona, despite the abundance of sophisticated but competing models \citep{henoux95}.
I briefly list a few suggestions that, however, all require further elaboration; another recent
summary has been given by \citet{drake03b}.

{\it Stratification of the atmosphere.} \citet{mewe97}  proposed that the abundance features observed
in active stars are due to the different scale heights of different ions in a hydrostatic 
coronal loop, depending on mass and charge. As a result, the ion distribution in a magnetic
loop is inhomogeneous and the line-to-continuum ratio depends on the scale height of a specific
ion.  After flares, the settling to equilibrium distributions is 
expected to occur on time scales of hours.
This model cannot, however, explain the high Ne abundances in active stars  because of the mass 
dependence of the proposed stratification: Ne has a mass intermediate between O and Mg!

{\it Coronal equilibrium.} All determinations of element abundances strongly rely on the assumption of collisional equilibrium.
The thermalization time is usually short enough to justify this assumption. If the observed coronal 
emission is, however, driven by a sequence of small flares (Sect.~\ref{obsstochflare}) then a number of further
conditions would have to apply, which are difficult to assess. Whether such effects could 
change the measured abundances is currently unknown (see also the discussion in
\citealt{huenemoerder03}).

{\it ``Anomalous flares''.} High Ne abundances are occasionally seen on the Sun during ``Ne rich'' flares \citep{schmelz93}. 
\citet{shemi91} argued that the high photoionization cross section of Ne makes it prone
to efficient ionization by X-ray irradiation of the chromosphere during flares,
thus making Ne behave like a low-FIP element in the solar corona. While this model explains the Ne enrichment,
it does not address the apparent underabundance of the low-FIP element Fe. 

{\it Evaporation.} The peak metallicities observed in large flares seem to be bounded by $Z \la Z_{\odot}$.
Assuming that most population I stars in the solar neighborhood have photospheric abundances
similar to the Sun, then the coronal abundance increase suggests that new chromospheric/photospheric
material is supplied to the corona, most likely by the chromospheric evaporation process
\citep{ottmann96, mewe97, guedel99}. 
The selective low-FIP element enhancement during some large flares \citep{guedel99, osten00, audard01a}
is further compatible with a chromospheric evaporation  model presented by
\citet{wang96} in which the evaporation induces upward drifts of electrons and
protons. These particles then efficiently drag chromospheric ions (i.e., preferentially low-FIP elements)
in an ambipolar diffusion process, while they leave neutral (preferentially high-FIP) elements behind. 

{\it Stellar evolution.} \citet{schmitt02b} found N/C abundance ratios that are enhanced by
factors of up to 40 in
Algol and in a sample of (sub-)giants. They attributed this anomaly to an enrichment of N from mass transfer
in Algol, and from dredge-up of N in the evolving giants. This has been quantitatively
worked out in the context of mass loss evolution and convective dredge-up 
by \citet{drake03a} for Algol. The N enrichment and C depletion then essentially come from nuclear processing through the CN cycle
in the secondary star, followed by strong mass loss from the outer convection zone and
dredge-up of N enriched material.

{\it He enrichment.} An increased abundance of He relative to H due to some fractionation process in the 
chromosphere would increase the
continuum level and therefore decrease the line-to-continuum ratios, thus leading to
an apparent underabundance of  the metals \citep{drake98}.  Helium 
enhancements by factors of a few are required, however \citep{rodono99}, up to levels 
where He would be the most abundant element of the plasma \citep{covino00}. FIP-related trends are
not explained by this model.

{\it Non-thermal emission.} I mention here another possibility that may lower all abundances in general, and in particular
the Fe abundance if measured from the Fe~K complex at 6.7~keV. If the continuum level is enhanced
by an additional power-law component due to impact of non-thermal electrons in the chromosphere,
then the equivalent width of the line is below predictions, simulating a ``depleted Fe abundance'' \citep{vilhu93}.
Observationally, quite high signal-to-noise ratios and good spectral
resolution may be required to rule out non-thermal contributions. Such a ``non-thermal'' model is 
interesting because appreciable suppression of the overall metallicity is found particularly
in stars that are strong non-thermal radio emitters \citep{guedel02c}.

There is another coincidence between the appearance of non-thermal electrons and element
abundance anomalies. The change from an inverse-FIP bias to a normal solar-type FIP effect
occurs at quite high activity levels, close to the empirical saturation limit for
main-sequence stars \citep{guedel02c}. At the same time, non-thermal radio
emission drops sharply (much more rapidly than the X-rays). If electrons continuously propagate into the chromosphere
at a modest rate without inducing strong evaporation, then a downward-pointing electric field
should build up. This field should tend to trap chromospheric ions, i.e., predominantly the
low-FIP elements, at low levels while neutrals, i.e., predominantly high-FIP elements, 
are free to drift into the corona. As radio emission disappears in lower-activity stars,
the low-FIP element suppression disappears, and a solar-type FIP effect may build up, for whatever
physical reasons, in analogy to the solar case.

\section{X-ray emission in the context of stellar evolution}

X-ray emission offers easy access to stellar evolution studies because
magnetic activity is governed by various stellar parameters such 
as convection zone depth and  rotation that change gradually as a star evolves.
Evolutionary studies have been based either on nearby, modestly-sized but well-defined samples of
stars or on  large statistical samples from  open clusters. I will first 
describe some principal results from the first approach and then concentrate on
open cluster studies.

\subsection{Main-sequence stars}\label{msevol}

The X-ray luminosity of field F-G stars clearly decays monotonically with age.
Studies of young open clusters (Sect.~\ref{clusters}) indicate
a slow decay during the initial few 100~Myr of a solar-like star, with $L_X$ scaling
approximately like the inverse of the age \citep{patten96}. In contrast, somewhat older field stars
show a steep decay  toward higher ages.
For nearby F-G main-sequence stars for which approximate ages have been derived mostly from their
surface Li abundance, moving group membership, or rotation (once sufficiently
converged toward a mass-dependent value), the X-ray luminosity decays like
\begin{equation}\label{tempage}
L_X \approx (3\pm1)\times 10^{28} t^{-1.5\pm 0.3}\quad {\rm [erg~s^{-1}]}
\end{equation}
where the age $t$ is given in Gyr \citep{maggio87, guedel97a}. The same trend is found in
open clusters (see Sect.~\ref{clusters} and Fig.~\ref{clusterage} below).
It is quite certain that this decay law reflects the
slowing of the rotation rate with age rather than some intrinsic dynamo ageing
\citep{hempelmann95}. It must therefore derive directly from the combination of 
the rotation-age relationship (e.g., \citealt{soderblom93}) and the
rotation-activity dependence (Sect.~\ref{rotation}).

\begin{figure} 
\centerline{
\resizebox{0.85\textwidth}{!}{\includegraphics{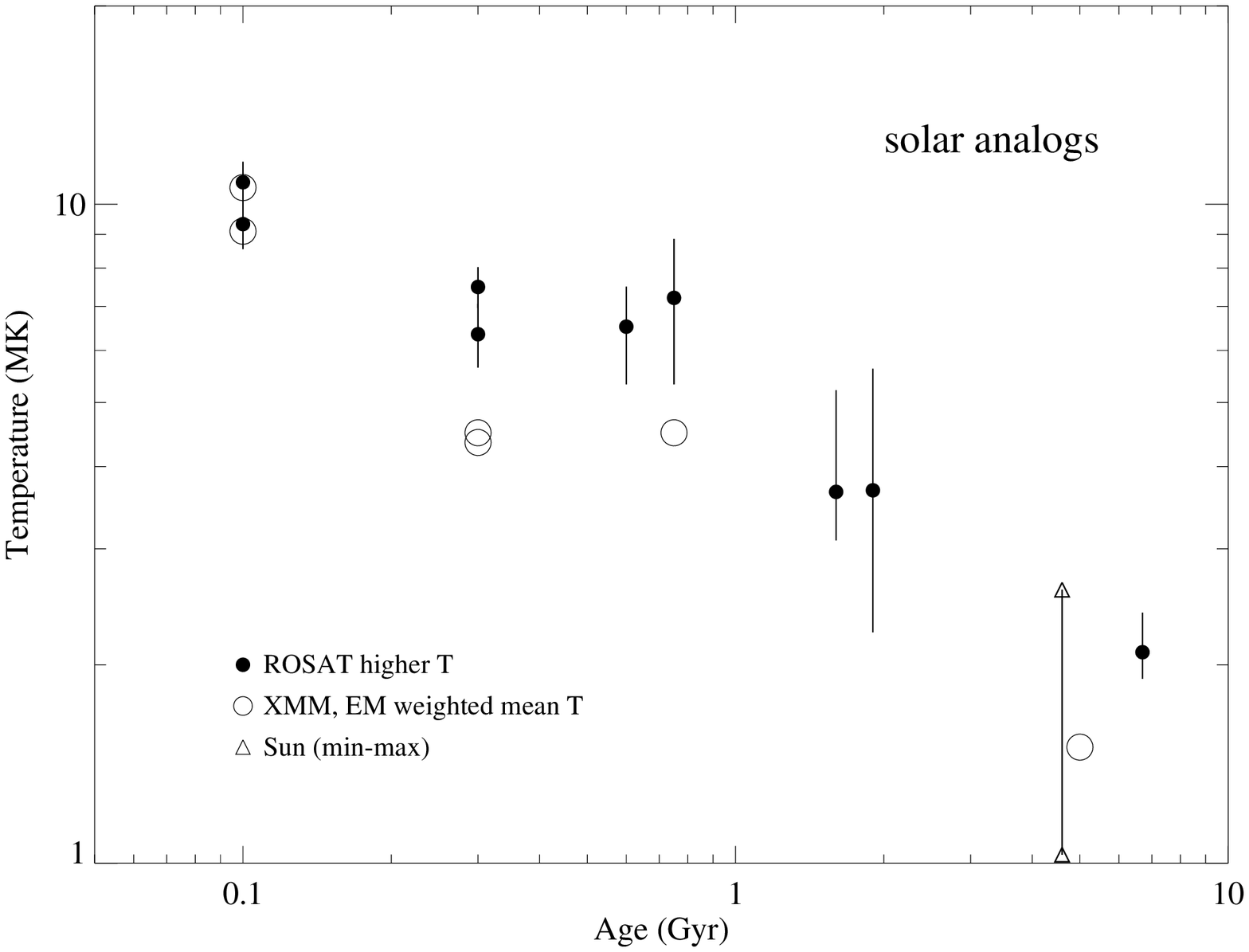}}
}
\caption{Decay of coronal temperature with age for solar-type stars. The temperatures of 
the hotter component in the G star sample of \citet{guedel97a} and \citet{guedel98} are shown (filled circles). Solar data are from 
\citet{peres00} and illustrate the range from activity minimum to maximum (triangles). The open circles
represent the EM-weighted logarithmic average  of $T$ as calculated from EMDs \citep{telleschi04} or
from multi-$T$ spectral fits (for $\alpha$ Cen at $t$ = 5~Gyr, \citealt{raassen03a}).}\label{agedecay}
\end{figure}

As the star ages and its overall luminosity decreases, the EM
distribution shifts to cooler temperatures, with a rapid decrease in particular
of the hot plasma component - see Fig.~\ref{agedecay} (and Sect.~\ref{demresults} and \ref{templx}). 
A possible cause of the temperature decrease are the
less efficient coronal magnetic interactions in less active stars given their lower magnetic filling factors,
and consequently a less efficient heating, or a lower rate of flares that produce high-temperature  
plasma  \citep{guedel97a, guedel97c}.

It is reassuring that the Sun and its near-twin, $\alpha$ Cen A, behave very much alike,  both in terms of 
coronal temperature and $L_X$  \citep{mewe95, mewe98a, drake97}. The similarity between $\alpha$ Cen A
and the Sun in spectral type, rotation period and age provides a convenient approximation  to the ``Sun as a star''
\citep{golub82, ayres82}. Seen from this perspective, the full-disk DEM has cooled to a distribution between 
1--5 MK with a peak at 3~MK \citep{mewe98a, raassen03a}, representative of a star in which
a mixture of active regions, quiet regions, and coronal  holes prevail 
(\citealt{orlando00, peres00}, see Sect.~\ref{constituents}).

When we move to lower-mass stars, the picture changes gradually.
A progressively larger fraction of stars is found at relatively high activity levels toward lower masses,
in particular among M dwarfs. This is a consequence of the smaller spin-down rate for less massive stars 
\citep{fleming88, fleming95, giampapa96}: {\it low-mass stars stay active for a longer time.} This may
be related to a different dynamo being in operation in late M dwarfs where the radiative core is 
small or absent altogether \citep{giampapa96}.
The more rapid spin-down of more massive dwarfs is, in turn, illustrated by the paucity of extremely active G and early 
K stars in the immediate solar vicinity; the nearest examples are found at distances of 15--30~pc
with ages of no more than $\approx 100$~Myr (e.g., the K0~V rapid rotator AB Dor [\citealt{vilhu93}] or
the young solar analogs EK Dra [\citealt{guedel95}] and 47 Cas B [\citealt{guedel98}]). For M dwarfs, we find a 
clear ageing  trend only when we look at much larger
age ranges: their X-ray luminosity significantly decays in concert with their metallicity from young to old disk
population stars (\citealt{fleming95}; metallicity generally decreases with increasing population age).  
This trend continues toward the oldest, halo population M dwarfs that reveal significantly softer X-ray spectra  
than young and old disk stars (\citealt{micela97a}; see also \citealt{barbera93}).

\subsection{Giants}\label{giants}

\begin{figure} 
\centerline{
\resizebox{0.7\textwidth}{!}{\includegraphics{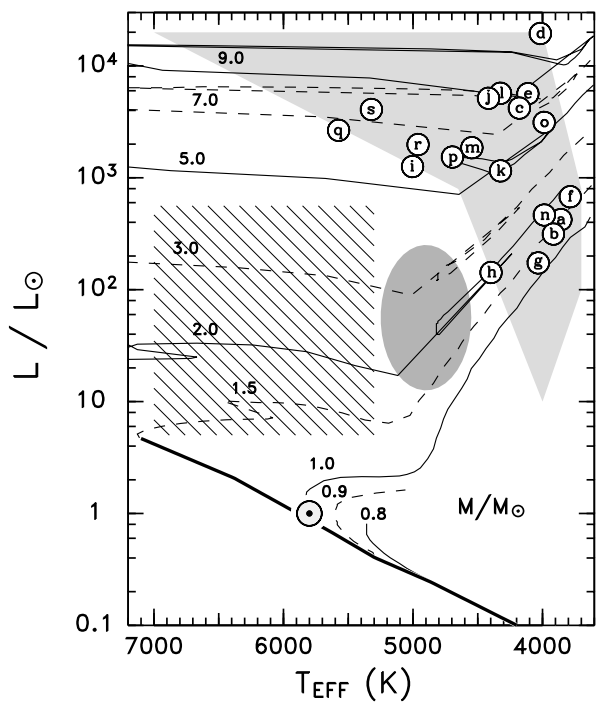}}
}
\caption{Main features of the giant HRD.  The hatched region schematically illustrates the
Hertzsprung gap area, while the gray oval region marks the area of the ``clump stars''. The polygonal
region in the upper right corner shows  the approximate range of the cool-wind stars. The original
dividing line is the lower left  hand edge of this area. Individual circles mark the loci of
``hybrid'' stars. Evolutionary tracks are given for a variety of masses 
(figure courtesy of T. Ayres, after  \citealt{ayres04}).}\label{gianthrd}
\end{figure}

The giant and supergiant area of the HRD is more complicated
because evolutionary tracks are running close to each other and partly overlap.
An overview of the relevant area  is shown in Fig.~\ref{gianthrd}.
As cool stars evolve from the main sequence to the giant branch, the deepening of the 
convection zone may temporarily increase the coronal magnetic activity level \citep{maggio90}.
In general, however, the X-ray activity of stars with masses $\la 1.5M_{\odot}$ further 
decreases as they move redward in the HRD \citep{pizzolato00} 
because these stars begin their evolution toward the giant branch as slow rotators on the main sequence, 
and the increasing radius further slows  their rotation rate.

The evolution of magnetic activity is different for more massive stars because they start their evolution off
the MS as rapidly rotating O, B, and A stars. As convection suddenly sets in, a dynamo begins
to operate in the stellar interior, and a magnetized wind starts braking the star in the region of early G-type
giants and supergiants \citep{huensch96, schroeder98}. These ``first crossing stars'' develop maximum X-ray activity
among G giants ($L_X \la 10^{31}$~erg~s$^{-1}$, \citealt{maggio90, micela92, scelsi04}). The break in  rotation 
period is located significantly blueward of the location where X-rays weaken \citep{gondoin87}.
This is because the convection zone deepens and, possibly, differential rotation strengthens as the star evolves 
redward; both trends help strengthen the dynamo against the slowing rotation  \citep{gondoin99, pizzolato00}.  
As a consequence, one finds
no unique rotation-activity relation for the complete ensemble of giants,
and some of the more rapidly rotating late giants keep extreme activity levels in X-rays \citep{gondoin02,
gondoin03a, gondoin03b, gondoin03c, gondoin04b}.
Only at $B-V \ga 0.9-1$ do we find  a gradual decrease of $L_X$, up to $B-V \approx 1.1$ ($\approx$~K1~III). Toward
later spectral types,  X-ray emission drops to very low values. 

The X-ray emission of early F giants and subgiants 
is deficient with respect to transition region line fluxes (\citealt{simon89, ayres95, ayres98}; Fig.~\ref{fluxflux}). 
\citet{simon89} therefore
argued in favor of acoustic coronal heating for the warmer stars that show no appreciable magnetic braking
by a wind, and solar-like magnetic heating for the cooler stars in which magnetic braking is effective. In contrast, 
\citet{ayres98} proposed very extended ($L \approx R_*$) coronal loops with long ($\approx 1$~day) filling durations 
and  cooling times that favor a redistribution of the energy into the transition region where it is efficiently radiated. 
Such loops may be remnants of global magnetospheres as proposed in different contexts for hotter MS stars, 
coexisting with a growing solar-type corona that is generated by a dynamo in the deepening convection zone.
The systematic behavior is less clear in supergiants given their smaller number statistics;
the outstanding early-type example among single Hertzsprung gap supergiants is $\alpha$ Car that shows an appreciable
$L_X$ despite its shallow convection  zone (F0~II, log\,$L_X = 29.8$, \citealt{maggio90}).

As giants evolve, and at least the less massive examples undergo a He-flash, they gather in the 
``clump'' region of the HRD, at roughly $0.95 \le {\rm B-V} \le 1.2$ (K giants, see Fig.~\ref{gianthrd}). Again,
many of these sources are detected in X-rays, albeit at relatively weak levels, with $L_X$ of a few times
$10^{27}$~erg~s$^{-1}$ \citep{schroeder98}). Strong flaring activity before and after the He-flash
testifies to the ability of the dynamo to survive the internal reconfiguration \citep{schroeder98, ayres99, ayres01a}.

In this context, the Capella binary is particularly interesting  since it contains a
primary clump giant (G8~III) and a presumably co-eval Hertzsprung gap star (G1~III) of about
equal mass ($\approx 2.6M_{\odot}$). The Fe\,{\sc xxi} $\lambda 1354$ lines show considerable
variability on timescales of several years in the clump giant, but not in the Hertzsprung gap star
\citep{johnson01, linsky98}. The former authors suggested that long-term variability, for example
induced by dynamo cycles, occurs
only in the clump phase where a deep convection zone has built up. 

Another example is the quadruplet of the nearly co-eval Hyades giants
($\gamma$, $\delta$, $\epsilon$, and $\theta^1$ Tau). The four stars are found at very similar
locations in the HRD, but show a spread in $L_X$ over orders of magnitude \citep{micela88, stern92c}. 
A possible explanation is, apart from contributions by companions, that the high-activity stars are first 
crossers while the others  have already evolved to the clump area \citep{collura93}.

\subsection{Dividing lines}\label{dividing}

From ultraviolet observations of chromospheric and transition region lines in giants, \citet{linsky79} proposed a 
dividing line in the giant and supergiant  area of the HRD roughly at V--R = 0.7--0.8, separating stars with 
chromospheres and hot transition regions to the left from stars that exhibit exclusively chromospheric lines, 
to the right. A steep gradient in transition region fluxes is found when one moves from spectral type G towards
the dividing line, located roughly between K2~III and G~Ib \citep{haisch90c}.  The absence of
warm material in the later-type stars suggested an absence of coronal material as well, which was soon
confirmed based on {\it Einstein} observations \citep{ayres81a, haisch82, maggio90}.
{\it ROSAT}  observations deepened this conclusion, converging to a more ``perpendicular''
dividing line at spectral type $\approx$ K3 between luminosity classes II--IV \citep{haisch91b}.

The dividing line roughly corresponds to the ``onset'' 
of massive cool winds toward M giants and supergiants \citep{reimers96}. This coincidence may be a consequence
of  the cooler stars carrying predominantly open magnetic field lines that produce stellar winds
somewhat similar  to solar coronal holes \citep{linsky79}. Plasma that resides in open magnetic
regions is necessarily cool in such stars because the escape temperature is much smaller than on the Sun, 
and this leads to small Alfv\'en-speed scale heights and 
thus to strong winds via Alfv\'en-wave reflection \citep{rosner91}. Why cooler stars should show 
predominantly open field lines is unexplained. A 
shift from an $\alpha\omega$ dynamo to a turbulent dynamo in
cooler stars \citep{gondoin87, gondoin99} could possibly change the magnetic field structure. 

An $\alpha^2$ dynamo that relies on convection close to the
surface but not on rotation may indeed be a favorable option given the slow rotation 
of red giants \citep{ayres03a}. In this case, one expects that smaller-scale loops are generated,
and this has two consequences (e.g., \citealt{rosner95, schrijver96}): First, the larger thermal
pressure in low-lying hot loops will open them up  if the magnetic field
strength is not sufficiently high, thus draining the available energy into a stellar wind. 
Second,  \citet{antiochos86a} proposed that the low
gravity of the cooler and therefore, larger stars to the right of the dividing line  allows for new
static cool loop solutions with $T < 10^5$~K and small heights (after \citealt{martens82, antiochos86b}), 
while hot X-ray loops with heights less
than the (large!) pressure scale height at $10^5$~K grow unstable and are therefore
not radiating. In this model, then, the hot magnetic corona
is replaced by a cooler but extended transition region. Any hot material detected from stars to the
right of the dividing line or from ``hybrid stars'' (see below) may  entirely
be due to short-term flaring \citep{kashyap94}. 
The above two instabilities apply to warmer stars as well: Because the magnetic loop size
is restricted by the convection zone depth which is small for F stars, there is 
a regime in that spectral range where the above arguments apply similarly, i.e., hot coronae
will become unstable \citep{schrijver96}.

Even if a magnetically confined X-ray corona is present in giants, its X-ray emission may be 
efficiently  absorbed by the stellar wind so that the dearth of coronal emission in M giants 
may only be apparent \citep{maggio90}. Since several non-coronal giants have been found to maintain transition 
regions up to $10^5$~K, \citet{ayres97} proposed that coronal loops are in fact submerged in the 
cool chromospheric and CO layers and are thus absorbed. The reason for an extended, not magnetically
trapped chromosphere is rooted in the 40-fold larger pressure scale height on red giants
such as Arcturus compared to the Sun. The model of ``buried'' coronae would again be aided if the dynamo in these stars favored
short, low-lying loops \citep{rosner95}. 
This view has gained strong support from the recent detection of very weak
X-ray emission from the K1~III giant Arcturus, with $L_X \approx 1.5\times 10^{25}$~erg~s$^{-1}$  \citep{ayres03a};
Arcturus  was previously thought to be X-ray dark 
\citep{ayres81a, ayres91}. \citet{ayres03a} also reported the presence of transition region lines that are
absorbed by overlying cooler material that acts like a ``cool absorber''.

\citet{huensch96} compared the dividing line with evolutionary models and concluded that it is actually
nearly parallel to the evolutionary tracks in the giant domain. They  suggested that the drop of
X-ray activity toward cooler stars is gradual and is a direct function of mass. In  this picture, the coolest
M giants with masses $\la 2M_{\odot}$ and ages around $10^9$~yr are X-ray faint simply because they had spun down on 
the MS before and are thus no longer able to drive a dynamo \citep{ayres81b, 
haisch82, huensch96, schroeder98} while the warmer giants, descendants of $M > 2M_{\odot}$ main-sequence
stars, have retained more angular momentum.

\subsection{Hybrid stars}\label{hybrid}

The situation is somewhat less clear for evolved supergiants and bright giants (luminosity class II) given 
the small statistics at hand, but there is now evidence against a clear dividing line in this
region of the HRD \citep{reimers96}. 
While the wind dividing line sharply bends to
the left in the HRD as most bright giants and supergiants show strong winds (Fig.~\ref{gianthrd}), the so-called 
 ``hybrid stars'' are formally right of a vertical dividing line extended from the giant
region, but they show indications both of cool winds and $\approx 10^5$~K transition region
material \citep{hartmann80}. After first X-ray detections \citep{brown91, haisch92}, it  soon
became evident that X-ray emitting hybrids are common across the regime of 
bright giants and supergiants of luminosity class I--II 
\citep{reimers92, reimers96, huensch96b}. The co-existence of hot coronae and cool
winds in hybrids renders them  pivotal for understanding the
physics of stellar atmospheres in this region of the HRD.

The X-ray luminosities of hybrids are relatively modest, with $L_X = 10^{27} - 10^{30}$~erg~s$^{-1}$ 
\citep{reimers96, ayres04};
like other luminosity-class II stars, they are X-ray deficient compared to giants or main-sequence stars
(\citealt{ayres95, reimers96, ayres04}; Fig.~\ref{fluxflux}), but at the same time they show very high 
coronal temperatures. The X-ray deficiency syndrome of 
supergiants is probably the same as that of F-type giants  and subgiants (\citealt{simon89}, 
Sect.~\ref{giants}). Indeed, the transformation from X-ray deficient to ``solar-like'' stars occurs at
progressively later spectral types as one moves to higher luminosity classes, eventually 
encompassing almost the complete cool half of the supergiant HRD \citep{ayres95}. 
The co-existence of winds and coronae may be the result of the very rapid evolution
of hybrid stars: The X-ray emission may compete with
cool-wind production as the dynamo activity persists. In the picture of \citet{huensch96}, then, hybrids
are X-ray sources because their masses are $>2M_{\odot}$, i.e., they start out as
fast rotators on the main-sequence and keep their dynamo while developing strong
winds. The X-ray dividing line defined by spun-down low-mass stars does
simply not reach up to these luminosity classes.

\subsection{Evolution of X-ray emission in open stellar clusters}\label{clusters}

\subsubsection{Overview.} Open clusters have become instrumental for the study of stellar coronae and their long-term evolution
for several reasons: i)  They provide large samples of nearly co-eval stars spread
over a broad mass range that encompasses all types of cool MS stars and possibly brown  dwarfs,
in the statistical proportion dictated by the processes of star formation; ii) their ages are rather 
well known from their distribution on the HRD; ii) their surface chemical 
composition is very likely to be constant across the stellar sample. Open clusters are therefore
ideal objects with which to test theories of stellar evolution and, in particular,
systematic dependences between {\it rotation, activity, and age}. While some of the issues have already been covered in the 
section on rotation (Sect.~\ref{rotation}), the present section emphasizes specific evolutionary effects and sample studies 
made with open clusters.

Table~\ref{clustertab} summarizes open cluster studies in the literature, also listing - if available - 
median X-ray luminosities for the spectral classes F, G, K, and M. The colors were not defined identically
by all authors, but the large spread of $L_X$ and the statistical uncertainties will make this a rather
negligible problem. Also included are clusters in star-forming regions. Because these stars
are not yet on the main sequence, comparing them with MS clusters based on color may be misleading.
For the sake of definition, I have lumped together all T Tau stars from such samples in the ``K star'' column
unless more explicit information was available. Several of the given ages (mostly from the Lausanne open cluster
database) must be treated as tentative.

\begin{table}
\begin{minipage}{1.0\textwidth}
\caption{Open cluster studies: Luminosities and bibliography}
\label{clustertab}       
\scriptsize
\begin{tabular}{lllllll}
\hline\noalign{\smallskip}
Cluster      & Age (Myr)    & \multicolumn{4}{c}{Median $L_X$ for spectral class}                               &    References$^a$    \\
             &              & F                    &  G                   &   K                &  M                    &   \\
\noalign{\smallskip}\hline\noalign{\smallskip}
NGC 1333     &    $<1$      &	...		   &   ...		  &    ...	       &       ...	       & 1     \\
$\rho$ Oph   &    0.1-1     &	...		   &   ...		  &    30.4	       &       ...	       & 2     \\ 
Serpens      &    1         &	...		   &   ...		  &    ...	       &       ...	       & 3     \\ 
NGC 2264     &  $\approx 1$ &	...		   &   ...		  &    30.4	       &       ...	       & 4     \\ 
Orion        & 0.1-10       &   ...                &   ... 	          &    29.8	       & 	    ...        & 5     \\
NGC 2024     & 0.3-few      &	...		   &   ...		  &    ...	       &       ...	       & 6     \\
Taurus WTTS  & 0.1-10       &   ...                &   30.2	          &    29.9	       &	    29.1       & 7     \\
ChaI  WTTS   & 0.1-10       &   ...                &   ... 	          &    29.8	       &	    ...        & 8     \\
IC 348       & 1-6	    &   ...                &   ... 	          &    29.15	       &	    ...        & 9     \\ 
R CrA        & 1.5-7	    &   ...                &   ... 	          &    ... 	       &	    ...        & 10    \\
Upper Sco-Cen& 5-6	    &   ...                &   ... 	          &    ... 	       &	    ...        & 11    \\
$\eta$ Cha   & 8	    &   ...                &   ... 	          &    ... 	       &	    ...        & 12    \\
TW Hya       & 10-30 	    &   ...                &   ... 	          &    29.6	       &	    ...        & 13     \\
Tucanae      & 10-30	    &   ...                &   ... 	          &    29.8	       &	    ...        & 14     \\
IC 2602      & 32	    &	...		   &   29.8		  &    29.75	       &       29.15	       & 15     \\
NGC 2547     & 36           &	...                &   ...		  &    ...	       &       ...	       & 16     \\
IC 4665      & 43	    &	...                &   ...		  &    ...	       &       ...	       & 17     \\
IC 2391      & 46	    &	...		   &   29.75		  &    ...	       &       ...	       & 18     \\
$\alpha$ Per & 50	    &	29.52		   &   29.68		  &    29.57	       &       28.86	       & 19     \\
NGC 2451 B   & 50           &	29.81              &   29.82		  &    29.71	       &       ...	       & 20     \\
NGC 2451 A   & 50-80        &	29.23              &   29.46		  &    $\approx 29$      &       29.29	       & 20     \\
Blanco 1     & 63	    &	29.45		   &   29.41		  &    29.11	       &       29.06	       & 21     \\
NGC 2422     & 73	    &   ...                &   $\approx 29.0$	  &    ...	       &       ...	       & 22     \\
Pleiades     & 100 	    &	29.20		   &   29.25		  &    29.2	       &       28.85	       & 23     \\
NGC 2516     & 110 	    &	29.40		   &   29.16		  &    29.08	       &       28.72	       & 24     \\
Stock 2      & 170	    &	...		   &   ...		  &    ...	       &       ...	       & 25     \\
NGC 1039     & 180	    &	...		   &   ...		  &    ...	       &       ...	       & 26     \\
Ursa Major   & 300          &	$\approx 28.2$     &   $\approx 28.2$	  &    ...	       &       ...	       & 27     \\
NGC 6475     & 300	    &	(29.25)		   &   (29.44)		  &    (29.5)	       &       ...	       & 28     \\
NGC 3532     & 310	    &	...		   &   ...		  &    ...	       &       ...	       & 29     \\
NGC 6633     & 430          &  $\approx 28.7$	   &  $\approx 28.5$	  &    ...	       &       ...	       & 30     \\
Coma Ber     & 450	    &	28.87		   &   ...		  &    ...	       &       ...	       & 31     \\
IC 4756      & 500	    &	...		   &   ...		  &    ...	       &       ...	       & 32     \\
NGC 6940     & 720	    &	...		   &   ...		  &    ...	       &       ...	       & 33     \\
Praesepe     & 730	    &	...		   &   29.02		  &    28.25	       &       28.3	       & 34     \\
Hyades       & 790	    &	28.7		   &   29.03		  &    28.19	       &       28.21	       & 35     \\
NGC 752      & 1100	    &	...		   &   ...		  &    ...	       &       ...	       & 36     \\
IC 4651      & 1100  	    &	...		   &   ...		  &    ...	       &       ...	       & 37     \\
M67          & 2600 	    &	...		   &   ...		  &    ...	       &       ...	       & 38     \\
NGC 188      & 4300  	    &	...		   &   ...		  &    ...	       &       ...	       & 39     \\
\noalign{\smallskip}\hline
\end{tabular}\\
\footnotetext{\scriptsize NOTE: For PMS clusters, the numbers refer to WTTS if separately available from the literature. 
              If given as a class, they are listed in the K star column. 
	     $^a$~REFERENCES AND NOTES: Reference from which quoted values were taken are given in {\bf boldface.} 
	     Approximate cluster ages are from the Lausanne open cluster database (WEBDA) or from the cited references.
             1   \citet{preibisch97b}, \citet{preibisch03a}, \citet{getman02};  
	     2   \citet{montmerle83}, \citet{koyama94}, \citet{casanova95}, \citet{kamata97}, \ {\bf \citet{grosso00}}, \citet{grosso01},
	         \citet{imanishi01a}, \citet{imanishi01b}, \citet{imanishi02}, \citet{imanishi03};  
	     3   \citet{preibisch98a}, \citet{preibisch03b};  
             4   \citet{simon85}, \citet{flaccomio99}, {\bf \citet{flaccomio00}};  
             5   \citet{gagne94}, \citet{gagne95b}, \citet{pravdo95}, \citet{yamauchi96}, \citet{garmire00}, \citet{tsujimoto02},
	         \citet{feigelson02a},  \citet{feigelson02b}, 
	     	 \citet{feigelson03}, \citet{flaccomio03a}, {\bf \citet{flaccomio03b}};
	     6   \citet{freyberg95, skinner03};  
	     7   \citet{feigelson87}, \citet{walter88}, \citet{strom90}, \citet{strom94}, 
	         \citet{damiani95a}, \citet{damiani95b}, \citet{neuhauser95a}, 
	         \citet{carkner96}, \citet{skinner97}, {\bf \citet{stelzer01}};  
	     8   \citet{feigelson89}, {\bf \citet{feigelson93}}, \citet{lawson96};  
	     9   \citet{preibisch96}, \citet{preibisch01}, {\bf \citet{preibisch02}}; dominated by M dwarf sample;     
	     10  \citet{koyama96}, \citet{walter97};     
	     11  \citet{walter94}, \citet{sciortino98};  
	     12  \citet{mamajek99}, \citet{mamajek00};  
	     13  \citet{hoff98}, \citet{jensen98}, \citet{kastner99}, {\bf \citet{stelzer00b}};  
	     14  {\bf \citet{stelzer00b}};
	     15  {\bf \citet{randich95b}};     
	     16  \citet{jeffries98b};			
	     17  \citet{giampapa98};   
	     18  \citet{patten93}, {\bf \citet{patten96}}, \citet{simon98};			
	     19  {\bf \citet{randich96a}}, \citet{prosser96};  
	     20  {\bf \citet{huensch03}}, B biased toward too high $L_X$;     
	     21  \citet{micela99b}, {\bf \citet{pillitteri03}};     
	     22  {\bf \citet{barbera02}};      
	     23  \citet{caillault85}, \citet{micela85},  \citet{schmitt93a}, \citet{stauffer94},
	     	   \citet{gagne95a}, \citet{hodgkin95}, {\bf \citet{micela96}}, \citet{micela99a},
	     	   \citet{stelzer01}, \citet{krishnamurthi01}, \citet{daniel02}, \citet{briggs03};	       
	     24  \citet{dachs96}, \citet{jeffries97}, \citet{micela00}, \citet{sciortino01},
	     	   \citet{harnden01}, {\bf \citet{damiani03}};   
	     25  \citet{sciortino00};	       
	     26  \citet{simon00};      
	     27  \citet{walter84b}, {\bf \citet{schmitt90b}};  
	     28  {\bf \citet{prosser95}}, \citet{james97}; sample is X-ray selected - true median should be lower;     
	     29  \citet{simon00}, \citet{franciosini00};       
	     30  \citet{briggs00}, {\bf \citet{harmer01}}; upper limit to median given;   
	     31  \citet{randich96b};   
	     32  \citet{randich98}, \citet{briggs00};	       
	     33  \citet{belloni97};    
	     34  \citet{randich95a}, \citet{randich96b}, \citet{barrado98}, 
	     	   {\bf \citet{franciosini04}};	       
	     35  \citet{stern81}, \citet{micela88}, \citet{stern92c}, \citet{stern94}, 
	     	   {\bf \citet{pye94}}, {\bf \citet{stern95b}}, \citet{reid95}, \citet{stelzer01}, {\bf G. Micela, priv. comm. (2004)};        
	     36  \citet{belloni96};    
	     37  \citet{belloni98a};   
	     38  \citet{belloni93}, \citet{belloni98b};        
	     39  \citet{belloni98b}.    
	     }\normalsize
\end{minipage}
\end{table}

\begin{figure} 
\rotatebox{90}{
\begin{minipage}{1.0\textheight} 
\vbox{\vskip -0.3truecm\hbox{ 
\hskip -0.3truecm\resizebox{0.43\textheight}{!}{\includegraphics{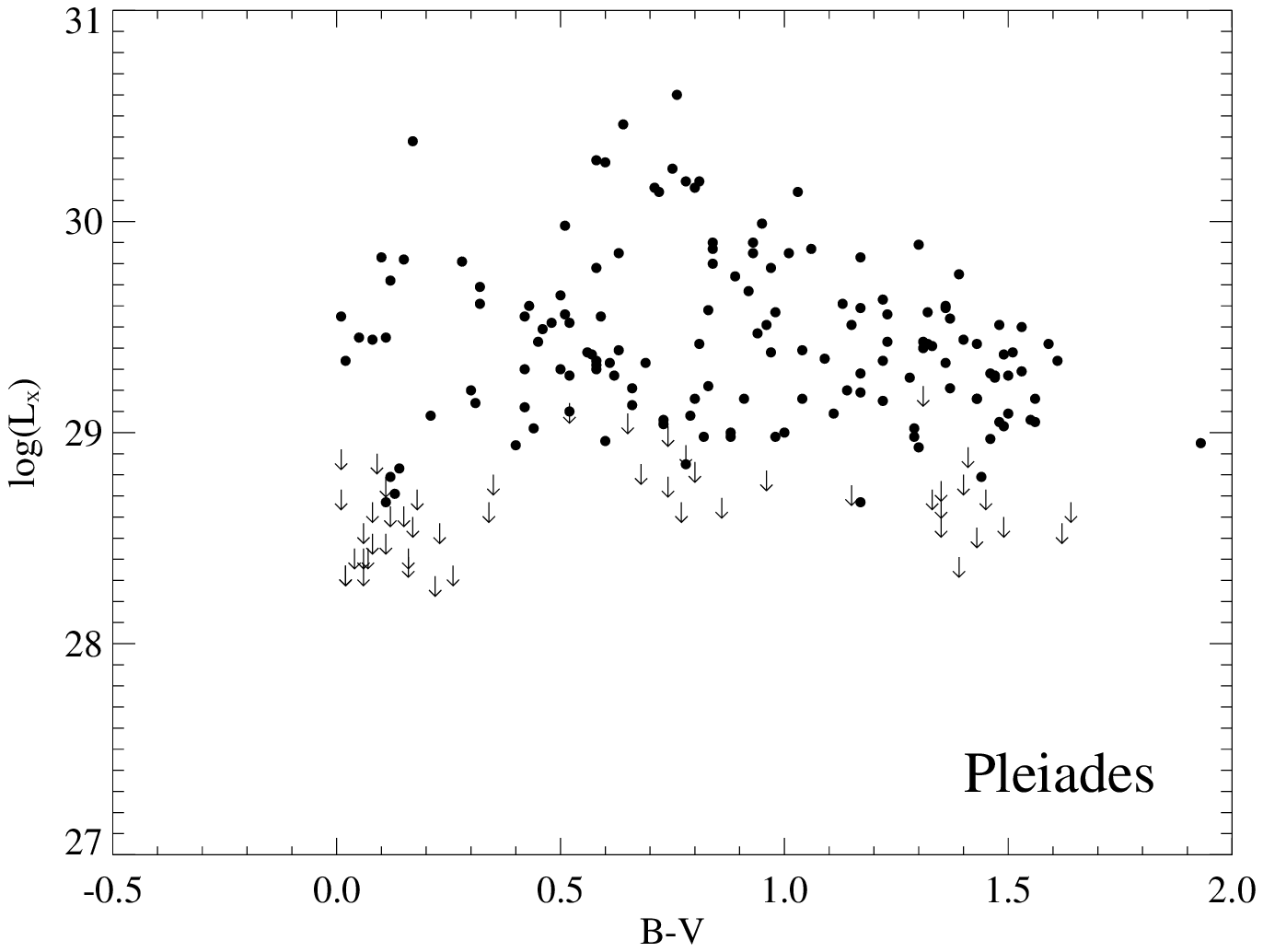}}
\hskip -0.3truecm\resizebox{0.43\textheight}{!}{\includegraphics{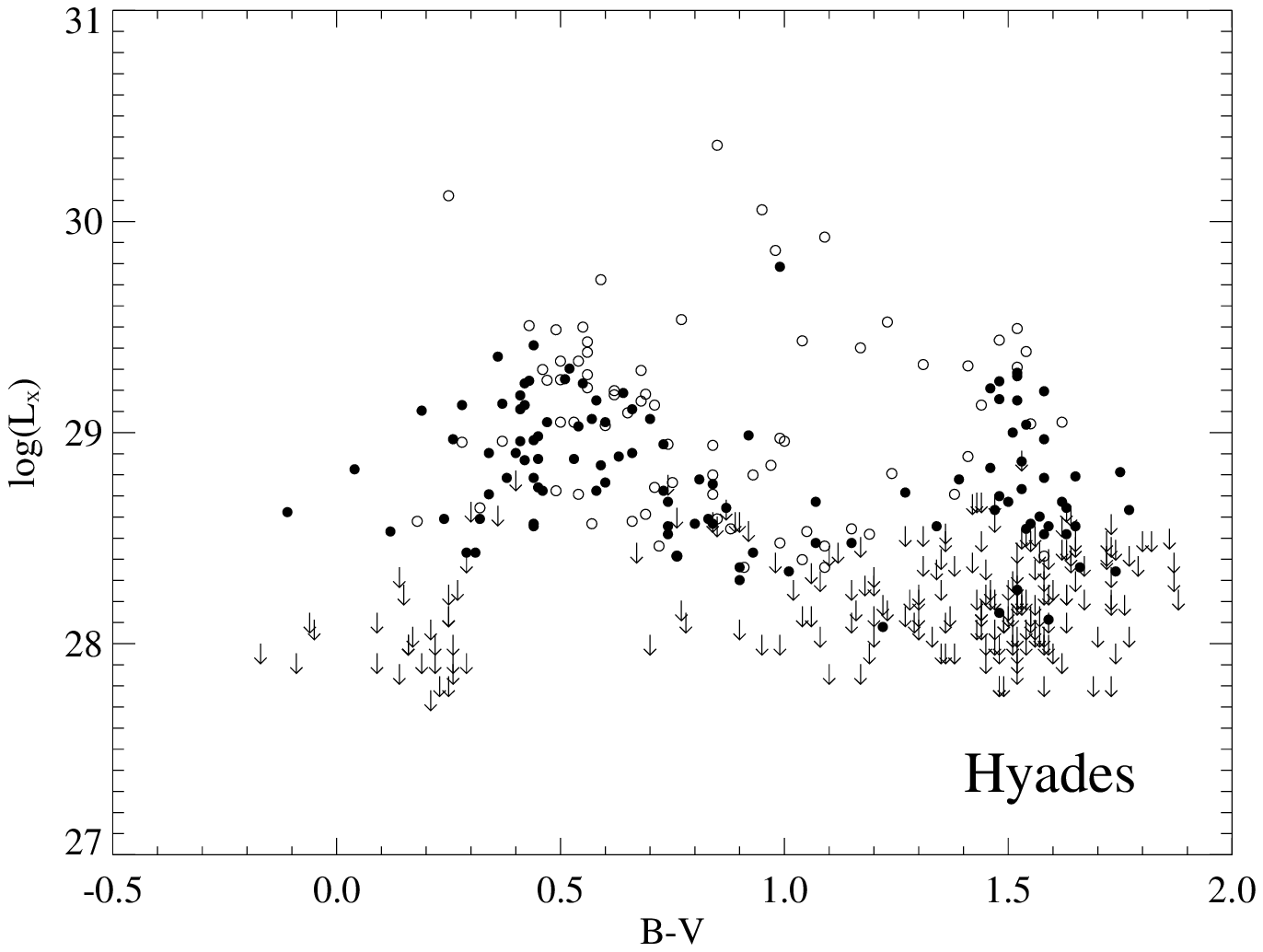}}
}
\hbox{ 
\hskip -0.3truecm\resizebox{0.43\textheight}{!}{\includegraphics{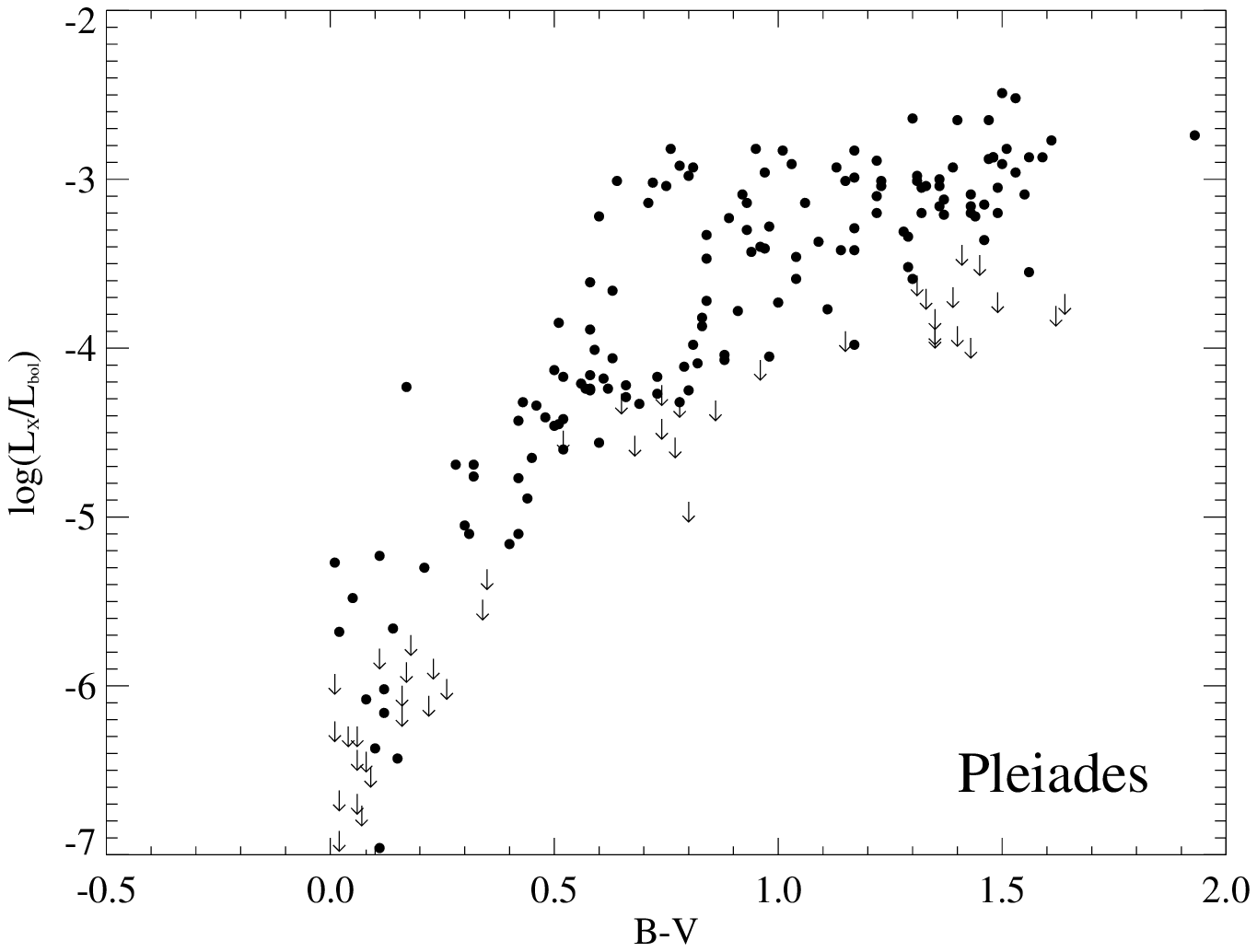}}
\hskip -0.3truecm\resizebox{0.43\textheight}{!}{\includegraphics{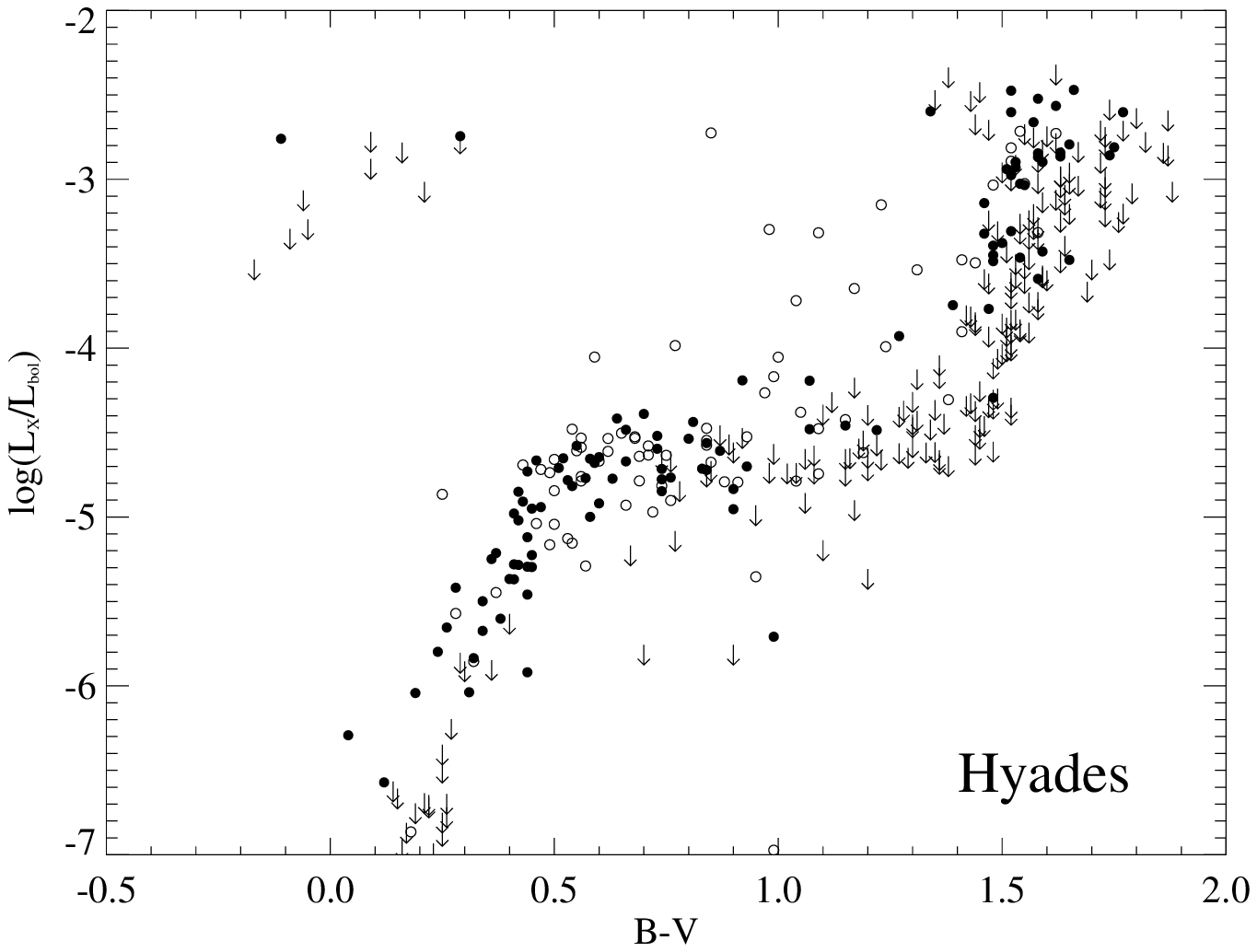}}
}
}
\vskip -12.5truecm\hskip 16.6truecm\begin{minipage}{3cm}
\caption{Evolution of X-ray luminosities in open clusters. The figures on the  left refer to the
Pleiades (data are from \citealt{stauffer94}), those on the right  to the Hyades
(data from \citealt{stern95b}). The upper plots show $L_X$ as a function of $B-V$, while
the lower panels show $L_X/L_{\rm bol}$.\label{clusterevol}} 
\end{minipage}
\end{minipage} 
}
\end{figure}

Figure~\ref{clusterevol} shows the distribution of $L_X$ and $L_X/L_{\rm bol}$ as a 
function of $B-V$ for the Pleiades (age $\approx 100$~Myr) and the Hyades  (age $\approx 700-800$~Myr). The older cluster 
shows distinctly lower median $L_X$ and $L_X/L_{\rm bol}$ for most spectral classes, with the exception of
M dwarfs. These systematics are clearly related to evolution  and rotation \citep{caillault85, micela85}, but we 
now also see a strong dependence of the evolution on $B-V$ and therefore mass. Let us briefly
walk across the available $B-V$ range.

As is the case for field stars (Sect.~\ref{astars}), A stars and early F-type stars are usually 
detected at very low luminosities or not at all, and in most other cases, there is a strong suspicion that later-type companions
are responsible for the observed level of X-ray emission \citep{micela85, micela88, micela96, schmitt90b, stauffer94, randich95b, 
randich96a, briggs03}.
F stars that are thought to be genuine X-ray emitters not only fall short by one order of magnitude of the maximum $L_X/L_{\rm bol}$ 
level attained by cooler stars, but their X-ray properties also show relatively little evolution in time, at least among younger clusters
with ages up to a few 100~Myr  \citep{caillault85, micela90, patten96, randich96a}. As 
discussed in  Sect.~\ref{rotation}, the decreasing X-ray efficiency toward earlier spectral types is most likely related to
the decreasing depth of the convection zone and the consequent limited ability of the dynamo to cover the 
star with a strong magnetic corona.

The most luminous stars in young clusters are found among G stars. These are the earliest stars that  reach up to
a saturation limit of $L_X/L_{\rm bol} \approx 10^{-3}$ (\citealt{stauffer94, randich96a}, 
Fig.~\ref{clusterevol}). At the same time, they also show the
largest spread in X-ray activity. An evolutionary decay in $L_X$ is well seen beyond the Pleiades' age: 
Hyades G stars are no longer found at the saturation level.
The X-ray activity evolution of G stars thus proceeds on time scales of $\approx$100~Myr. 

Moving now to K dwarfs, similar systematics are seen except that the upper bound of $L_X$ continuously drops toward later spectral types 
(Fig.~\ref{clusterevol}), which is a direct consequence of the decreasing maximum $L_X$ allowed by the saturation boundary 
$L_X/L_{\rm bol} \approx 10^{-3}$. The evolution seems to be slowest among M dwarfs that are still close to the saturation 
level in the Hyades. As already mentioned above for field stars, the slower spin-down keeps their magnetic activity at high levels 
during at least several 100~Myr.

\subsubsection{Rotation-age-activity relations} 

Open clusters offer a great amount of information on rotation-activity and activity-age relations. Although initial reports 
failed to find such dependencies  \citep{caillault85, micela85}, the best-defined  rotation-activity 
relations in fact now derive from open cluster studies and include the saturation 
plateau and stars in the  supersaturation regime (\citealt{randich00b}, Fig.~\ref{activityrotation}).

Such investigations have converged to a comprehensive basic picture of MS coronal evolution (e.g., 
\citealt{stern95b, patten96}).
As the stars arrive on the main sequence, their initial rotation periods may largely vary, probably as a consequence of 
different star-disk coupling and disk-dispersal histories. Somewhat counter-intuitively, X-ray luminosities
remain  narrowly confined at predictable levels during this phase. This is because most 
stars rotate rapidly enough to keep them in the saturation regime where $L_X$ is controlled essentially by $L_{\rm bol}$
rather than by rotation. Stars thus typically enter the rotation-activity diagram (Fig.~\ref{activityrotation}) 
spread over a range in the saturation regime. There is often some additional intrinsic scatter that is
possibly related to long-term, cyclic modulation (\citealt{patten96}, but see \citealt{micela96, simon98}).

The rotation rate of stars declines steadily as they lose angular momentum via  a magnetized wind.   
Magnetic braking occurs at the highest rate for the most rapid rotators, because they produce the 
strongest magnetic fields and presumably winds. The stars eventually drop out of the
saturation regime  and enter the standard rotation-activity branch, but not all
at the same time given their spread of initial rotation periods. This is probably the cause of the wide spread 
seen in $P$ and $L_X$ below the saturation regime at any given moment in time, i.e., for a given cluster, as long as the 
cluster is relatively young \citep{stauffer94}.  

Because the spin-down rate is higher for higher-mass stars, the
X-ray activity of the earliest-type stars decays first. This leads to the characteristic development of the $L_X$ vs. $B-V$ 
diagram where the G-K regime is ``hollowed out'' from the left on times scales of $20-50$~Myr for G 
stars and $\ga 75$~Myr to a few 100~Myr 
for K stars, while M dwarfs remain at high activity levels for much longer (\citealt{patten96, james97}, Fig.~\ref{clusterevol}). 
Incidentally, the distribution of $L_X$ levels may then be relatively flat across a wide range of $B-V$ in the age period between 
the Pleiades  and the Hyades.
At the same time, the $L_X/L_{\rm bol}$ ratio systematically increases toward larger $B-V$ (Fig.~\ref{clusterevol}). At the age of 
$\alpha$ Per or the Pleiades, G, K, and M dwarfs still all reach up to the saturation limit \citep{hodgkin95, prosser96, 
randich96a, micela99a}, while at the ages of the Hyades, this is true only for late-K and M dwarfs  \citep{reid95, stern95b}. The break
point at which saturation is reached thus moves to progressively larger $B-V$ as the cluster ages. 

\begin{figure} 
\centerline{\resizebox{1.\textwidth}{!}{\includegraphics{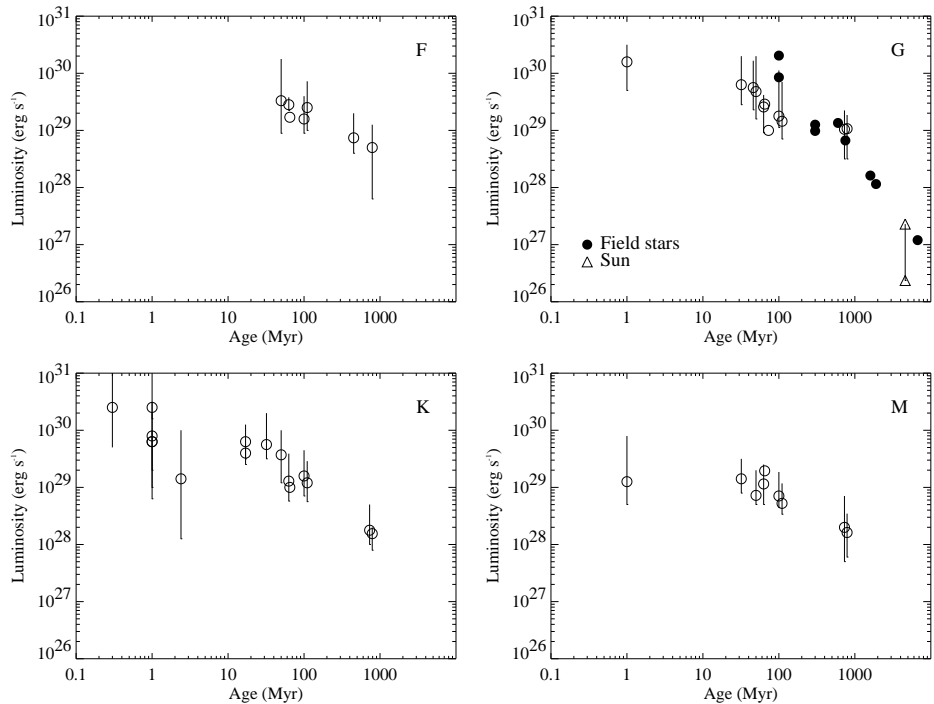}}}
\caption{The evolution of the median X-ray luminosities of F, G, K, and M stars derived from various open 
clusters (from Table~\ref{clustertab}: $\rho$ Oph, NGC 2264, Orion, Taurus, Cha~I, IC 348, TW Hya, Tucanae, IC 2602, IC 2391,
$\alpha$ Per, Blanco 1, NGC 2451 A, NGC 2422, Pleiades, NGC 2516, Coma Ber, Praesepe, and Hyades). 
The error bars give the approximate 1$\sigma$ scatter (if available from the literature). 
The panel for G stars also includes the field G stars shown in Fig.~\ref{agedecay} (filled circles).
The TTS samples from star-forming regions are lumped together in the K star panel, except for
the Taurus region for which G, K, and M stars are given separately. The ages of the star-forming regions
are only characteristic averages as they are typically spread between 0.1--10~Myr.}\label{clusterage}
\end{figure}

The different braking histories lead to different characteristic decays of the X-ray luminosities for 
various spectral ranges, as illustrated in Fig.~\ref{clusterage}. 
The error bars refer to the $\pm 1\sigma$ spread read  off the
luminosity functions published in the literature. They are only approximate and are probably dominated
by uncertainties from small-number statistics. As we see from Fig.~\ref{clusterage}, 
the $L_X$ of all spectral classes decays
beyond $100$~Myr although this decay is clearly slowest for M dwarfs.  As a cluster reaches the 
Hyades' age ($\approx 700-800$~Myr), the rotation  periods of both G and K dwarfs have mostly converged to relatively low values 
with little spread, and the $L_X$ values are consequently also expected to have 
dropped significantly below the saturation limit, with small statistical spread, while M dwarfs are now in a regime of rapid 
braking, increasing their scatter in $L_X$ as they settle at lower rotation rates \citep{micela88, stern92c, stern94, stern95b}. 

\begin{figure} 
\centerline{\hbox{
\resizebox{0.5\textwidth}{!}{\rotatebox{270}{\includegraphics{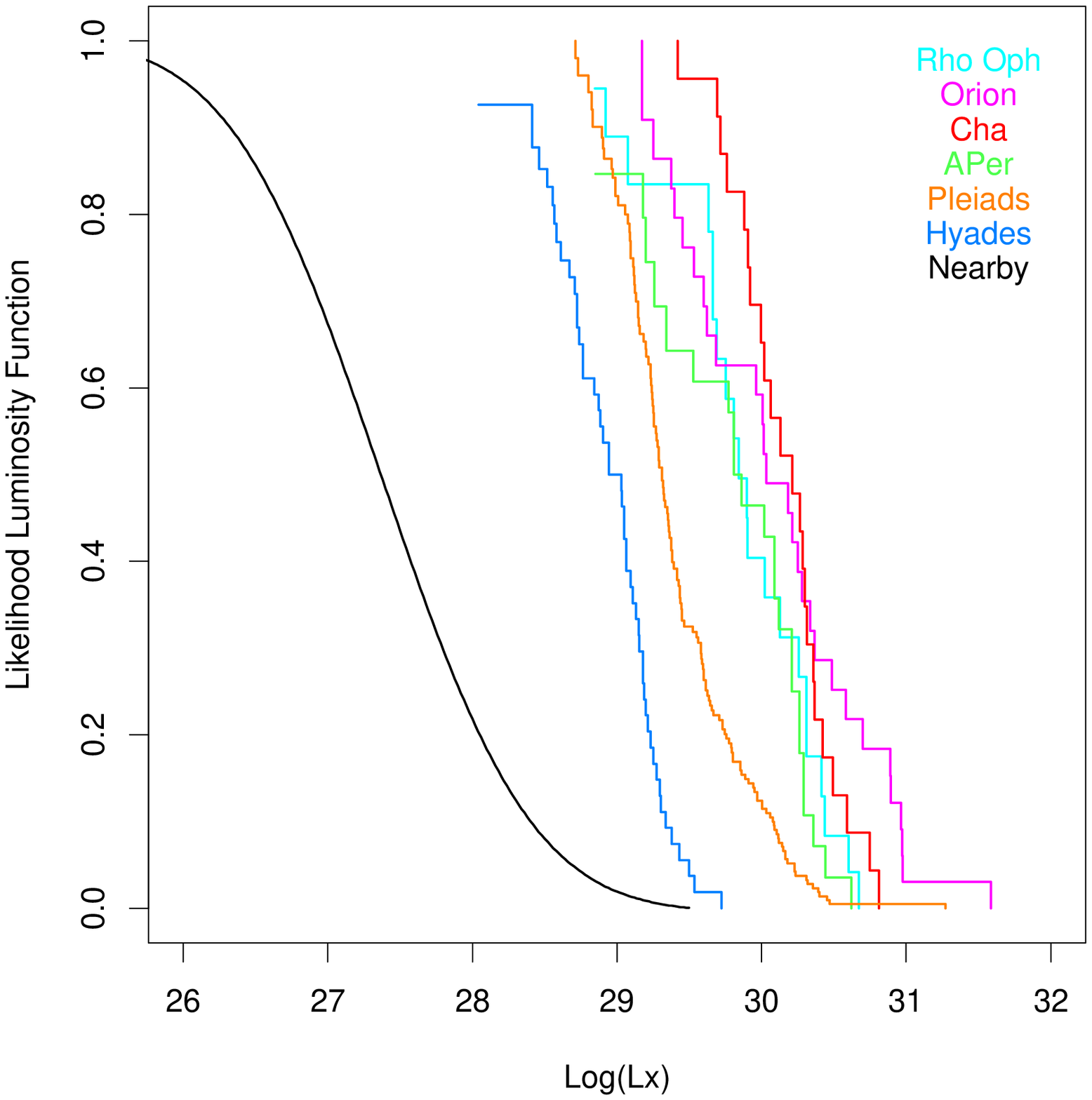}}}
\resizebox{0.5\textwidth}{!}{\rotatebox{270}{\includegraphics{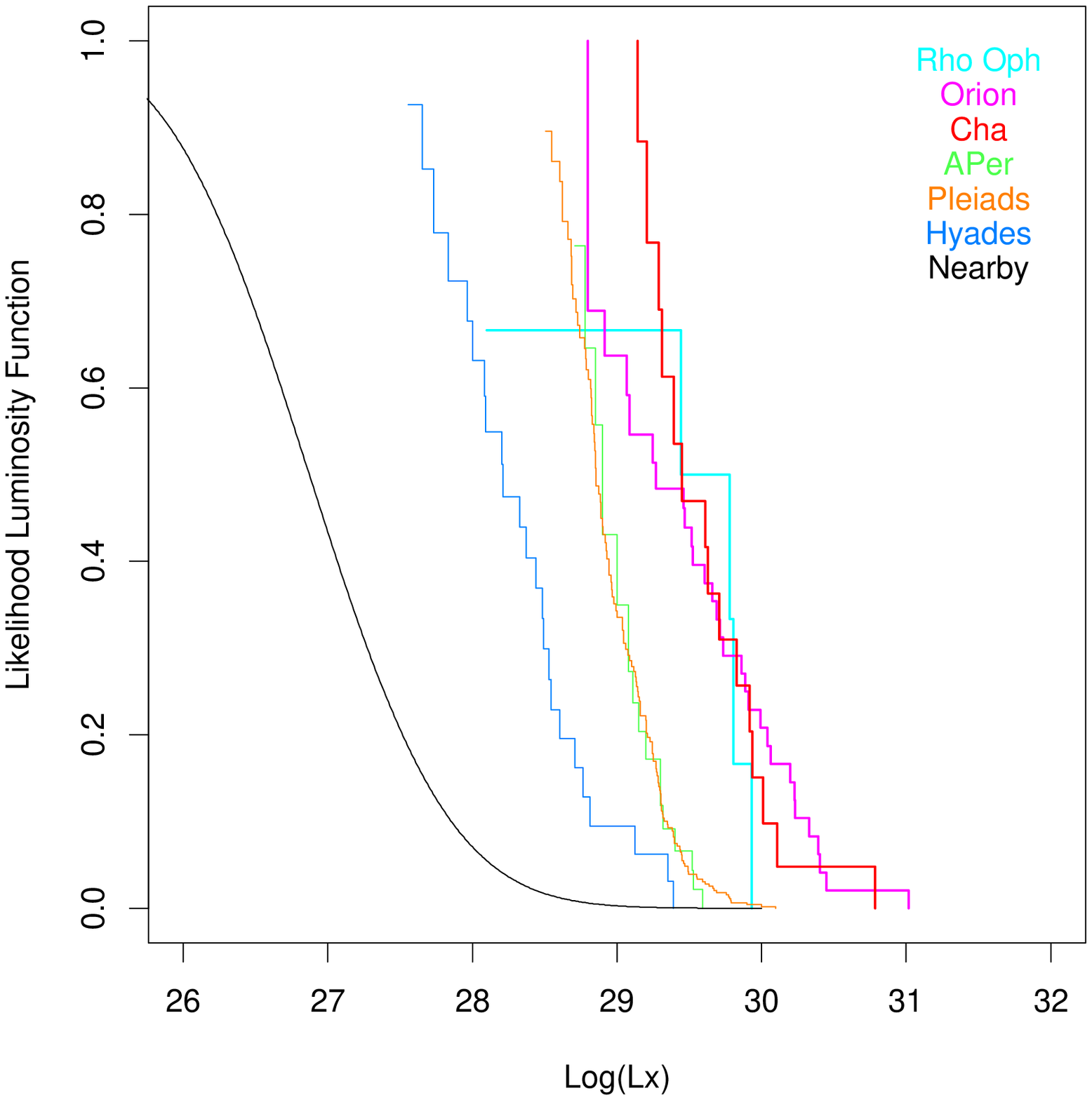}}}
}
}
\caption{Luminosity functions of (left) stars with masses of $\approx 0.5-1M_{\odot}$, representative
of G--K stars,   and (right) with masses of  $\approx 0.25-0.5M_{\odot}$, representative of M-type stars,  
for various star-forming regions and open clusters. The median luminosity decreases with increasing age
(figures courtesy of G. Micela; references used: \citealt{pye94, stern95b, randich96a, flaccomio03c}).}\label{lumfunction}
\end{figure}

The overall trend agrees nicely with the decay law found from field stars, as is illustrated for the G star panel where the 
field stars from Fig.~\ref{agedecay} are overplotted as filled circles.
This scenario is alternatively illustrated by X-ray luminosity functions
for various spectral ranges; the sample median and the spread of
X-ray luminosities as a function of spectral type can then be followed as a function of age (e.g., \citealt{micela90}),
as shown in Fig.~\ref{lumfunction}.

At this point, we recognize the close interrelationship between the {\it activity-age} relationship discussed here
(see also Sect.~\ref{msevol} and Eq.~\ref{tempage}), 
the {\it temperature-activity} relation  introduced in Sect.~\ref{templx} (Fig.~\ref{temperaturelx}, Eq.~\ref{TLx}),
and the {\it temperature-age} relation (Sect.~\ref{msevol}, Fig.~\ref{agedecay}). {\it They are all an expression
of the coupling between coronal activity and the rotation-induced internal dynamo.}

\subsubsection{Binaries}

A different scenario seems to apply to late-type binary stars even if they are not tidally locked. 
\citet{pye94} and \citet{stern95b} found components in K- and M-type binaries (but not F- and G-type binaries)
in the Hyades to be overluminous by an order of magnitude when compared to
single stars. The explanation for this effect is unclear but could involve a less rapid braking in the PMS stage as the
circumstellar disks were disrupted, thus producing very rapid rotators that have still not spun down at the age of the 
Hyades \citep{pye94}. A reanalysis by \citet{stelzer01}, however, questioned a significant difference between
singles and binaries.

\subsubsection{Co-eval clusters}

Intercomparisons between clusters of the same age have great\-ly helped confirm the above evolutionary picture. Similar
trends for $L_X$ and $L_X/L_{\rm bol}$ are 
generally found for co-eval clusters (e.g., \citealt{giampapa98}). A notable exception was discussed 
by \citet{randich95a}, \citet{randich96b}, and \citet{barrado98} who compared the nearly co-eval Hyades, Coma Berenices, and 
Praesepe clusters, finding significant $L_X$ deficiencies in the latter cluster but agreement between the former two. The 
cause of the discrepancy is not clear, but could be related to the recent finding that Praesepe shows a spatial segregation
of activity levels, which may be the result of a merger of two non-coeval clusters (\citealt{franciosini04} and references therein).
Similar discrepant cases include Stock 2 \citep{sciortino00} and possibly IC 4756 \citep{randich98},
NGC 3532 \citep{franciosini00}, and NGC 6633 \citep{briggs00, harmer01}.  Another important influence may come from
different metallicities in clusters. \citet{jeffries97}, \citet{harnden01}, \citet{sciortino01}, and \citet{damiani03} 
compared NGC 2516 with the slightly younger Pleiades and found F stars to be {\it more luminous} and G/K stars less
luminous in the former. 
\citet{jeffries97} suspected that the rotational history of the F stars is different owing to different convective turnover times
given NGC 2516's lower metallicity. This explanation, however, seems to be ruled out by recent spectroscopy that indicates
solar metallicity for this cluster. Slightly different rotation period distributions may be present for G and K stars perhaps due to  
a small age difference (see \citealt{damiani03} and references therein). A metallicity-related effect on X-ray radiation has 
also been discussed for  the metal-rich cluster Blanco 1 that reveals an X-ray excess for M dwarfs \citep{micela99b, pillitteri03}.

\subsubsection{Toward older clusters}

An extension of cluster studies toward higher ages suffers both from the small number of
clusters that are still bound, thus implying larger typical distances, and from the low X-ray luminosities of old 
main-sequence stars. Only close binaries that are kept in  rapid rotation by tidal forces, or giants have typically been 
detected in such clusters, with no less interesting results (e.g., \citealt{belloni96}, \citealt{belloni97}, and
\citealt{belloni98b} for binaries in NGC~752, NGC~6940, and M~67, respectively; see Table~\ref{clustertab}).

\subsubsection{Toward younger clusters}

Clusters significantly younger than $\alpha$ Per connect to
 pre-main-sequence (PMS) evolution  and the phase of circumstellar-disk dispersal. For example, \citet{jeffries98b} found
the X-ray luminosities of all G and K stars in the young (14~Myr; $\approx 35$~Myr according to other sources) cluster NGC~2547
to be a factor of two below the saturation 
limit of $L_X/L_{\rm bol} \approx 10^{-3}$,
suggesting that all stars have rotation  periods $> 3$~d. Possibly, their circumstellar
disks have only recently, at an unusually late point in time, been dispersed and the stars have not yet spun up from
the disk-controlled  slow rotation. The preceding evolution of disk environments is thus obviously of pivotal
importance for cluster development, as we will discuss in the next section.

\section{X-ray coronae and star formation}\label{starformation}

Modern theory of star formation together with results from comprehensive observing
programs have converged to a picture in which a forming low-mass star evolves through various stages
with progressive clearing of a contracting circumstellar envelope. In its ``class 0''
stage (according to the mm/infrared classification scheme), the majority of the future
mass of the star still resides in the contracting molecular envelope. ``Class I'' protostars
have essentially accreted their final mass while still being deeply embedded in an
envelope and surrounded by a thick circumstellar disk. Jets and outflows may be driven
by these optically invisible ``infrared stars''. Once the envelope is dispersed, the
stars  enter their ``classical T Tauri'' stage (CTTS, usually belonging to class II) with excess H$\alpha$
line emission if they are still surrounded by a massive   
circumstellar disk; the latter results in an infrared excess. ``Weak-lined
T Tauri stars'' (also ``naked T Tauri stars'', \citealt{walter86}; usually with class III characteristics) have lost most of 
their disk and are dominated by photospheric light \citep{walter88}. 

Consequently, X-ray emission from
these latter stars has unequivocally been attributed to solar-like coronal activity \citep{feigelson81a, feigelson81b,
feigelson89, walter84a, walter88}, an assertion that is less clear for earlier PMS classes. 
Arguments in favor of solar-like coronal activity in {\it all} TTS include i) typical electron temperatures of 
order $10^7$~K that require magnetic confinement, ii) a ``solar-like'' correlation with chromospheric
emission, iii) the presence of flares, and iv) rotation-activity relations that are similar to those  in more evolved
active stars (e.g., \citealt{feigelson81a, walter84a, walter88, bouvier90, damiani95b}). The solar analogy may,
on the other hand, not hold for the emission excess of optical lines in CTTS; this excess is uncorrelated with
X-rays \citep{bouvier90} but that seems to relate to the accretion process.

Because this review is primarily concerned with coronal X-rays that are - in the widest sense - solar-like,
the discussions in the following sections are confined to low-mass stars in nearby star-forming regions. For a broader
view of X-rays in the star-formation process, I refer the reader to the review by \citet{feigelson99} and references therein.

\subsection{T Tauri stars}\label{ttau}

\subsubsection{Overview}
Because most of the low-mass CTTS and WTTS are fully convective, there has
been a special interest in their X-ray behavior and in  their rotation-activity relations. 
Quite some debate has unfolded on whether CTTS and WTTS belong to the same parent population if
their X-ray luminosities are compared. CTTS have been found to be, on average, less luminous than WTTS in particular in Taurus
\citep{strom94, damiani95a, neuhauser95a, stelzer01}, but other authors have found the two samples to be 
indistinguishable in various other star-forming regions (\citealt{feigelson89, feigelson93, strom90, casanova95, 
lawson96, grosso00, flaccomio00, preibisch01, preibisch02, feigelson02a, getman02}, but see contradicting result 
for Orion by \citealt{flaccomio03b}). 

It is plausible that WTTS are stronger X-ray sources than CTTS because the latters' coronae could be absorbed by an overlying wind
(\citealt{walter81c}, see also \citealt{stassun04}), their coronal activity could  somehow be inhibited by the process of mass 
accretion onto the stellar surface \citep{damiani95b}, or because WTTS produce more efficient dynamos given their
generally higher measured rotation rates \citep{neuhauser95a}, 
although most PMS are in the saturation regime \citep{flaccomio03b}. 
However, there are also a number of arguments against any intrinsic X-ray difference between WTTS and CTTS.  
CTTS are usually identified optically, while WTTS are inconspicuous at those wavelengths
but are typically selected from X-ray surveys \citep{feigelson87} where many of them stand out given their 
rapid rotation, hence their bias toward strong X-rays \citep{feigelson89, preibisch96}. Several authors \citep{gagne94, damiani95a,
preibisch01, preibisch02, getman02, feigelson02a, feigelson03} also tested X-ray activity against infrared excess, but no 
distinction was found, suggesting that the observed X-ray emission does not directly relate to the presence 
of massive disks.

\begin{figure} 
\centerline{\resizebox{0.85\textwidth}{!}{\includegraphics{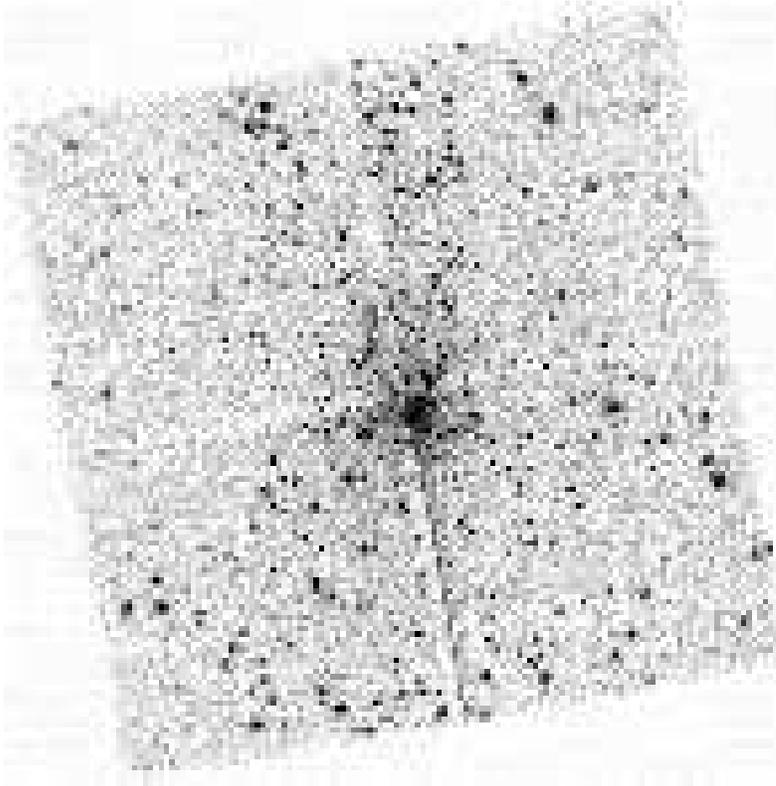}}}
\caption{The Orion Nebula Cluster as seen in X-rays by {\it Chandra}. North is up \citep{feigelson02a}.}\label{ONC}
\end{figure}

This issue has seen some, albeit still not full, clarification with recent work by \citet{flaccomio03a, flaccomio03b, flaccomio03d}
who studied various ranges of stellar mass. It probably does matter 
whether stars are distinguished by an indicator for the {\it presence} of a disk (such as their IR classification) 
or by an indicator of active {\it accretion} (e.g., H$\alpha$ emission). There is no one-to-one correspondence between these indicators (see also 
\citealt{preibisch02}). The debate remains open; Flaccomio et al.'s work  seems to suggest that either indicator 
points at CTTS being less X-ray luminous than WTTS in a given mass range, at least for the considered star forming clusters. 
\citet{stassun04} argued, from a reconsideration of the Orion X-ray samples, that the suppression of X-ray luminosity in 
a subsample of stars, accompanied by increased hardness, is in fact only apparent and is due to the increased attenuation 
of softer  photons by magnetospheric accretion columns in actively accreting stars.

\subsubsection{X-ray luminosity and age}
 
\begin{figure} 
\centerline{\resizebox{0.85\textwidth}{!}{\rotatebox{270}{\includegraphics{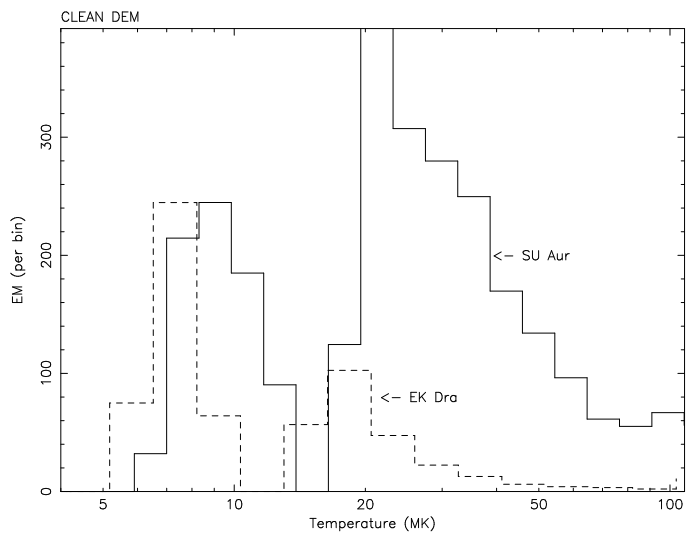}}}}
\caption{Comparison of EMDs of the CTTS SU Aur (solid histogram) and the ZAMS solar
analog EK Dra (dashed histogram; figure courtesy of S. Skinner, after \citealt{skinner98}).}\label{suaur}
\end{figure}

Spectroscopic evidence and EM distributions in TTS point to an analogy to young
solar analogs. It appears that both the EM and the hot temperatures are progressively
more enhanced as one moves from the ZAMS into the pre-main sequence regime of CTTS 
(Fig.~\ref{suaur}, \citealt{skinner98}). 
The overall X-ray levels of pre-main sequence TTS also  fit into the general
picture of declining X-ray activity with increasing age (\citealt{walter88, feigelson89, feigelson93, gagne94, gagne95b, 
damiani95a, stelzer01}, Figs.~\ref{clusterage}, \ref{lumfunction}), although 
this effect is not directly - or not only - coupled with rotation. While WTTS may spin up toward the
main sequence and reach X-ray saturation, they at the same time contract and decrease in 
$L_{\rm bol}$; the X-ray luminosity  then peaks around an age of 1~Myr and subsequently  slowly decays \citep{neuhauser95a, damiani95b, feigelson03,
flaccomio03c} whereas $L_X/L_{\rm bol}$ remains saturated for all TTS during their descent along the Hayashi track 
\citep{flaccomio03b, flaccomio03a}.
It is important to mention that CTTS and WTTS properties do not provide reliable age indicators per se. CTTS 
and WTTS in a star-forming region may have the same age while disk/envelope dispersal histories were different,
although WTTS do tend to dominate the final pre-main sequence episodes
(see, e.g., \citealt{walter88, feigelson93, lawson96, alcala97, stelzer01}).

\subsubsection{X-ray luminosity, saturation, and rotation} 

Rotation may therefore be a more pivotal parameter.
In this context, activity-rotation relationships are
particularly interesting because the rotation history is strongly coupled with the
presence of accretion disks, probably by magnetic coupling. CTTS generally rotate
slowly, with rotation periods typically of $P \approx 4-8$~days; only once the stars have lost their massive
inner accretion disks will the star spin up to rotation periods of typically one to a few days \citep{bouvier93}.
Many T Tau stars, in particular WTTS, are therefore in the saturation regime  
\citep{bouvier90, strom90, strom94, gagne94, gagne95b, casanova95, flaccomio00}. While often 
quoted as supportive of the same type of magnetic activity as in normal stars, this result is, at hindsight,
rather surprising because PMS have no significant, stable nuclear energy source.
The saturation level therefore does not seem to  relate to the total nuclear energy production.

A normal rotation dependence of $L_X$ can be found as well, in particular in the Taurus sample.
In this case, excellent agreement is found with the behavior of more evolved stars, 
again suggesting that essentially the same type of magnetic dynamo and coronal activity are
at work \citep{bouvier90, damiani95b, neuhauser95a, stelzer01}. However, in other star-forming regions, such as the Orion Nebula 
Cluster, $L_X$ appears to be independent of $P_{\rm rot}$ up to quite long periods of 30~d, the stars
essentially all residing in a saturated regime despite considerable scatter in $L_X/L_{\rm bol}$. 

\citet{feigelson03} discussed these results
in terms of various dynamo theories. On the other hand, \citet{flaccomio03c} have studied rotation and convection
for various PMS age ranges. Convection zone parameters
indicate that most PMS stars should indeed be in the saturation regime, as observed, for example,
in the Orion Nebula Cluster. The interesting point is, 
however, that the very youngest stars show a suppressed ``saturation''  level, by as much as an order of magnitude
(e.g., $L_X/L_{\rm bol} \approx 10^{-4}$ for Orion and $\rho$ Oph, \citealt{feigelson03, skinner03, grosso00}). Whereas \citet{feigelson03} 
hypothesized that a less efficient distributed dynamo is in effect, \citet{flaccomio03c}
speculated that the disk is somehow inhibiting strong coronal activity during the first few Myr, which 
then led them to find a characteristic (inner-) disk dispersal time of 1--2~Myr.

\subsubsection{The widely dispersed ``field WTTS'' samples}
 
The easy identification of WTTS in X-ray surveys has led to a rather controversial issue related to
the dispersal of star-formation clusters and TTS evolution. In general, the location of 
TTS relative to the cloud cores  provides interesting insight into the star formation history in and around 
a molecular cloud. \citet{feigelson87}, \citet{walter88}, \citet{feigelson89}, and \citet{strom90} were first to point to a rather 
large overpopulation of WTTS in Taurus by factors of 2 to 10 compared with CTTS, probably constituting the long-sought post-T Tau 
population evolving toward the ZAMS. An estimated WTTS/CTTS ratio of order 10 is then consistent with a disk dispersal time of
a few Myr \citep{neuhauser95c, feigelson96}. If star formation is cut off in time, the ratios
may even be higher; \citet{walter94} found a ratio of 40 in Upper Sco-Cen where star formation 
has ceased about 2~Myr ago.

Nevertheless, it came as quite a surprise when further associated WTTS were discovered tens of degrees away from the Taurus, 
Chamaeleon, Orion, and Lupus molecular clouds \citep{walter88, neuhauser95b, sterzik95, wichmann96, alcala95, alcala97, krautter97}. 
They show no spatial correlation with the present molecular clouds, which suggests that they have either drifted 
away from their place of formation \citep{neuhauser95b}, or that they are the
products of star formation in now dispersed molecular clouds \citep{walter88} or local high-velocity, turbulent cloudlets 
\citep{feigelson96}. The latter explanation is attractive because some of the samples are as young
as $10^6$~yr despite their large distance from the (present-day) ``parent cloud'' \citep{alcala97}. 

However, the evolutionary stage of these samples has been the subject of considerable debate. On the one
hand, \citet{neuhauser97b} estimated their ages 
at typically $\le 30$~Myr, including a moderately-sized population of widely distributed ZAMS G stars.
On the other hand,  \citet{micela93}, \citet{briceno97}, and \citet{briceno99} used  star-count models, statistics of X-ray detected
M dwarfs, and Li measurements to argue
that the dispersed WTTS do not relate to the ``missing post-T  Tau stars'' but mostly have a larger, 
near-ZAMS age of  $20-100$~Myr. This would agree with the near-ZAMS status of X-ray selected field stars reported by
\citet{micela97b} that were, however, not drawn from the sample in question.
Such an age population would also solve the problem of the ``missing'' ZAMS population that is  expected from standard star
formation models. A likely explanation for the potential misclassification of young field stars as
WTTS may relate to the use of low-resolution spectroscopy for Li measurements \citep{briceno97, favata97d, micela97b, martin99}. 
If this revised  age classification is correct, then, however, the problem with the ``missing post-T Tau'' stars at younger 
ages of $2-10$~Myr reopens. 

\citet{covino97}, \citet{alcala98}, and \citet{alcala00}  reconsidered and discussed this 
issue for the  Chamaeleon and the Orion regions, arriving at intermediate conclusions.
The distributed ``WTTS'' population appears to consist of at least 50\% genuinely young ($<5$~Myr) WTTS somewhat concentrated 
toward the molecular clouds, plus a smaller, widely distributed population of unrelated older, possibly near-ZAMS field stars.
There is little evidence for a genuine post-T Tau population, indicating that the star formation process
in a given cloud occurs on time scales $< 10$~Myr and is not continuous, as assumed in some of the population models
(see also discussion in \citealt{favata97d} and the counter-arguments in \citealt{feigelson96}). The entire issue
remains under debate; see, for example, \citet{alcala00} and references therein for a recent assessment
discussing Li abundance, the location of the stars on the HRD, and their spatial distribution.

An argument supporting an intermediate age of a few $10^7$~yrs for these stars at the interface 
between PMS and ZAMS may be
their large-scale spatial correlation with the young Gould Belt structure which also contains several 
star formation regions in the solar vicinity \citep{krautter97, wichmann97}. The ages of many foreground Orion
``WTTS'' are indeed compatible with this hypothesis \citep{alcala98} - see Sect.~\ref{largescale}.

\subsubsection{Flares}

X-ray flares  have given clear evidence for
underlying magnetic fields not only in WTTS \citep{walter84a} but also in CTTS \citep{feigelson81a, 
walter84a}. They strongly support the presence of solar-like closed coronae in the broadest sense. From an
energetics point of view, they play a very important role: As much as half of
the emitted X-ray energy, if not more, may be due to  strong flares \citep{montmerle83}, and many
TTS are nearly continuously variable probably also owing to flares \citep{mamajek00, feigelson02a, preibisch02, skinner03}.
Examples with extreme luminosities and temperatures up to 100~MK have been reported 
(\citealt{feigelson81a, montmerle83, preibisch95, skinner97, tsuboi98, tsuboi00, imanishi02}, see Sect.~\ref{flares}).
\citet{stelzer00a} systematically studied flares in TTS, comparing them with flares
in the Pleiades and the Hyades clusters. They found that TTS flares tend to reach higher luminosities
and temperatures than their main-sequence equivalents, and the flare rate above a given limit
is also higher. The most extreme flares are found on CTTS, a possible hint at star-disk magnetic interactions 
during flares although this is at variance with suggestions made by \citet{montmerle00} (Sect.~\ref{protoflares}). 
The high activity level found for  CTTS is echoed by the work of \citet{skinner03} who noted a large variability fraction among
the hottest and most  absorbed sources in NGC~2024.

\subsubsection{The circumstellar environment}

The high-energy emission related to the
luminous X-rays  may have considerable impact on the circumstellar environment
and on the entire surrounding molecular clouds because it changes the ionization balance and induces
chemical reactions in molecular material \citep{montmerle83, casanova95, 
kastner99}. Recently, \citet{feigelson02b} speculated that the elevated rate of large flares in young solar analogs
in the Orion Nebula Cluster may be applicable to the young Sun. The increased activity may  possibly explain
the production of chondrules by flash-melting and isotopic anomalies in meteorites
by high proton fluxes.

\subsubsection{Accretion-driven X-ray emission?}

An entirely different model for X-ray production in CTTS  was put forward by 
\citet{kastner02} based on observations of the CTTS TW Hya. This star has been conspicuous by producing  
luminous radiation with $L_X = 2\times 10^{30}$~erg~s$^{-1}$ which is, however, very
soft, with a best-fit temperature of $T \approx 3$~MK \citep{hoff98, kastner99}. While unusual for active coronal sources, the temperature
is compatible with shock-induced X-ray emission at the base of magnetically funneled accretion flows. 
Explicit density measurements using He-like triplets of O\,{\sc vii} and Ne\,{\sc ix} indeed suggest 
very high densities of
$n_e = 10^{12}-10^{13}$~cm$^{-3}$, densities that are not seen at these temperatures in any coronal source. \citet{stelzer04}
have supported this view and further argued that the low C and Fe abundances relate to their being grain-forming elements,
that is, C and Fe have condensed out in the circumstellar disk or cloud. In the light of 
similarly extreme abundance anomalies in other stars (Sect.~\ref{composition}), this model necessarily remains 
tentative at this time. 

The role of accretion columns for the X-ray production has further been discussed by \citet{stassun04}. 
They found that actively accreting stars in the Orion Nebula Cluster on average show less luminous but also
harder X-ray emission than non-accreting stars, suggesting that accretion columns may attenuate the X-rays
while the intrinsic X-ray emission is similar in both samples, i.e., not related to accretion but to rotation.
A time-dependent effect of this sort has possibly been seen in the CTTS XZ Tau \citep{favata03}. To 
what extent TW Hya is an exception presently remains  open.

\begin{figure} 
\centerline{
\hbox{
\resizebox{0.53\textwidth}{!}{\includegraphics{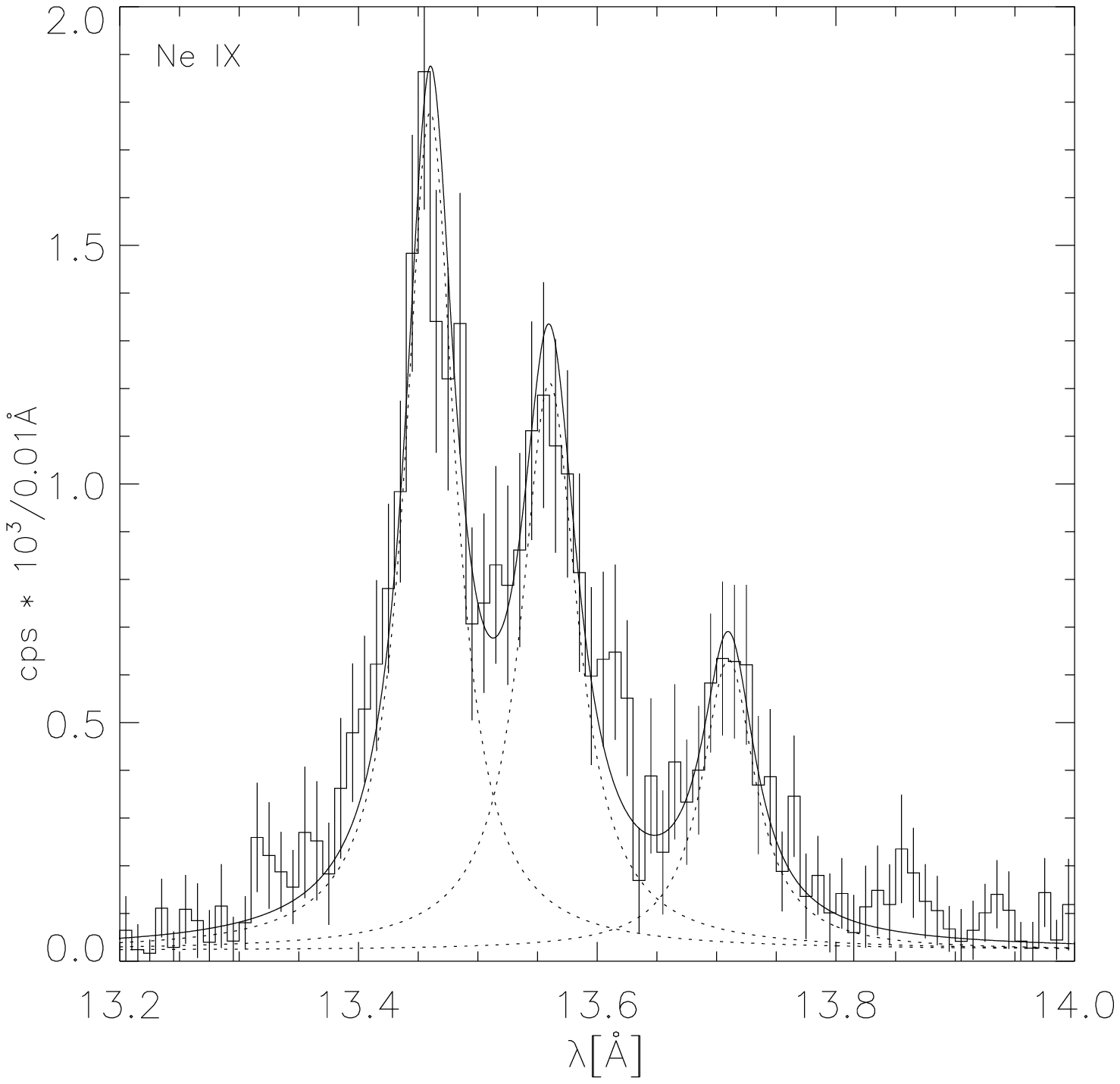}}
\hskip -0.4truecm\resizebox{0.53\textwidth}{!}{\includegraphics{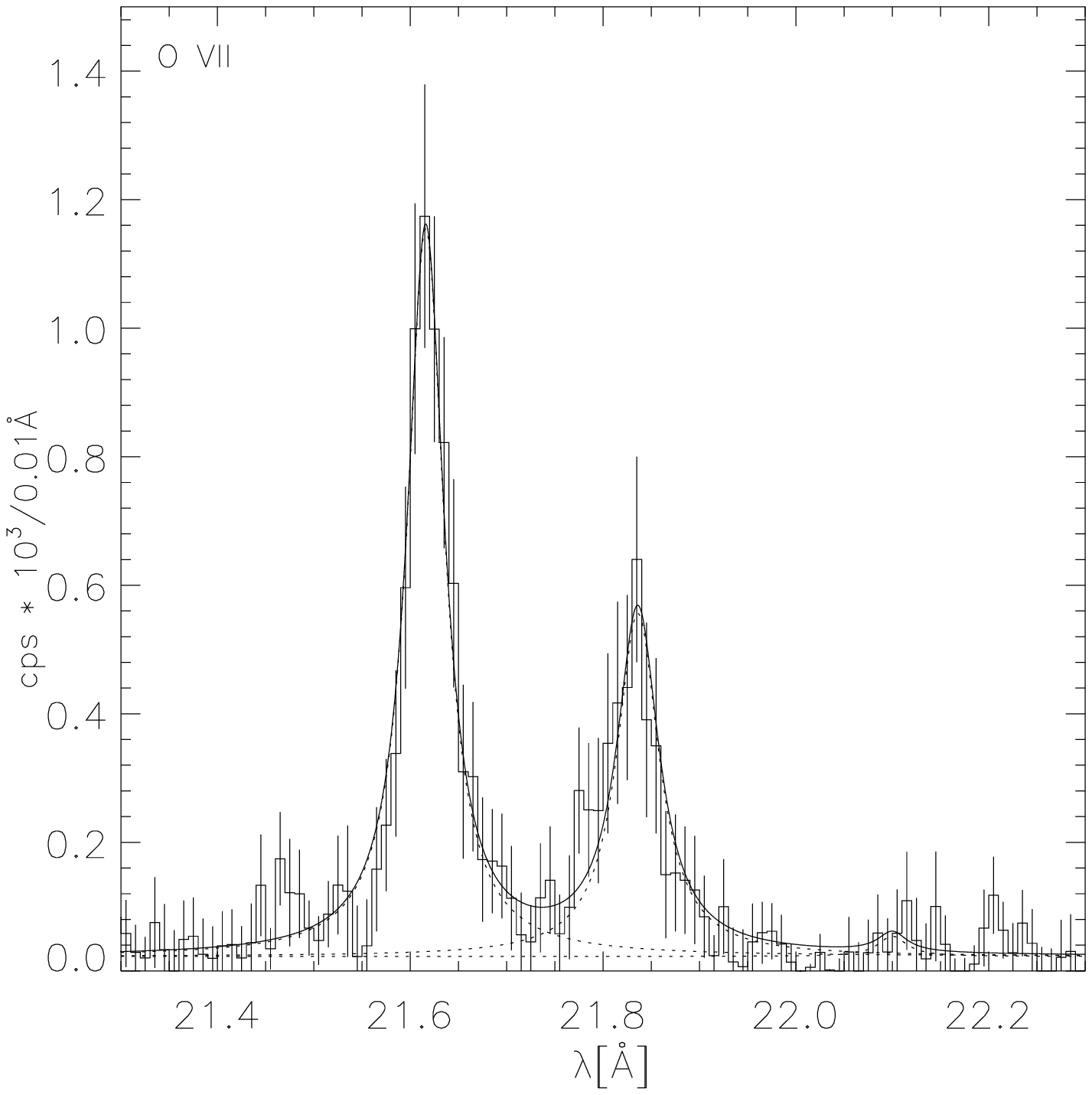}}
}
}
\caption{He-like Ne\,{\sc ix}  (left) and O\,{\sc vii} triplets (right) of the nearby CTTS TW Hya observed
by {\it XMM-Newton}. The forbidden lines at 13.7~\AA\ and 22.1~\AA, respectively, are largely suppressed, indicating high electron densities 
(figures courtesy of B. Stelzer, after \citealt{stelzer04}).}\label{twhya}
\end{figure}

\subsection{Protostars}

\subsubsection{Overview}

X-rays from embedded (``infrared'') protostars are strongly attenuated and require access to
relatively hard photons. The protostellar survey by \citet{carkner98} listed only about a dozen X-ray 
class I protostars detected at that time, with luminosities of order $10^{30}$~erg~s$^{-1}$, and no class 0 
objects. First tentative evidence for heavily absorbed hard emission was reported for $\rho$ Oph 
\citep{koyama94, casanova95}, followed by several unambiguous detections of very hard 
class I sources with temperatures of up to 7~keV in the R CrA cloud \citep{koyama96, neuhauser97a}, $\rho$ Oph \citep{kamata97, grosso01}, 
Orion \citep{ozawa99}, Taurus (\citealt{skinner97}, see also \citealt{carkner98}), NGC~1333 \citep{preibisch97b, preibisch98b}, 
and Serpens \citep{preibisch98a}. Their luminosities apparently correspond to classical 
``saturation'' ($L_X/L_{\rm bol} \approx 10^{-3}$, \citealt{ozawa99}), but extreme levels of $L_X \approx (6-18)\times
10^{32}$~erg~s$^{-1}$ have been reported \citep{preibisch98a}.

Class I protostars have now become accessible in larger numbers thanks to {\it Chandra}'s and {\it XMM-Newton}'s
hard-band sensitivity \citep{imanishi01a, preibisch01, preibisch02, preibisch03b, getman02}. Their measured characteristic 
temperatures are very high, of order 20--40~MK \citep{tsujimoto02, imanishi01a}. Some of these values may, however, be biased 
by strongly absorbed (``missing'') softer components in particular in spectra with limited signal-to-noise ratios. It is correspondingly 
difficult to characterize the $L_X$ values in traditional soft X-ray bands for comparison with other stellar sources.

\subsubsection{Flares and magnetic fields}\label{protoflares}

\begin{figure} 
\centerline{\resizebox{0.70\textwidth}{!}{\includegraphics{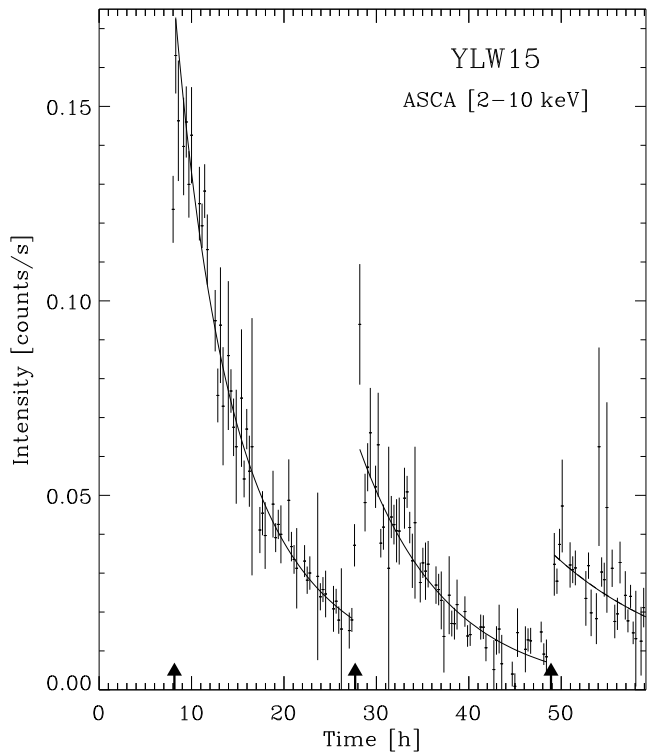}}}
\caption{Triple X-ray flare on the protostar YLW~15, observed with {\it ASCA}  
(figure courtesy of N. Grosso, after \citealt{montmerle00}).}\label{tripleflare}
\end{figure}

 Direct evidence for magnetic processes and perhaps truly coronal emission came with the
detection of flares.  Many of these events are exceedingly large, with total soft X-ray energies of up to
$\approx 10^{37}$~erg (Table~\ref{flaretable}; \citealt{koyama96, kamata97, grosso97, ozawa99, imanishi01a}, 
and \citealt{preibisch03a}).  \citet{imanishi03} conducted a systematic comparative analysis of flares from class I--III 
stars. They interpreted rise times, decay times and temperatures within the framework of MHD models (\citealt{shibata02},
Sect.~\ref{mhd}, Fig.~\ref{shibata}). 
The magnetic fields tend to become stronger toward the typically hotter class I flaring sources.
Such flares realistically require large volumes, in fact to an extent that star-disk magnetic fields become 
a possibility for the flaring region (\citealt{grosso97} for YLW~15 in $\rho$ Oph). 

\begin{figure} 
\centerline{\resizebox{0.85\textwidth}{!}{\includegraphics{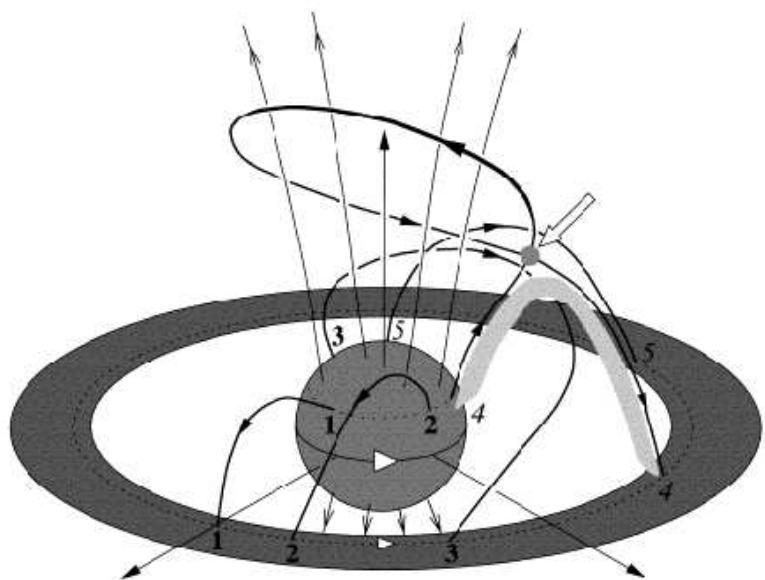}}}
\caption{Sketch illustrating a star-disk magnetic-field model in which field lines wind up and reconnect 
 (line labeled `5') because the star rotates faster than the inner edge of the disk
 (figure courtesy of T. Montmerle, after \citealt{montmerle00}).}\label{disk}
\end{figure}

\citet{tsuboi00} reported on
quasi-periodic flare events in YLW 15 that occurred three times in sequence, separated by  about 1 day 
(Fig.~\ref{tripleflare}). If magnetic  fields indeed connect the star with the inner border of
the circumstellar disk, then they may periodically ignite
flares each time the field lines have become sufficiently stretched due to the difference in rotation rates of
the star and the disk (Fig.~\ref{disk}). This scenario was computed  by \citet{hayashi96}; their MHD simulations
showed extensive episodic heating and large
plasmoids detaching from the star-disk magnetic fields. \citet{montmerle00} expanded this view
qualitatively to suggest that star-disk magnetic flares should be common in protostars through winding-up  
star-disk magnetic fields because stellar and inner-disk rotation rates have
not synchronized at that age. But because the same fields will eventually brake the star to disk-synchronized
rotation in TTS, the large star-disk flares should then cease to occur, and the X-ray activity becomes
related exclusively to the stellar corona.

\subsubsection{The stellar environment}

\begin{figure} 
\centerline{\resizebox{0.85\textwidth}{!}{\includegraphics{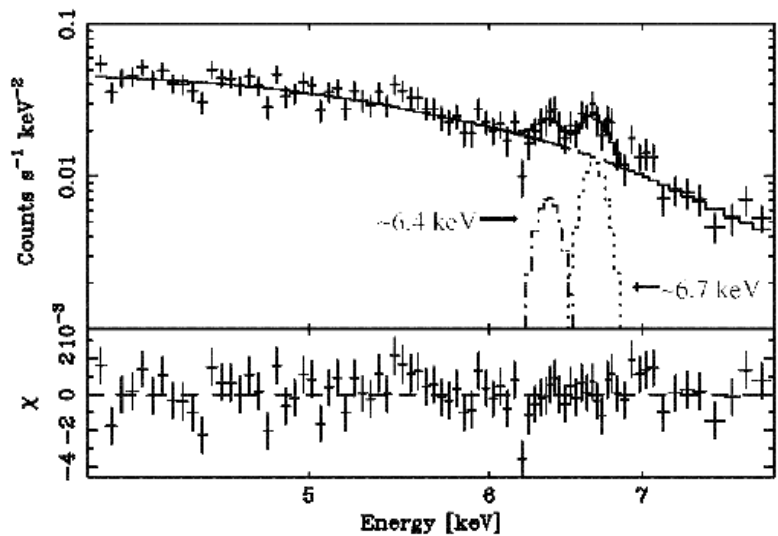}}}
\caption{{\it Chandra} CCD spectrum of the protostar YLW 16A observed during  a large flare. The  spectral
line feature at 6.7~keV  refers to the Fe\,{xxv} complex, whereas the line at  6.4~keV
is a fluorescent line of cold Fe (figure courtesy of K. Imanishi, after \citealt{imanishi01a}).}\label{fluorescence}
\end{figure}

X-ray emission in protostars and strong flaring in particular may have far-reaching consequences
for the evolution of the stars themselves, but also for the environment in which planets
form. First, X-rays efficiently ionize the molecular environment, which may lead to modifications
in the accretion rate, for example via the magnetic Balbus-Hawley instability (see \citealt{montmerle00}
for a discussion). While X-ray flares may thus induce accretion events, the latter will tend to quench
strong magnetic activity, in turn decreasing disk ionization. The effect of flare irradiation of 
disks may indeed have been seen explicitly: \citet{imanishi01a} detected, during
a giant flare in YLW 16A ($\rho$ Oph), strong Fe fluorescence line at 6.4 keV that is possibly induced
by X-ray irradiation of a circumstellar disk
(Fig.~\ref{fluorescence}; see also \citealt{koyama96}). Second, strong, frequent flares and disk ionization may also 
be of fundamental importance  for the generation of jets \citep{hayashi96}, spallation reactions in solids in the 
circumstellar disk, and the  formation of planets, a subject beyond the scope of the present review (see, e.g., 
\citealt{feigelson99, feigelson02b}, and the extensive review of this subject by
\citealt{glassgold00}).

\subsubsection{``Class 0'' objects}

Knowing that stars are already extremely active at the deeply embedded class I stage, the interest
in the magnetic behavior of class 0 objects is obvious, but the strong photoelectric absorption makes
detection experiments extremely challenging. \citet{tsuboi01} reported a possible detection of a class 0 object
in Orion, with properties surprisingly similar to more evolved class I--III objects, such as
an X-ray luminosity of $L_X \approx 
2\times 10^{30}$~erg~s$^{-1}$, but additional detections and confirmations are badly needed. Specific 
searches have, so far, given a null result (T. Montmerle 2004, private communication).

\subsection{Young brown dwarfs}\label{youngBD}

\begin{figure}
\centerline{
\vbox{\centerline{\resizebox{0.75\textwidth}{!}{\includegraphics{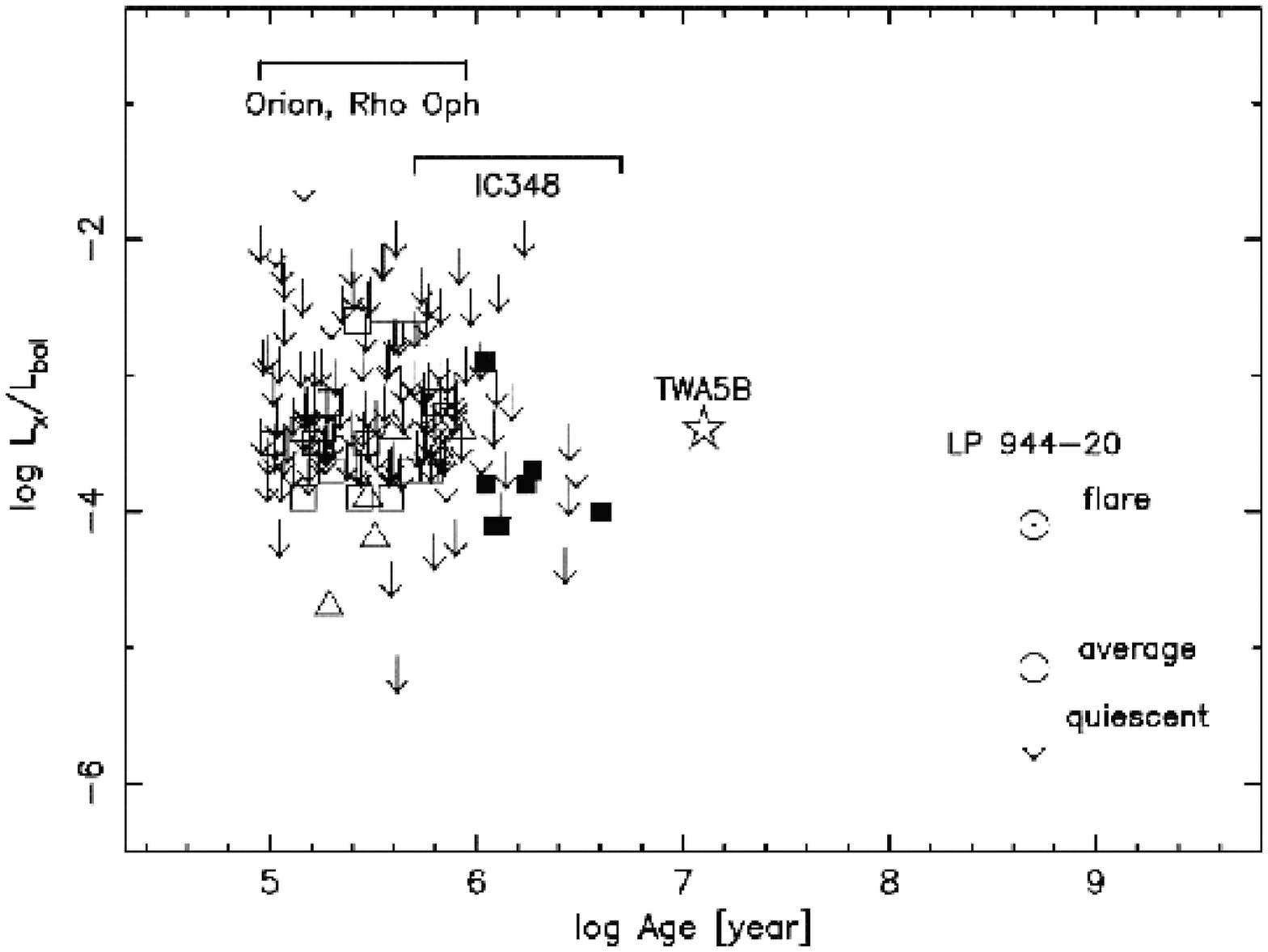}}} 
      \centerline{\resizebox{0.75\textwidth}{!}{\includegraphics{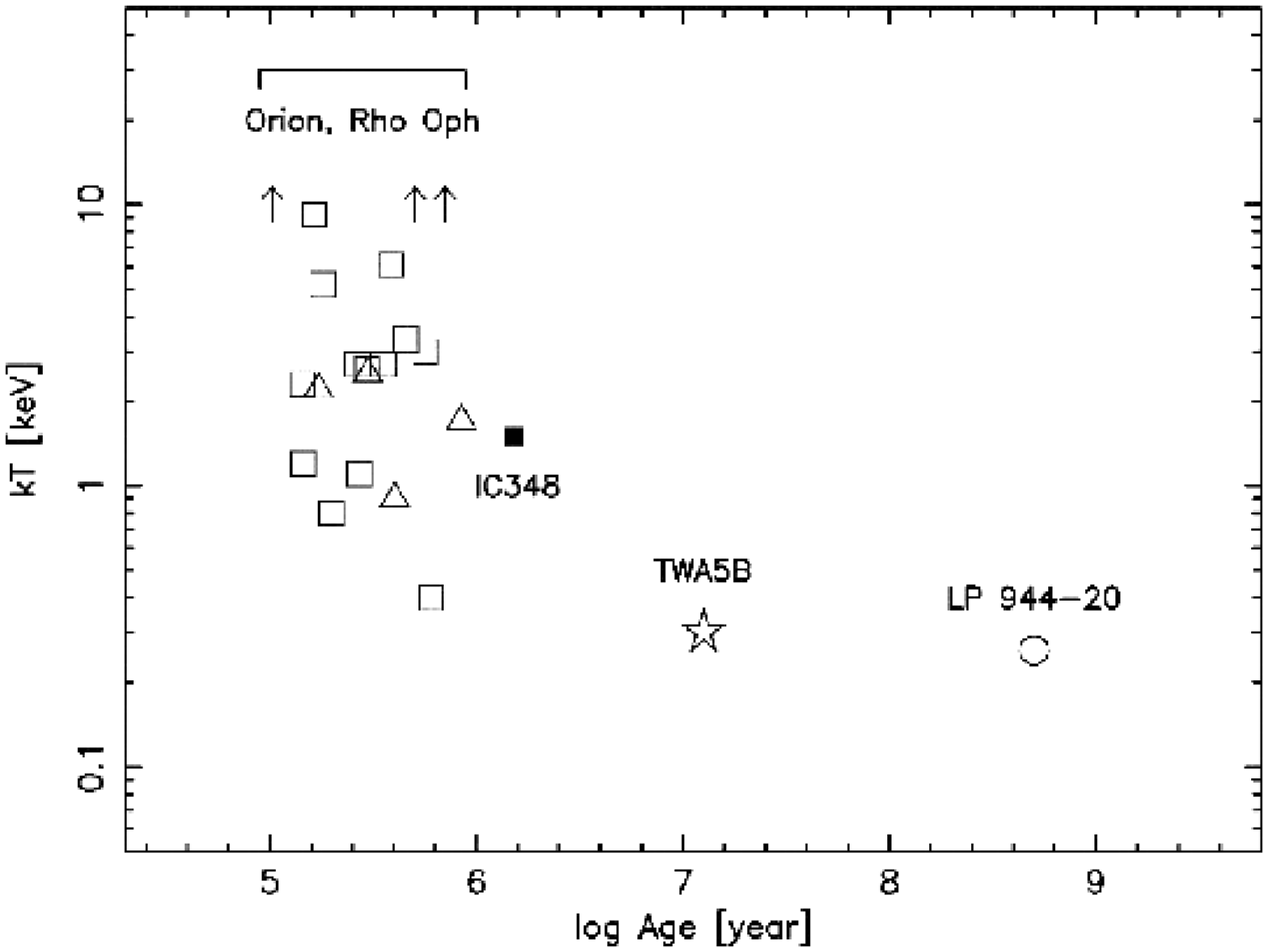}}}
}
}
\caption{X-ray characteristics of brown dwarfs as a function of age. {\it Top panel:} log\,$L_X/L_{\rm bol}$;
{\it bottom panel:} characteristic electron temperature. Key to the symbols:
open squares: Orion Nebula Cluster; triangles: $\rho$ Oph; filled squares: IC 348.
All observations were obtained by {\it Chandra} (figure courtesy of Y. Tsuboi, after \citealt{tsuboi03}).}\label{BDplot}
\end{figure}

Brown dwarfs (BD) were first detected as faint X-ray sources  in star forming regions such as Cha~I, Taurus,
and  $\rho$ Oph \citep{neuhauser98, neuhauser99, comeron00}. They correspond to spectral types beyond M6 at this age.
The list of X-ray detections of BDs and candidates is now rapidly growing thanks to sensitive observations with 
{\it Chandra} and {\it XMM-Newton}. Considerable numbers have been reported from the Orion Nebula Cluster 
(\citealt{garmire00, feigelson02a, flaccomio03b}), $\rho$ Oph 
\citep{neuhauser99, imanishi01b}, IC 348 \citep{preibisch01, preibisch02}, Taurus (\citealt{mokler02}; \citealt{neuhauser99}
using {\it ROSAT}), and  the $\sigma$ Ori cluster \citep{mokler02}. The latter authors and \citet{tsuboi03}
provided a summary of all measurements and put 
the X-ray properties in a wider context. Although alternative origins of the X-rays are possible 
such as primordial magnetic fields or star-disk fields (see \citealt{comeron00}), all X-ray properties suggest thermal
emission from a solar-like corona. Specifically, $L_X$ is typically found to follow the saturation law ($L_X/L_{\rm bol} 
\approx 10^{-4} - 10^{-3}$) of more massive stars, implying a general decrease toward later spectral types, without any evident break. 
In absolute terms, $L_X$ reaches up to a few times $10^{28}$~erg~s$^{-1}$. Young BDs reveal coronal temperatures
typically exceeding 1~keV, similar to TTS. A number of these objects have also been found to flare \citep{imanishi01b, 
feigelson02a}. It is only beyond $10^7$~yr that the X-ray emission from BDs decays (see also Sect.~\ref{BD}) - 
again seemingly  similar to main-sequence stars (Fig.~\ref{BDplot}).

No thorough study of relations between accretion disk signatures and X-rays is available at this time.
For $\rho$ Oph, \citet{imanishi01b} found no relation between $L_X$  and $K$ band luminosity excess. However,
\citet{tsuboi03} suggested that the lack of X-ray detections among stars with very large H$\alpha$ equivalent widths 
indicates, as in CTTS, increased accretion at the cost of strong X-ray emission.

The similar behavior of young BDs and TTS is perhaps not entirely surprising, whatever the internal dynamo mechanism is. 
At this young age, both classes of stars are descending the Hayashi track, are fully convective, have surface temperatures
like cool main-sequence dwarfs, and derive most of their energy from gravitational contraction, while they lack a significant 
central nuclear energy source. Young BDs are not yet aware of their fate to fail as stars.

\section{Young populations in the solar neighborhood}\label{largescale}

Stellar coronae have also proven practical in investigations of (nearby) galactic structure.
Large statistical samples of stellar X-ray sources are interesting 
to compare with galactic population models, for example to uncover ``excess populations'' of 
active stars or to assess the stellar contribution to the apparently diffuse galactic 
soft X-ray background. Sophisticated models take into account the statistical galactic 
distribution of stellar populations (in terms of scale heights, space densities, etc.) as a function 
of their color, evolutionary status, and binarity, together with effects of extinction 
\citep{favata92, guillout96}. I will only touch upon this field and mention some of the recent key results. 

When observations of large samples were compared with predictions from galactic population 
models, an excess of nearby ``yellow'' (F, G, K) stars  was found \citep{favata88}.
Later investigations interpreted them as predominantly young ($<1$~Gyr), near-ZAMS main-sequence stars \citep{sciortino95}. 
RS CVn binaries  may contribute as well, but their number is not sufficient 
to account for the observed excess \citep{favata95a, favata95b}. A young population is
also conspicuous in the {\it ROSAT} All-Sky Survey, providing evidence for a considerable population of
such stars in the solar neighborhood. Population studies have further been used to 
derive constraints on the stellar birthrate. \citet{micela93} found that the observed source counts 
are compatible with a nearly constant galactic stellar birthrate and exclude a rapid decline during 
the first few Gyr of the  galaxy's life. 

\citet{guillout98a} and \citet{guillout98b} undertook a large correlation study between the {\it ROSAT}
All-Sky Survey and the Tycho Catalog. They found rather compelling evidence for a concentration of luminous
and therefore probably young stars along a great circle on the sky that corresponds to the Gould Belt. 
This local galactic, disk-like structure  in the solar vicinity spans over a few hundred parsec 
and comprises a number of the prominent nearby star-forming regions. The detected population
of X-ray stars appears to correspond to the low-mass population of this structure.

\section{Long-term variability and stellar magnetic cycles}

Records of sunspot numbers back over several hundred years show a near-cyclic modulation that has turned out to
be a central challenge for dynamo theories. The magnetic activity period between two successive spot maxima of approximately
11 years expresses itself most beautifully in soft X-rays, more so than in any longer wavelength range 
(see Fig.~\ref{solarminmax}). The X-ray luminosity variation between maximum and minimum has been quoted variably 
as ranging between a factor of about ten \citep{aschwanden94, acton96} as measured by {\it Yohkoh} and {\it GOES}, to factors 
of 20--200 \citep{kreplin70} as measured with {\it SOLRAD}, although the spectral bandpasses clearly matter as well for a 
closer comparison. Converted to the {\it ROSAT} bandpass, \citet{hempelmann96} estimated variations by a factor of 10, whereas 
\citet{ayres96} found a somewhat more modest factor of 4 as extrapolated from far-UV data. 

The Mount Wilson HK project \citep{baliunas95} has collected a continuous data stream
of the chromospheric H\&K line flux diagnostic for many stars over several decades. This stupendous observing
project has now clearly demonstrated that many stars show magnetic activity cycles somewhat similar to the Sun's.  A subset
of stars appear to lack such cycles, however, and very active stars tend to exhibit an irregular rather than
a cyclic mode of variability \citep{hempelmann96}. 

The sensitive response of the coronal luminosity to changes in the surface magnetic field 
should make us believe that cycles are easily seen in X-ray active stars. Surprisingly, evidence is
still tentative at the time of  writing. Magnetic X-ray cycles have eluded detection because no appropriate program
has been carried out for sufficiently long periods; however, there may also be interesting physical reasons for 
a lack of detections. Two strategies toward detecting X-ray cycles have been followed, the first observing
young open clusters or field stars with various satellites to obtain statistical information on long-term variability,
and the second approach concentrating on dedicated long-term ``monitoring'' of select field stars.   

\subsection{Clusters and field star samples}

Many open clusters or field star samples were observed by {\it Einstein}, {\it EXOSAT}, and {\it ROSAT}, and they obtain new visits again
with {\it XMM-Newton} and {\it Chandra}. One of the principal results of comparative studies over time
scales of 10~years has in fact been a suspicious absence of strong long-term variability. Most of these observing
programs reported variations of no more than a factor of two for the large majority of stars. Such results apply to
samples of active field dMe stars \citep{pallavicini90a}, dM dwarfs \citep{marino00}, dF--dK stars \citep{marino02}, 
volume-limited samples of nearby stars \citep{schmitt95, fleming95}, old disk and halo stars \citep{micela97a}, 
RS CVn binaries \citep{dempsey93a}, T Tau stars \citep{sciortino98, gagne94, gagne95b, 
grosso00}  and, most notably, open clusters such as the Pleiades \citep{gagne95a, micela96, marino03b} 
and the Hyades \citep{stern94, stern95b}. Much of the observed variability is statistically consistent with shorter-term
flare-like fluctuations or variability due to slow changes in active regions \citep{ambruster87}.
In some cases, there is a modest excess of long-term over short-term variations (e.g., \citealt{gagne95a} for the Pleiades,
\citealt{marino02} for F-K field stars) but the evidence remains marginal. 

These investigations were put on a solid statistical footing by comparing short-term with long-term
variability of a large sample of active binaries \citep{kashyap99}. Again, most of the variability was
identified to occur on time scales $\la 2$~yr although there is marginal evidence for
excess variability on longer terms (see also \citealt{marino00}). The latter could then be 
attributed to cyclic variability although the ratio of maximum to minimum luminosity would be bounded by a factor 
of 4, much less than in the case of the Sun. Overall, such results may be taken as 
evidence for the operation of a distributed dynamo producing relatively unmodulated small-scale magnetic fields 
\citep{kashyap99, drake96}. If this is the case, then it would support a model
in which not only the latest M dwarfs, but also earlier-type magnetically active stars are prone 
to a turbulent dynamo \citep{weiss93}.

The long-term variability in stellar coronae was further quantified with a large {\it ROSAT} sample of observations by \citet{micela03}.
They modeled the $\approx 10$~yr X-ray light curve collected from {\it Yohkoh} full-disk observations (Fig.~\ref{solarminmax}) 
by transforming it to the {\it ROSAT} bandpass. Samples of stellar snapshot observations
can then be compared with the distribution of the solar amplitude variations
for a given time separation. In the solar case,  statistical variation within a factor of two is
due to short-term variability, while another factor of three to four is contributed
by the solar cycle. Turning to the stellar sample, the authors concluded that there is
some likelihood for {\it inactive} stars to have magnetic X-ray cycles somewhat similar to 
the Sun's, whereas the most active stars tend to be less variable on long time scales.

\citet{hempelmann96} attacked the problem by measuring surface flux $F_X$ for a number of
stars with known Ca H\&K activity cycles. They then compared the excess flux with 
predictions from a rotation-activity relation based on the known  Ca H\&K cycle phase, 
finding tentative indications that the X-ray flux varies in concert with
Ca along the cycle, but again, the significance is low.

\subsection{Case studies}

\begin{figure} 
\centerline{\resizebox{0.85\textwidth}{!}{\rotatebox{270}{\includegraphics{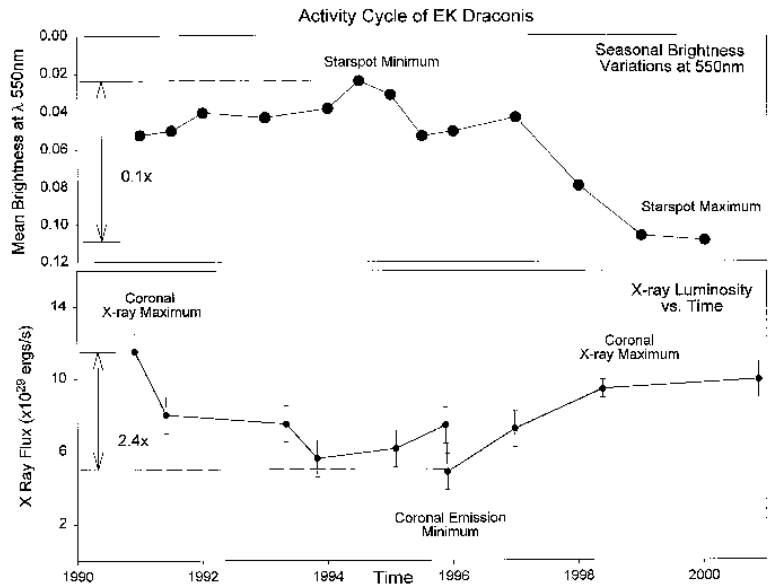}}}}
\caption{Illustration of a tentative identification of a magnetic cycle in the corona of EK Dra (G0~V)
during one full photospheric magnetic cycle period (about 10~yr). The upper panel gives brightness offset,
maximum brightness referring to smallest starspot coverage.  The lower panel shows X-ray luminosity
 measurements (figure courtesy of E. Guinan).}\label{cycle_ek}
\end{figure}

Obviously, dedicated long-term programs are in order to obtain more sensitive results on cyclic 
behavior. First {\it tentative} evidence from such a program was reported by \citet{dorren95}. The star under
scrutiny, EK Draconis, is a Pleiades-age solar analog near X-ray saturation and is therefore not the
type of star for which we expect regular cycles. The evidence for the latter is, however, clear and compelling,
with a cycle period of about ten years in optical photometry (measuring the spot coverage) and
partly in Ca\,{\sc ii} and Mg\,{\sc ii} lines \citep{dorren94, dorren95}. This cycle has now been followed 
over two full periods \citep{guedel03c}. The star has obtained regular coverage
with X-ray satellites for over ten years. Initial results were presented in \citet{dorren95}.
The attempt to detect an X-ray cycle is illustrated in Fig.~\ref{cycle_ek}. While there is 
a suggestive anti-correlation between X-ray flux and photospheric brightness where the star is
brightest at its activity minimum, the evidence must be considered as tentative. The snapshots
were partly obtained with different detectors, and the variation is rather modest, i.e., within a 
factor of two, illustrating the challenges of such observing programs.

A more recent report by \citet{hempelmann03} indicated a correlation between X-ray luminosity and 
the Ca H\&K S index for the two inactive stars 61 Cyg A and B. Both show chromospheric modulations
on time scales of about 10 years, one being regular and the other irregular, and the X-ray fluxes vary in concert
during the 4~1/2 years of coverage. A gradual X-ray modulation was also seen during a time span 2.5 years
in HD~81809, although there seems to be a phase shift by about 1 year with respect to the
Ca cycle \citep{favata04}. The evidence in these two
examples, while promising, should again be regarded
as preliminary given that only 30--50\% of the cycle has been covered, and the variations
in 61 Cyg are - so far - again within a factor of only $\approx$ 2.

\section{Outlook}

Stellar coronal  X-ray astronomy  enters its fourth decade now, with no lack of wonderful instrumentation
available to pursue its goals. Its - so far - short history started with feeble detections of exotic sources in the pioneering
seventies; the eighties consolidated our view of X-ray coronae and brought them into a perspective of 
stellar evolution within which we have placed our Sun. The nineties saw rapid progress with increased
sensitivities and routine access to spectroscopy, permitting the systematic study of numerous
physical processes. And finally, the present decade has brought high-resolution
X-ray spectroscopy and high-resolution X-ray imaging, paving the way to deep studies of coronal structure,
heating, and evolution. Where should we go from here?

Challenging questions wait to be answered. The following summary of important issues and recommendations
represents my current view:

{\it How are coronae heated?} The most fundamental question of our subject has started with the very
recognition of the presence of hot plasma around the Sun -- and it remains unanswered! The subject has
not been reviewed in this paper because it is a predominantly solar and plasma-theoretical issue. However,
stellar astronomy can shed more light on the problem by making use of what has been called the
``solar-stellar connection'': through observations of systematic trends over a wide parameter range, while we
assume that the basic physical processes remain the same in all stellar systems and the Sun. Stars offer access
to plasmas as hot as 100~MK. Coronae are observed across a range of stellar gravities, magnetic field strengths,
convection zone depths, and rotation periods. The systematic studies of stellar X-ray activity across the HRD 
have played a key role in the arguments for or against acoustic heating. High-resolution spectroscopy offers 
new possibilities by studying abundances and the form of DEMs, both being the result of heating processes 
occurring across the face of a star. The new satellites help us address these issues more systematically.

{\it How are coronae structured?} The new, refined observational facilities enhance our diagnostic capabilities
that will lead to conclusive insights into the structure of magnetically confined coronae, 
magnetic field generation and distribution, and eventually will help us to 
locate and identify the dynamo mechanism. It is still unclear whether in certain stars $\alpha\omega$ 
or $\alpha^2$ dynamos are in operation. Structure information may provide the necessary boundary
conditions as has been discussed in this review, for example in the context of giants. To understand the
global structure of a magnetic corona, however, we have to start recognizing {\it structure distribution}. 
Many spectroscopically determined observables such as ``density'', ``temperature'', ``abundance'', or ``opacity'' 
are the product of radiation from a plethora of coronal features that are, in fact, {\it distributed} in these 
parameters. Our X-ray spectra provide only a highly degenerate view of a true corona, with many possible 
realizations for one given observation. The use of painstakingly and accurately determined singular 
spectroscopic parameters may produce entirely misleading results in our modeling efforts; a 
value for a density or an element abundance derived from simple line-flux ratios may not reflect the physical 
conditions of the magnetic features we aim to understand. If we are to
succeed in meaningfully describing entire stellar coronae based on spectroscopic data in  the future, 
we must cope with the challenge of modeling {\it statistical distributions} of physical parameters across
various coronal features.
Additionally, new coronal structure reconstruction methods such as X-ray Doppler imaging, to mention at 
least one, hold promise to gain complementary insight into this problem.

{\it What is the physics behind flares?} Although the detailed plasma physical mechanisms of flares can probably only
be studied on the Sun, stellar flares offer access to quite extreme situations that may provide 
important boundary conditions for flare theories. Many characteristics of stellar coronae 
also suggest contributions of flares. Flare-like processes may be a fundamental mechanism to 
heat astrophysical plasmas in general. The underlying physical mechanisms certainly deserve more, in-depth
observational and theoretical studies.

{\it How does magnetic activity evolve?} Although a rough outline of rotation-age-activity relations has
been established, many details  remain unclear.  How is magnetic activity controlled at early stages
of evolution? Where are the magnetic fields generated in protostars or their environments, and where 
and how is magnetic energy released? Do  circumstellar accretion disks influence it? Is the accretion process itself
important for the generation or suppression of X-ray activity? Do star-disk magnetic fields contribute
significantly to X-ray emission? Do they regulate the rotational history of a star that defines 
the starting conditions for stars settling on the main sequence?

{\it How do magnetic fields and coronal X-rays interact with the stellar environment?} It has become 
clear that X-rays and magnetic fields do not only weakly interact with surrounding
planets via a stellar wind and some magnetospheric processes; in young stellar systems, the high levels
of X-ray emission may directly alter accretion disk properties by ionization, possibly controlling
accretion events and governing, together with associated ultraviolet radiation, the disk and envelope 
chemistry. Magnetic fields may play a fundamental  role in redistributing matter or in making accretion 
disks unstable. Coronal research turns into ``interplanetary research'' when dealing with forming young stellar systems.

Many of these aspects are being addressed by present-day observations and theory. But we need to
take bolder steps as well, in particular in terms of more sophisticated instrumentation. Spectroscopy is still at
an infant stage, essentially separating emission lines but hardly resolving their profiles. With a resolving power
exceeding $\approx 3000$, rotational broadening, turbulent motions, or bulk plasma motions (e.g., during
flares) could become routinely detectable in many stellar sources. ``Doppler imaging'' then would start really 
imaging stellar coronae, possibly even for different temperatures, thus producing a 3-D thermal structure model
of a stellar corona. The required resolving powers are within reach of current detector technology, for example 
in the cryogenic domain.

An ultimate goal will be producing spatially resolved images of nearby coronal stars through X-ray interferometric 
techniques, i.e., the equivalent of what radio astronomy has achieved through very long baseline interferometry.
Instead of using a baseline of the Earth's diameter as in conventional radio VLBI, two X-ray telescopes separated 
by 1.5~m would  - in principle - 
reach a resolution of 0.3 milliarcsecond at 20~\AA, sufficient to resolve the diameter of the corona of $\alpha$~Cen into 
about 30 resolution elements, a resolution that can be compared with early solar X-ray images. 
First X-ray interferometric laboratory experiments are indeed promising \citep{cash00}.

Many of our questions started in the solar system. I consider it a privilege that stellar astronomy 
is in the position of having an example for close scrutiny near-by. There is hope that further exploiting the
solar-stellar connection, i.e., carefully comparing detailed solar studies with investigations of stars,
will solve many of the remaining outstanding questions.

\begin{acknowledgement}
It is a pleasure to thank the editors of this journal, Profs. L. Woltjer and M.~C.~E. Huber, for inviting me to write 
this review, and for their thoughtful comments on the manuscript. I am grateful to many of my colleagues for helpful advice, 
the contributions of figures, and answers to my persistent questions.  Kaspar Arzner, Marc Audard, 
and Kevin Briggs critically commented on the text. I extend special thanks to them and to Stephen Skinner and
Ton Raassen for discussions on our subject over the past many years. I owe much to the late Rolf Mewe who
was a prolific source of inspiration and motivation in all aspects of coronal physics and spectroscopy. 
Discussions with Roberto Pallavicini, Antonio Maggio, and Giovanni Peres helped clarify issues on coronal loops. 
I acknowledge permission from several colleagues to use their previously published figures or data, in particular:
Drs. Marc Audard, Tom Ayres, Nancy Brickhouse, Kevin Briggs, Jeremy Drake, Fabio Favata, Eric Feigelson, Thomas Fleming, 
Elena Franciosini, Mark Giampapa, Nicolas Grosso, Edward Guinan, Kensuke Imanishi, 
Moira Jardine, Antonietta
Marino, Giusi Micela, Thierry Montmerle, Jan-Uwe Ness, Rachel Osten, Ton Raassen, Sofia Randich, Karel Schrijver,
Kazunari Shibata, Marek Siarkowski, Stephen Skinner, Beate Stelzer, Alessandra Telleschi, Yohko Tsuboi, and 
Takaaki Yokoyama. The figures are reproduced with the permission of the publishers.
I am particularly indebted to  M. Audard, T. Ayres, N. Brickhouse, 
G. Micela, J.-U. Ness, R. Osten, M. Siarkowski, K. Shibata, A. Telleschi, and T. Yokoyama for their generous help by
preparing entirely new figures for the present article, and Doris Lang for her editorial support. Several previously unprocessed and unpublished data sets or related results are shown in this paper, 
namely from {\it XMM-Newton}, {\it Chandra}, {\it GOES}, {\it Yohkoh}, and {\it TRACE}. {\it XMM-Newton} is an ESA science mission with instruments and 
contributions directly funded by ESA Member States and the USA (NASA). The {\it GOES}
soft X-ray data shown in Fig.~\ref{goeslight}  are from the Space Environment Center, Boulder, CO,
of the US National Oceanic and Atmospheric Administration (NOAA). The solar X-ray images in Fig.~\ref{solarminmax} were taken by the {\it Yohkoh}
mission of ISAS, Japan, which is operated by international collaboration of Japanese, US, and UK scientists under the
support of ISAS, NASA, and SERC, respectively. Figs.~\ref{traceloop} and \ref{2R_Trace} were obtained by the  Transition
Region and Coronal Explorer, {\it TRACE}, which is a mission of the Stanford-Lockheed Institute for Space Research (a joint program of 
the Lockheed-Martin  Advanced Technology Center's Solar and Astrophysics Laboratory and Stanford's Solar Observatories Group),
and part of the NASA Small Explorer program. Use was made of the Lausanne open cluster database at 
http://obswww.unige.ch/webda/, and of the SIMBAD database, operated at CDS, Strasbourg, France. 
General stellar X-ray astronomy research at PSI has been supported by the Swiss National
Science Foundation under projects 2100-049343.96, 20-58827.99, and 20-66875.01.
\end{acknowledgement}

\end{document}